\documentclass{cambridge6A}
\usepackage[numbers]{natbib}

\usepackage{lipsum}

  \usepackage{rotating}
  \usepackage{floatpag}
  \rotfloatpagestyle{empty}

 \usepackage{amsmath}
  \usepackage{amsthm,amssymb,units,amscd}
  \usepackage{graphicx}

\usepackage{bm}
\usepackage{txfonts,dsfont,tikz}	
\usepackage{epstopdf} 
\DeclareGraphicsExtensions{.tif,.pdf,.eps,.png,.jpg,.jpeg,.mps} 

\usepackage{mathrsfs,url}

\usepackage{bbm}
\usepackage{comment}

\usepackage{subeqnarray}

\usepackage[active]{srcltx}
\usepackage{amsfonts}
\usepackage{epsfig,here}

 \usepackage{makeidx}
 \makeindex

\newcommand{\beq}{\begin{quote}}
\newcommand{\enq}{\end{quote}}

\newcommand{\be}{\begin{equation}}
\newcommand{\ee}{\end{equation}}

\def\ket#1{\left|#1\right>}
\def\bra#1{\langle#1\vert}
\def\exptt#1{\langle#1\rangle}
\def\expt(#1#2){\langle #2 \vert #1 \vert #2 \rangle}

\begin{document}

  \title[In search of the ontology of quantum mechanics]
    {MEANING OF THE WAVE FUNCTION}

  \author{Shan Gao\\[3\baselineskip]}

  \frontmatter
  \maketitle

\newpage
 \vspace*{10cm}

\textit{\large This book is dedicated to Erwin Schr\"{o}dinger, who introduced the wave function, discovered the equation named after him, and had been attempting to find the ontology of quantum mechanics throughout the rest of his life.}\\ 

\setcounter{tocdepth}{2}
  \tableofcontents

\newpage
\thispagestyle{empty}
 \vspace*{10cm}

\hfill{Erwin with his psi can do}

\hfill{Calculations quite a few.}

\hfill{But one thing has not been seen:}

\hfill{Just what does psi really mean?}

\hfill\textit{--- Erich H\"uckel, translated by Felix Bloch}

  \listoffigures

  \mainmatter


  \chapter*{Preface}

\vspace*{1cm}

The meaning of the wave function has been a hot topic of debate since the early days of quantum mechanics. Recent years have witnessed a growing interest in this long-standing question.\footnote{See, e.g. Pusey, Barrett and Rudolph (2012), Ney and Albert (2013), and Gao (2014a, 2015b).} Is the wave function ontic, directly representing a state of reality, or epistemic, merely representing a state of (incomplete) knowledge, or something else? If the wave function is not ontic, then what, if any, is the underlying state of reality? If the wave function is indeed ontic, then exactly what physical state does it represent? 

In this book, I aim to make sense of the wave function in quantum mechanics and find the ontological content of the theory. The book can be divided into three parts. The first part addresses the question of the nature of the wave function (Chapters 1-5). After giving a comprehensive and critical review of the competing views of the wave function, I present a new argument for the ontic view in terms of protective measurements. In addition, I also analyze the origin of the wave function by deriving the free Schr\"{o}dinger equation. The second part analyzes the ontological meaning of the wave function (Chapters 6, 7). I propose a new ontological interpretation of the wave function in terms of random discontinuous motion of particles, and give two main arguments supporting this interpretation. 
The third part investigates whether the suggested quantum ontology is complete in accounting for our definite experience and whether it needs to be revised in the relativistic domain (Chapters 8, 9).


The idea of random discontinuous motion of particles came to my mind when I was a postgraduate at the Institute of Electronics, Chinese Academy of Sciences in 1993. I am happy that finally it has a more logical and satisfying formulation in this book. 
During the past twenty years, I have benefited from interactions and discussions with many physicists and philosophers of physics who care about the way the world really is. They are: Steve Adler, Guido Bacciagaluppi, Jeremy Butterfield, Tian Yu Cao, Ze-Xian Cao, Eli Cohen, Lajos Di\'{o}si, Bernard d'Espagnat, Arthur Fine, Shelly Goldstein, Guang-Can Guo, Bob Griffiths, Basil Hiley, Richard Healey, Jenann Ismael, Adrian Kent,  Vincent Lam, Tony Leggett, Matt Leifer, Peter Lewis, Chuang Liu, David Miller, Owen Maroney, Wayne Myrvold, Philip Pearle, Roger Penrose, Matt Pusey, Huw Price, Alastair Rae, Dean Rickles, Abner Shimony, Max Schlosshauer, Lee Smolin, Antoine Suarez, Hans Westman, Ken Wharton, Ling-An Wu, Jos Uffink, Lev Vaidman, and H. Dieter Zeh, among others. I thank them all deeply.


I would like to thank Steve Adler, Guido Bacciagaluppi, Eli Cohen, Vincent Lam, David Miller, Owen Maroney, and Dean Rickles for reading through some parts of an early draft of this manuscript and providing helpful suggestions for improving it. I also wish to express my warm thanks to all participants of First iWorkshop on the Meaning of the Wave Function and Quantum Foundations Workshop 2015, which were both hosted by \emph{International Journal of Quantum Foundations} and in which the main ideas of this book were discussed.\footnote{See http://www.ijqf.org/ for more information about the workshops.} 
I thank Simon Capelin of Cambridge University Press for his kind support as I worked on this project, and the referees who gave helpful suggestions on how the work could best serve its targeted audience. 

During the writing of this book, I have been assisted by research funding from the Ministry of Education of the People's Republic of China, Chinese Academy of Sciences, and the Institute for the History of Natural Sciences, Chinese Academy of Sciences. Some parts of this book were written when I taught \emph{The Philosophy of Quantum Mechanics} to the postgraduates at the University of Chinese Academy of Sciences. I thank the International Conference Center of the University for providing comfortable accommodation. 

Finally, I am deeply indebted to my parents, Qingfeng Gao and Lihua Zhao, my wife Huixia and my daughter Ruiqi for their unflagging love and support; this book would have been simply impossible without them. Moreover, they have never let me forget the true values of life.

\hfill Shan Gao

\hfill \emph{Beijing} 

\hfill May 2016


\chapter{Quantum mechanics and experience}

Quantum mechanics is an extremely successful physical theory due to its accurate empirical predictions.
The core of the theory, which is contained in various nonrelativistic quantum theories, is the Schr\"{o}dinger equation\index{Schr\"{o}dinger equation} and the Born rule\index{Born rule}.\footnote{An apparent exception is collapse theories (Ghirardi\index{Ghirardi, GianCarlo}, 2016). In these theories, however, the additional collapse term in the revised Schr\"{o}dinger equation\index{Schr\"{o}dinger equation} is so tiny for microscopic systems that it can be ignored in analyzing the ontological status and meaning of the wave function.} The Schr\"{o}dinger equation\index{Schr\"{o}dinger equation} governs the time evolution of the wave function assigned to a physical system, and the Born rule\index{Born rule} connects the wave function with the probabilities of possible results of a measurement on the system. 
In this chapter, I will introduce the core of quantum mechanics, especially the connections of its mathematical formalism with experience\index{quantum mechanics!and experience}. The introduction is not intended to be complete but enough for the later analysis of the meaning of the wave function and the ontological content of quantum mechanics.

\section{The mathematical formalism}

The mathematical formalism of quantum mechanics is mainly composed of two parts. The first part assigns a mathematical object, the so-called wave function or quantum state, to a physical system appropriately prepared at a given instant.\footnote{It is worth noting that although all quantum theories assign the same wave function to an isolated physical system, different quantum theories, such as no-collapse theories and collapse theories, may assign different wave functions to a non-isolated physical system. The assignment, which depends on the concrete laws of motion\index{laws of motion} in the theory, does not influence the ontological status and meaning of the wave function.}
The second part specifies how the wave function evolves with time. The evolution of the wave function is governed by the Schr\"{o}dinger equation\index{Schr\"{o}dinger equation}, whose concrete form is determined by the properties of the system and its interactions with environment.

There are two common representations for the wave function: the Hilbert space\index{Hilbert space} representation\index{wave function!Hilbert space representation of} and the configuration space\index{configuration space} representation\index{wave function!configuration space\index{configuration space} representation of}, which have their respective advantages. According to the Hilbert space\index{Hilbert space} representation\index{wave function!Hilbert space representation of}, the wave function is an unit vector or state vector in a Hilbert space\index{Hilbert space}, usually denoted by $\ket{\psi(t)}$ with Dirac's bracket notation. The Hilbert space\index{Hilbert space} is a complete vector space with scalar product, and its dimension and structure depend on the particular system. For example, the Hilbert space\index{Hilbert space} associated with a composite system is the tensor product of the Hilbert spaces associated with the systems of which it is composed.\footnote{Similarly, the Hilbert space\index{Hilbert space} associated with independent properties is the tensor product of the Hilbert spaces associated with each property.}

This structure of the Hilbert space\index{Hilbert space} can be seen more clearly from the configuration space\index{configuration space} representation\index{wave function!configuration space\index{configuration space} representation of}. The configuration space\index{configuration space} of an $N$-body quantum system has $3N$ dimensions, and each point in the space can be specified by an ordered $3N$-tuple, where each group of three coordinates are position coordinates of each subsystem in three-dimensional space. The wave function of the system is a complex function on this configuration space\index{configuration space},\footnote{To be consistent with convention, I will also say ``the wave function of a physical system", but it still means ``the wave function assigned to a physical system".} and it can be written as $\psi(x_1,y_1,z_1,..., x_N,y_N,z_N,t)$, where $x_i,y_i, z_i$ are coordinates of the $i$-th subsystem in the $3N$-dimensional configuration space\index{configuration space}. Moreover, the wave function is normalized, namely the integral of the modulus squared of the wave function over the whole space is one. When the $N$ subsystems are independent, the whole wave function can be decomposed as the product of the wave functions of the $N$ subsystems, each of which lives in three-dimensional space.

For an $N$-body quantum system, there are also a $3N$-dimensional space and wave functions on the space for other properties of the system besides position. For example, the momentum space of an $N$-body system is a $3N$-dimensional space parameterized by $3N$ momentum coordinates, and the momentum wave function is a complex function on this space. Here the Hilbert space\index{Hilbert space} representation\index{wave function!Hilbert space representation of} is more convenient. Every measurable property or observable of a physical system is represented by a Hermitian operator on the Hilbert space\index{Hilbert space} associated with the system, and the wave functions for different properties such as position and momentum may be transformed into each other by considering the relationship between the corresponding operators of these properties.

The second part of the mathematical formalism of quantum mechanics specifies how the wave function assigned to a physical system evolves with time. The time evolution of the wave function, $\ket{\psi(t)}$, is governed by the Schr\"{o}dinger equation\index{Schr\"{o}dinger equation}

\begin{equation}
i \hbar {\partial \ket{\psi(t)} \over  \partial t}=H \ket{\psi(t)}, 
\end{equation}

\noindent where $\hbar$ is Planck's constant divided by $2\pi$, and $H$ is the Hamiltonian operator that depends on the energy properties of the system. The time evolution is linear and unitary in the sense that the Hamiltonian is independent of the evolving wave function and it keeps the normalization of the wave function unchanged. The concrete forms of the Hamiltonian and the Schr\"{o}dinger equation\index{Schr\"{o}dinger equation} depend on the studied system and its interactions with other systems in the environment. For example, the wave function of an electron evolving in an external potential obeys the following Schr\"{o}dinger equation\index{Schr\"{o}dinger equation}:

\begin{equation}
i\hbar{\partial \psi(x,y,z,t) \over \partial t}=[-{\hbar^2 \over 2m}\nabla^2 +V(x,y,z,t)]\psi(x,y,z,t),
\label{Sch}
\end{equation}

\noindent where $\psi(x,y,z,t)$ is the wave function of the electron, $m$ is the mass of the electron, and $V(x,y,z,t)$ is the external potential. 

\section{The Born rule\index{Born rule}}

What is the empirical content of quantum mechanics\index{quantum mechanics!empirical content of}? Or how does the wave function assigned to a physical system relate to the results of measurements on the system? The well-known connection rule is the Born rule\index{Born rule}, which has been precisely tested by experiments. It says that a (projective) measurement of an observable $A$ on a system with the wave function $\ket{\psi}$ will randomly obtain one of the eigenvalues of $A$, and the probability of obtaining an eigenvalue $a_i$ is given by $|\langle a_i | \psi \rangle|^2$, where $\ket{a_i}$ is the eigenstate corresponding to the eigenvalue $a_i$. 

The Born rule\index{Born rule} can also be formulated in the language of configuration space\index{configuration space}. It says that the integral of the modulus squared of the wave function over a certain region of the configuration space\index{configuration space} associated with a property of a physical system gives the probability of the measurement of the property of the system obtaining the values inside the region. For example, for a physical system whose wave function is $\psi(x,y,z,t)$, $|\psi(x,y,z,t)|^2dxdydz$ represents the probability of a position measurement on the system obtaining a result between $(x,y,z)$ and $(x+dx,y+dy,z+dz)$, and $|\psi(x,y,z,t)|^2$ is the corresponding probability density in position $(x,y,z)$. Similarly, for an $N$-body system whose wave function is $\psi(x_1,y_1,z_1,..., x_N,y_N,z_N,t)$, $|\psi(x_1,y_1,z_1,..., x_N,y_N,z_N,t)|^2$ represents the probability density that a position measurement on the first subsystem obtains result $(x_1,y_1,z_1)$ and a position measurement on the second subsystem obtains result $(x_2,y_2,z_2)$ ... and a position measurement on the $N$-th subsystem obtains result $(x_N,y_N,z_N)$.

The Born rule\index{Born rule} provides a probabilistic connection between the wave function and the results of measurements. However, it may be not the only connection rule, as the involved measurements are only one kind of measurements, the projective measurements\index{projective measurements}. In order to know whether there are other possible connections between quantum mechanics and experience\index{quantum mechanics!and experience}, we need to analyze measurements in more detail.

A measurement is an interaction between a measured system and a measuring device. It can be described by using the standard von Neumann\index{von Neumann, John} procedure. Suppose the wave function of the measured system is $\ket{\psi}$ at a given instant $t=0$, and the initial wave function of the pointer of a measuring device at $t=0$ is a Gaussian wavepacket\index{Gaussian wavepacket} of very small width $w_0$ centered in initial position $x_0$, denoted by $|\phi(x_0)\rangle$.
The total Hamiltonian of the combined system can be written as

\begin{equation}
H = H_S + H_D + H_I, \label{H_full}
\end{equation}

\noindent  where $H_S$ and $H_D$ are the free Hamiltonian\index{Hamiltonian!free}s of the measured system and the measuring device, respectively, and $H_I$ is the interaction Hamiltonian\index{Hamiltonian!interaction} coupling the measured system to the measuring device, which can be further written as 

\begin{equation}
H_I = g(t)PA,
\label{H_int}
\end{equation} 

\noindent where $P$ is the momentum of the pointer of the measuring device, $A$ is the measured observable, and $g(t)$ represents the time-dependent coupling strength of the interaction, which is a smooth function normalized to $\int dt g(t)=1$ during the measurement interval $\tau$, and $g(0)=g(\tau)=0$. 

It has been known that there are different types of measurements, depending on the interaction strength and time and whether the measured system is appropriately protected etc. The most common measurements are projective measurements\index{projective measurements} involved in the Born rule\index{Born rule}. For a projective measurement, the interaction $H_I$  is of very short duration and so strong that it dominates the rest of the Hamiltonian, and thus the effect of the free Hamiltonian\index{Hamiltonian!free}s of the measuring device and the measured system can be neglected. Then the state of the combined system at the end of the interaction can be written as

\begin{equation}
\ket{t=\tau} = e^{-{i\over\hbar} P A } \ket{\psi}  \ket{\phi(x_0)}.
\end{equation}

\noindent By expanding $\ket{\psi}$  in the eigenstates of $A$, $\ket{a_i}$, we obtain
 
\begin{equation}
\ket{t=\tau} = \sum_{i} e^{-{i\over\hbar} P a_i } c_i \ket{a_i} \ket{\phi(x_0)},
\end{equation}

\noindent where $c_i$ are the expansion coefficients. The exponential term shifts the center of the pointer by $a_i$:

\begin{equation}
\ket{t=\tau} = \sum_{i} c_i \ket{a_i} \ket{\phi(x_0+a_i)}.
\label{ENS}
\end{equation}

\noindent  This is an entangled state\index{entangled states}, where the eigenstates of $A$ with eigenvalues $a_i$ are correlated to the measuring device states in which the pointer is shifted by these eigenvalues $a_i$.

The Born rule\index{Born rule} tells us (and we also know by experience) that the result of this projective measurement is one of the eigenvalues of the measured observable, say $a_i$, with probability $|c_i|^2$. However, we still don't know whether this entangled superposition is the final state of the combining system after the measurement.\footnote{In other words, it is still unknown how the wave function evolves during a projective measurement. In standard quantum mechanics\index{quantum mechanics!standard formulation of}, which is formulated by Dirac\index{Dirac, Paul A. M.} (1930) and von Neumann\index{von Neumann, John} (1932), it is assumed that after a projective measurement of an observable the entangled superposition formed by the Schr\"{o}dinger\index{Schr\"{o}dinger, Erwin} evolution collapses to one of the eigenstates of the observable that corresponds to the result of the measurement. This assumption is called the collapse postulate\index{quantum mechanics!standard formulation of!collapse postulate in}. For a helpful introduction of \index{quantum mechanics!standard formulation of}standard quantum mechanics for philosophers see Ismael\index{Ismael, Jenann} (2015).}
The appearance of the definite result seems apparently incompatible with the superposed state. This is the notorious measurement problem\index{measurement problem}. I will try to solve this problem in Chapter 8.


\section{A definite connection with experience\index{protective measurements!as a definite connection with experience}}

It is not surprising that since the interaction between the measured system and the measuring device is very strong during a projective measurement, the measurement disturbs the measured system and changes its wave function greatly. This is not a good measurement. A good measurement is required not to disturb the state of the measured system so that it can measure the realistic properties of the system. This is only possible for projective measurements\index{projective measurements} when the initial state of the measured system is an eigenstate of the measured observable. In this case, the final state of the combining system is not an entangled state but a product state such as:

\begin{equation}
\ket{t=\tau} = \ket{a_i} \ket{\phi(x_0+a_i)}.
\label{PS}
\end{equation}

\noindent According to the Born rule\index{Born rule}, this projective measurement obtains a definite result $a_i$.


A general way to make a good measurement is to protect the measured state from being changed when the  measurement is being made. A universial protection scheme is via the quantum Zeno\index{protective measurements!Zeno-type}\index{quantum Zeno effect} effect (Aharonov\index{Aharonov, Yakir}, Anandan\index{Anandan, Jeeva S.} and Vaidman\index{Vaidman, Lev}, 1993).\footnote{Another protection scheme is to introduce a protective potential such that the measured wave function of a quantum system is a  nondegenerate energy eigenstate of the Hamiltonian of the system with finite gap to neighboring energy eigenstates (Aharonov\index{Aharonov, Yakir} and Vaidman\index{Vaidman, Lev}, 1993). By this scheme, the measurement of an observable is required to be weak and adiabatic.}  
Let us see how this can be done\index{protective measurements!introduction of}. We make projective measurements\index{projective measurements} of an observable $O$, of which the measured state $\ket{\psi}$ is an nondegenerate eigenstate, a large number of times which are dense in a very short measurement interval $[0,\tau]$. For example, $O$ is measured in $[0,\tau]$ at times $t_n=(n/N)\tau, n = 1,2, . . . ,N$, where $N$ is an arbitrarily large number.  At the same time, we make the same projective measurement of an observable $A$ in the interval $[0,\tau]$ as in the last section, which is described by the interaction Hamiltonian (\ref{H_int}). 


As noted before, since the interaction $H_I$  is of very short duration and so strong that it dominates the rest of the Hamiltonian, the effect of the free Hamiltonian\index{Hamiltonian!free}s of the measuring device and the measured system can be neglected. Then the branch of the state of the combined system after $\tau$, in which each projective measurement of $O$ results in the state of the measured system being in $\ket{\psi}$, is given by

\begin{eqnarray}
\ket{t=\tau} &=& \ket{\psi}\bra{\psi}e^{-{i\over\hbar}{\tau\over N}H(t_N)}... \ket{\psi}\bra{\psi}e^{-{i\over\hbar}{\tau\over N}H(t_2)}
\ket{\psi}\bra{\psi}e^{-{i\over\hbar}{\tau\over N}H(t_1)}\ket{\psi}\ket{\phi(x_0)}\nonumber \\ 
&=&\ket{\psi}\bra{\psi}e^{-{i\over\hbar}{\tau\over N}g(t_N)P A}... \ket{\psi}\bra{\psi}e^{-{i\over\hbar}{\tau\over N}g(t_2)P A}
\ket{\psi}\bra{\psi}e^{-{i\over\hbar}{\tau\over N}g(t_1)P A}\ket{\psi}\ket{\phi(x_0)},\nonumber \\
\label{}
\end{eqnarray}

\noindent where $\ket{\phi(x_0)}$ is the initial wave function of the pointer of the measuring device, which is supposed to be a Gaussian wavepacket\index{Gaussian wavepacket} of very small width centered in initial position $x_0$.

Thus in the limit of $N \rightarrow \infty$, we have

\begin{eqnarray}
\ket{t=\tau} = \ket{\psi}e^{-{i\over\hbar}\int_{0}^{\tau}g(t)\expt(A{\psi})Pdt}\ket{\phi(x_0)}
= \ket{\psi}\ket{\phi(x_0+\exptt{A})},
\label{PMZ}
\end{eqnarray}

\noindent where $\exptt{A} \equiv \expt(A{\psi})$ is the expectation value\index{protective measurements!expectation values as results of} of $A$ in the measured state $\ket{\psi}$. Since the modulus squared of the amplitude of this branch approaches one when $N \rightarrow \infty$, this state will be the state of the combined system after $\tau$.\footnote{It is worth noting that the possible collapse of the wave function resulting from the projective measurements\index{projective measurements} of $O$ does not influence this result. The reason is that the probability of the measured state collapsing to another state different from $\ket{\psi}$ after each projective measurement of $O$ is proportional to $1/N^2$, and thus the sum of these probabilities is proportional to $1/N$ after $\tau$ and approaches zero when $N \rightarrow \infty$. Moreover, since the pointer of a  measuring device may be a microscopic system, whose shift can be further read out by another measuring device, the effect of the possible collapse of the wave function resulting from the projective measurements\index{projective measurements} of $A$ can also be ignored.} It can be seen that after the measurement, the measuring device state and the system state are not entangled, and the pointer of the measuring device is shifted by the expectation value $\exptt{A}$.\footnote{Note that after the measurement the pointer wavepacket does not spread, and the width of the wavepacket is the same as the initial width. This ensures that the pointer shift can represent a valid measurement result.} 

This demonstrates that for an arbitrary state of a quantum system at a given instant, we can protect the state from being changed via the quantum Zeno\index{protective measurements!Zeno-type}\index{quantum Zeno effect} effect,
and a projective measurement of an observable, which is made at the same time, yields a definite measurement result, the expectation value\index{protective measurements!expectation values as results of} of the observable in the measured state. 
Such measurements have been called protective measurements\index{protective measurements!introduction of}\index{protective measurements} (Aharonov\index{Aharonov, Yakir} and Vaidman\index{Vaidman, Lev}, 1993; Aharonov\index{Aharonov, Yakir}, Anandan\index{Anandan, Jeeva S.} and Vaidman\index{Vaidman, Lev}, 1993; Vaidman\index{Vaidman, Lev}, 2009).

In fact, it can be shown that if the measured state is not changed during a projective measurement, then the result must be the expectation value of the measured observable in the measured state. In this case, the evolution of the state of the combined system is

\begin{equation}
\ket{\psi(0)}\ket{\phi(0)} \rightarrow \ket{\psi(t)}\ket{\phi(t)}, t>0,
\end{equation}

\noindent where $\ket{\phi(0)}$ and $\ket{\phi(t)}$ are the states of the measuring device at instants 0 and $t$, respectively, $\ket{\psi(0)}$ and $\ket{\psi(t)}$ are the states of the measured system at instants 0 and $t$, respectively, and $\ket{\psi(t)}$ is the same as $\ket{\psi(0)}$ up to a phase factor during the measurement interval $[0,\tau]$. The interaction Hamiltonian\index{Hamiltonian!interaction} is still given by (\ref{H_int}). Then by Ehrenfest's theorem\index{Ehrenfest's theorem} we have

\begin{equation}
{d \over dt}\bra{\psi(t)\phi(t)}X\ket{\psi(t)\phi(t)} = g(t)\bra{\psi(0)}A\ket{\psi(0)},
\end{equation}

\noindent where $X$ is the pointer variable. This further leads to

\begin{equation}
\bra{\phi(\tau)}X\ket{\phi(\tau)}- \bra{\phi(0)}X\ket{\phi(0)}= \bra{\psi(0)}A\ket{\psi(0)},
\end{equation}

\noindent which means that the shift of the center of the pointer of the measuring device is the expectation value\index{protective measurements!expectation values as results of} of the measured observable in the measured state. This clearly demonstrates  that the result of a measurement which does not disturb the measured state is the expectation value\index{protective measurements!expectation values as results of} of the measured observable in the measured state.

Since the wave function can be reconstructed from the expectation value\index{protective measurements!expectation values as results of}s of a sufficient number of observables, the wave function of a single quantum system can be measured by a series of protective measurements\index{protective measurements}. Let the explicit form of the measured state at a given instant $t$ be $\psi(x)$, and the measured observable $A$ be (normalized) projection operators on small spatial regions $V_n$ having volume $v_n$:

\begin{equation}
A= 
\begin{cases} 
{1\over{v_n}},& \text{if $x \in V_n$,}
\\
0,&\text{if $x \not\in V_n$.} 
\end{cases}
\label{OA}
\end{equation}

\noindent  A protective measurement of $A$ then yields 

\begin{equation}
\exptt{A} = {1\over {v_n}} \int_{V_n}|\psi(x)|^2 dv,
\end{equation}

\noindent  which is the average of the density $\rho(x) = |\psi(x)|^2$ over the small region $V_n$. Similarly, we can measure another observable $B ={\hbar \over{2mi}} (A\nabla + \nabla A)$. The measurement yields

\begin{equation}
\exptt{B} ={1\over {v_n}} \int_{V_n}{\hbar \over{2mi}}(\psi^* \nabla \psi - \psi  \nabla \psi^* )dv = {1\over {v_n}} \int_{V_n}j(x)dv.
\end{equation}

\noindent This is the average value of the flux density $j(x)$ in the region $V_n$. Then when $v_n \rightarrow 0$ and after performing measurements in sufficiently many regions $V_n$ we can measure $\rho(x)$ and $j(x)$ everywhere in space.
Since the wave function $\psi(x,t)$ can be uniquely expressed by $\rho(x,t)$ and $j(x,t)$ (except for an overall phase factor), the whole wave function of the measured system at a given instant can be measured by protective measurements\index{protective measurements}.

Protective measurements provide a definite, direct connection between the wave function assigned to a physical system and the results of measurements on the system, and the connection is only determined by the linear Schr\"{o}dinger\index{Schr\"{o}dinger, Erwin} evolution.\footnote{Note that besides the wave function there are also state-independent quantities such as $m$ (mass) and $Q$ (charge) in the Schr\"{o}dinger equation\index{Schr\"{o}dinger equation}, and the measurement of such a quantity will obtain a definite result. This is also a definite, direct connection between the mathematical formalism of quantum mechanics and experience\index{quantum mechanics!and experience}.} As I will argue later in this book, although this connection seems less well-known, it will be extremely important for understanding the meaning of the wave function and searching for the ontology of quantum mechanics\index{quantum mechanics!ontology of}.


\chapter{The wave function: ontic vs epistemic}



The mathematical formalism of quantum mechanics and its connections with experience\index{quantum mechanics!and experience} are the starting point of our search for the ontology\index{quantum mechanics!ontology of} of quantum mechanics. The theory admits the mind-independent existence of a physical world, which contains various physical systems. Moreover, it assigns a wave function to each appropriately prepared physical system, and states the relationship of the wave function with the results of measurements on the system.
But the theory does not tell us what a physical system ontologically is, and especially, what the connection of the wave function with the ontic state of the system is.
Does the wave function directly represent the ontic state of a physical system? Or does it merely represent the state of (incomplete) knowledge about the ontic state of the system? In short, is the wave function ontic or epistemic? We need to first answer this question in order to find the ontological content of quantum mechanics. But wait, is there such a thing like the ontic state of a physical system such as an atom?

\section{There is an underlying reality}

There is an argument for a negative answer to the above question, which is due to Bohr\index{Bohr, Niels}\index{quantum mechanics!anti-realist views of}. It is based on the concepts of entanglement\index{quantum entanglement} and nonseparability\index{nonseparability} between system and device during a measurement (Bohr\index{Bohr, Niels}, 1948; Faye\index{Faye, Jan}, 2014). Bohr\index{Bohr, Niels} believed that atoms are real, but he would not attribute intrinsic and measurement-independent state properties to atomic objects (Faye\index{Faye, Jan}, 1991). Moreover, he did not regard the wave function as a description of something physically real. According to Bohr\index{Bohr, Niels}, the distinction between the measured system and the measuring device is a necessary condition for a measurement to reveal information about the properties of the system. Concretely speaking, in order to measure the properties of the measured system, it must be assumed that the system possesses an independent state, which is in principle distinguishable from the state of the measuring device with which it interacts. However, a quantum-mechanical treatment of the interaction between the measured system and the measuring device will make the very distinction ambiguous, since it requires that the combining system must be described by a single inseparable entangled state. Therefore, according to Bohr\index{Bohr, Niels}, the impossibility of ``separating the behaviour of the objects from their interaction with the measuring instruments" in quantum mechanics ``implies an ambiguity in assigning conventional attributes to atomic objects" (Bohr\index{Bohr, Niels}, 1948; see also Camilleri\index{Camilleri, Kristian} and Schlosshauer\index{Schlosshauer, Maximilian}, 2015). 


It can be argued that there are several loopholes in the above argument. First of all, it is possible that microscopic objects\index{microscopic objects} have some properties that cannot be measured. Thus the unmeasurability of certain properties does not necessarily exclude the existence of these properties. \index{quantum mechanics!anti-realist views of}
Next, the above argument only applies to certain kinds of measurements such as projective measurements\index{projective measurements}, which lead to inevitable entanglement\index{quantum entanglement} and nonseparability\index{nonseparability} between the measured system and the measuring device. It is indeed true that such a measurement disturbs the measured system and changes its state, and thus the result does not reflect the  properties of the measured system even if they exist. However, the argument does not apply to protective measurements\index{protective measurements}, during which there is no entanglement\index{quantum entanglement} and nonseparability\index{nonseparability}. Thirdly, even for projective measurements\index{projective measurements}, the above argument does not hold true when the measured system is in an eigenstate of the measured observable. In this case, there is no entanglement\index{quantum entanglement} and nonseparability\index{nonseparability} between the measured system and the measuring device during the measurement either. Finally, if quantum mechanics is universal,\footnote{It seems that Bohr\index{Bohr, Niels} did not exclude the application of quantum mechanics to any system (Faye\index{Faye, Jan}, 2014).} then the above argument also applies to macroscopic objects\index{macroscopic objects} including measuring devices. Therefore, if the argument is valid, then measuring devices will have no properties either. This contradicts the presupposition that measuring devices can obtain definite results, and in particular, the pointers of measuring devices have definite positions.

Certainly, even if there are no convincing arguments, one may also assume that a microscopic object has no  properties, e.g. the properties that determine the result of a measurement on it. It has been suggested that the behaviour of microscopic objects\index{microscopic objects} falls under no law and they do not properly admit of direct description (Timpson\index{Timpson, Christopher G.}, 2008; Fuchs\index{Fuchs, Christopher A. }, 2011). 
This view seems reasonable when the measurements of microscopic objects\index{microscopic objects} yield only random results. But even for projective measurements\index{projective measurements}, when the measured system is in an eigenstate of the measured observable, the measurement result is not random, but definite.
In this case, this view cannot explain the definite measurement result.
In fact, even for projective measurements\index{projective measurements} whose results are random, this view cannot explain the probabilities of different measurement results either.
In addition, according to Timpson\index{Timpson, Christopher G.} (2008), this view cannot provide an explanation of why macroscopic objects\index{macroscopic objects} have the kinds of physical properties that they do in virtue of the properties their constituents possess, since it does not ascribe properties to microscopic objects\index{microscopic objects} along with laws describing how they behave. 


Besides this explanatory deficit problem, it can be argued that the above view has a more serious inconsistency problem. An observation of the position of the pointer of a measuring device obtains a definite result (without disturbing it), and the result indicates that the pointer has a definite position. This is the very reason why we assume macroscopic objects\index{macroscopic objects} including the pointers of measuring devices have definite properties. Now if we can measure a microscopic object without disturbing it and also obtain a definite result as for protective measurements\index{protective measurements}, then why cannot we ascribe properties to these objects? It will be inconsistent if we do not ascribe properties to microscopic objects\index{microscopic objects} in this case.\footnote{Similarly, when considering the existence of protective measurements\index{protective measurements}, if classical mechanics can attribute intrinsic and observation-independent state properties to macroscopic objects\index{macroscopic objects}, then quantum mechanics can also attribute intrinsic and observation-independent state properties to microscopic objects\index{microscopic objects}.} In fact, if quantum mechanics is universal and can be applied to any physical system, then microscopic objects\index{microscopic objects} will have the same ontological status as macroscopic objects\index{macroscopic objects}. Therefore, if macroscopic objects\index{macroscopic objects} have properties (that can be measured by a measuring device), so do microscopic objects\index{microscopic objects}. Note that denying macroscopic objects\index{macroscopic objects} have  properties that we can experience will slip into solipsism\index{solipsism} for all practical purposes (see also Norsen\index{Norsen, Travis}, 2016). \index{quantum mechanics!anti-realist views of}

Last but not least, it is worth noting that whether or not microscopic objects\index{microscopic objects} have properties related to their states, it is arguable that they at least have some state-independent properties such as mass and charge. This is clearly indicated by the Schr\"{o}dinger equation\index{Schr\"{o}dinger equation} that governs the evolution of microscopic objects\index{microscopic objects}, in which there are quantities $m$ and $Q$. Moreover, these properties also determine the results of the measurements on them, which are definite and involve no quantum randomness.

\section{The $\psi$-epistemic view\index{wave function!epistemic view of}}

I have argued that there is an underlying reality in the sense that microscopic objects\index{microscopic objects} such as atoms, like our familiar macroscopic objects\index{macroscopic objects}, also have state properties or an underlying ontic state (which can be measured). However, it is still possible that the wave function, the key mathematical object of quantum mechanics, does not directly represent the underlying reality. Concretely speaking, the wave function assigned to a physical system may not represent the ontic state of the system, but merely represent a state of (incomplete) knowledge -- an epistemic state -- about the ontic state of the system. There are indeed some heuristic arguments against the $\psi$-ontic view\index{wave function!ontic view of} and supporting the (realist) $\psi$-epistemic view\index{wave function!epistemic view of}.\footnote{It is worth noting that there are two different $\psi$-epistemic view\index{wave function!epistemic view of}s. One is the realist $\psi$-epistemic view\index{wave function!epistemic view of}, which will be discussed below. The other is the operationalist $\psi$-epistemic view\index{wave function!epistemic view of!operationalist, \emph{see also} QBism}, which regards the wave function of a quantum system as representing a state of incomplete knowledge about which outcome will occur if a measurement is made on the system. A further development of this view is QBism\index{QBism} (Caves\index{Caves, Carlton}, Fuchs and Schack\index{Schack, R\"{u}diger}, 2002; Fuchs\index{Fuchs, Christopher A. }, 2011), according to which the Born probabilities are subjective even for probability one. This subjective Bayesian account of quantum mechanical probability has been debated by Timpson\index{Timpson, Christopher G.} (2008) and Stairs (2011). As argued in the last section, this view is also plagued by the explanatory deficit problem\index{QBism!explanatory deficit problem of} etc (see Marchildon\index{Marchildon, Louis}, 2004; Timpson\index{Timpson, Christopher G.}, 2008).} In the following, I will examine these arguments.


\subsection{Multidimensionality\index{multidimensionality}}

The first sign of the non-reality of the wave function\index{wave function!non-reality of} is that the space it lives on is not three-dimensional but multidimensional. If a physical system consists of $N$ subsystems, then the space in which its wave function is defined, namely the configuration space\index{configuration space} of the system, will be $3N$-dimensional. 

When Schr\"{o}dinger\index{Schr\"{o}dinger, Erwin} first introduced the wave function for a two-body system, he already worried about its reality:

\begin{quote}
The direct interpretation of this wave function of six variables in three-dimensional space meets, at any rate initially, with difficulties of an abstract nature. (Schr\"{o}dinger\index{Schr\"{o}dinger, Erwin}, 1926a, p.39)
\end{quote}

\noindent Einstein\index{Einstein, Albert} expressed the same doubt:

\begin{quote}
The field in a many-dimensional coordinate space does not smell like something real. (Einstein\index{Einstein, Albert}, 1926)
\end{quote}
\noindent Later, after failing to develop a satisfactory ontology for the wave function, Schr\"{o}dinger\index{Schr\"{o}dinger, Erwin} also conceded this point when writing to Einstein\index{Einstein, Albert}:

\begin{quote}
I am long past the stage where I thought that one can consider the w-function as somehow a direct description of reality. (Schr\"{o}dinger\index{Schr\"{o}dinger, Erwin}, 1935a)
\end{quote}

\noindent Interestingly, the multidimensionality of configuration space\index{configuration space} also made Bohm doubt the reality of the wave function\index{wave function!reality of} in the initial phase of formulating his theory:

\begin{quote}
While our theory can be extended formally in a logically consistent way by introducing the concept of a wave in a $3N$-dimensional space, it is evident that this procedure is not really acceptable in a physical theory, and should at least be regarded as an artifice that one uses provisionally until one obtains a better theory in which everything is expressed once more in ordinary three-dimensional space. (Bohm,\index{Bohm, David} 1957, p.117)
\end{quote}

However, although the multidimensionality of configuration space\index{configuration space} may pose a serious difficulty for wave function realism\index{wave function!realism} (Albert\index{Albert, David Z.}, 1996, 2013), which regards the wave function as a physical field in configuration space\index{configuration space}, it does not lead to difficulties when interpreting the wave function as a representation of a property of particles in three-dimensional space (Monton\index{Monton, Bradley}, 2002, 2013; Lewis,\index{Lewis, Peter J.} 2013, 2016; Gao\index{Gao, Shan}, 2014b).\footnote{I will analyze these two interpretations of the wave function in detail in Chapters 6 and 7.} Therefore, the above analysis does not consititute a decisive argument supporting the $\psi$-epistemic view\index{wave function!epistemic view of} and against the $\psi$-ontic view\index{wave function!ontic view of}. 

\subsection{Collapse of the wave function}

Besides the multidimensional form of the wave function, the laws that govern the evolution of the wave function seem to also suggest its non-reality. 

Recall that there are two distinct evolution laws in \index{quantum mechanics!standard formulation of}standard quantum mechanics. When a physical system is not being measured, its wave function evolves continuously according to the Schr\"{o}dinger equation\index{Schr\"{o}dinger equation}. On the other hand, according to the collapse postulate\index{quantum mechanics!standard formulation of!collapse postulate in}, if a (projective) measurement is made on the system, the original wave function will instantaneously and discontinuously be updated by the wave function corresponding to the measurement result. It has been a hot topic of debate how to explain or explain away the collapse of the wave function\index{wave function!collapse}. 

It seems that the $\psi$-epistemic view\index{wave function!epistemic view of} may provide a natural explanation of the collapse of the wave function\index{wave function!collapse} (Leifer\index{Leifer, Matthew S.}, 2014a). If the wave function does not directly represent a state of reality but merely represent a state of incomplete knowledge about reality, namely if the wave function is not ontic but epistemic, then it seems that the collapse of the wave function\index{wave function!collapse} can be readily explained as the effect of acquiring new information, no more mysterious than the updating of a classical probability distribution when new data is obtained. 
For example, in Schr\"{o}dinger\index{Schr\"{o}dinger, Erwin}'s cat thought experiment\index{Schr\"{o}dinger's cat thought experiment}\index{Schr\"{o}dinger's cat thought experiment!$\psi$-epistemic explanation of}, the cat may be definitely dead or alive before we observe it, and the superposition of dead and alive cats we assign to it may simply reflect the fact that we do not know the actual state of the cat. Then after we observe the cat and know whether it is dead or alive, the superposition will naturally be updated by the state corresponding to the dead or alive cat. 



There is evidence that Einstein\index{Einstein, Albert} once supported this explanation of the collapse of the wave function\index{wave function!collapse} (see Fine\index{Fine, Arthur}, 1993, 1996 for a more careful analysis). In a letter to Heitler, he criticized Heitler's notion that the observer plays an important role in the process of wavefunction collapse:

\begin{quote}
[I advocate] that one conceives of the $\psi$-function [i.e., wavefunction] only as an incomplete description of a real state of affairs, where the incompleteness of the description is forced by the fact that observation of the state is only able to grasp part of the real factual situation. Then one can at least escape the singular conception that observation (conceived as an act of consciousness) influences the real physical state of things; the change in the $\psi$-function through observation then does not correspond essentially to the change in a real matter of fact but rather to the alteration in \emph{our knowledge} of this matter of fact. (emphasis in original) (Einstein\index{Einstein, Albert}, 1948)
\end{quote}

If this simple, intuitive explanation of the collapse of the wave function\index{wave function!collapse} is indeed valid, then it strongly suggests that the wave function is not real. However, this explanation cannot be wholly correct. Consider a quantum system being in the following superposed state:

\begin{equation}
\ket{\psi} = \sum_{i} {c_i \ket{a_i}},
\end{equation}

\noindent where $\ket{a_i}$ are the eigenstates of an arbitrary observable $A$, and $c_i$ are the expansion coefficients.
The Born rule\index{Born rule} tells us that the result of a projective measurement of $A$ is one of the eigenvalues of $A$, say $a_i$, with probability $|c_i|^2$.
The collapse postulate\index{quantum mechanics!standard formulation of!collapse postulate in} in \index{quantum mechanics!standard formulation of}standard quantum mechanics further says that after the measurement the original superposition instantaneously and discontinuously collapses to the corresponding eigenstate $\ket{a_i}$.
The explanation of the collapse of the wave function\index{wave function!collapse} provided by the $\psi$-epistemic view\index{wave function!epistemic view of} is as follows.
Before  the measurement, the observable $A$ of the system has a definite value $a_i$. The superposed state we assign to the system reflects our incomplete knowledge about the actual value of the observable.
Then after we measure the observable and know its actual value $a_i$, the superposed state will naturally be updated by the state corresponding to the value, $\ket{a_i}$.

Since this explanation of wavefunction collapse is supposed to hold true for the measurement of every observable at any time, it must assume that all observables defined for a quantum system have definite values at all times, which are their eigenvalues, independently of any measurement context, and moreover, measurements also reveal these pre-existing values. As a result, the functional relations between commuting observables must hold for the values assigned to them in order to avoid conflicts with the predictions of quantum mechanics. 
This obviously violates the Kochen-Specker theorem\index{Kochen-Specker theorem}, which asserts the impossibility of assigning values to all observables whilst, at the same time, preserving the functional relations between them  (in a Hilbert space\index{Hilbert space} of dimension $d \geqslant 3$) (Kochen and Specker, 1967\index{Kochen, Simon B.}\index{Specker, Ernst}).\footnote{Here I ignore the finite precision loophole of the Kochen-Specker theorem\index{Kochen-Specker theorem}\index{Kochen-Specker theorem!finite precision loophole of}, which allows non-contextual hidden-variables theories, but which is widely regarded as physically implausible (Bartlett and Kent\index{Kent, Adrian}, 2004). Besides, it is worth noting that for a quantum system all observables can have their expectation values in the state of the system at all times (Gao\index{Gao, Shan}, 2015b). The Kochen-Specker theorem\index{Kochen-Specker theorem!assumptions of} does not prohibit this, since the definite values being expectation values violates the production rule, which is one of the key assumptions of the theorem (Kochen and Specker, 1967\index{Kochen, Simon B.}\index{Specker, Ernst}).} 

Today, most $\psi$-epistemists already abandon the above explanation of wavefunction collapse. However, a satisfactory explanation consistent with the $\psi$-epistemic view\index{wave function!epistemic view of} is still wanting. 
In order that a $\psi$-epistemic model is consistent with the predictions of quantum mechanics, the underlying ontic state also needs to be changed during the measurement process in general. For example, in Spekkens\index{Spekkens, Robert W.}'s toy model (Spekkens\index{Spekkens, Robert W.}, 2007), even a measurement of an eigenstate of the measured observable also causes change of the underlying ontic state. 
Therefore, even if denying the reality of the wave function\index{wave function!reality of} and wavefunction collapse, the $\psi$-epistemic view\index{wave function!epistemic view of} also needs to explain why the underlying ontic state is changed during the measurement process. This is still a big challenge to the $\psi$-epistemic view\index{wave function!epistemic view of}.\footnote{In Spekkens\index{Spekkens, Robert W.}'s (2007) toy model, this change is explained by the requirement of the so-called knowledge balance principle, which, roughly speaking, states that at most half of the information needed to specify the ontic state can be known at any given time. But again, the problem turns to explaining why there is this knowledge balance principle. For further discussion see Leifer\index{Leifer, Matthew S.} (2014a).}



To sum up, I have argued that the collapse of the wave function\index{wave function!collapse} cannot simply be explained as a process of updating information about the ontic state of the measured system. This result provides a support for the $\psi$-ontic view\index{wave function!ontic view of}, not for the $\psi$-epistemic view\index{wave function!epistemic view of}.





\subsection{Indistinguishability of nonorthogonal states\index{nonorthogonal states}}

Although some features of quantum mechanics seem to also suggest the non-reality of the wave function\index{wave function!non-reality of}, a more careful analysis shows that this is not the case. In the following, I will take the indistinguishability of nonorthogonal states\index{nonorthogonal states!indistinguishability of} as a typical example.

At first sight, the $\psi$-epistemic view\index{wave function!epistemic view of} may provide a very natural explanation of the impossibility of distinguishing between nonorthogonal states\index{nonorthogonal states} with certainty. The usual argument is as follows. If the wave function represents an ontic state, then two nonorthogonal states\index{nonorthogonal states} will correspond to distinct ontic states, and thus it seems puzzling that we cannot measure the difference between them. On the other hand, if the wave functions are epistemic states and represented by probability distributions that have support over some set of ontic states, then two nonorthogonal states\index{nonorthogonal states} will overlap and may correspond to the same ontic state. Thus it is quite understandable that we cannot perfectly distinguish them by a measurement. Take again Spekkens\index{Spekkens, Robert W.}'s (2007) toy model as an example. In the model, two nonorthogonal states\index{nonorthogonal states} $|x+)$ and $|y+)$ overlap on the ontic state $(+1, +1)$, which will be occupied by the system half the time whenever $|x+)$ or $|y+)$ is prepared. When this happens, there is nothing about the ontic state of the system that could possibly tell us whether $|x+)$ or $|y+)$ was prepared. Therefore, we cannot distinguish between these two prepared nonorthogonal states\index{nonorthogonal states} at least half the time. The overlap of the two epistemic states explains their indistinguishability.\index{nonorthogonal states!indistinguishability of}

In order to justify the above $\psi$-epistemic explanation of the indistinguishability of nonorthogonal states\index{nonorthogonal states}, we need more than just that the probability distributions corresponding to two nonorthogonal states\index{nonorthogonal states} should have nonzero overlap. We also need that the overlap should be equal to the overlap of the two nonorthogonal states\index{nonorthogonal states} (i.e. the modulus squared of the inner product of these two states). That is: when measuring a physical system prepared in one of these two states, the probability of obtaining the result corresponding to the projector of the other state should be equal to the overlap between the probability distributions corresponding to the two states. Such models are called maximally $\psi$-epistemic models\index{psi-epistemic models}. Only in these models, can the difficulty of distinguishing nonorthogonal states\index{nonorthogonal states} be completely and quantitatively explained by the difficulty of distinguishing the corresponding epistemic states, and thus the $\psi$-epistemic explanation of the indistinguishability of nonorthogonal states\index{nonorthogonal states} can be wholly satisfying (Maroney\index{Maroney, Owen J. E.}, 2012).\index{nonorthogonal states!indistinguishability of}

However, it has been shown that this simple, intuitive understanding of the indistinguishability of nonorthogonal states\index{nonorthogonal states} cannot go through (Leifer\index{Leifer, Matthew S.} and Maroney\index{Maroney, Owen J. E.}, 2013; Barrett\index{Barrett, Jonathan} et al, 2014\index{Cavalcanti, Eric G.}; Leifer\index{Leifer, Matthew S.}, 2014b; Branciard\index{Branciard, Cyril}, 2014). For example, it proves that the maximally $\psi$-epistemic models\index{psi-epistemic models}, in which the overlap of the probability distributions is large enough to explain fully the indistinguishability of nonorthogonal states\index{nonorthogonal states}, must make different predictions from quantum mechanics for Hilbert space\index{Hilbert space} dimension $d \geq 3$ (Barrett\index{Barrett, Jonathan}\index{Cavalcanti, Eric G.} et al, 2014). As we will see in the next section, the $\psi$-ontology theorems\index{psi-ontology theorems} will further show that $\psi$-epistemic models\index{psi-epistemic models} must be severely constrained if they are to reproduce the predictions of quantum mechanics.\index{nonorthogonal states!indistinguishability of}



\subsection{The eigenvalue-eigenstate half link\index{quantum mechanics!standard formulation of!eigenvalue-eigenstate half link in}}

In the last part of this section, I will also discuss a heuristic argument against the $\psi$-epistemic view\index{wave function!epistemic view of} and supporting the $\psi$-ontic view\index{wave function!ontic view of}. The argument is based on the so-called eigenvalue-eigenstate half link\index{quantum mechanics!standard formulation of!eigenvalue-eigenstate half link in}, which says that when a physical system is in an eigenstate of an observable, the system has an observation-independent property with value being the eigenvalue corresponding to the eigenstate (Monton\index{Monton, Bradley}, 2006, 2013).

Here is the argument given by Leifer\index{Leifer, Matthew S.} (2014a). First of all, according to the eigenvalue-eigenstate half link\index{quantum mechanics!standard formulation of!eigenvalue-eigenstate half link in}, the observables of which the wave function is an eigenstate are properties of a physical system with values being the corresponding eigenvalues. Next, the wave function of a physical system is uniquely determined by the set of observables of which it is an eigenstate. In fact, every wave function is uniquely determined by a single observable; $\ket{\psi}$ is an eigenstate of the observable $\ket{\psi}\bra{\psi}$ with eigenvalue $+1$ and it is the only state in the $+1$ eigenspace of $\ket{\psi}\bra{\psi}$ (up to a global phase). Then the argument goes like this: if a physical system has a set of properties, and those properties uniquely determine its wave function, then the wave function is also real, representing a property of the system.

This argument deserves careful examination. 
To begin with, if one denies the eigenvalue-eigenstate half link\index{quantum mechanics!standard formulation of!eigenvalue-eigenstate half link in}, then one can certainly refute the argument.
However, it is worth noting that unlike the problematic eigenvalue-eigenstate link\index{quantum mechanics!standard formulation of!eigenvalue-eigenstate link in}, the eigenvalue-eigenstate half link\index{quantum mechanics!standard formulation of!eigenvalue-eigenstate half link in} seems more reasonable.
The former states that if and only if a physical system is in an eigenstate of an observable, the system has a property with value being the eigenvalue corresponding to the eigenstate, and it restricts the scope of possible  properties of a physical system. While the latter removes the words ``and only if" and thus avoids this unnecessary restriction. 
I will analyze the basis of the eigenvalue-eigenstate half link\index{quantum mechanics!standard formulation of!eigenvalue-eigenstate half link in} in more detail in Chapter 4.

Next, when assuming the validity of the eigenvalue-eigenstate half link\index{quantum mechanics!standard formulation of!eigenvalue-eigenstate half link in}, it is indeed surprising how this weak link can be so strong as to be able to derive the reality of the wave function\index{wave function!reality of}. 
The crux of the matter is whether  the eigenvalue-eigenstate half link\index{quantum mechanics!standard formulation of!eigenvalue-eigenstate half link in} really implies that the observables of which the wave function is an eigenstate are properties of a physical system.
It can be argued that the answer is negative.
The reason is that the eigenvalue-eigenstate half link\index{quantum mechanics!standard formulation of!eigenvalue-eigenstate half link in} is overused here; 
the link only says that an observable is a property of a physical system when the system is in an eigenstate of the observable, and it does not say that the observable is a property of the system in all cases, e.g. when the system is not in an eigenstate of the observable. For example, according to the eigenvalue-eigenstate half link\index{quantum mechanics!standard formulation of!eigenvalue-eigenstate half link in}, the observable $\ket{\psi}\bra{\psi}$ is a property of a physical system when the system is in the state $\ket{\psi}$. However, the link does not say that the observable $\ket{\psi}\bra{\psi}$ is still a  property of the system when the system is in a state other than $\ket{\psi}$. Only if the latter is true can the above argument go through.

In fact, this understanding of the eigenvalue-eigenstate half link\index{quantum mechanics!standard formulation of!eigenvalue-eigenstate half link in} is still not very accurate. 
What the link really says is that when a physical system is in an eigenstate of an observable,  the system has a  property represented by the eigenvalue associated with the eigenstate. 
This property  is certainly not the observable itself, although we may say that it is the observable possessing the corresponding eigenvalue.
Therefore, although  the wave function $\ket{\psi}$ can be uniquely determined by the observable $\ket{\psi}\bra{\psi}$, since the observable is not a  property of the system, this has no implication on the reality of the wave function\index{wave function!reality of}.
On the other hand, even though the observable $\ket{\psi}\bra{\psi}$ possessing an eigenvalue +1 is a  property of the system, this property cannot uniquely determine  the wave function $\ket{\psi}$; the wave function is a very complex object, while some property having value +1 contains little information.

However, this analysis may not persuade Penrose\index{Penrose, Roger}. He said:

\begin{quote}
One of the most powerful reasons for rejecting such a subjective
viewpoint concerning the reality of $|\psi\rangle$ comes from the
fact that whatever $|\psi\rangle$ might be, there is always---in
principle, at least---a {\it primitive measurement\/} whose {\bf YES}
space consists of the Hilbert-space ray determined by $|\psi\rangle$.
The point is that the physical state $|\psi\rangle$ (determined by
the ray of complex multiples of $|\psi\rangle$) is {\it uniquely\/}
determined by the fact that the outcome {\bf YES}, for this state, is
{\it certain}.  No other physical state has this property.  For any
other state, there would merely be some probability, short of
certainty, that the outcome will be {\bf YES}, and an outcome of {\bf
NO} might occur. Thus, although there is no measurement which will
tell us what $|\psi\rangle$ actually {\it is}, the physical state
$|\psi\rangle$ is uniquely determined by what it asserts must be the
result of a measurement that {\it might\/} be performed on it. (Penrose\index{Penrose, Roger}, 1994, p.314)
\end{quote}

Penrose\index{Penrose, Roger}'s argument is different from the above argument in that it is not based on the eigenvalue-eigenstate half link\index{quantum mechanics!standard formulation of!eigenvalue-eigenstate half link in}. His reason for the reality of the wave function\index{wave function!reality of} is that the wave function $|\psi\rangle$ is  uniquely determined by the fact that the result of the measurement of the projector $\ket{\psi}\bra{\psi}$ on this state is certain, and other wave functions do not have this property. 
This seems to be a good reason, but it in fact assumes too much. Concretely speaking, it assumes that the Born probabilities are not epistemic, but objective, intrinsic to a single measurement process. 
This already rejects the $\psi$-epistemic view\index{wave function!epistemic view of}, according to which the Born probabilities are at least partly epistemic. Therefore, Penrose\index{Penrose, Roger}'s argument is not a valid argument supporting the $\psi$-ontic view\index{wave function!ontic view of} either.

\section{$\psi$-ontology theorems\index{psi-ontology theorems}}

I have analyzed several heuristic arguments for the $\psi$-epistemic view\index{wave function!epistemic view of} and $\psi$-ontic view\index{wave function!ontic view of}. The analysis shows that these arguments do not constitute a decisive proof of either view. In order to obtain a definite result, we need to find a general, rigorous approach to determine whether the wave function is ontic or epistemic. In this section, I will introduce one such approach, the ontological models framework\index{ontological models framework}\index{psi-ontology theorems}, and two $\psi$-ontology theorems\index{psi-ontology theorems} based on this framework.

\subsection{The ontological models framework\index{ontological models framework}}

The ontological models framework\index{ontological models framework} provides a rigorous approach to address the question of the nature of the wave function (Spekkens\index{Spekkens, Robert W.}, 2005; Harrigan\index{Harrigan, Nicholas} and Spekkens\index{Spekkens, Robert W.}, 2010). 
It has two fundamental assumptions. \index{ontological models framework!two assumptions of}
The first assumption is about the existence of the underlying state of reality. It says that if a quantum system is prepared such that quantum mechanics assigns a pure state to it, then after preparation the system has a well-defined set of physical properties or an underlying ontic state, which is usually represented by a mathematical object, $\lambda$. This assumption is necessary for the analysis of the ontological status of the wave function, since if there are no any underlying ontic states, it will be meaningless to ask whether or not the wave functions describe them. 

Here a strict $\psi$-ontic/epistemic distinction can be made\index{wave function!ontic/epistemic distinction}. In a $\psi$-ontic ontological model, the ontic state of a physical system determines its wave function uniquely, and thus the wave function represents a  property of the system. While in a $\psi$-epistemic ontological model, the ontic state of a physical system can be compatible with different wave functions, and the wave function represents a state of incomplete knowledge -- an epistemic state -- about the actual ontic state of the system. Concretely speaking, the wave function corresponds to a probability distribution $p(\lambda|P)$ over all possible ontic states when the preparation is known to be $P$, and the probability distributions corresponding to two different wave functions can overlap.\footnote{Note that it is possible that two wave functions are compatible with the same ontic state but the probability distributions corresponding to the two different wave functions do not overlap, although the probability of this situation occurring is zero. I will not consider this undetectable difference between the two formulations of the $\psi$-epistemic view in my following discussion.} 

In order to investigate whether an ontological model is consistent with the empirical predictions of quantum mechanics, we also need a rule of connecting the underlying ontic states with the results of measurements. This is the second assumption of the ontological models framework\index{ontological models framework}, which says that when a measurement is performed, the behaviour of the measuring device is only determined by the ontic state of the system, along with the physical properties of the measuring device. More specifically, the framework assumes that for a projective measurement $M$, the ontic state $\lambda$ of a physical system determines the probability $p(k|\lambda,M)$ of different results $k$ for the measurement $M$ on the system.\footnote{This specific assumption is not necessarily a consequence of the second assumption. For further discussion see Chapter 3.} The consistency with the predictions of quantum mechanics then requires the following relation: $\int{d\lambda p(k|\lambda, M)p(\lambda|P)} = p(k|M, P)$, where $p(k|M, P)$ is the Born probability of $k$ given $M$ and $P$. A direct inference of this relation is that different orthogonal states correspond to different ontic states.

In recent years, there have appeared several no-go theorems which attempt to refute the $\psi$-epistemic view\index{wave function!epistemic view of} within the ontological models framework\index{ontological models framework}. These theorems are called $\psi$-ontology theorems\index{psi-ontology theorems}, including the Pusey-Barrett-Rudolph theorem\index{Pusey-Barrett-Rudolph theorem}, the Colbeck\index{Colbeck, Roger}-Renner\index{Renner, Renato} theorem\index{Colbeck-Renner theorem}, and Hardy's theorem\index{Hardy's theorem} (Pusey,\index{Pusey, Matthew F.} Barrett and\index{Barrett, Jonathan} Rudolph,\index{Rudolph, Terry} 2012; Colbeck and Renner, 2012; Hardy,\index{Hardy, Lucien} 2013).\footnote{Note that the early no-go theorems for hidden variables, such as Bell's theorem\index{Bell's theorem} and the Kochen-Specker theorem\index{Kochen-Specker theorem}, are not $\psi$-ontology theorems\index{psi-ontology theorems}, since they do not explicitly addresses the $\psi$-ontic/epistemic distinction. For a comprehensive review of $\psi$-ontology theorems\index{psi-ontology theorems} and related work see Leifer\index{Leifer, Matthew S.} (2014a).} The key assumption of the $\psi$-epistemic view\index{wave function!epistemic view of} is that there exist two nonorthogonal states\index{nonorthogonal states} which are compatible with the same ontic state (i.e. the probability distributions corresponding to these two nonorthogonal states\index{nonorthogonal states} overlap).\footnote{In other words, when these two nonorthogonal states\index{nonorthogonal states} are prepared, there is a non-zero probability that the prepared ontic states are the same.} A general strategy of these $\psi$-ontology theorems\index{psi-ontology theorems} is to prove the consequences of this assumption are inconsistent with the predictions of quantum mechanics (under certain auxiliary assumptions). In the following, \index{psi-ontology theorems}I will introduce two typical $\psi$-ontology theorems\index{psi-ontology theorems}. 

\subsection{Pusey-Barrett-Rudolph theorem\index{Pusey-Barrett-Rudolph theorem}}

The first $\psi$-ontology theorem is the Pusey-Barrett-Rudolph theorem\index{Pusey-Barrett-Rudolph theorem} (Pusey,\index{Pusey, Matthew F.} Barrett and\index{Barrett, Jonathan} Rudolph,\index{Rudolph, Terry} 2012). Its basic proof strategy is as follows. 
Assume there is a nonzero probability that $N$ nonorthogonal states\index{nonorthogonal states} $\ket{\psi_i}$ ($i$=1, ... , $N$) are compatible with the same ontic state $\lambda$. The ontic state $\lambda$ determines the probability $p(k|\lambda,M)$ of different results $k$ for the measurement $M$. Moreover, there is a normalization relation for any $N$ result measurement: $\sum_{i=1}^{N}p(k_i|\lambda,M)=1$. Now if an $N$ result measurement satisfies the condition that the first state gives zero Born probability to the first result and the second state gives zero Born probability to the second result and so on, then there will be a relation $p(k_i|\lambda,M)=0$ for any $i$, which leads to a contradiction.

The task is then to find whether there are such nonorthogonal states\index{nonorthogonal states} and the corresponding measurement\index{psi-ontology theorems}. Obviously there is no such a measurement for two nonorthogonal states\index{nonorthogonal states} of a physical system, since this will permit them to be perfectly distinguished. However, such a measurement does exist for four nonorthogonal states\index{nonorthogonal states} of two copies of a physical system (Pusey,\index{Pusey, Matthew F.} Barrett\index{Barrett, Jonathan} and Rudolph,\index{Rudolph, Terry} 2012). The four nonorthogonal states\index{nonorthogonal states} are the following product states: $\ket{0} \otimes \ket{0}$, $\ket{0} \otimes \ket{+}$,$\ket{+} \otimes \ket{0}$ and $\ket{+} \otimes \ket{+}$, where $\ket{+}= {1 \over \sqrt{2}}(\ket{0} + \ket{1})$. The corresponding measurement is a joint measurement of the two systems, which projects onto the following four orthogonal states:

\begin{eqnarray}\label{}
\ket{\phi_1}&=&\tfrac{1}{\sqrt{2}}(\ket{0}\otimes\ket{1}+\ket{1}\otimes\ket{0}), \nonumber\\
\ket{\phi_2}&=&\tfrac{1}{\sqrt{2}}(\ket{0}\otimes\ket{-}+\ket{1}\otimes\ket{+}), \nonumber\\
\ket{\phi_3}&=&\tfrac{1}{\sqrt{2}}(\ket{+}\otimes\ket{1}+\ket{-}\otimes\ket{0}), \nonumber\\
\ket{\phi_4}&=&\tfrac{1}{\sqrt{2}}(\ket{+}\otimes\ket{-}+\ket{-}\otimes\ket{+}),
\end{eqnarray}

\noindent where $\ket{-}={1 \over \sqrt{2}}(\ket{0}-\ket{1})$. This proves that the four nonorthogonal states\index{nonorthogonal states} are ontologically distinct. In order to further prove the two nonorthogonal states\index{nonorthogonal states} $\ket{0}$ and $\ket{+}$ for one system are ontologically distinct, a preparation independence assumption\index{Pusey-Barrett-Rudolph theorem!preparation independence assumption of} is needed, which says that multiple systems can be prepared such that their ontic states are uncorrelated. Under this assumption, a similar proof for every pair of nonorthogonal states\index{nonorthogonal states} can also be found, which requires more than two copies of a physical system (Pusey,\index{Pusey, Matthew F.} Barrett\index{Barrett, Jonathan} and Rudolph,\index{Rudolph, Terry} 2012).

\subsection{Hardy's theorem\index{Hardy's theorem}}

The proof of the Pusey-Barrett-Rudolph theorem\index{Pusey-Barrett-Rudolph theorem} requires an analysis of multiple copies of the system in question. Hardy's theorem\index{Hardy's theorem} improves this by pertaining to a single copy of the system in question (Hardy,\index{Hardy, Lucien} 2013). 
The price it needs to pay is to resort to assumptions about how dynamics is represented in an ontological model. In contrast, the Pusey-Barrett-Rudolph theorem\index{Pusey-Barrett-Rudolph theorem}\index{psi-ontology theorems} only involves prepare-and-measure experiments.

Hardy's theorem\index{Hardy's theorem}  can be illustrated with a simple example (Leifer\index{Leifer, Matthew S.}, 2014a). Assume there are two nonorthogonal states\index{nonorthogonal states} $\ket{\psi_1}$ and ${1 \over \sqrt{2}}(\ket{\psi_1}+\ket{\psi_2})$, which are compatible with the same ontic state $\lambda$ as required by the $\psi$-epistemic view\index{wave function!epistemic view of}. Consider a unitary evolution which leaves $\ket{\psi_1}$ invariant but changes ${1 \over \sqrt{2}}(\ket{\psi_1}+\ket{\psi_2})$ to its orthogonal state ${1 \over \sqrt{2}}(\ket{\psi_1}-\ket{\psi_2})$. Since two orthogonal states correspond to different ontic states,\footnote{Note that in order to prove this result and Hardy's theorem\index{Hardy's theorem}, it is not necessary to resort to the stronger assumption that the ontic state determines the probability for measurement results (which is needed to prove the Pusey-Barrett-Rudolph theorem\index{Pusey-Barrett-Rudolph theorem}); rather, one only needs to assume that the ontic state determines whether the probability is zero or nonzero. This weaker assumption is called possibilistic completeness assumption (Hardy,\index{Hardy, Lucien} 2013).} the original ontic state $\lambda$ must be changed by the unitary evolution. How to derive a contradiction then? If assuming that the unitary evolution that leaves $\ket{\psi_1}$ invariant also leaves the underlying ontic state $\lambda$ invariant, then there will be a contradiction. In other words, under this assumption we can prove that the two nonorthogonal state $\ket{\psi_1}$ and ${1 \over \sqrt{2}}(\ket{\psi_1}+\ket{\psi_2})$ are ontologically distinct. 

This is the simplest example of Hardy's theorem\index{Hardy's theorem}. The above auxiliary assumption is called ontic indifference assumption\index{Hardy's theorem!ontic indifference assumption of}. One strong motivation for this assumption is locality. When $\ket{\psi_1}$ and $\ket{\psi_2}$ are two spatially separated states prepared in regions 1 and 2 respectively, it seems reasonable to assume that the local evolution of the ontic state in region 2 does not influence the ontic state in region 1. Interestingly and surprisingly, even if the ontic indifference assumption holds only for a single pure state, Hardy's theorem\index{Hardy's theorem} can also be proved (Hardy,\index{Hardy, Lucien} 2013; Patra\index{Patra, Manas K.}, Pironio\index{Pironio, Stefano} and Massar\index{Massar, Serge}, 2013).


\chapter{The nomological view\index{wave function!nomological view of}}

The ontological models framework\index{ontological models framework} provides a rigorous approach to address the question of whether the wave function is ontic or epistemic. 
However, as noted by the proponents of this framework (Harrigan\index{Harrigan, Nicholas} and Spekkens\index{Spekkens, Robert W.}, 2010), there could exist realist interpretations of quantum mechanics that are not suited to it. This is indeed the case. For example, Bohm's theory\index{Bohm's theory} is just an exception.\footnote{Bohm's theory\index{Bohm's theory} is a realistic alternative to \index{quantum mechanics!standard formulation of}standard quantum mechanics initially proposed by de Broglie\index{de Broglie, Louis} (1928) and later rediscovered and developped by Bohm (\index{Bohm, David}1952) (see also Bohm and Hiley\index{Hiley, Basil J.}, 1993; Holland\index{Holland, Peter R.}, 1993; D\"{u}rr\index{D\"{u}rr, Detlef} and Teufel\index{Teufel, Stefan}, 2009; D\"{u}rr\index{D\"{u}rr, Detlef}, Goldstein\index{Goldstein, Sheldon} and Zangh\`{i}\index{Zangh\`{i}, Nino}, 2012; Goldstein\index{Goldstein, Sheldon}, 2013).  The theory is also called the de Broglie-Bohm theory or the pilot wave theory in the literature. I will use the appellation ``Bohm's theory\index{Bohm's theory}" throughout this book.}
The reason is that the ontological models framework\index{ontological models framework} and Bohm's theory\index{Bohm's theory} have different assumptions about the connection between the underlying ontic state and the probabilities of measurement results; the former assumes that the ontic state determines the probability of a measurement result, while the latter as a deterministic theory assumes that  the ontic state completely determines the measurement result, and  the (epistemic) probability of a measurement result is determined by the initial condition of the universe. 
Therefore, the $\psi$-ontology theorems\index{psi-ontology theorems} such as the Pusey-Barrett-Rudolph theorem\index{Pusey-Barrett-Rudolph theorem}, which are based on the ontological models framework\index{ontological models framework}, do not apply to Bohm's theory\index{Bohm's theory} even though their auxiliary assumptions can be avoided (see also Feintzeig\index{Feintzeig, Benjamin}, 2014; Gao\index{Gao, Shan}, 2014b; Drezet\index{Drezet, Aurelien}, 2015). In other words, these theorems do not require that the wave function should be ontic in Bohm's theory\index{Bohm's theory}. 
Since Bohm's theory\index{Bohm's theory} clearly rejects the $\psi$-epistemic view\index{wave function!epistemic view of}, how can it interpret the wave function if not assuming the $\psi$-ontic view\index{wave function!ontic view of}? Interestingly, there is a third option, the nomological view\index{wave function!nomological view of}. 
In this chapter, I will introduce this view of the wave function and give a critical analysis of it.

\section{The effective wave function\index{Bohm's theory!effective wave function in}}

According to Bohm's theory\index{Bohm's theory}, a complete realistic description of a quantum system is provided by the configuration defined by the positions of its particles together with its wave function. The Bohmian law of motion is expressed by two equations: a guiding equation\index{Bohm's theory!guiding equation of} for the configuration of particles and the Schr\"{o}dinger equation\index{Schr\"{o}dinger equation}, describing the time evolution of the wave function which enters the guiding equation\index{Bohm's theory!guiding equation of}. The law can be formulated as follows:
              
\begin{equation}
{{dQ(t)} \over {dt}}=v^{\Psi(t)}(Q(t)),
\label{GEf}
\end{equation}

\begin{equation}
i\hbar {\partial \Psi(t) \over \partial t}=H\Psi(t),
\label{Scf}
\end{equation}

\noindent where $Q(t)$ denotes the spatial configuration of particles, $\Psi(t)$ is the wave function at time $t$, and $v$ equals to the velocity of probability density in \index{quantum mechanics!standard formulation of}standard quantum mechanics.\footnote{Note that there are two somewhat different formulations of Bohm's theory\index{Bohm's theory!two formulations of}, in one of which the guiding equation\index{Bohm's theory!guiding equation of} is second-order as Bohm originally formulated, and in the other the guiding equation\index{Bohm's theory!guiding equation of} is first-order. Here I introduce the first-order formulation of Bohm's theory\index{Bohm's theory}, which is usually called Bohmian mechanics (Goldstein\index{Goldstein, Sheldon}, 2013). See Belousek (2003) for a comparison of these two formulations. In addition, it is worth noting that there are also other velocity formulas with nice properties, including Galilean symmetry, and yielding theories that are empirically equivalent to \index{quantum mechanics!standard formulation of}standard quantum mechanics and to Bohm's theory\index{Bohm's theory} (Deotto and Ghirardi\index{Ghirardi, GianCarlo}, 1998), although the Bohmian choice is arguably the simplest.}
Moreover, it is assumed that at some initial instant $t_0$, the epistemic probability of the configuration, $\rho(t_0)$, is given by the Born rule\index{Born rule}: $\rho(t_0)=|\Psi(t_0)|^2$. This is the so-called quantum equilibrium hypothesis\index{Bohm's theory!quantum equilibrium hypothesis of}, which, together with the law of motion, ensures the empirical equivalence between Bohm's theory\index{Bohm's theory} and \index{quantum mechanics!standard formulation of}standard quantum mechanics.

The status of the above equations is different depending on whether one considers the physical description of the universe as a whole or of a subsystem thereof. Bohm's theory\index{Bohm's theory} starts from the concept of a universal wave function\index{Bohm's theory!universal wave function in} (i.e. the wave function of the universe), figuring in the fundamental law of motion for all the particles in the universe. That is, $Q(t)$ describes the configuration of all the particles in the universe at time $t$, and $\Psi(t)$ is the wave function of the universe at time $t$, guiding the motion of all particles taken together. To describe subsystems of the universe, the appropriate concept is the effective wave function\index{Bohm's theory!effective wave function in} in Bohm's theory\index{Bohm's theory}.

The effective wave function\index{Bohm's theory!effective wave function in} is the Bohmian analogue of the usual wave function in \index{quantum mechanics!standard formulation of}standard quantum mechanics. It is not primitive, but derived from the universal wave function\index{Bohm's theory!universal wave function in}\index{wave function!universal|see{Bohm's theory}} and the actual spatial configuration of all the particles ignored in the description of the respective subsystem (D\"{u}rr\index{D\"{u}rr, Detlef}, Goldstein\index{Goldstein, Sheldon} and Zangh\`{i}\index{Zangh\`{i}, Nino}, 1992). The effective wave function\index{Bohm's theory!effective wave function in} of a subsystem can be defined as follows. Let $A$ be a subsystem of the universe including $N$ particles with position variables $x=(x_1,x_2,...,x_N)$. Let $y=(y_1,y_2,...,y_M)$ be the position variables of all other particles not belonging to $A$. Then the subsystem $A$'s conditional wave function at time $t$ is defined as the universal wave function\index{Bohm's theory!universal wave function in} $\Psi_t(x, y)$ evaluated at $y = Y(t)$:

\begin{equation}
\psi_t^A(x)=\Psi_t(x, y)|_{y=Y(t)}.
\end{equation}

\noindent If the universal wave function\index{Bohm's theory!universal wave function in} can be decomposed in the following form:

\begin{equation}\label{EF}
\Psi_t(x, y)=\varphi_t(x)\phi_t(y)+\Theta_t(x, y),
\end{equation}

\noindent where $\phi_t(y)$ and $\Theta_t(x, y)$ are functions with macroscopically disjoint supports, and $Y(t)$ lies within the support of $\phi_t(y)$, then $\psi_t^A(x)=\varphi_t(x)$ (up to a multiplicative constant) is $A$'s effective wave function\index{Bohm's theory!effective wave function in} at $t$. It can be seen that the temporal evolution of $A$'s particles is given in terms of $A$'s conditional wave function in the usual Bohmian way, and when the conditional wave function is  $A$'s effective wave function\index{Bohm's theory!effective wave function in},  it also obeys a Schr\"{o}dinger\index{Schr\"{o}dinger, Erwin} dynamics of its own. This means that the effective descriptions of subsystems are of the same form of the law of motion as given above. This is a satisfactory result.

\section{The universal wave function\index{Bohm's theory!universal wave function in} as law}

It has been a hot topic of debate how to interpret the wave function in Bohm's theory\index{Bohm's theory}. An influential view is the nomological interpretation of the wave function, which was originally suggested by D\"{u}rr\index{D\"{u}rr, Detlef}, Goldstein\index{Goldstein, Sheldon} and Zangh\`{i}\index{Zangh\`{i}, Nino} (1997).\footnote{See also Goldstein\index{Goldstein, Sheldon} and Teufel\index{Teufel, Stefan} (2001) and Goldstein\index{Goldstein, Sheldon} and Zangh\`{i}\index{Zangh\`{i}, Nino} (2013).} They argued that the universal wave function\index{Bohm's theory!universal wave function in} or the wave function of the universe has a law-like character, that is, it is more in the nature of a law than a concrete physical reality. In their own words,

\begin{quote}
The wave function of the universe is not an element of physical reality. We propose that the wave function belongs to an altogether different category of existence than that of substantive physical entities, and that its existence is nomological rather than material. We propose, in other words, that the wave function is a component of a physical law rather than of the reality described by the law. (D\"{u}rr\index{D\"{u}rr, Detlef}, Goldstein\index{Goldstein, Sheldon} and Zangh\`{i}\index{Zangh\`{i}, Nino}, 1997, p. 10)
\end{quote}

The reasons to adopt this nomological view\index{wave function!nomological view of} of the wave function come from the unusual kind of way in which Bohm's theory\index{Bohm's theory} is formulated, and the unusual kind of behavior that the wave function undergoes in the theory.
First of all, although the wave function affects the behavior of the configuration of the particles, which is expressed by the guiding equation\index{Bohm's theory!guiding equation of} (\ref{GEf}), there is no back action of the configuration upon the wave function. The evolution of the wave function is governed by the Schr\"{o}dinger equation\index{Schr\"{o}dinger equation} (\ref{Scf}), in which the actual configuration $Q(t)$ does not appear. Since a physical entity is supposed to satisfy the action-reaction principle, the wave function cannot describe a physical entity in Bohm's theory\index{Bohm's theory}.

Next, the wave function of a many-particle system, $\psi(q_1, . . . , q_N)$, is defined not in our ordinary three-dimensional space, but in the $3N$-dimensional configuration space\index{configuration space}, the set of all hypothetical configurations of the system. Thus it seems untenable to view the wave function as directly describing a physical field. I have discussed such worries in the last chapter. In fact, the sort of physical field the wave function is supposed to describe is even more abstract. Since two wave functions such that one is a (nonzero) scalar multiple of the other are physically equivalent,  what the wave function describes is not even a physical field at all, but an equivalence class of physical fields. Moreover, Bohm's theory\index{Bohm's theory} regards identical particles such as electrons as unlabelled, so that the configuration space\index{configuration space} of $N$ such particles is not the familiar high dimensional space, like $R^{3N}$, but is the unfamiliar high-dimensional space $^{N}R^3$ of $N$-point subsets of $R^3$. This space has a nontrivial topology, which may naturally lead to the possibilities of bosons and fermions. But it seems odd as a fundamental space in which a physical field exists.

Thirdly, and more importantly, the wave function in Bohm's theory\index{Bohm's theory} plays a role that is analogous to that of the Hamiltonian in classical Hamiltonian mechanics\index{Hamiltonian mechanics} (Goldstein\index{Goldstein, Sheldon} and Zangh\`{i}\index{Zangh\`{i}, Nino}, 2013). To begin with, both the classical Hamiltonian\index{Hamiltonian!classical} and the wave function live on a high dimensional space. The wave function is defined in configuration space\index{configuration space}, while the classical Hamiltonian\index{Hamiltonian!classical} is defined in phase space: a space that has twice as many dimensions as configuration space\index{configuration space}. 
Next, there is a striking analogy between the  guiding equation\index{Bohm's theory!guiding equation of} in Bohm's theory\index{Bohm's theory} and the Hamiltonian equations in classical mechanics. The guiding equation\index{Bohm's theory!guiding equation of} can be written as:

\begin{equation}
{{dQ} \over {dt}}=der(log\psi),
\label{GE1}
\end{equation}

\noindent where the symbol $der$ denotes some sort of derivative. Similarly, the Hamiltonian equations can be written in a compact way as:

\begin{equation}
{{dX} \over {dt}}=der (H),
\label{HE}
\end{equation}

\noindent  where $der (H)$ is a suitable derivative of the Hamiltonian. Moreover, it is also true that both $log \psi$ and $H$ are normally regarded as defined only up to an additive constant. Adding a constant to $H$ doesn't change the equations of motion. Similarly, when multiplying the wave function by a scalar, which amounts to adding a constant to its $log$, the new wave function is physically equivalent to the original one, and they define the same velocity for the configuration in the equations of motion in Bohm's theory\index{Bohm's theory}.
Since the classical Hamiltonian\index{Hamiltonian!classical} is regarded not as a description of some physical entity, but as the generator of time evolution in classical mechanics, by the above analogy it seems natural to assume that the wave function is not a description of some physical entity either, but a similar generator of the equations of motion in Bohm's theory\index{Bohm's theory}.

These analyses suggest that one should think of the wave function as describing a law and not as some sort of concrete physical reality in Bohm's theory\index{Bohm's theory}. However, it seems that there is a serious problem with this nomological view\index{wave function!nomological view of} of the wave function. The wave function of a quantum system typically changes with time, but laws are supposed not to change with time. Moreover, we can prepare the wave function of a quantum system and control its evolution, but laws are not supposed to be things that we can prepare and control. 
This problem indeed exists for the effective wave function\index{Bohm's theory!effective wave function in} of a subsystem of the universe, but it may not exist for the wave function of the universe, only which deserves to be interpreted nomologically (Goldstein\index{Goldstein, Sheldon} and Zangh\`{i}\index{Zangh\`{i}, Nino}, 2013). The wave function of the universe is certainly not controllable. And it may not be dynamical either. This can be illustrated by the Wheeler\index{Wheeler, John A.}-DeWitt\index{DeWitt, Bryce S.} equation\index{Wheeler-DeWitt equation}, which is the fundamental equation for the wave function of the universe in canonical quantum cosmology:

\begin{equation}
H\Psi(q)=0,
\label{WD}
\end{equation}

\noindent where $\Psi(q)$ is the wave function of the universe, $q$ refers to 3-geometries, and $H$ is the Hamiltonian constraint which involves no explicit time-dependence. Unlike the Schr\"{o}dinger equation\index{Schr\"{o}dinger equation}, the Wheeler\index{Wheeler, John A.}-DeWitt\index{DeWitt, Bryce S.} equation\index{Wheeler-DeWitt equation} has on one side, instead of a time derivative of $\Psi$, simply 0, and thus its natural solutions are time-independent. 
Moreover, the wave function of the universe may be unique. Although the Wheeler\index{Wheeler, John A.}-DeWitt\index{DeWitt, Bryce S.} equation\index{Wheeler-DeWitt equation} presumably has a great many solutions, when supplemented with additional natural conditions such as the Hartle-Hawking boundary condition, the solution may become unique. Such uniqueness also fits nicely
with the conception of the wave function as law.

Whether the wave function of the universe is a stationary function, uniquely obeying some constraints and hence resembling the classical Hamiltonian\index{Hamiltonian!classical} is still unknown, since the final theory of quantum gravity\index{quantum gravity} is not yet available.
What we can do now is to examine whether the effective wave function\index{Bohm's theory!effective wave function in}s of subsystems also have a tenable physical explanation under the nomological view\index{wave function!nomological view of} of the universal wave function\index{Bohm's theory!universal wave function in}.  As we will see in the next section, the answer to this question will have implications for the  nomological view\index{wave function!nomological view of} of the wave function, as well as for the ontology of Bohm's theory\index{Bohm's theory}\index{Bohm's theory!ontology of}.

\section{A critical analysis}

What is the physical meaning of the effective wave function\index{Bohm's theory!effective wave function in} in Bohm's theory\index{Bohm's theory}? 
If the wave function of the universe is nomological, then the ontology of Bohm's theory\index{Bohm's theory}\index{Bohm's theory!ontology of} will consist only in particles and their positions. As a consequence, the effective wave function\index{Bohm's theory!effective wave function in} of a subsystem of the universe, which is not nomological in general, must be ontologically explained by these particles and their positions. Moreover, it is uncontroversial that the effective wave function\index{Bohm's theory!effective wave function in} of a subsystem does not supervene on the distribution of the system's particles' positions. For instance, for the electron in the hydrogen atom, there are countably many real-valued wave functions corresponding to different energy eigenstates of the electron, but they may all describe a particle that is at rest in the same position at all times. Therefore, if the ontology of Bohm's theory\index{Bohm's theory}\index{Bohm's theory!ontology of} consists only in particles and their positions, then the effective wave function\index{Bohm's theory!effective wave function in} of a subsystem must encode the influences of the particles which are not part of the subsystem. 

This line of reasoning is also supported by the analysis of Esfeld\index{Esfeld, Michael} et al (2013). According to these authors, the effective wave function\index{Bohm's theory!effective wave function in} of a subsystem encodes the non-local influences of other particles on the subsystem via the non-local law of Bohm's theory\index{Bohm's theory}. For example, in the double-slit experiment with one particle at a time, the particle goes through exactly one of the two slits, and that is all there is in the physical world. There is no field or wave that guides the motion of the particle and propagates through both slits and undergoes interference. The development of the position of the particle (its velocity and thus its trajectory) is determined by the positions of other particles in the universe, including the particles composing the experimental setup, and the non-local law of Bohm's theory\index{Bohm's theory} can account for the observed particle position on the screen (Esfeld\index{Esfeld, Michael} et al, 2013).\footnote{See also Dorato\index{Dorato, Mauro} (2015) for a recent  evaluation of this view.}

In the following, I will argue that  the effective wave function\index{Bohm's theory!effective wave function in} of a subsystem of the universe does not encode the influences of other particles on the subsystem, and thus the nomological view\index{wave function!nomological view of} of the wave function seems problematic. First of all, consider the simplest case in which the universal wave function\index{Bohm's theory!universal wave function in} factorizes so that

\begin{equation}
\Psi_t(x, y)=\varphi_t(x)\phi_t(y).
\end{equation}

\noindent Then $\psi_t^A(x)=\varphi_t(x)$ is subsystem $A$'s effective wave function\index{Bohm's theory!effective wave function in} at $t$. This is the first example considered by D\"{u}rr\index{D\"{u}rr, Detlef}, Goldstein\index{Goldstein, Sheldon} and Zangh\`{i}\index{Zangh\`{i}, Nino} (1992) in explaining the effective wave function\index{Bohm's theory!effective wave function in}. In this case, it is uncontroversial that subsystem $A$ and its environment are independent of each other, and the functions $\varphi_t(x)$ and $\phi_t(y)$ describe subsystem $A$ and its environment, respectively. Thus, the effective wave function\index{Bohm's theory!effective wave function in} of subsystem $A$ is independent of the particles in the environment, and it does not encode the non-local influences of these particles. 
Note that even although the universal wave function is time-independent, the effective wave functions of subsystem $A$ and its environment may be both time-dependent, and thus it is arguable that they cannot be interpreted nomologically.
\footnote{Moreover, this simplest case seems to also pose a difficulty for the dispositionalist interpretation of Bohm's theory\index{Bohm's theory!dispositionalist interpretation of} suggested by Esfeld\index{Esfeld, Michael} et al (2013). If the universal wave function\index{Bohm's theory!universal wave function in} represents the disposition of motion of all particles in the universe, then when the universal wave function\index{Bohm's theory!universal wave function in} factorizes, the effective wave function\index{Bohm's theory!effective wave function in} of each subsystem will also represent the disposition of motion of the particles of the subsystem, and thus Belot\index{Belot, Gordon}'s (2012) objections will be valid in this case.}

Next, consider the general case in which there is an extra term in the factorization of the universal wave function\index{Bohm's theory!universal wave function in}, which is denoted by (\ref{EF}). In this case, the effective wave function\index{Bohm's theory!effective wave function in} of subsystem A is determined by both the universal wave function\index{Bohm's theory!universal wave function in} and the positions of the particles in its environment (via a measurement-like process). If $Y(t)$ lies within the support of $\phi_t(y)$, $A$'s effective wave function\index{Bohm's theory!effective wave function in} at $t$ will be $\varphi_t(x)$. If $Y(t)$ does not lie within the support of $\phi_t(y)$, $A$'s effective wave function\index{Bohm's theory!effective wave function in} at $t$ will be not $\varphi_t(x)$. For example, suppose $\Theta_t(x, y)=\sum_n{f_n(x)g_n(y)}$, where $g_i(y)$ and $g_j(y)$ are functions with macroscopically disjoint supports for any $i \neq j$, then if $Y(t)$ lies within the support of $g_i(y)$, $A$'s effective wave function\index{Bohm's theory!effective wave function in} at $t$ will be $f_i(x)$. 
It can be seen that the role played by the particles in the environment is only selecting which function the effective wave function\index{Bohm's theory!effective wave function in} of subsystem A is, while each selected function is independent of the particles in the environment and completely determined by the universal wave function\index{Bohm's theory!universal wave function in}. 

Therefore, it seems that the effective wave function\index{Bohm's theory!effective wave function in} of a subsystem of the universe does not encode the influences of other particles in the universe  in general cases. When the effective wave function\index{Bohm's theory!effective wave function in}\index{wave function!effective|see{Bohm's theory}} of a subsystem has been selected, the other particles in the universe will have no influences on the particles of the subsystem. For example, in the double-slit experiment with one particle at a time, the development of the position of the particle does not depend on the positions of other particles in the universe (if only the positions of these particles select the same effective wave function\index{Bohm's theory!effective wave function in} of the particle during the experiment, e.g. $Y(t)$ has been within the support of $\phi_t(y)$ during the experiment).

Since it is arguable that the nomological view\index{wave function!nomological view of} of the wave function implies that the effective wave function\index{Bohm's theory!effective wave function in} of a subsystem of the universe encodes the influences of other particles in the universe, the above result seems to pose a threat to the view. Moreover, the result also suggests that the ontology of Bohm's theory\index{Bohm's theory}\index{Bohm's theory!ontology of} consist in not only Bohmian particle\index{Bohm's theory!Bohmian particles in}s and their positions, but also the wave function.

\chapter{Reality of the wave function}



I have analyzed the competing views of the wave function, including the $\psi$-epistemic view\index{wave function!epistemic view of}, the $\psi$-ontic view\index{wave function!ontic view of} and the nomological view\index{wave function!nomological view of}. 
Which interpretation is true, then? 
Although there are already several $\psi$-ontology theorems\index{psi-ontology theorems}, a definite  answer to this question is still unavailable.
On the one hand, auxiliary assumptions are  required to prove these $\psi$-ontology theorems\index{psi-ontology theorems}, e.g. the preparation independence assumption for the Pusey-Barrett-Rudolph theorem\index{Pusey-Barrett-Rudolph theorem} (Pusey,\index{Pusey, Matthew F.} Barrett\index{Barrett, Jonathan} and Rudolph,\index{Rudolph, Terry} 2012) and the ontic indifference assumption for Hardy's theorem\index{Hardy's theorem} (Hardy,\index{Hardy, Lucien} 2013). It thus seems to be impossible to completely rule out the $\psi$-epistemic view\index{wave function!epistemic view of} without auxiliary assumptions.\footnote{Indeed, by removing the assumptions of these $\psi$-ontology theorems\index{psi-ontology theorems}, explicit $\psi$-epistemic models\index{psi-epistemic models} can be constructed to reproduce the statistics of (projective) quantum measurements in Hilbert spaces of any dimension (Lewis et al\index{Lewis, Peter G.}, 2012; Aaronson\index{Aaronson, Scott} et al, 2013).}
On the other hand, as noted before, the ontological models framework\index{ontological models framework}, on which these $\psi$-ontology theorems\index{psi-ontology theorems} are based, is not very general\index{ontological models framework!limitations of}. For example, the framework does not apply to deterministic theories such as Bohm's theory\index{Bohm's theory}. Thus the $\psi$-ontology theorems\index{psi-ontology theorems}, even if their  auxiliary assumptions can be avoided, cannot rule out the nomological view\index{wave function!nomological view of} of the wave function either. 

But this is not the end of the story. 
In this chapter, I will extend the ontological models framework\index{ontological models framework} by introducing protective measurements\index{protective measurements}, and give a new argument for the $\psi$-ontic view\index{wave function!ontic view of} in terms of protective measurements\index{protective measurements}, first in the extended ontological models framework\index{ontological models framework!extended}\index{ontological models framework} and then beyond the framework (see also Gao, 2015b). 
The argument does not rely on auxiliary assumptions, and it also applies to deterministic theories.

\section{Ontological models framework extended}

In order to obtain a definite answer to the question of the nature of the wave function, the ontological models framework\index{ontological models framework} must be amended and extended.

The first limitation\index{ontological models framework!limitations of} of the ontological models framework\index{ontological models framework} is that it does not apply to deterministic theories. 
This limitation can be readily removed by assuming that the ontic state of a physical system determines the probabilities for different results of a projective measurement on the system \emph{only} for indeterministic theories, and for deterministic theories the ontic state of a physical system (and the ontic state of the measuring device) determine the result of a projective measurement on the system.
In this way, the ontological models framework\index{ontological models framework} can be amended to apply to deterministic theories. 
However, since the result of a projective measurement is random, the additional connection between the ontic state and the measurement result for deterministic theories will have little use in addressing the question of the nature of the wave function in these theories.

The second limitation\index{ontological models framework!limitations of} of the ontological models framework\index{ontological models framework} is that the framework only consider conventional projective measurements\index{projective measurements}. This is not beyond expectations, as these measurements are most well-known and have been once regarded as the only type of quantum measurements. However, it has been known that there are in fact other types of quantum measurements, one of which is the relatively less-known protective measurements\index{protective measurements} (Aharonov\index{Aharonov, Yakir} and Vaidman,\index{Vaidman, Lev} 1993; Aharonov\index{Aharonov, Yakir}, Anandan\index{Anandan, Jeeva S.} and Vaidman,\index{Vaidman, Lev} 1993; see also Section 1.3). 
During a protective measurement, the measured state is protected by an appropriate mechanism such as via the quantum Zeno\index{protective measurements!Zeno-type} effect, so that it neither changes nor becomes entangled with the state of the measuring device. In this way, such protective measurements\index{protective measurements} can measure the expectation values of observables on a single quantum system, even if the system is initially not in an eigenstate of the measured observable, and the wave function of the system can also be measured as expectation values of a sufficient number of observables.

Protective measurements are distinct from projective measurements\index{projective measurements} in that a protective measurement always obtains a definite result, while a projective measurement in general obtains a random result with certain probability in accordance with the Born rule\index{Born rule}. 
As a consequence, the ontological models framework\index{ontological models framework} will be greatly extended by including protective measurements\index{protective measurements}. The framework assumes as its second assumption that when a measurement is performed, the behaviour of the measuring device is only determined by the ontic state of the system, along with the physical properties of the measuring device. 
For a projective measurement,  this assumption means the ontic state of a physical system determines the probabilities for different results of the projective measurement on the system (for indeterministic theories). Similarly, for a protective measurement, this assumption will mean that the ontic state  of a physical system determines the definite result of the protective measurement on the system.\footnote{One may think that the result of a protective measurement is determined not by the ontic state of the measured system, but by the protection procedure, or in other words, what a protective measurement measures is not the ontic state of the measured system but the protection procedure, such as the protection potential for an adiabatic\index{protective measurements!adiabatic-type} protective measurement (see, e.g. Rovelli\index{Rovelli, Carlo}, 1994). However, it has been argued that this understanding is not right (Aharonov\index{Aharonov, Yakir}, Anandan\index{Anandan, Jeeva S.} and Vaidman\index{Vaidman, Lev}, 1996). The main reason is that for infinite number of various protective procedures which are all characterized by having the same wave function, the protective measurements\index{protective measurements} of the same observable will always yield the same results. For further discussion see Section 4.3.} 
Note that this inference for protective measurements\index{protective measurements} is independent of the origin of the Born probabilities, and it applies to both deterministic theories and indeterministic theories.

As we will see immediately, by extending the ontological models framework\index{ontological models framework}, protective measurements\index{protective measurements} will provide more resources for proving the reality of the wave function\index{wave function!reality of}. 

\section{A new proof in terms of protective measurements\index{protective measurements}}

Since the wave function can be measured from a single physical system by a series of protective measurements\index{protective measurements!and the reality of the wave function}, it seems natural to assume that the wave function refers directly to the ontic state of the system. Several authors, including the discoverers of protective measurements\index{protective measurements}, have given similar arguments supporting this implication of protective measurements\index{protective measurements} for the ontological status of the wave function (Aharonov\index{Aharonov, Yakir} and Vaidman\index{Vaidman, Lev}, 1993; Aharonov\index{Aharonov, Yakir}, Anandan\index{Anandan, Jeeva S.} and Vaidman\index{Vaidman, Lev}, 1993; Anandan\index{Anandan, Jeeva S.}, 1993; Dickson\index{Dickson, Michael}, 1995; Gao\index{Gao, Shan}, 2013d, 2014b; Hetzroni\index{Hetzroni, Guy} and Rohrlich\index{Rohrlich, Daniel}, 2014). However, these analyses are not very rigorous and also subject to some objections (Unruh\index{Unruh, William G.}, 1994; Rovelli\index{Rovelli, Carlo}, 1994; Uffink\index{Uffink, Jos}, 1999, 2013; Dass and Qureshi\index{Dass, N. D. Hari}\index{Qureshi, Tabish}, 1999; Schlosshauer\index{Schlosshauer, Maximilian} and Claringbold, 2014).\footnote{See Gao\index{Gao, Shan} (2014b) for a brief review of these objections.} It is still debatable whether protective measurements\index{protective measurements} imply the reality of the wave function\index{wave function!reality of}. In the following, I will give a new, rigorous argument for $\psi$-ontology in terms of protective measurements\index{protective measurements} in the extended ontological models framework\index{ontological models framework!extended}\index{ontological models framework} (see also Gao\index{Gao, Shan}, 2015b)\index{protective measurements!and the reality of the wave function}.

I first use the proof strategy of the existing $\psi$-ontology theorems\index{psi-ontology theorems}, namely first assuming that two nonorthogonal wave functions are compatible with the same ontic state, and then proving the consequences of this assumption are inconsistent with the predictions of quantum mechanics. The argument is as follows. For two different wave functions such as two nonorthogonal states\index{nonorthogonal states}, select an observable whose expectation values in these two states are different. For example, consider a spin half particle. The two nonorthogonal states\index{nonorthogonal states} are $\ket{0}$ and  $\ket{+}$, where $\ket{+}= {1 \over \sqrt{2}}(\ket{0} + \ket{1})$, and $\ket{0}, \ket{1}$ are eigenstates of spin in the z-direction. As Aharonov\index{Aharonov, Yakir}, Anandan\index{Anandan, Jeeva S.} and Vaidman\index{Vaidman, Lev} (1993) showed, a spin state can be protected by a magnetic field in the direction of the spin. Let $B_0, B_+$ be protecting fields for the states $\ket{0}$, $\ket{+}$, respectively, and let the measured observable  be $P_0=\ket{0}\bra{0}$. Then the protective measurements\index{protective measurements} of this observable on these two nonorthogonal states\index{nonorthogonal states} yield results 1 and 1/2, respectively. Although these two nonorthogonal states\index{nonorthogonal states} need different protection procedures, the protective measurements\index{protective measurements} of the observable on the two (protected) states are the same, and the results of the measurements are different with certainty. 
If these two (protected) wave functions are compatible with the same ontic state $\lambda$, then according to the extended ontological models framework\index{ontological models framework!extended}\index{ontological models framework}, the results of the protective measurements\index{protective measurements} of the observable on these two states will be the same. 
This leads to a contradiction. Therefore, two (protected) wave functions correspond to different ontic states.\footnote{This result is not surprising, since two (protected) wave functions of a single system can be distinguished with certainty by protective measurements\index{protective measurements}.} By assuming that whether an unprotected state or a corresponding protected state is prepared, the probability distribution of the ontic state $\lambda$ is the same, which may be called preparation noncontextuality assumption (Spekkens\index{Spekkens, Robert W.}, 2005; Leifer\index{Leifer, Matthew S.}, 2014a),\footnote{According to Leifer\index{Leifer, Matthew S.} (2014a), ``Preparation noncontextuality says that if there is no difference between two preparation procedures in terms of the observable statistics they predict, i.e. they are represented by the same quantum state, then there should be no difference between them at the ontological level either, i.e. they should be represented by the same probability." Due to the existence of protective measurements, however, this definition is vague, since even though two preparation procedures are represented by the same wave function, they may have different observable statistics. it is arguable that a more appropriate definition of preparation noncontextuality is that if two preparation procedures are represented by the same wave function, then there should be no difference between them at the ontological level, in other words, the same wave function corresponds to the same probability distribution of the ontic state. I use this definition of preparation noncontextuality in my argument.} we can further reach the conclusion that two (unprotected) wave functions also correspond to distinct ontic states. In other words, the wave function represents the ontic state of a single  system.\index{protective measurements!and the reality of the wave function}

A similar argument can also be given in terms of realistic protective measurements\index{protective measurements}. A realistic protective measurement cannot be performed on a single quantum system with absolute certainty. For a realistic protective measurement of an observable $A$, there is always a small probability to obtain a result different from $\exptt{A}$. In this case, according to the ontological models framework\index{ontological models framework}, the probabilities for different results will be determined by the ontic state of the measuring device and the realistic measuring condition such as the measuring time, as well as by the ontic state of the measured system.\footnote{Note that this applies only to indeterministic theories. Similarly, the probabilities for different results of a realistic projective measurement will be also determined by the ontic state of the measuring device and the measuring time, as well as by the ontic state of the measured system. As we will see later, the existing $\psi$-ontology theorems\index{psi-ontology theorems} will be difficult or even impossible to prove for realistic projective measurements\index{projective measurements}.} Now consider two  (protected)  wave functions, and select an observable whose expectation values in these two states are different. Then we can perform the same realistic protective measurements\index{protective measurements} of the observable on these two states. The overlap of the probability distributions of the results of these two measurements can be arbitrarily close to zero when the realistic condition approaches the ideal condition (In the limit, each probability distribution will be a Dirac $\delta-$function localized in the expectation value of the measured observable in the measured state, and it will be determined only by the ontic state of the measured system). If there exists a non-zero probability $p$ that these two wave functions correspond to the same ontic state $\lambda$, then since the same $\lambda$ yields the same probability distribution of measurement results under the same measuring condition according to the ontological models framework\index{ontological models framework}, the overlap of the probability distributions of the results of protective measurements\index{protective measurements} of the above observable on these two states will be not smaller than $p$. Since $p>0$ is a determinate number, this leads to a contradiction.\footnote{Note that it is indeed true that for any given realistic condition one can always assume that there exists some probability $p$ that the two measured wave functions correspond to the same ontic state $\lambda$. However, the point is that if the unitary dynamics of quantum mechanics is valid, the realistic condition can always approach the ideal condition arbitrarily closely, and thus the probability $p$ must be arbitrarily close to zero, which means that any $\psi$-epistemic model with finite overlap probability $p$ is untenable. Certainly, our argument will be invalid if quantum mechanics breaks down when reaching certain realistic condition.} Therefore, two (protected) wave functions correspond to different ontic states, and so do two (unprotected) wave functions by the preparation noncontextuality assumption.\index{protective measurements!and the reality of the wave function}

The above argument, like the existing $\psi$-ontology theorems\index{psi-ontology theorems}, is also based on an auxiliary assumption, the preparation noncontextuality assumption this time.\footnote{Since the above argument only considers individual quantum systems and makes no appeal to entanglement\index{quantum entanglement}, it avoids the preparation independence assumption for multiple systems used by the Pusey-Barrett-Rudolph theorem\index{Pusey-Barrett-Rudolph theorem} (Pusey,\index{Pusey, Matthew F.} Barrett\index{Barrett, Jonathan} and Rudolph,\index{Rudolph, Terry} 2012).} However, the argument can be further improved to avoid this auxiliary assumption. The key is to notice that the result of a protective measurement depends only on the measured observable and the ontic state of the measured system. If the result is also determined by other factors such as the ontic state of the measuring device or the protection setting, then the result may be different for the same measured observable and wave function. This will contradict the predictions of quantum mechanics, according to which the result of a protective measurement is always the expectation value of the measured observable in the measured wave function. Now consider  two  (unprotected) wave functions, and select an observable whose expectation values in these two states are different. The results of the protective measurements\index{protective measurements} of the observable on these two states are different with certainty. 
If these two wave functions are compatible with the same ontic state $\lambda$, then according to the above analysis, the results of the protective measurements\index{protective measurements} of the observable on these two states will be the same. 
This leads to a contradiction. Therefore, two different wave functions correspond to different ontic states.\index{protective measurements!and the reality of the wave function}

There is also a direct argument for $\psi$-ontology in terms of protective measurements\index{protective measurements}, which is not based on auxiliary assumptions either. As argued above, the result of a protective measurement is determined only by the measured observable and the ontic state of the measured system. Since the measured observable also refers to the measured system, this further means that the result of a protective measurement, namely the expectation value of the measured observable in the measured wave function, is determined only by the properties of the measured system. Therefore, the expectation value of the measured observable  in the measured wave function is also a property of the measured system. Since a wave function can be constructed from the expectation values of a sufficient number of observables, the wave function also represents the property of a single quantum system.\index{protective measurements!and the reality of the wave function}

 \section{With more strength}

The above arguments for $\psi$-ontology, like the other $\psi$-ontology theorems\index{psi-ontology theorems}, are based on the second assumption of the ontological models framework\index{ontological models framework}, according to which when a measurement is performed, the behaviour of the measuring device is determined by the ontic state of the measured system (along with the physical properties of the measuring device) immediately before the measurement, whether the ontic state of the measured system changes or not during the measurement. This is a simplified assumption, and it may be not valid in general. A more reasonable assumption is that the ontic state of the measured system may be disturbed and thus evolve in a certain way during a measurement, and the behaviour of the measuring device is determined by the total evolution of the ontic state of the system during the measurement, not simply by the initial ontic state of the system. 
For example, for a projective measurement it is the total evolution of the ontic state of the measured system during the measurement that determines the probabilities for different results of the measurement. 
Certainly, if the measuring interval is extremely short and the change of the ontic state of the measured system is continuous, then the ontic state will be almost unchanged during the measurement, and thus the original simplified assumption will be still valid. However, if the change of the ontic state of the measured system is not continuous but discontinuous, then even during an arbitrarily short time interval the ontic state may also change greatly, and the original simplified assumption will be wrong.

It seems that the existing $\psi$-ontology theorems\index{psi-ontology theorems} such as the Pusey-Barrett-Rudolph theorem\index{Pusey-Barrett-Rudolph theorem} will be invalid under the above new assumption. The reason is that under this assumption, even if two nonorthogonal states\index{nonorthogonal states} correspond to the same ontic state initially, they may correspond to different evolution of the ontic state, which may lead to different probabilities of measurement results. Then the proofs of the $\psi$-ontology theorems\index{psi-ontology theorems} by reduction to absurdity cannot go through. 
However, it can be seen that the above arguments for $\psi$-ontology in terms of protective measurements can still go through under the new assumption. 

For a protective measurement, there are two sources which may interfere with the spontaneous evolution of the ontic state of the measured system: one is the protection procedure, and the other is the measuring device. 
However, no matter how they influence the evolution of the ontic state of the measured system, they cannot generate the definite result of the protective measurement, namely the expectation value of the measured observable in the measured wave function, since they contain no information about the measured wave function.\footnote{In other words, the properties of the protection setting and the measuring device and their time evolution do not determine the measured wave function.}  
The measuring device only contains information about the measured observable, and it does not contain information about the measured wave function. Compared with the measuring device, the protection procedure ``knows" less. 
The protection procedure is either a protective potential or a Zeno measuring device. In each case, the protection procedure contains no information about both the measured observable and the measured wave function.\footnote{Certainly, the measurer who does the protective measurement knows more information than that contained in the measuring device and protection procedure. Besides the measured observable, the measurer also knows the measured wave function is one of infinitely many known states (but she needs not know which one the measured wave function is). In the case of protective potential, the measurer knows that the measured wave function is one of infinitely many nondegenerate discrete energy eigenstate of the Hamiltonian of the measured system. In the case of Zeno protection, the measurer knows that the measured wave function is one of infinitely many nondegenerate eigenstates of an observable. Note that this permits the possibility that the measurer can cheat us by first measuring which one amongst these infinitely many states the measured wave function is (e.g. by measuring the eigenvalue of energy for the case of protective potential) and then calculating the expectation value and outputing it through a device. Then the result will have no implications for the reality of the wave function. But obviously this is not a protective measurement.}   
Thus, if the information about the measured wave function is not contained in the measured system, then the result of a protective measurement cannot be the expectation value of the measured observable in the measured wave function.  

On the other hand, as noted before, if the result of a protectie measurement is also determined by the ontic state of the measuring device or the protection procedure due to their influences on the evolution of the ontic state of the measured system, then the result may be different for the same measured observable and the same measured wave function. This  contradicts the predictions of quantum mechanics, according to which the result of a protective measurement is always the expectation value of the measured observable in the measured wave function. 

Therefore, the definite result of a protective measurement, namely the expectation value of the measured observable in the measured wave function, is determined by the spontaneous evolution of the ontic state of the measured system during the measurement. Since the spontaneous evolution of the ontic state of the measured system is an intrinsic property of the system independent of the protective measurement, the expectation value of the measured observable in the measured wave function is also a property of the system. This then proves the reality of the wave function, which can be constructed from the expectation value\index{protective measurements!expectation values as results of}s of a sufficient number of observables.

In the following, I will present a more detailed analysis of how a protective measurement obtains the expectation value of the measured observable in the measured wave function. 
The analysis may help understand the above result by further clarifing the roles the measured system, the protection procedure, and the measuring device play in a protective measurement. 

By a projective measurement on a single quantum system, one obtains one of the eigenvalues of the measured observable, and the expectation value of the observable can only be obtained as the statistical average of eigenvalues for an ensemble of identically prepared systems. 
Thus it seems surprising that a protective measurement can obtain the expectation value of the measured observable directly from a single quantum system. 
In fact, however, this result is not as surprising as it seems to be. 
The key point is to notice that the pointer shift rate at any time during a projective measurement is proportional to the expectation value of the measured observable in the measured wave function at the time. Concretely speaking, for a projective measurement of an observable $A$, whose interaction Hamiltonian\index{Hamiltonian!interaction} is given by the usual form $H_I = g(t)PA$, where $g(t)$ is the time-dependent coupling strength of the interaction, and $P$ is the conjugate momentum of the pointer variable, the pointer shift rate at each instant $t$ during the measurement is:

\begin{equation}
{d\exptt{X} \over dt} = g(t)\exptt{A},
\end{equation}

\noindent where $X$ is the pointer variable, $\exptt{X}$ is the center of the pointer wavepacket at instant $t$, and $\exptt{A}$ is the expectation value of the measured observable $A$ in the measured wave function at instant $t$.  
This pointer shift rate formula indicates that at any time during a projective measurement, the pointer shift after an infinitesimal time interval is proportional to the expectation value of the measured observable in the measured wave function at the time.  
As is well known, however, since the projective measurement changes the wave function of the measured system greatly, and especially it also results in the pointer wavepacket spreading greatly, the point shift after the measurement does not represent the actual measurement result, and it cannot be measured either. 
Moreover, even if the point shift after the measurement represents the actual measurement result (e.g. for collapse theories), the result is not definite but random, and it is not the expectation value of the measured observable in the initial measured wave function either. 

Then, how to make the expectation value of the measured observable  in the measured wave function, which is hidden in the process of a projective measurement, visible in the final measurement result?  
This requires that the pointer wavepacket should not spread considerably during the measurement so that the final pointer shift is qualified to represent the measurement result, and moreover, the final pointer shift should be also definite. 
A direct way to satisfy the requirement is to protect the measured wave function from changing as a protective measurement does. Take the Zeno protection scheme as an example. 
We make frequent projective measurements\index{projective measurements} of an observable $O$, of which the measured state $\ket{\psi}$ is a nondegenerate eigenstate, in a very short measurement interval $[0,\tau]$. For instance, $O$ is measured in $[0,\tau]$ at times $t_n=(n/N)\tau, n = 1,2, . . . ,N$, where $N$ is an arbitrarily large number.  At the same time, we make the same projective measurement of an observable $A$ in the interval $[0,\tau]$ as above. Different from the derivation given in Section 1.3, here I will calculate the post-measurement state in accordance of the order of time evolution. This will let us see the process of protective measurement more clearly.\footnote{Note that in the derivation given in Section 1.3, the measurement result of a protective measurement, namely the expectation value of the measured observable in the measured wave function, is already contained in the measurement operator which describes the measurement procedure. But this does not imply what the measurement measures is not the property of the measured system, but the property of the measurement procedure such as the protection procedure (cf. Combes et al, 2015).  Otherwise, for example, diseases will exist not in patients, but in doctors or expert systems for disease diagnosis.}  

The state of the combined system immediately before $t_1=\tau/N$ is given by

\begin{eqnarray}
  e^{-{i\over\hbar}{\tau\over N}g(t_1)P A}\ket{\psi}\ket{\phi(x_0)} &=& \sum_{i} c_i \ket{a_i} \ket{\phi(x_0+{\tau\over N}g(t_1)a_i)}\nonumber \\
 &=& \ket{\psi}\ket{\phi(x_0+ {\tau\over N}g(t_1)\exptt{A})} \nonumber \\ 
& & +{\tau\over N}g(t_1)(A-\exptt{A}) \ket{\psi} \ket{\phi'(x_0+{\tau\over N}g(t_1)\exptt{A})}\nonumber \\ & & + O({1 \over N^2}),
\label{}
\end{eqnarray}

\noindent where $\ket{\phi(x_0)}$ is the pointer wavepacket centered in initial position $x_0$, $\ket{a_i}$ are the eigenstates of $A$, and $c_i$ are the expansion coefficients. Note that the second term in the r.h.s of the formula is orthogonal to the measured state $\ket{\psi}$. Then the branch of the state of the combined system after $t_1=\tau/N$, in which the projective measurement of $O$ results in the state of the measured system being in $\ket{\psi}$, is given by 

\begin{eqnarray}
  \ket{\psi}\bra{\psi}e^{-{i\over\hbar}{\tau\over N}g(t_1)P A}\ket{\psi}\ket{\phi(x_0)} 
= \ket{\psi}\ket{\phi(x_0+ {\tau\over N}g(t_1)\exptt{A})} + O({1 \over N^2}).
\label{}
\end{eqnarray}

\noindent Thus after $N$ such measurements and in the limit of $N \rightarrow \infty$, the branch of the state of the combined system, in which each projective measurement of $O$ results in the state of the measured system being in $\ket{\psi}$, is

\begin{eqnarray}
\ket{t=\tau} = \ket{\psi}\ket{\phi(x_0+\int_{0}^{\tau}g(t)dt\exptt{A})}= \ket{\psi}\ket{\phi(x_0+\exptt{A})}.
\label{}
\end{eqnarray}

\noindent Since the modulus squared of the amplitude of this branch approaches one when $N \rightarrow \infty$, this state will be the state of the combined system after the protective measurement.


By this derivation, it can be clearly seen that the role of the protection procedure is not only to protect the measured wave function from the change caused by the projective measurement, but also to prevent the pointer wavepacket from the spreading caused by the projective measurement. 
As a result, the pointer shift after the measurement can represent a valid measurement result, and moreover, it is also definite, being natually the expectation value of the measured observable in the \emph{initial} measured wave function.  

In fact, since the width of the pointer wavepacket keeps unchanged during the above protective measurement, and the pointer shift rate at any time during the measurement is proportional to the expectation value of the measured observable in the measured wave function at the time,\footnote{Since the pointer shift is always continuous and smooth during a protective measurement, it is arguable that the evolution of the ontic state of the measured system (which determines the pointer shift) is  also continuous. Then for an ideal situation where the protective measurement is instantaneous, the ontic state of the measured system will be unchanged after the measurement and my previous arguments for $\psi$-ontology in terms of protective measurements\index{protective measurements} will be still valid. Note that the evolution of the position of the pointer as its ontic state may be discontinuous in an $\psi$-epistemic model. However, the range of the position variation is limited by the width of the pointer wavepacket, which can be arbitrarily small in principle. Thus such discontinuous evolution cannot be caused by the evolution of the ontic state of the measured system, whether continuous or discontinuous.} which is the same as the initial measured wave function, we can obtain the final measurement result at any time during the protective measurement (when the time-dependent coupling strength is known). This indicates that the result of a protective measurement is determined by the initial ontic state of the measured system, not by the evolution of the ontic state of the system during the measurement, whether spontaneous or not. Thus the second, simplified assumption of the ontological models framework is still valid for protective measurements, so do my previous arguments for the reality of the wave function based on this assumption.



It has been conjectured that the result of a protective measurement is determined not by the ontic state of the measured system but by the protection procedure, which may lead to a certain evolution of the ontic state of the system that may generate the measurement result (Combes et al, 2015). If this is true, then protective measurements will have no implications for the reality of the wave function. 
However, as I have argued in the beginning of this section, this conjecture cannot be correct. The essential reason is that the protection procedure does not ``know" the measured wave function, and thus it cannot generate the measurement result, the expectation value of the measured observable in the measured wave function.\footnote{Note that in the $\psi$-epistemic models given by Combes et al (2015), it is implicitly assumed that the protection procedure knows the measured wave function. Thus it is not surprising that the models can reproduce the predictions of quantum mechanics for protective measurements.}  
In addition, the above analysis clearly shows that the result of a protective measurement is generated not by the protection procedure.  
The expectation value of the measured observable in the measured wave function is already hidden in the process of the projective measurement, and what the protection procedure does is to make it visible in the final measurement result by keeping the measured wave function unchanged.  

In conclusion, the above analysis of how a protective measurement influences the measured system and obtains its result does not refute but strengthen my previous arguments for $\psi$-ontology in terms of protective measurements.\index{protective measurements!and the reality of the wave function}

\section{A weaker criterion of reality\index{criterion of reality}}

The first assumption of the ontological models framework\index{ontological models framework} is that if a quantum system is prepared such that quantum mechanics assigns a pure state to it, then after preparation the system has a well-defined set of physical properties (Harrigan\index{Harrigan, Nicholas} and Spekkens\index{Spekkens, Robert W.}, 2010; Pusey,\index{Pusey, Matthew F.} Barrett\index{Barrett, Jonathan} and Rudolph,\index{Rudolph, Terry} 2012). The $\psi$-ontology theorems\index{psi-ontology theorems}, including the above arguments in terms of protective measurements\index{protective measurements}, are all based on this assumption. If one drops this assumption as anti-realists would like to do, then one can still restore the (non-realist) $\psi$-epistemic view\index{wave function!epistemic view of} or assume another non-realist view. In this section, I will give a stronger proof of the reality of the wave function\index{wave function!reality of} based not on this realistic assumption but on a weaker criterion of reality\index{criterion of reality}. The analysis is beyond the ontological models framework\index{ontological models framework}.\index{protective measurements!and the reality of the wave function}\index{criterion of reality!and reality of the wave function}

A well-known criterion of reality\index{criterion of reality} is the EPR criterion of reality\index{criterion of reality!EPR}, which says that  ``If, without in any way disturbing a system, we can predict with certainty ({i.e.} with probability equal to unity) the value of a physical quantity, then there exists an element of physical reality\index{wave function!reality of} corresponding to this physical quantity." (Einstein\index{Einstein, Albert}, Podolsky and Rosen\index{Rosen, Nathan}\index{Podolsky, Boris}, 1935).\footnote{Note that Hetzroni\index{Hetzroni, Guy} and Rohrlich\index{Rohrlich, Daniel} (2014) gave an argument for $\psi$-ontology based on protective measurements\index{protective measurements} and the EPR criterion of reality\index{criterion of reality!EPR}, and Gao\index{Gao, Shan} (2014b) improved the argument by revising the criterion of reality\index{criterion of reality} so that it can also be applied to realistic protective measurements\index{protective measurements}.} 
The main difficulty of applying this criterion of reality\index{criterion of reality} is to determine whether the measured system is disturbed during a measurement. 
Since we don't know the ontic state of the measured system and its dynamics during a measurement before our analysis using the criterion of reality\index{criterion of reality}, the requirement of ``without in any way disturbing a system" in the criterion seems difficult or even impossible to justify. In addition, disturbing the measured system does not necessarily exclude the possibility that the measurement result reflects the property of the measured system. The disturbance may not influence the parts of the ontic state of the measured system which generate the measurement result (see, e.g.  Spekkens\index{Spekkens, Robert W.}, 2007 for an example). 

Here, based on the analysis given in the last section, I suggest an improved criterion of reality\index{criterion of reality!a new} that may avoid the above problems. 
It is that if a measurement of a physical quantity on a system obtains a definite result, which is denoted by the value of a pointer variable after the measurement, and during the measurement the pointer shift rate is also determined by the value, then the measurement result reflects a physical property of the measured system. This criterion of reality provides a direct link from the mathematical quantities in a realistic theory to the properties of a physical system via experience. By using it to analyze the ontological content of a theory, we need not care about the underlying ontic state of a physical system and its possible dynamics during a measurement. Thus the analysis will be simpler.  

It can be seen that this suggested criterion of reality can be directly applied to classical mechanics. Moreover, due to the existence of protective measurements\index{protective measurements}, it can also be applied to quantum mechanics to analyze the ontological status of the wave function. 
As I have pointed out in the last section, a protective measurement on a physical system will obtain a definite result, namely the expectation value of the measured observable in the measured wave function, and during the measurement the pointer shift rate is also determined by the expectation value. 
Thus, according to the suggested criterion of reality\index{criterion of reality}, the expectation value of the measured observable in the measured wave function is a physical property of the measured system. Since the wave function can be constructed from the expectation values of a sufficient number of observables, the measured wave function also represents a physical property of the measured system. This proves the reality of the wave function\index{wave function!reality of} in quantum mechanics.\index{criterion of reality!and reality of the wave function}


Since the suggested criterion of reality\index{criterion of reality!a new} does not necessarily require that a quantum system have properties, it is weaker than the realistic assumption of the ontological models framework\index{ontological models framework}. Even though some people refuse to attribute properties to quantum systems, they may well accept this criterion of reality\index{criterion of reality}. On the one hand, this criterion of reality\index{criterion of reality} can be perfectly applied to classical mechanics, and one can use it to get the anticipant ontological content of the theory. On the other hand, people usually think that this criterion of reality\index{criterion of reality} cannot be applied to quantum mechanics in general (although it can be applied to the measurements of the eigenstates of the measured observable), and thus it does not influence the anti-realist views of the theory\index{quantum mechanics!anti-realist views of}. However, the existence of protective measurements\index{protective measurements} must be a surprise for these people. It will be interesting to see whether some anti-realists will reject this criterion of reality\index{criterion of reality} due to the existence of protective measurements\index{protective measurements}.\index{protective measurements!and the reality of the wave function}\index{criterion of reality!and reality of the wave function}

Certainly, one can also restore the (non-realist) $\psi$-epistemic view\index{wave function!epistemic view of} by rejecting the suggested criterion of reality\index{criterion of reality!a new}. However, there is a good reason why this is not a good choice. it is arguable that a reasonable, universal criterion of reality\index{criterion of reality}, which may provide a plausible link between theory and reality via experience, is useful or even necessary for realistic theories. The criterion of reality\index{criterion of reality} is not necessarily complete, being able to derive all ontological content of a theory, which seems to be an impossible task. However, we can at least derive the basic ontological content of a realistic theory by using this criterion of reality\index{criterion of reality}. If one admits the usefulness and universality of such a criterion of reality\index{criterion of reality}, then the similarity between classical measurements and protective measurements\index{protective measurements} will require that if one assumes a realist view of classical mechanics, admitting the ontological content of the theory derived from the suggested criterion of reality\index{criterion of reality!a new}, then one must also admit the ontological content of quantum mechanics derived from this criterion of reality\index{criterion of reality}, such as the reality of the wave function\index{wave function!reality of}. The essential point is not that the suggested criterion of reality\index{criterion of reality!a new} must be true, but that if we accept the usefulness and universality of such a  criterion of reality\index{criterion of reality} and apply it to classical mechanics and macroscopic objects\index{macroscopic objects} to derive the anticipant classical ontology, we should also apply it to quantum mechanics and microscopic objects\index{microscopic objects} to derive the quantum ontology, no matter how strange it is. Otherwise we will have to divide the world into a quantum part and a classical part artificially, and we will not have a unified world view as a result.\index{protective measurements!and the reality of the wave function}\index{criterion of reality!and reality of the wave function}

\chapter{Origin of the Schr\"{o}dinger equation\index{Schr\"{o}dinger equation}}






I have argued that the wave function in the Schr\"{o}dinger equation\index{Schr\"{o}dinger equation} of quantum mechanics represents the physical state of a single system. In this chapter, I will provide additional evidence supporting this conclusion by analyzing the origin of the Schr\"{o}dinger equation\index{Schr\"{o}dinger equation}. 

Many quantum mechanics textbooks provide a heuristic derivation of the Schr\"{o}dinger equation\index{Schr\"{o}dinger equation} (see, e.g. Schiff\index{Schiff, Leonard}, 1968;  Landau and Lifshitz\index{Landau Lev  D.}\index{Lifshitz, Evgeny M.}, 1977; Greiner\index{Greiner, Walter}, 1994).
It begins with the assumption that the state of a free microscopic particle has the form of a plane wave $e^{i(kx-\omega t)}$. When combining with the de Broglie\index{de Broglie, Louis} relations\index{de Broglie relations} for momentum and energy $p=\hbar k$ and $E=\hbar \omega$, this state becomes $e^{i(px-Et)/\hbar}$. Then it uses the nonrelativistic energy-momentum relation\index{energy-momentum relation}\index{energy-momentum relation!nonrelativistic} $E=p^2/2m$ to obtain the free Schr\"{o}dinger equation\index{Schr\"{o}dinger equation!free}. Lastly, this equation is generalized to include an external potential, and the end result is the Schr\"{o}dinger equation\index{Schr\"{o}dinger equation}. 

In the following sections, I will show that the heuristic derivation of the free Schr\"{o}dinger equation\index{Schr\"{o}dinger equation!free}\index{Schr\"{o}dinger equation!free!derivation of} can be made more rigorous by resorting to spacetime translation invariance\index{laws of motion!spacetime translation invariance of} and relativistic invariance\index{laws of motion!relativistic invariance of}.
Spacetime translation gives the definitions of momentum and energy, and spacetime translation invariance\index{laws of motion!spacetime translation invariance of} entails that the state of a free quantum system with definite momentum and energy assumes the plane wave form $e^{i(px-Et)}$. Moreover, the relativistic transformations of the generators of space translation and time translation further determine the relativistic energy-momentum relation\index{energy-momentum relation!relativistic}\index{energy-momentum relation}, whose nonrelativistic approximation is $E=p^2/2m$. 
Although the requirements of these invariances are already well known, an explicit and complete derivation\index{Schr\"{o}dinger equation!free!derivation of} of the free Schr\"{o}dinger equation\index{Schr\"{o}dinger equation!free} using them seems still missing in the literature.\footnote{Note that several authors have claimed that the free Schr\"{o}dinger equation\index{Schr\"{o}dinger equation!free} can be derived in terms of Galilean invariance and a few other assumptions (Musielak and Fry\index{Musielak, Zdzislaw E.}\index{Fry, J. L.}, 2009a, 2009b). But the derivation is arguably problematic (Gao\index{Gao, Shan}, 2014e). In addition, there are also attempts to derive the Schr\"{o}dinger equation\index{Schr\"{o}dinger equation} from Newtonian mechanics\index{Newtonian mechanics}, a typical example of which is Nelson\index{Nelson, Edward}'s stochastic mechanics (Nelson\index{Nelson, Edward}, 1966). It has been pointed out that Nelson\index{Nelson, Edward}'s stochastic mechanics is not equivalent to quantum mechanics (Grabert, H\"{a}nggi and Talkner, 1979; Wallstrom\index{Wallstrom, Timothy C.}, 1994). Moreover, Nelson\index{Nelson, Edward} himself also showed that there is an empirical difference between the predictions of quantum mechanics and his stochastic mechanics when considering quantum entanglement\index{quantum entanglement} and nonlocality (Nelson\index{Nelson, Edward}, 2005). But it seems still possible to obtain quantum mechanics ``as a statistical mechanics canonical ensemble average of classical variables obeying classical dynamics." (Pearle\index{Pearle, Philip}, 2005)  Adler's (2004)\index{Adler, Stephen L.} trace dynamics is an excellent example. In the theory, it is shown that under plausible assumptions, thermodynamic averages leads to the Schr\"{o}dinger equation, while fluctuations around the averages leads to a stochastic modification of the Schr\"{o}dinger equation, which may naturally explain the collapse of the wave function and the Born rule.} 
The new analysis may not only answer why the physical state of a single system is described by a wave function, but also help answer why the linear nonrelativistic time evolution of the wave function is governed by the  Schr\"{o}dinger equation\index{Schr\"{o}dinger equation}. 


\section{Spacetime translation invariance}

It is well known that the laws of motion\index{laws of motion} that govern the time evolution of an isolated system satisfies spacetime translation invariance\index{laws of motion!spacetime translation invariance of}.\footnote{This is due to the homogeneity of space and time. The homogeneity of space ensures that the same experiment performed at two different places gives the same result, and the homogeneity in time ensures that the same experiment repeated at two different times gives the same result.} 
In this section, I will analyze how the requirement of spacetime translation invariance\index{laws of motion!spacetime translation invariance of} restricts the possible forms of the laws of motion\index{laws of motion}. For the sake of simplicity, I will mainly analyze one-dimensional motion.
The physical state of an isolated system is assumed to be represented by a general analytic function with respect to both $x$ and $t$, $\psi(x,t)$.\footnote{It is arguable that $\psi(x,t)$ is the most general scalar representation of the physical state of a system. As we will see later, however, the equation that governs the time evolution of the state will restrict the possible forms of $\psi(x,t)$.} A space translation operator can be defined as 

\begin{equation}
T(a)\psi(x,t)=\psi(x-a,t).
\label{}
\end{equation}

\noindent It means translating rigidly the state of the system, $\psi(x,t)$, by an infinitesimal amount $a$ in the positive $x$ direction.\footnote{There are in general two different pictures of translation: active transformation and passive transformation. The active transformation corresponds to displacing the studied system, and the passive transformation corresponds to moving the coordinate system. Physically, the equivalence of the active and passive pictures is due to the fact that moving the system one way is equivalent to moving the coordinate system the other way by an equal amount. Here I will analyze spacetime translations in terms of active transformations.} $T(a)$ can be further expressed as 

\begin{equation}
T(a)=e^{-ia\hat{p}}, 
\label{}
\end{equation}

\noindent where $\hat{p}$ is the generator of space translation.\footnote{In order to differentiate the momentum and energy eigenvalues from the momentum and energy operators, I add a hat to the momentum and energy operators as usual. But I omit the hat for all other operators in this book. In addition, for convenience of later discussion I introduce the imaginary unit \emph{i} in the expression. This does not influence the validity of the following derivation.} By expanding $\psi(x-a,t)$  in order of $a$, we can further get 

\begin{equation}
\hat{p}=-i{ \partial \over \partial x}. 
\label{}
\end{equation}

\noindent Similarly, a time translation operator can be defined as 

\begin{equation}
U(t)\psi(x,0)=\psi(x,t). 
\label{}
\end{equation}

\noindent And it can also be expressed as $U(t) = e^{-it\hat{E}}$, where 

\begin{equation}
\hat{E}=i{\partial \over \partial t}
\label{}
\end{equation}

\noindent  is the generator of time translation. In order to know the laws of motion\index{laws of motion}, we need to find the concrete manifestation of $\hat{E}$ for a physical system, which means that we need to find the evolution equation of state:

\begin{equation}
i{\partial \psi(x,t) \over \partial t} =H\psi(x,t),
\label{Sch0}
\end{equation}

\noindent  where $H$ is a to-be-determined operator that depends on the properties of the studied system, and it is also called the generator of time translation.\footnote{Similarly I also introduce the imaginary unit \emph{i} in the equation for convenience of later discussion.} In the following analysis, I assume $H$ is a linear operator independent of the evolved state, namely the evolution is linear, which is an important presupposition in my derivation\index{Schr\"{o}dinger equation!free!derivation of} of the free Schr\"{o}dinger equation\index{Schr\"{o}dinger equation!free}. 

Let us now see the implications of spacetime translation invariance\index{laws of motion!spacetime translation invariance of} for the laws of motion\index{laws of motion}. First of all, time translational invariance requires that $H$ have no time dependence, namely $dH/dt=0$. This can be demonstrated as follows (see also Shankar\index{Shankar, Ramamurti}, 1994, p.295). Suppose an isolated system is in state $\psi_0$ at time $t_1$ and evolves for an infinitesimal time $\delta t$. The state of the system at time $t_1+\delta t$, to first order in $\delta t$, will be  

\begin{equation}
\psi (x,t_1+\delta t)=[I-i\delta t H(t_1)]\psi_0.
\label{}
\end{equation}

\noindent  If the evolution is repeated at time $t_2$, beginning with the same initial state, the state at $t_2+\delta t$ will be

\begin{equation}
\psi (x,t_2+\delta t)=[I-i\delta t H(t_2)]\psi_0 .
\label{}
\end{equation}

\noindent  Time translational invariance requires the outcome state should be the same:

\begin{equation}
\psi (x,t_2+\delta t)-\psi (x,t_1+\delta t)=i\delta t [H(t_1)-H(t_2)]\psi_0=0.
\label{}
\end{equation}

\noindent  Since the initial state $\psi_0$ is arbitrary, it follows that $H(t_1)=H(t_2)$. Moreover, since $t_1$ and $t_2$ are also arbitrary, it follows that $H$ is time-independent, namely $dH/dt=0$. It can be seen that this result relies on the linearity of time evolution. If $H$ depends on the state, then obviously we cannot obtain $dH/dt=0$ because the state is time-dependent. But we still have $H(t_1,\psi_0)=H(t_2,\psi_0)$, which means that the state-dependent $H$ also satisfies time translational invariance.

Secondly, space translational invariance requires $[T(a),U(t)]=0$, which further leads to $[\hat{p},\hat{E}]=0$ and $[\hat{p},H]=0$. This can be demonstrated as follows (see also Shankar\index{Shankar, Ramamurti}, 1994, p.293). Suppose at $t = 0$ two observers $A$ and $B$ prepare identical isolated systems at $x = 0$ and $x = a$, respectively. Let $\psi(x,0)$ be the state of the system prepared by $A$. Then $T(a)\psi(x,0)$ is the state of the system prepared by B, which is obtained by translating (without distortion) the state  $\psi(x,0)$ by an amount $a$ to the right. The two systems look identical to the observers who prepared them. After time $t$, the states evolve into  $U(t)\psi(x,0)$ and  $U(t)T(a)\psi(x,0)$. Since the time evolution of each identical system at different places should appear the same to the local observers, the above two systems, which differed only by a spatial translation at $t = 0$, should differ only by the same spatial translation at future times. Thus the state $U(t)T(a)\psi(x,0)$  should be the translated version of $A$'s system at time $t$, namely we have $U(t)T(a)\psi(x,0)=T(a)U(t)\psi(x,0)$. This relation holds true for any initial state $\psi(x,0)$, and thus we have $[T(a),U(t)]=0$, which says that space translation operator and time translation operator are commutative. Again, it can be seen that the linearity of time evolution is an important presupposition of this result. If $U(t)$ depends on the state, then the space translational invariance will only lead to $U(t, T\psi)T(a)\psi(x,0)=T(a)U(t,\psi)\psi(x,0)$, from which we cannot obtain $[T(a),U(t)]=0$. 

When $dH/dt=0$, the solutions of the evolution equation Eq.(\ref{Sch0}) assume the basic form

\begin{equation}
\psi(x,t)=\varphi_E(x)e^{-iEt},
\label{Sch1}
\end{equation}

\noindent and their linear superpositions, where $E$ is an eigenvalue of $H$, and $\varphi_E(x)$ is an eigenfunction of $H$ and satisfies the time-independent equation: 

\begin{equation}
H\varphi_E(x)=E\varphi_E(x). 
\label{}
\end{equation}

\noindent Moreover, the commutative relation $[\hat{p},H]=0$ further implies that $\hat{p}$ and $H$ have common eigenfunctions. Since the eigenfunction of $\hat{p} = -i{ \partial \over \partial x}$ is $e^{ipx}$ (except a normalization factor),
where $p$ is the eigenvalue, the basic solutions of the evolution equation Eq.(\ref{Sch0}) for an isolated system assume the form $e^{i(px-Et)}$, which represents the state of an isolated system with definite properties $p$ and $E$. 
In quantum mechanics, $\hat{p}$ and $\hat{E}$, the generators of space translation and time translation, are also called momentum operator and energy operator, respectively, and $H$ is called the Hamiltonian of the system. Correspondingly, $e^{i(px-Et)}$ is the eigenstate of both momentum and energy, and $p$ and $E$ are the corresponding momentum and energy eigenvalues, respectively. Then the state $e^{i(px-Et)}$ describes an isolated system (e.g. a free electron) with definite momentum $p$ and energy $E$.

\section{The energy-momentum relation\index{energy-momentum relation}}

The  energy-momentum relation\index{energy-momentum relation} can be further determined by considering the relativistic transformations of the generators of space translation and time translation. The operator $\hat{P}_{\mu}=(\hat{E}/c, -\hat{{p}})=i({1 \over c}{\partial \over \partial t}, \nabla)$ is a four-vector operator. In order that its eigenvalue equation holds in all inertial frames, its eigenvalues must transform as a four-vector too. In other words,  every eigenvalue of the four-vector operator $\hat{P}_{\mu}$, $(E/c, -{p})$, is also a four-vector. 
Since the dot product of two four-vectors is Lorentz invariant (a Lorentz scalar), we can form a Lorentz  scalar $p^2-E^2/c^2$ with the four-vector  $(E/c, -{p})$. Then the energy-momentum relation\index{energy-momentum relation} is:

\begin{equation}
E^2=p^2c^2+E_0^2,
\label{EM1}
\end{equation}

\noindent where $p$ and $E$ are the momentum and energy of a microscopic particle, respectively, and $E_0$ is the energy of the particle when its momentum is zero, called the rest energy of the particle.\footnote{For other derivations of the energy-momentum relation\index{energy-momentum relation!derivation of}\index{energy-momentum relation} see Sonego\index{Sonego, Sebastiano} and Pin\index{Pin, Massimo} (2005) and references therein.} By defining $m=E_0/c^2$ as the (rest) mass of the particle, we can further obtain the familiar energy-momentum relation\index{energy-momentum relation}

\begin{equation}
E^2=p^2c^2+m^2 c^4.
\label{EM2}
\end{equation}

\noindent In the nonrelativistic domain, this energy-momentum relation\index{energy-momentum relation!nonrelativistic} reduces to $E=p^2/2m$.

\section{Derivation of the Schr\"{o}dinger equation\index{Schr\"{o}dinger equation}}

Since the operators $\hat{E}$ and $\hat{p}$ have common eigenfunctions for an isolated system, the relation between their eigenvalues $E$ and  $p$ or the energy-momentum relation\index{energy-momentum relation} implies the corresponding operator relation between $\hat{E}$ and $\hat{p}$. In the nonrelativistic domain, the operator relation is $\hat{E}=\hat{p}^2/2m$ for an isolated system. Then we can obtain the free Schr\"{o}dinger equation\index{Schr\"{o}dinger equation!free}:

\begin{equation}
i{\partial \psi(x,t) \over \partial t}=-{1 \over 2m}{{\partial^2 \psi(x,t)}\over \partial x^2}.
\label{FSch}
\end{equation}

Here it needs to be justified that the only parameter $m$ in this equation assumes real values; otherwise the existence of the imaginative unit $i$ in the equation will be an illusion and the equation will be distinct from the free Schr\"{o}dinger equation\index{Schr\"{o}dinger equation!free}. Since velocity assumes real values, this is equivalent to proving that momentum or the eigenvalue of the generator of space translation assumes real values, namely that the generator of space translation itself is Hermitian. This is indeed the case. Since the space translation operator $T(a)$ preserves the norm of the state: $\int_{-\infty}^{\infty} \psi^{\ast}(x,t)\psi(x,t)dx=\int_{-\infty}^{\infty} \psi^{\ast}(x-a,t)\psi(x-a,t)dx$, $T(a)$ is unitary, satisfying $T^{\dagger}(a)T(a)=I$. Thus the generator of space translation, $\hat{p}$, which is defined by $T(a)=e^{-ia\hat{p}}$, is Hermitian. 

In addition, it is worth noting that the reduced Planck constant $\hbar$ with dimension of action is missing in the above free Schr\"{o}dinger equation\index{Schr\"{o}dinger equation!free}. However, this is in fact not a problem. The reason is that the dimension of $\hbar$ can be absorbed in the dimension of $m$. For example, we can stipulate the dimensional relations as $p=1/L$, $E=1/T$ and $m=T/L^2$, where $L$ and $T$ represent the dimensions of space and time, respectively (see Duff, Okun and Veneziano, 2002 for a more detailed analysis). Moreover, the value of $\hbar$ can be set to the unit of number $1$ in principle. Thus the above equation is essentially the free Schr\"{o}dinger equation\index{Schr\"{o}dinger equation!free} in quantum mechanics.

When assuming the time evolution due to interaction is still linear, we can further obtain the Schr\"{o}dinger equation\index{Schr\"{o}dinger equation} under an external potential $V(x,t)$:

\begin{equation}
i{\partial \psi(x,t) \over \partial t}=-{1 \over 2m}{{\partial^2 \psi(x,t)}\over \partial x^2}+V(x,t)\psi(x,t).
\label{Sch}
\end{equation}

\noindent The concrete form of the potential in a given situation is determined by a theory of interactions, such as the nonrelativistic approximation of interacting quantum field theory\index{quantum field theory}. 

The Schr\"{o}dinger equation\index{Schr\"{o}dinger equation} for one-body systems can also be extended to many-body systems. For simplicity, consider a free two-body system containing two subsystems $m_1$ and $m_2$. When ignoring the interaction between the two subsystems, the Schr\"{o}dinger equation\index{Schr\"{o}dinger equation} that governs the evolution of this system will be

\begin{equation}
i{\partial [\psi_1(x_1,t)\psi_2(x_2,t)] \over \partial t}=-[{1 \over 2m_1}{{\partial^2 }\over \partial x_1^2}+{1 \over 2m_2}{{\partial^2 }\over \partial x_2^2}][\psi_1(x_1,t)\psi_2(x_2,t)]. 
\label{FSch2}
\end{equation}

\noindent wherer $x_1$ and $x_2$ are the coordinates of the two subsystems, respectively, and $\psi_1(x_1,t)$ and $\psi_2(x_2,t)$ are their wave functions, respectively. When considering the interaction between the two subsystems and assuming the time evolution due to interaction is still linear, the interaction will form an entangled state of the whole system which is defined in a six-dimensional configuration space\index{configuration space}, and the free Schr\"{o}dinger equation\index{Schr\"{o}dinger equation!free} that governs the evolution of this  interacting two-body system will be

\begin{equation}
i{\partial \psi(x_1,x_2,t) \over \partial t}=-[{1 \over 2m_1}{{\partial^2 }\over \partial x_1^2}+{1 \over 2m_2}{{\partial^2 }\over \partial x_2^2}]\psi(x_1,x_2,t)+V_{12}(x_1,x_2,t)\psi(x_1,x_2,t),
\label{FSch2i}
\end{equation}

\noindent where the entangled wave function $\psi(x_1,x_2,t)$ describes the whole two-body system, and the potential energy term $V_{12}(x_1,x_2,t)$ describes the interaction between its two subsystems. 

\section{Further discussion}

I have derived the free Schr\"{o}dinger equation\index{Schr\"{o}dinger equation!free} based on an analysis of spacetime translation invariance\index{laws of motion!spacetime translation invariance of} and relativistic invariance\index{laws of motion!relativistic invariance of}. The new analysis may not only make the Schr\"{o}dinger equation\index{Schr\"{o}dinger equation} in quantum mechanics more logical and understandable, but also help understand the origin of the complex and multi-dimensional wave function.

As noted before, the free Schr\"{o}dinger equation\index{Schr\"{o}dinger equation!free} is usually derived in quantum mechanics textbooks by analogy and correspondence with classical physics.
There are at least two mysteries in this heuristic derivation. First of all, even if the behavior of microscopic particles likes wave and thus a wave function is needed to describe them, it is unclear why the wave function must assume a complex form. Indeed, when Schr\"{o}dinger\index{Schr\"{o}dinger, Erwin} invented his equation, he was also puzzled by the inevitable appearance of the imaginary unit ``$i$"  in the equation. Next, one doesn't know why there are the de Broglie\index{de Broglie, Louis} relations\index{de Broglie relations} for momentum and energy and why the nonrelativistic energy-momentum relation\index{energy-momentum relation!nonrelativistic} is $E=p^2/2m$.

According to the analysis given in the previous sections, the key to unveiling these mysteries is to analyze spacetime translation invariance\index{laws of motion!spacetime translation invariance of} of laws of motion\index{laws of motion}\index{Schr\"{o}dinger equation!free!derivation of}. Spacetime translation gives the definitions of momentum and energy in quantum mechanics. The momentum operator $\hat{p}$ is defined as the generator of space translation, and it is Hermitian and its eigenvalues are real. Moreover, the form of the momentum operator is uniquely determined by its definition, which turns out to be $\hat{p}=-i{ \partial / \partial x}$, and its eigenfunctions are $e^{ipx}$ , where $p$ is the corresponding real eigenvalue. Similarly, the energy operator $\hat{E}$ is defined as the generator of time translation, and its universal form is $\hat{E}=i{\partial /\partial t}$. But the concrete manifestation of this operator for a physical system, denoted by $H$ and called the Hamiltonian of the system, is determined by the concrete situation. 

Fortunately, for an isolated system, the form of $H$, which determines the evolution equation of state, can be fixed for linear evolution by the requirements of spacetime translation invariance\index{laws of motion!spacetime translation invariance of} and relativistic invariance\index{laws of motion!relativistic invariance of}. Concretely speaking, time translational invariance requires that $dH/dt=0$, and this implies that the solutions of the evolution equation $i{\partial \psi(x,t) / \partial t} =H\psi(x,t)$ are $\varphi_E(x)e^{-iEt}$ and their superpositions, where $\varphi_E(x)$ is the eigenfunction of $H$. Moreover, space translational invariance requires $[\hat{p},H]=0$. This means that $\hat{p}$ and $H$ have common eigenfunctions, and thus $\varphi_E(x)=e^{ipx}$. Therefore, $e^{i(px-Et)}$ and their superpositions are  solutions of the evolution equation for an isolated system, where $e^{i(px-Et)}$ represents the state of the system with  momentum $p$ and energy $E$. In other words, the state of an isolated system (e.g. a free electron) with definite momentum and energy assumes the plane wave form $e^{i(px-Et)}$. Furthermore, the relation between $p$ and $E$ or the energy-momentum relation\index{energy-momentum relation} can be determined by considering  the relativistic transformation of the generators of space translation and time translation, and in the nonrelativistic domain it is $E=p^2/2m$. Then we can obtain the Hamiltonian of an isolated system, $H=\hat{p}^2/2m$, and the free Schr\"{o}dinger equation\index{Schr\"{o}dinger equation!free}, Eq.(\ref{FSch}). 

Finally, I emphasize again that the linearity of time evolution is an important presupposition in the above derivation\index{Schr\"{o}dinger equation!free!derivation of} of the free Schr\"{o}dinger equation\index{Schr\"{o}dinger equation!free}. It is only for linear evolution that spacetime translation invariance\index{laws of motion!spacetime translation invariance of} of laws of motion\index{laws of motion} can help determine the precise form of the equation of motion for isolated systems. It is possible that the free evolution equation also contains nonlinear evolution terms. However, although nonlinear time evolution can also satisfy spacetime translation invariance\index{laws of motion!spacetime translation invariance of}, the invariance requirement cannot help determine the precise form of the nonlinear evolution equation. Nonlinear time evolution, if it exists, must have an additional physical origin. I will discuss this issue in Chapter 8.


\chapter{The ontology of quantum mechanics (I)}






I have argued in the previous chapters that the wave function in quantum mechanics represents the physical state of a single system.
The next question is: What physical state does the wave function represent? We must answer this question in order to know the ontology of\index{quantum mechanics!ontology of} quantum mechanics.

Unfortunately, like the nature of the wave function, the ontological meaning of the wave function has also been a hot topic of debate since the early days of quantum mechanics. Today it is still unclear what ontic state the wave function represents in the realistic alternatives to quantum mechanics, such as Bohm's theory\index{Bohm's theory}, Everett's theory\index{Everett's theory}, and collapse theories\index{collapse theories}\index{quantum mechanics!alternatives to, \emph{see} Bohm's theory, Everett's theory, and collapse theories}. 
It can be expected that the $\psi$-ontology theorems\index{psi-ontology theorems}, which says that the wave function is ontic, may have further implications for the ontological meaning of the wave function.
The reason is that these theorems say that the ontic state of a physical system, which is represented by  the wave function, has certain efficacy during a measurement, and a further analysis of the efficacy of the ontic state may help find what the ontic state really is.

According to the existing $\psi$-ontology theorems\index{psi-ontology theorems} such as the Pusey-Barrett-Rudolph theorem\index{Pusey-Barrett-Rudolph theorem}, which are based on an analysis of projective measurements\index{projective measurements}, the efficacy of the ontic state of a physical system is to determine the probabilities for different results of a projective measurement on the system.
It seems that such efficacy says little about what the ontic state of a physical system is (see also Dorato\index{Dorato, Mauro} and Laudisa\index{Laudisa, Federico}, 2014). Moreover, whether  the efficacy exists or not also depends on the solutions to the measurement problem\index{measurement problem}, e.g. it does not exist in deterministic theories such as Bohm's theory\index{Bohm's theory}. 
In contrast, my arguments for $\psi$-ontology in terms of protective measurements\index{protective measurements} says something different.
According to these arguments, the efficacy of the ontic state of a physical system is to determine the definite result of a protective measurement on the system, not probabilities. 
This direct, definite link is obviously stronger than the above indirect, probabilistic link.
Moreover, the efficacy exists in any realist theory consistent with the predictions of quantum mechanics, independently of the solutions to the measurement problem\index{measurement problem}.
Therefore, the efficacy revealed by protective measurements\index{protective measurements} may tell us something about the underlying ontology.

For a quantum system whose wave function is $\psi(x)$ at a given instant, we can measure the density $|\psi(x)|^2$ in each position $x$ in space by a protective measurement (see Section 1.3). In other words, the density $|\psi(x)|^2$ as part of the ontic state has efficacy to shift the pointer of the measuring device and yield the  result of the protective measurement. Then, what density is the density $|\psi(x)|^2$?
Since a measurement must always be realized by certain physical interaction between the measured system and the measuring device, the density must be, in the first place, the density of certain interacting charge. For example, if the measurement is realized by an electrostatic interaction between the measured system (with charge $Q$) and the measuring device, then the density multiplied by the charge of the system, namely $|\psi(x)|^2Q$, will be charge density. It is such concrete properties that have the actual efficiencies during a measurement.

In this chapter, I will analyze the existence and origin of the charge distribution\index{charge distribution!origin of}\index{charge distribution!existence of} of a quantum system. As we will see, the analysis will help unveil the deeper ontological meaning of the wave function.

\section{Schr\"{o}dinger\index{Schr\"{o}dinger, Erwin!charge density hypothesis of}'s charge density hypothesis}

In quantum mechanics, an electron has an electric charge represented by $-e$ in the potential term of the Schr\"{o}dinger equation\index{Schr\"{o}dinger equation}, $-e\varphi \psi(x,t)$, where $\psi(x,t)$ is the wave function of the electron, and $\varphi$ is an external electric scalar potential. An intriguing question is: how is the charge of the electron distributed in space?
We can measure the total charge of an electron by electromagnetic interaction and find it in a certain region of space. Thus it seems that the charge of an electron must exist in space with a certain distribution. When Schr\"{o}dinger\index{Schr\"{o}dinger, Erwin} introduced the wave function and founded his wave mechanics in 1926, he also suggested an answer to this question. Schr\"{o}dinger\index{Schr\"{o}dinger, Erwin} assumed that the charge of an electron is distributed in the whole space, and the charge density in position $x$ at instant $t$ is $-e|\psi(x,t)|^2$, where $\psi(x,t)$ is the wave function of the electron. In the following, I will give a detailed historical and logical analysis of Schr\"{o}dinger\index{Schr\"{o}dinger, Erwin!charge density hypothesis of}'s charge density hypothesis.

In his paper on the equivalence between wave mechanics and matrix mechanics (Schr\"{o}dinger\index{Schr\"{o}dinger, Erwin}, 1926b), Schr\"{o}dinger\index{Schr\"{o}dinger, Erwin} suggested that it might be possible to give an extraordinarily visualizable and intelligible interpretation of the intensity and polarization of radiation by assuming the wave function, which was then called mechanical field scalar, is the source of the radiation. In particular, he assumed that the charge density of an electron as the source of radiation is given by the real part of $-e\psi \partial\psi^*/\partial t$, where $\psi$ is the wave function of the electron. In his third paper on wave mechanics (Schr\"{o}dinger\index{Schr\"{o}dinger, Erwin}, 1926c), which deals with perturbation theory and its application to the Stark effect, Schr\"{o}dinger\index{Schr\"{o}dinger, Erwin} noted in an addendum in proof that the correct charge density of an electron was given by $-e|\psi|^2$. Then in his fourth paper on wave mechanics and his 1927 Solvay report (Schr\"{o}dinger\index{Schr\"{o}dinger, Erwin}, 1926d, 1928), Schr\"{o}dinger\index{Schr\"{o}dinger, Erwin} further showed how this gives rise to a sensible notion of charge density for several electrons, each contribution being obtained by integrating over the other electrons. Concretely speaking, for a many-body system, select one subsystem and keep the coordinates of the subsystem that describe its position fixed at a given position and integrate $|\psi|^2$ over all the rest of the coordinates of the system and multiply the charge of the subsystem, and do a similar thing for each subsystem, in each case fixing the selected subsystem at the same given position. Then the sum of all these partial results gives the charge density at the given position. 

At the 1927 Solvay conference, Born\index{Born, Max} posed an objection relating to quadrupole moments to Schr\"{o}dinger\index{Schr\"{o}dinger, Erwin!charge density hypothesis of}\index{Schr\"{o}dinger, Erwin!charge density hypothesis of!Born's objection to}'s charge density hypothesis (Bacciagaluppi\index{Bacciagaluppi, Guido} and Valentini\index{Valentini, Antony}, 2009, p.426). Born considered two microscopic particles with charge $e$ whose wave function is $\psi(x_1,x_2)$, where $x_1$ and $x_2$ stand for all the coordinates of the two particles. According to Schr\"{o}dinger\index{Schr\"{o}dinger, Erwin}, the charge density is

\begin{equation}
    \rho(x)=e\int|\psi(x,x_2)|^2dx_2+e\int|\psi(x_1,x)|^2dx_1\ .
\end{equation}

\noindent But the electric quadrupole moment
  \[
    e\int\!\int x_1x_2|\psi(x_1,x_2)|^2dx_1dx_2
  \]
cannot be expressed using the function $\rho(x)$. As a result, one cannot reduce the radiation of the quadrupole to the motion of a charge distribution\index{charge distribution} $\rho(x)$ in the usual three-dimensional space. Born then concluded that interpreting the quantity $|\psi|^2$ as charge density leads to difficulties in the case of quadrupole moments.

However, it can be seen by a more careful analysis that the above problem is not really a problem of Schr\"{o}dinger\index{Schr\"{o}dinger, Erwin!charge density hypothesis of}'s charge density hypothesis, but a problem of Schr\"{o}dinger\index{Schr\"{o}dinger, Erwin}'s interpretation of the wave function in terms of charge density. In fact, Schr\"{o}dinger\index{Schr\"{o}dinger, Erwin} also clearly realized this problem. As early as in his equivalence paper (Schr\"{o}dinger\index{Schr\"{o}dinger, Erwin}, 1926b), Schr\"{o}dinger\index{Schr\"{o}dinger, Erwin} already noticed the difficulty relating to the problem of several electrons, which lies in the fact that the wave function is a function in configuration space\index{configuration space}, not in real space. Although the charge distribution\index{charge distribution} in three-dimensional space can be consistently defined for an $N$-body system, it does not reflect all information encoded in the wave function of the system which lives in a $3N$-dimensional configuration space\index{configuration space}. Therefore, although the existence of charge distribution\index{charge distribution!existence of} may provide an approximate classical explanation for some phenomena of radiation, it cannot account for all experimental observations, e.g. as Born rightly pointed out, the motion of a charge distribution\index{charge distribution} cannot explain the radiation of the quadrupole.\index{Schr\"{o}dinger, Erwin!charge density hypothesis of!a new analysis of}

Besides this incompleteness problem for many-body systems, Schr\"{o}dinger\index{Schr\"{o}dinger, Erwin} (1928) also realized that the charge distribution\index{charge distribution} of a quantum system such as an electron cannot be purely classical either, because his equation does not include the usual Coulomb interaction between the distributions. In particular, there is no electrostatic self-interaction of the charge distribution\index{charge distribution} of a quantum system. Moreover, according to the Schr\"{o}dinger equation\index{Schr\"{o}dinger equation}, the interacting systems should be treated as a whole, whose wave function is defined in a multi-dimensional configuration space\index{configuration space}, and cannot be decomposed into a direct product of the wave functions of all interacting systems. This makes the interaction between two charged quantum systems more complex than the interaction between two classical charges.

Schr\"{o}dinger\index{Schr\"{o}dinger, Erwin}'s interpretation of the wave function in terms of charge density was latter investigated and extended by a few authors (see, e.g. Madelung\index{Madelung, Erwin}, 1926, 1927; J\'{a}nossy\index{J\'{a}nossy, Lajos}, 1962; Jaynes\index{Jaynes, Edwin T.}, 1973; Barut\index{Barut, Asim O.}, 1988).\footnote{It is worth noting that Wallace\index{Wallace, David} and Timpson\index{Timpson, Christopher G.}'s (2010) ``spacetime state realism" can be regarded as a generation of Schr\"{o}dinger\index{Schr\"{o}dinger, Erwin}'s interpretation in some sense (see also Wallace, 2012, chap. 8). Although this view may avoid the problems of wave function realism\index{wave function!realism} (Albert\index{Albert, David Z.}, 1996, 2013), it has the same problems as Schr\"{o}dinger\index{Schr\"{o}dinger, Erwin}'s interpretation. For a critical analysis of this view see Norsen\index{Norsen, Travis} (2016).}  
Due to the above problems, however, this semiclassical interpretation cannot be satisfactory in the final analysis. Moreover, although this fact does not imply the non-existence of the charge distribution\index{charge distribution} of an electron, the very limited success of the interpretation does not provide a convincing argument for its existence either. Presumably because people thought that the hypothetical charge distribution\index{charge distribution} of an electron cannot be directly measured and its existence also lacks a consistent physical explanation, Schr\"{o}dinger\index{Schr\"{o}dinger, Erwin!charge density hypothesis of}'s charge density hypothesis has been largely ignored.\index{Schr\"{o}dinger, Erwin!charge density hypothesis of!a new analysis of}\footnote{However, there is a modern variant of Schr\"{o}dinger\index{Schr\"{o}dinger, Erwin!charge density hypothesis of}'s charge density hypothesis, which has been called mass density ontology\index{GRW theory!mass density ontology of} (Ghirardi\index{Ghirardi, GianCarlo}, Grassi\index{Grassi, Renata} and Benatti, 1995; Ghirardi\index{Ghirardi, GianCarlo}, 1997, 2016). I will briefly discuss it later in Section 8.5.}

\section{Is an electron a charge cloud\index{charge cloud}?}

In order to answer the question of whether a quantum system such as an electron has a well-defined charge distribution\index{charge distribution} as Schr\"{o}dinger\index{Schr\"{o}dinger, Erwin} assumed, we need to first determine what exists in quantum mechanics.
According to the extended ontological models framework\index{ontological models framework!extended}\index{ontological models framework} or the suggested criterion of reality\index{criterion of reality!a new} (see Chapter 4),  
the definite result obtained by a protective measurement reflects a property of the measured quantum system.  
While what property the measured property is depends on the concrete interaction between the measured system and the measuring device during the protective measurement.  
In this section, I will analyze the existence of\index{charge distribution!existence of} the charge distribution\index{charge distribution} of a quantum system with the help of protective measurements.\index{Schr\"{o}dinger, Erwin!charge density hypothesis of!a new analysis of}

\subsection{Two simple examples}

Before my analysis of the charge distribution\index{charge distribution} of a quantum system, I will give two simple examples to explain how to determine whether a physical system has a well-defined charge distribution\index{charge distribution}.


\begin{center} 
\begin{figure}[h]\label{cmc}

\includegraphics[scale=0.39]{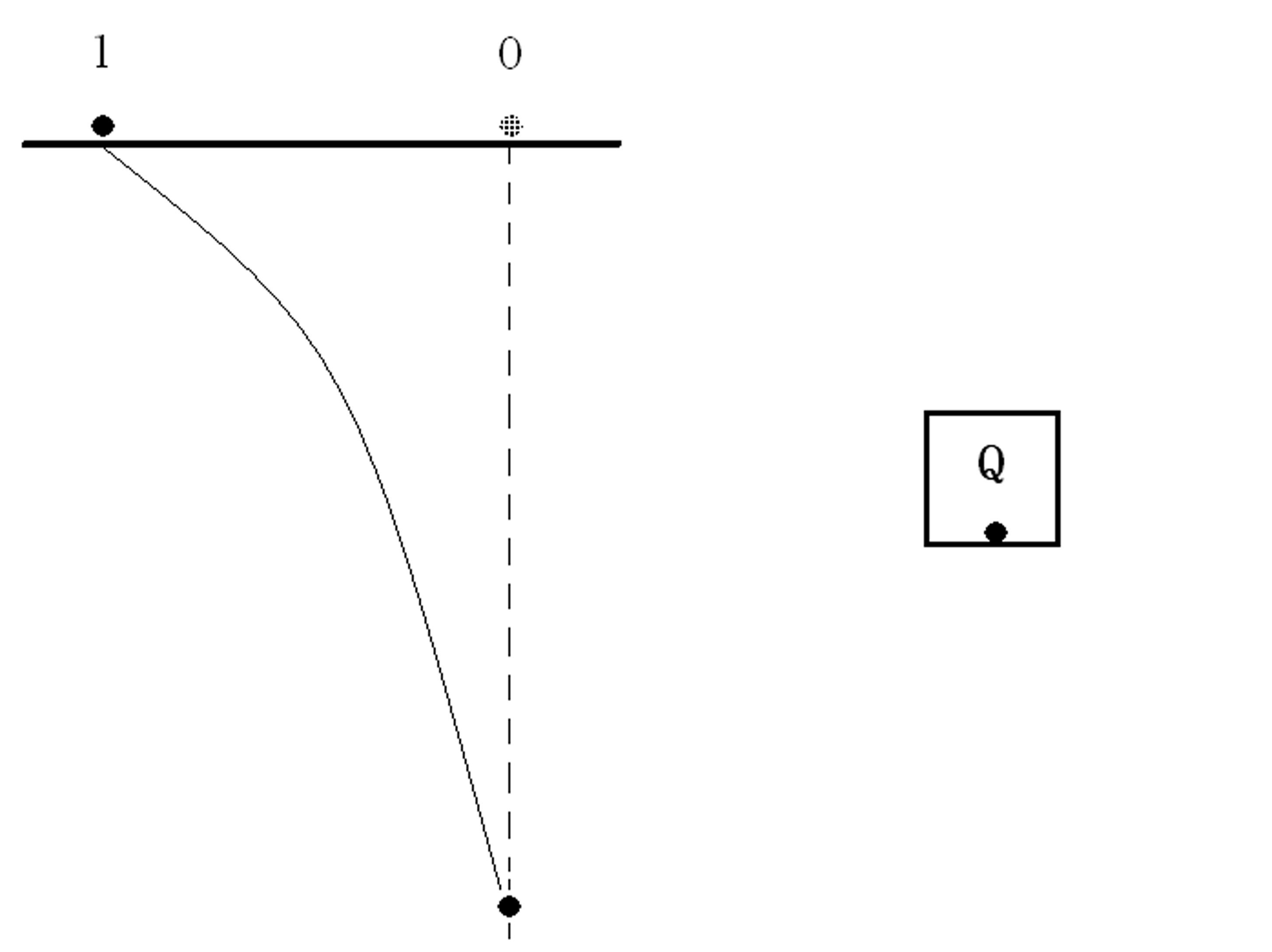}


\caption{Scheme of a non-disturbing measurement of the charge distribution of a classical system}

\end{figure}

\end{center} 

First of all, I will analyze the charge distribution\index{charge distribution} of a classical system. 
Consider a classical particle with charge $Q$, which is trapped in a small box. A measuring electron is shot along a straight line near the box, and then detected on a screen after passing by the box.  
According to Newton's laws of motion\index{laws of motion} and Coulomb's law, the deviation of the trajectory of the measuring electron is determined by the charge of the measured particle, as well as by the distance between the electron and the particle. 
If there were no charged particle in the box, the trajectory of the electron would be a straight line as denoted by position ``0" in Figure 6.1.  
Now the trajectory of the electron will be deviated by a definite amount as denoted by position ``1" in Figure 6.1. 
Then according to the second assumption of the ontological models framework\index{ontological models framework} or the suggested criterion of reality\index{criterion of reality!a new}, a simple analysis of the definite measurement result will tell us that the measured particle has a charge $Q$ in the box as its property, which has the efficacy to deviate the  measuring electron. Certainly, such ontological content of classical mechanics is already well-known.  However, this is because the theory was founded based on the classical ontology. Here I show that the ontology of the theory can be derived from its connections with experience and a connecting rule between experience and reality such as the suggested criterion of reality\index{criterion of reality!a new}. This approach to the ontology of a physical theory is universal, and it can be applied to all physical theories.

Here it may be necessary to further clarify the meaning of charge distribution\index{charge distribution} as a property of a physical system. As noted before, any physical measurement is necessarily based on certain interaction between the measured system and the measuring system.  Concretely speaking, the measuring system is influenced by the measured system through an interaction that depends on the measured property, and the definite change of the measuring system then reflects this property of the measured system (in accordance with the ontological models framework or the suggested criterion of reality\index{criterion of reality!a new}). 
For example, a position measurement must depend on the existence of certain position-dependent interaction between the measured system and the measuring system such as electrostatic interaction between two electric charges. The existence of an electrostatic interaction during a measurement, which is indicated by the deviation of the trajectory of the charged measuring system such as an electron, then tells us that the measured system also has a charge responsible for the interaction. Moreover, since the strength of the interaction relates to the distance between the two interacting systems, the measurement result may also reflect the charge distribution\index{charge distribution} of the measured system in space. In the above example, the definite deviation of the trajectory of the measuring electron will tell us that there exists a definite amount of charge in the box, and the extent of the deviation will further tell us the amount of the charge there.

\begin{center}
\begin{figure}[h]\label{pmc}

\includegraphics[scale=0.39]{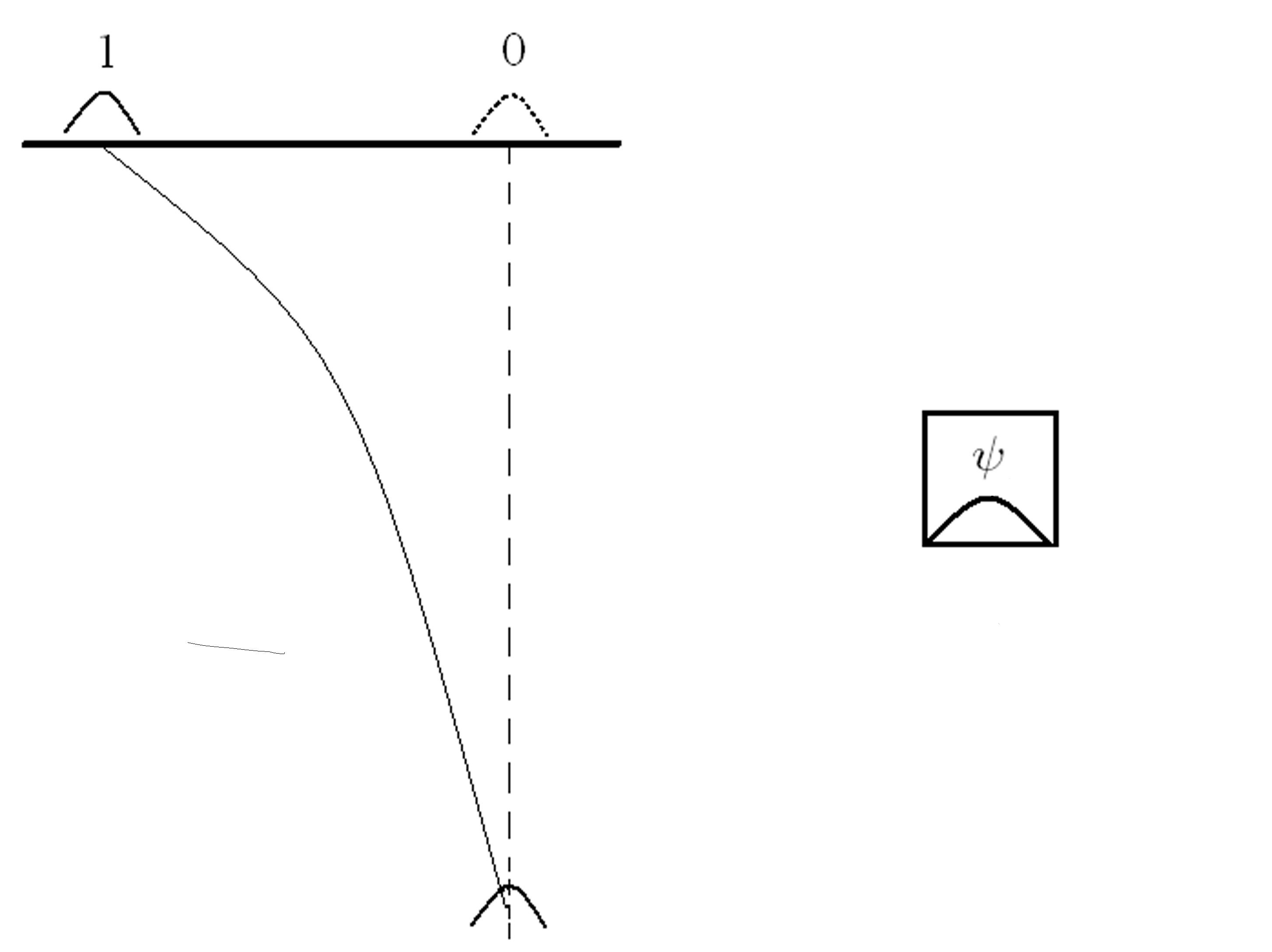}


\caption{Scheme of a projective measurement of the charge distribution of a quantum system}

\end{figure} 
\end{center} 

\vspace{2mm}

Secondly, I will analyze the charge distribution\index{charge distribution} of a quantum system being in a  position eigenstate.  
Consider a quantum system with charge $Q$ whose wave function is localized in a small box. A measuring electron, whose initial state is a Gaussian wavepacket\index{Gaussian wavepacket} narrow in both position and momentum, is shot along a straight line near the box. The electron is detected on a screen after passing by the box. 
According to the Schr\"{o}dinger equation\index{Schr\"{o}dinger equation} with an external Coulomb potential, the deviation of the trajectory of the electron wavepacket is determined by the charge of the measured particle in the box, as well as by the distance between the electron and the particle. 
If there were no charged quantum system in the box, then the trajectory of the electron wavepacket will be a straight line as denoted by position ``0" in Figure 6.2. Now, the trajectory of the electron wavepacket will be deviated by a definite amount as denoted by position ``1" in Figure 6.2. 
In an ideal situation where the size of the box can be ignored, this can be regarded as a conventional projective measurement of an eigenstate of the system's charge in the box. 
Then according to the ontological models framework\index{ontological models framework} or the suggested criterion of reality\index{criterion of reality!a new}, an analysis of the definite measurement result will tell us that the measured system has a charge $Q$ in the box as its property.

In general, when a quantum system is in an eigenstate of an observable, a projective measurement of the observable will obtain a definite result, namely the eigenvalue of the observable corresponding to the eigenstate. Then, similarly, the system has a property with value being the eigenvalue. 
This result is also called the eigenvalue-eigenstate half link\index{quantum mechanics!standard formulation of!eigenvalue-eigenstate half link in}, which says that if a system is in an eigenstate of an observable, the system has a property with value being the eigenvalue corresponding to the eigenstate (see Section 2.2). This link provides a very limited connection between quantum mechanics and reality.


\subsection{The answer of protective measurement}

I have analyzed the charge distribution\index{charge distribution}s of a classical system and a quantum system being in a position eigenstate. 
It is demonstrated that a classical charged particle has a well-defined charge distribution\index{charge distribution}; the charge is localized in the definite position in space where the particle is. Similarly, a charged quantum system being in a position eigenstate also has a well-defined charge distribution\index{charge distribution}; the charge is localized in a definite position in space, which is the position eigenvalue corresponding to the position eigenstate. 
Then, is there also a well-defined charge distribution\index{charge distribution} for a quantum system being in a superposition of position eigenstates? And if the answer is yes, what is the charge distribution\index{charge distribution} of the system? 

According to quantum mechanics, a projective position measurement of a superposition of position eigenstates will change the state greatly by entanglement and possible wavefunction collapse, and the measurement result is not definite but random with certain probability in accordance with the Born rule\index{Born rule}. Thus  neither the ontological models framework\index{ontological models framework} nor  the suggested criterion of reality\index{criterion of reality!a new} can be used to analyze the charge distribution\index{charge distribution}s of such superpositions when considering only projective measurements\index{projective measurements}. 
However, as I have argued before (in Chapter 4), both the ontological models framework\index{ontological models framework} and the suggested criterion of reality\index{criterion of reality!a new} can be applied to a general quantum state when considering protective measurements\index{protective measurements}; the definite result obtained by a protective measurement reflects a property of the measured system. 
In the following, I will demonstrate that protective measurements\index{protective measurements!and charge density hypothesis} can tell us that a quantum system has a well-defined charge distribution\index{charge distribution} in the same sense that classical measurements can tell us that a classical system has a charge distribution\index{charge distribution} and projective measurements\index{projective measurements} can tell us that a quantum system being in a position eigenstate has a charge distribution\index{charge distribution}.

Consider a quantum system with  charge $Q$ whose wave function is

\begin{equation}
\psi(x,t)=a\psi_1(x,t)+b\psi_2(x,t),
\end{equation}

\noindent where $\psi_1(x,t)$ and $\psi_2(x,t)$ are two normalized wave functions respectively localized in their ground states in two small identical boxes 1 and 2, and $|a|^2+|b|^2=1$. A measuring electron, whose initial state is a Gaussian wavepacket\index{Gaussian wavepacket} narrow in both position and momentum, is shot along a straight line near box 1 and perpendicular to the line of separation between the two boxes. The electron is detected on a screen after passing by box 1. Suppose the separation between the two boxes is large enough so that a  charge $Q$ in box 2 has no observable influence on the electron. Then if the system is in box 2, namely $|a|^2=0$, the trajectory of the electron wavepacket will be a straight line as denoted by position ``0" in Figure 6.3, indicating that there is no charge in box 1. If  the system is in box 1, namely $|a|^2=1$, the trajectory of the electron wavepacket will be deviated by a maximum amount as denoted by position ``1" in Figure 6.3, indicating that there is a charge $Q$ in box 1. As noted above, these two measurements are conventional projective measurements\index{projective measurements} of two eigenstates of the system's charge in box 1, and their results can tell us that the measured system has a well-defined charge distribution\index{charge distribution} in box 1 as its property. However, when $0< |a|^2 <1$, i.e. when the measured system is in a superposition of two eigenstates of its charge in box 1, such projective measurements\index{projective measurements} cannot obtain definite results and thus cannot tell us whether there is a well-defined charge distribution\index{charge distribution} in box 1.\index{protective measurements!and charge density hypothesis}

\begin{center} 
\begin{figure}[h]\label{ptmc}

\includegraphics[scale=0.39]{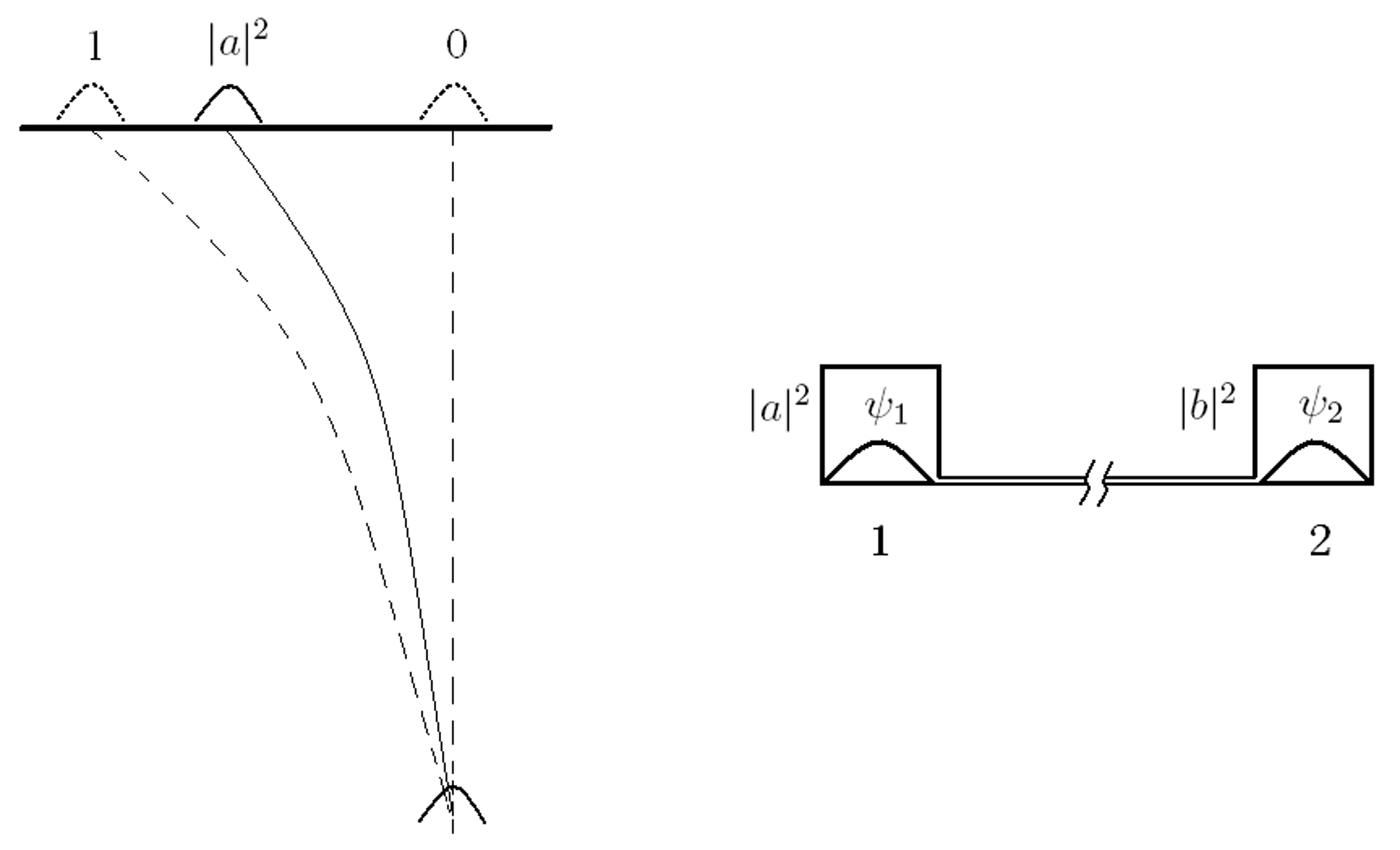}

\caption{Scheme of a protective measurement of the charge distribution of a quantum system}

\end{figure} 
\end{center} 

Now let us make a protective measurement\index{protective measurements!and charge density hypothesis} of the charge of the system in box 1 for the superposition $\psi(x,t)$.\footnote{An adiabatic\index{protective measurements!adiabatic-type}-type protective measurement can be realized as follows (Aharonov\index{Aharonov, Yakir}, Anandan\index{Anandan, Jeeva S.} and Vaidman\index{Vaidman, Lev}, 1993). Since the state $\psi(x,t)$ is degenerate with its orthogonal state $\psi^{\perp}(x,t)=b^*\psi_1(x,t)-a^*\psi_2(x,t)$, we first need an artificial protection procedure to remove the degeneracy, e.g. joining the two boxes with a long tube whose diameter is small compared to the size of the box. By this protection $\psi(x,t)$ will be a nondegenerate energy eigenstate. Then we need to realize the adiabatic\index{protective measurements!adiabatic-type} condition and the weakly interacting condition, which are required for a protective measurement. These conditions can be satisfied when assuming that (1) the measuring time of the electron is long compared to $\hbar/\Delta E$, where $\Delta E$ is the smallest of the energy differences between $\psi(x,t)$ and the other energy eigenstates, and (2) at all times the potential energy of interaction between the electron and the system is small compared to $\Delta E$. Then the measurement by means of the electron trajectory is a realistic protective measurement, and when the conditions approach ideal conditions, the measurement will be an (ideal) protective measurement with certainty. Note that weak measurements have been implemented in experiments (see, e.g. Lundeen\index{Lundeen, Jeff S.} et al, 2011), and it can be reasonably expected that adiabatic\index{protective measurements!adiabatic-type}-type protective measurements\index{protective measurements} can also be implemented in the near future with the rapid development of quantum technologies (Cohen, 2016).} 
Then the trajectory of the electron wavepacket is only influenced by the expectation value of the charge of the system in box 1, and thus the electron wavepacket will reach the definite position ``$|a|^2$" between ``0" and ``1"  on the screen as denoted in Figure 6.3. 
According to the extended ontological models framework\index{ontological models framework!extended} or the suggested criterion of reality\index{criterion of reality!a new}, this definite result of the protective measurement indicates that the measured system has a charge $|a|^2Q$ in box 1 as its property. \index{protective measurements!and charge density hypothesis}

This result can be generalized to an arbitrary superposition of position eigenstates. For a quantum system with charge $Q$ whose wave function is $\psi(x)$ at a given instant, we can make a protective measurement of the charge of the system in a small spatial region $V$ having volume $v$ near position $x$. This means to protectively measure the following observable:\index{protective measurements!and charge density hypothesis}

\begin{equation}
A= 
\begin{cases} 
{Q},& \text{if $x \in V$,}
\\
0,&\text{if $x \not\in V$.} 
\end{cases}
\end{equation}

\noindent The result of the measurement is

\begin{equation}
\exptt{A} = {Q}\int_{V}|\psi(x,t)|^2 dv.
\end{equation}
\noindent

\noindent  It indicates that the measured system has a charge ${Q}\int_{V}|\psi(x,t)|^2 dv$ in region $V$. Then when $v \rightarrow 0$ and after performing measurements in sufficiently many regions $V$, we can find that the measured system has a charge distribution\index{charge distribution} in the whole space, and the charge density  in each position $x$ is $|\psi(x)|^2Q$.\footnote{Similarly, we can protectively measure another observable $B ={\hbar \over{2mi}}(A\nabla + \nabla A)$. The measurements will tell us the measured system also has an electric flux distribution in space, and the electric flux density in position $x$ is $j_Q(x) ={\hbar Q\over{2mi}}(\psi^* \nabla \psi - \psi  \nabla \psi^* )$. These results can also be generalized to a many-body system.}\index{protective measurements!and charge density hypothesis}

To sum up, I have argued with the help of  protective measurements\index{protective measurements} that a quantum system has a well-defined charge distribution\index{charge distribution} in space, in exactly the same sense that a classical system has a well-defined charge distribution\index{charge distribution} in space. Moreover, it is shown that the charge of a charged quantum system is distributed throughout space, and the charge density in each position is equal to the modulus squared of the wave function of the system there multiplied by the charge of the system. 
Thus, visually speaking, a charged quantum system such as an electron is a charge cloud\index{charge cloud}.
This confirms Schr\"{o}dinger\index{Schr\"{o}dinger, Erwin}'s original charge density hypothesis.\index{protective measurements!and charge density hypothesis}

\section{The origin of charge density}

In this section, I will further investigate the physical origin of the charge distribution\index{charge distribution!origin of} of a quantum system such as an electron. As we will see, the answer may provide an important clue to the ontological meaning of the wave function.

As I have pointed out previously, there are at least two good motivations for investigating of the origin of the charge distribution\index{charge distribution} of a quantum system. First, although the charge distribution\index{charge distribution} can be consistently defined for a many-body system, the distribution contains no information about the entanglement\index{quantum entanglement} between the subsystems of the many-body system. This indicates that the charge distribution\index{charge distribution} is an incomplete manifestation of the underlying physical state and thus must have a deeper physical origin.\footnote{This conclusion is also supported by the seemingly puzzling fact that although each charged quantum system has a charge distribution\index{charge distribution} in space, the electrostatic interaction between two charged quantum systems is described not by certain charge density terms but by the potential terms in the Schr\"{o}dinger equation\index{Schr\"{o}dinger equation}. Since charge density (and electric flux density) are not a complete manifestation of the physical state of a two-body system, e.g. they do not contain the entanglement\index{quantum entanglement} between the sub-systems of the two-body system, they are not enough to describe the interaction between these two sub-systems when there is entanglement\index{quantum entanglement} between them in general cases. However, when there is no entanglement\index{quantum entanglement} between two quantum systems in special cases such as during a protective measurement, the charge density (and electric flux density) are enough to describe the interaction and can also be directly manifested, e.g. in the results of protective measurements\index{protective measurements}.}
Second, even for one-body systems the charge distribution\index{charge distribution} also has some puzzling features, e.g. the charge distribution\index{charge distribution} of a single electron has no electrostatic self-interaction. These puzzling features are in want of a reasonable explanation, which may be provided by the origin of the charge distribution\index{charge distribution}. In addition, the charge distribution\index{charge distribution} has two possible forms, and we need to determine which form is the actual form. Again, this is closely related to the physical origin of the distribution.\index{charge distribution!origin of}

Then, what kind of entity or process generates the charge distribution\index{charge distribution} of a quantum system in space, $|\psi(x,t)|^2Q$?
There are two possibilities. The  charge distribution\index{charge distribution} can be generated by either (1) a continuous charge distribution\index{charge distribution} with density $|\psi(x,t)|^2Q$ or (2) the motion of a discrete point charge $Q$ with spending time $|\psi(x,t)|^2dvdt$ in the infinitesimal spatial volume $dv$ around $x$ in the infinitesimal time interval $[t,t+dt]$.\footnote{Note that the expectation value of an observable at a given instant, such as $\exptt{A} = {Q}\int_{V}|\psi(x,t)|^2 dx$, is either the physical property of a quantum system at the precise instant (like the position of a classical particle) or the limit of the time-averaged property of the system at the instant (like the standard velocity of a classical particle). These two interpretations correspond to the above two possibilities. For the later, the observable assumes an eigenvalue at each instant, and its value spreads all eigenvalues during an infinitesimal time interval around the given instant. Moreover, the spending time in each eigenvalue is proportional to the modulus squared of the wave function of the system there. In this way, such ergodic motion may generate the expectation value of the observable (see also Aharonov\index{Aharonov, Yakir} and Cohen\index{Cohen, Eliahu}, 2014). I will discuss in the next chapter whether this picture of ergodic motion applies to properties other than position.} Correspondingly, the underlying physical entity is either a continuous field or a discrete particle. 
For the first possibility, the charge distribution\index{charge distribution} exists throughout space at the same time. For the second possibility, at every instant there is only a localized, point-like particle with the total charge of the system, and its motion during an infinitesimal time interval forms the effective charge distribution\index{charge distribution!effective}. Concretely speaking, at a particular instant the charge density of the particle in each position is either zero (if the particle is not there) or singular (if the particle is there), while the time average of the density during an infinitesimal time interval around the instant gives the effective charge density. Moreover, the motion of the particle is ergodic in the sense that the integral of the formed charge density in any region is equal to the expectation value of the total charge in the region.\index{charge distribution!origin of}

\subsection{Electrons are particles}

In the following, I will try to determine the existent form of the charge distribution\index{charge distribution!existent form of} of a quantum system. \index{random discontinuous motion of particles!arguments for}

If the charge distribution\index{charge distribution} of a quantum system is continuous in nature and exists throughout space at the same time,  then any two parts of the distribution (e.g. the two partial charges in box 1 and box 2 in the example discussed in the last section), like two electrons, will arguably have electrostatic interaction too.\footnote{The interaction will be described by an additional potential term in the Schr\"{o}dinger equation\index{Schr\"{o}dinger equation}. Moreover, the two parts of the distribution will be entangled, and their wave function will be defined in a six-dimensional configuration space\index{configuration space}.} The existence of such electrostatic self-interaction for individual quantum systems contradicts the superposition principle of quantum mechanics (at least for microscopic systems such as electrons). Moreover, the existence of the electrostatic self-interaction for the charge distribution\index{charge distribution} of an electron is incompatible with experimental observations either. For example, for the electron in the hydrogen atom, since the potential of the electrostatic self-interaction is of the same order as the Coulomb potential produced by the nucleus, the energy levels of hydrogen atoms would be remarkably different from those predicted by quantum mechanics and confirmed by experiments if there existed such electrostatic self-interaction. In contrast, if there is only a localized particle with charge at every instant, and the charge distribution\index{charge distribution} is effective, formed by the motion of the particle, then it is understandable that there exists no such electrostatic self-interaction for the effective charge distribution\index{charge distribution!effective}. This is consistent with the superposition principle of quantum mechanics and experimental observations.\index{charge distribution!existent form of}

Here is a further clarification of the above analysis. As noted before, the non-existence of electrostatic self-interaction for the charge distribution\index{charge distribution} of a single quantum system poses a puzzle. According to quantum mechanics, two charge distribution\index{charge distribution}s such as two electrons, which exist in space at the same time, have electrostatic interaction described by the potential term in the Schr\"{o}dinger equation\index{Schr\"{o}dinger equation}, but in the two-box example discussed in the last section, the two charges in box 1 and box 2 have no such electrostatic interaction. In fact, this puzzle does not depend so much on the actual existence of the charge distribution\index{charge distribution!existence of} as a property of a quantum system. It is essentially that according to quantum mechanics, the wavepacket $\psi_1$ in box 1 has electrostatic interaction with any test electron, so does the wavepacket $\psi_2$ in box 2, but these two wavepackets, unlike two electrons, have no electrostatic interaction. 

Facing this puzzle one may have two responses. The usual one is simply admitting that the non-existence of the self-interaction of the charge distribution\index{charge distribution} is a distinct feature of the laws of quantum mechanics, but insisting that the laws are what they are and no further explanation is needed. However, this response seems to beg the question and is unsatisfactory in the final analysis. A more reasonable response is to try to explain this puzzling feature, e.g. by analyzing its relationship with the existent form of the charge distribution\index{charge distribution!existent form of}. The charge distribution\index{charge distribution} has two possible existent forms after all. On the one hand, the non-existence of the self-interaction of the distribution may help determine which form is the actual one. For example, one form is inconsistent with this distinct feature, while the other form is consistent with it. On the other hand, the actual existent form of the charge distribution\index{charge distribution} may also help explain the non-existence of the self-interaction of the distribution.\index{charge distribution!existent form of}\index{random discontinuous motion of particles!arguments for}

This is just what the previous analysis has done. The analysis establishes a connection between the non-existence of the self-interaction of the charge distribution\index{charge distribution} of a quantum system and the actual existent form of the distribution. The reason why two wavepackets of an electron, each of which has part of the electron's charge, have no electrostatic interaction is that these two wavepackets, unlike two electrons, do not exist at the same time, and their charges are formed by the motion of a localized particle with the total charge of the electron. Since there is only a localized particle at every instant, there exists no electrostatic self-interaction of the effective charge distribution\index{charge distribution!effective} formed by the motion of the particle. In contrast, if the two wavepackets with charges, like two electrons, existed at the same time, then they would also have the same form of electrostatic interaction as two electrons.\footnote{For further discussion of this argument see Epilogue.}

To sum up, I have argued that the superposition principle of quantum mechanics requires that the charge distribution\index{charge distribution} of a quantum system such as an electron is effective; at every instant there is only a localized particle with the total charge of the system, while during an infinitesimal time interval around the instant the ergodic motion of the particle forms the effective charge distribution\index{charge distribution!effective} at the instant.\index{charge distribution!existent form of}

\subsection{The motion of a particle is discontinuous}


Which sort of ergodic motion then?  If the ergodic motion of a particle is continuous, then it can only form an effective charge distribution\index{charge distribution!effective} during a finite time interval. But, according to quantum mechanics, the charge distribution\index{charge distribution} is required to be formed by the ergodic motion of the particle during an infinitesimal time interval (\emph{not} during a finite time interval) around a given instant. Thus it seems that the ergodic motion of a particle cannot be continuous.\index{random discontinuous motion of particles!arguments for}
This is at least what the existing theory says. However, this argument may have a possible loophole. Although the classical ergodic models that assume continuous motion of particles are inconsistent with quantum mechanics due to the existence of finite ergodic time, they may be not completely refuted by experiments if only the ergodic time is extremely short. After all quantum mechanics is only an approximation of a more fundamental theory of quantum gravity\index{quantum gravity}, in which there may exist a minimum time interval such as the Planck time (see also Section 8.4.1). Therefore, we need to investigate the classical ergodic models more thoroughly.

First, consider an electron being in a momentum eigenstate. For this state the charge distribution\index{charge distribution} is even in the whole space at any time.
If the motion of the electron as a particle is continuous, moving with a finite speed, then it is obvious that the motion cannot generate a charge distribution\index{charge distribution} in the whole space during a finite time interval, whether the distribution is even or not. The reason is that during a finite time interval the particle can only move in a finite spatial region.
Thus it seems that only if the electron moves with an infinite speed at every instant, can it form the required charge distribution\index{charge distribution} in the whole space. But in this case, it seems already meaningless to say that the motion of the electron is continuous. \index{random discontinuous motion of particles!arguments for}

Next, consider an electron being in a superposition of two momentum eigenstates with opposite momenta. For this state the charge distribution\index{charge distribution} is cyclical in the whole space at all times.
Similarly, the continuous motion of the electron with a finite speed cannot generate the charge distribution\index{charge distribution} during any finite time interval.\footnote{Even if this is possible, it is also difficult to explain why the electron moves back and forth in space.} 
Moreover, even if the electron moves with an infinite speed at every instant, it can only form an even charge distribution\index{charge distribution} in the whole space, and it cannot form a cyclical charge distribution\index{charge distribution} in the whole space. Thus it seems that the continuous motion of the electron cannot form the required charge distribution\index{charge distribution} in this case.
This conclusion also holds true for general superpositions of momentum eigenstates which spread in the whole space.

Thirdly, consider an electron in a one-dimensional box in the first excited state (Aharonov\index{Aharonov, Yakir} and Vaidman\index{Vaidman, Lev}, 1993). \index{random discontinuous motion of particles!arguments for}
For this state the charge distribution\index{charge distribution} is symmetry relative to the center of the box, and the charge density is zero at the center of the box, as well as at the two ends of the box. 
Since the charge distribution\index{charge distribution} only exists in a finite spatial region, it seems that the continuous motion of the electron with a finite speed may generate the charge distribution\index{charge distribution} during a very short time interval.\footnote{Note that in Nelson\index{Nelson, Edward}'s stochastic mechanics, the electron, which is assumed to undergo Brownian motion, moves only within a region bounded by the nodes (Nelson\index{Nelson, Edward}, 1966). This ensures that the theory can be equivalent to quantum mechanics in a limited sense. Obviously this sort of motion is not ergodic and cannot generate the right charge distribution\index{charge distribution}. This conclusion also holds true for the motion of particles in Bohm's theory\index{Bohm's theory} (Bohm,\index{Bohm, David} 1952), as well as in some variants of stochastic mechanics and Bohm's theory\index{Bohm's theory} (Bell,\index{Bell, John S.} 1986b; Vink\index{Vink, Jeroen C.}, 1993; Barrett\index{Barrett, Jonathan}, Leifer\index{Leifer, Matthew S.} and Tumulka\index{Tumulka, Roderich}, 2005).}  
However, since the amount of time the electron spends around a given position is proportional to the charge density in the position, the electron can spend no time at the center of the box where the charge density is zero; in other words, it must move at infinite velocity at the center. 
Although the appearance of infinite velocities at an instant may be not a fatal problem (since the infinite potential is only an ideal situation), it seems difficult to explain why the electron speeds up at the node and where the infinite energy required for the acceleration comes from (Aharonov\index{Aharonov, Yakir} and Vaidman\index{Vaidman, Lev}, 1993).
 
Lastly, consider an electron in a superposition of two energy eigenstates in two boxes. In this case, even if the electron can move with infinite velocity, it cannot \emph{continuously} move from one box to another due to the restriction of box walls. Therefore, any sort of continuous motion cannot generate the required charge distribution\index{charge distribution}. One may still object that this is merely an artifact of the idealization of infinite potential. However, even in this ideal situation, the ergodic motion of the electron should also be able to generate the required charge distribution\index{charge distribution}; otherwise the model will be inconsistent with quantum mechanics. 

In view of these serious drawbacks of classical ergodic models and their inconsistency with quantum mechanics, I conclude that the ergodic motion of a particle cannot be continuous. If the motion of a particle is essentially discontinuous, then the particle can readily move throughout all regions where the wave function is nonzero during an arbitrarily short time interval around a given instant. Furthermore, when the probability density that the particle appears in each position is equal to the modulus squared of its wave function there at every instant, the discontinuous motion will be ergodic and can generate the right charge distribution\index{charge distribution}, for which the charge density in each position is proportional to the modulus squared of its wave function there. This will solve the above problems plagued by the classical ergodic models. The discontinuous ergodic motion requires no finite ergodic time. Moreover, a particle undergoing discontinuous motion can also ``jump" from one region to another spatially separated region, whether there is an infinite potential wall between them or not. Finally, discontinuous motion has no problem of infinite velocity. The reason is that no classical velocity and acceleration can be defined for discontinuous motion, and energy and momentum will require new definitions and understandings as in quantum mechanics. \index{random discontinuous motion of particles!arguments for}

In summary, I have argued that the ergodic motion of a particle is discontinuous, and the probability density that the particle appears in each position is equal to the modulus squared of its wave function there.

\subsection{An argument for random discontinuous motion\index{random discontinuous motion of particles}}

For the discontinuous motion of a particle, since quantum mechanics provides no further information about the position of the particle at each instant, it seems that the discontinuous motion should be also essentially random according to the theory. 
In the following, I will give a further argument for the existence of random discontinuous motion\index{random discontinuous motion of particles} of particles.

In order to know whether the motion of particles is random or not, we need to analyze the cause of motion. For example, if motion has no deterministic cause, then it will be  random and determined only by a probabilistic cause. This may be also the right way to find how particles move. Since motion involves change in position, if we can find the cause or instantaneous condition that determines the change,\footnote{The word ``cause" used here only denotes a certain instantaneous condition determining the change of position, which may appear in the laws of motion\index{laws of motion}. My following analysis is independent of whether the condition has a causal power or not.} we will be able to find how particles move in reality.\index{random discontinuous motion of particles!arguments for}

Consider the simplest states of motion of a free particle, for which the instantaneous condition determining the change of its position is constant during the motion. The instantaneous condition can be deterministic or indeterministic. That the instantaneous condition is deterministic means that it leads to a deterministic change of the position of the particle at each instant. That the instantaneous condition is indeterministic means that it only determines the probability density that the particle appears in each position in space at each instant. If the instantaneous condition is deterministic, then the simplest states of motion of the free particle will be continuous motion with constant velocity, for which the equation of motion is $x(t+dt)=x(t)+vdt$, where the deterministic instantaneous condition $v$ is a constant.\footnote{This deterministic instantaneous condition has been often called intrinsic velocity (Tooley\index{Tooley, Michael}, 1988).} 
On the other hand, if the instantaneous condition is indeterministic, then the simplest states of motion of the free particle will be random discontinuous motion\index{random discontinuous motion of particles} with even position probability distribution; at each instant the probability density that the particle appears in every position is the same.\index{random discontinuous motion of particles!arguments for}

In order to know whether the instantaneous condition is deterministic or not, we need to determine which sort of simplest states of free motion are the solutions of the equation of free motion in quantum mechanics (i.e. the free Schr\"{o}dinger equation\index{Schr\"{o}dinger equation!free}).\footnote{I have derived this equation of free motion from a few fundamental physical principles in Chapter 5. This makes the argument given here more complete.} According to the analysis given in the previous subsections, the momentum eigenstates of a free particle, which are the solutions of the free Schr\"{o}dinger equation\index{Schr\"{o}dinger equation!free}, describe the ergodic motion of the particle with even position probability distribution in space. Therefore, the simplest states of free motion with a constant probabilistic instantaneous condition are the solutions of the equation of free motion in quantum mechanics. On the other hand, it can be seen that the simplest states of free motion with a constant deterministic instantaneous condition are the solutions of the equation of free motion in classical mechanics, and not the solutions of the equation of free motion in quantum mechanics.

When assuming that the instantaneous condition determining the position change of a particle is always deterministic or indeterministic for any state of motion, the above result then implies that motion, whether it is free or forced, has no deterministic cause, and thus it is random and discontinuous,  determined only by a probabilistic cause. This argument for random discontinuous motion\index{random discontinuous motion of particles} may be improved by further analyzing this seemingly reasonable assumption, but I will leave this for future work.\index{random discontinuous motion of particles!arguments for}

\section{Further discussion}

Historically, it is Schr\"{o}dinger\index{Schr\"{o}dinger, Erwin} who first assumed the existence of the charge distribution\index{charge distribution!existence of} of an electron in space in 1926. According to his charge density hypothesis, the charge of a quantum system is distributed throughout space, and the charge density in each position is equal to the modulus squared of the wave function of the system there multiplied by the charge of the system.
Schr\"{o}dinger\index{Schr\"{o}dinger, Erwin}'s purpose was not to simply assume the existence of the charge distribution\index{charge distribution} of a quantum system, but to interpret the wave function of the system in terms of its charge distribution\index{charge distribution}. This is the first attempt to give an ontological interpretation of the wave function.


In the previous sections, I have re-examined Schr\"{o}dinger\index{Schr\"{o}dinger, Erwin!charge density hypothesis of}'s charge density hypothesis. 
It is argued that although Schr\"{o}dinger\index{Schr\"{o}dinger, Erwin}'s ontological interpretation of the wave function in terms of charge density meets serious problems and is unsatisfactory, this does not imply that the charge distribution\index{charge distribution} of an electron does not exist. 
Moreover, I have argued with the help of protective measurements\index{protective measurements} that a quantum system has a well-defined charge distribution\index{charge distribution}, and the charge density in each position is equal to the modulus squared of the wave function of the system there multiplied by the charge of the system.
This confirms Schr\"{o}dinger\index{Schr\"{o}dinger, Erwin}'s original charge density hypothesis. 

In order to explain the puzzling behaviours of the charge distribution\index{charge distribution} of a quantum system, I have also investigated the physical origin of the distribution.
It is argued that the charge distribution\index{charge distribution} of a quantum system is effective, that is, it is formed by the ergodic motion of a localized particle with the total charge of the system. 
Visually speaking, the ergodic motion of a particle will form a particle ``cloud" extending throughout space (during an infinitesimal time interval around a given instant), and the density of the cloud in each position, which represents the probability density that the particle appears there, is $|\psi(x,t)|^2$, where $\psi(x,t)$ is the wave function of the particle. For a charged particle such as an electron, the cloud will be a charged cloud, and the density $|\psi(x,t)|^2$, when multiplied by the charge of the particle, will be the charge density of the cloud. This picture of ergodic motion of a particle may explain some puzzling behaviours of the charge distribution\index{charge distribution} of a quantum system such as the non-existence of electrostatic self-interaction for the distribution.

\begin{center} 
\begin{figure}[h]\label{tec}

\includegraphics[scale=0.39]{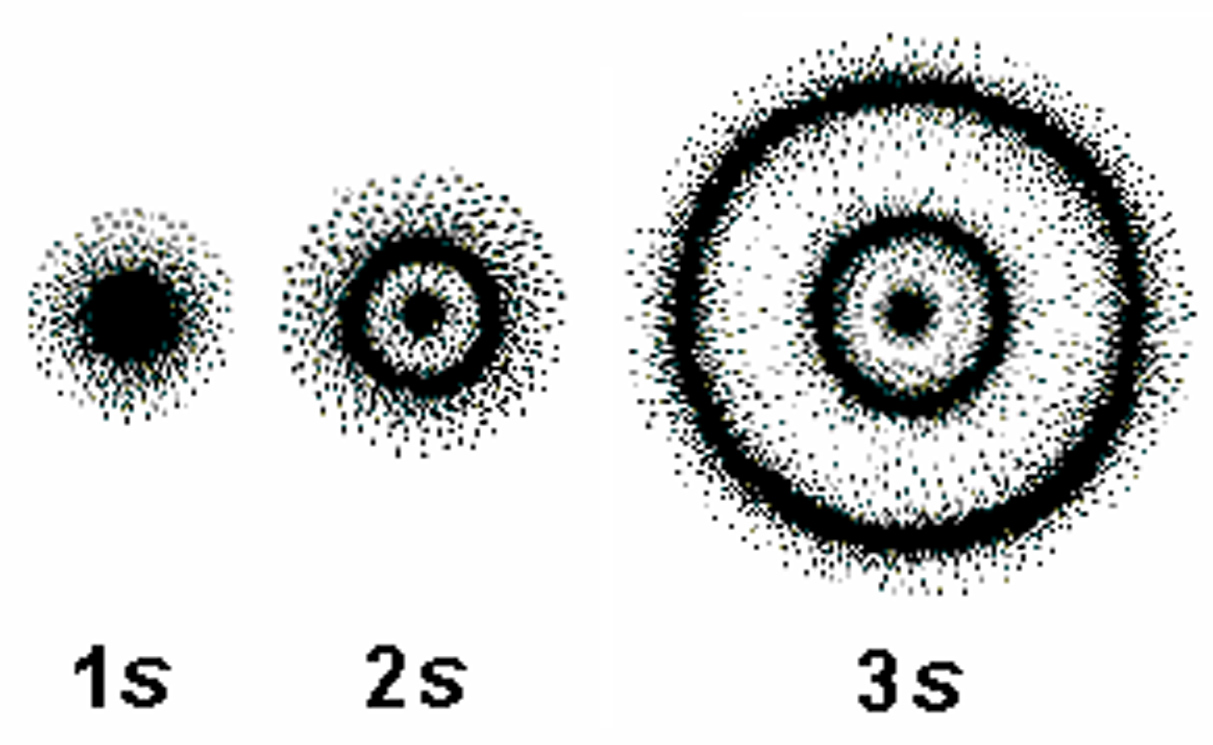}


\caption{Three electron clouds in a hydrogen atom}

\end{figure}

\end{center} 

Although the charge distribution\index{charge distribution} can be consistently defined for a many-body system, the distribution contains no information about the entanglement\index{quantum entanglement} between its subsystems.
In order to further solve this incompleteness problem, we need to extend the above picture of ergodic motion of a particle for one-body systems to many-body systems.
At a given instant, an $N$-body quantum system can be represented by a point in a $3N$-dimensional configuration space\index{configuration space}. During an infinitesimal time interval around the instant, the representative point performs ergodic motion in the configuration space\index{configuration space}, which is also random and discontinuous, and forms a cloud there. 
Then, similar to the single particle case, the representative point is required to spend in each volume element in the configuration space\index{configuration space} a time that is proportional to the modulus squared of the wave function of the system there. In other words, the density of the cloud in the configuration space\index{configuration space} is $\rho(x_1,x_2,...x_N,t)=|\psi(x_1, x_2, ...x_N, t)|^2$, where $\psi(x_1, x_2, ...x_N, t)$ is the wave function of the system. 
Since such ergodic motion in the configuration space\index{configuration space} contains entanglement\index{quantum entanglement} between the subsystems, its existence will solve the incompleteness problem for the charge distribution\index{charge distribution} of a many-body system.\footnote{I will give a detailed analysis of quantum entanglement\index{quantum entanglement} in Chapters 7 and 9.}

Here appears an intriguing question.
Are there $N$ particles in three-dimension space for an $N$-body quantum system? Or there is only one particle in configuration space\index{configuration space} for an $N$-body quantum system?
According to my previous analysis, the ontology for a one-body quantum system is a particle with the mass and charge of the system, which undergoes ergodic motion in three-dimension space. 
If this analysis is valid, it does require that the ontology for many-body quantum systems is also discrete particle, not continuous field in configuration space\index{configuration space}; otherwise when the wave function of a many-body system is a product state of the wave functions of one-body systems, the ontology for one-body systems is not particle.
However, the analysis does not require that the ontology for $N$-body quantum systems should be $N$ particles in three-dimension space, not one particle in configuration space\index{configuration space}. 
In order to answer the above question, we need to further analyze the nature of configuration space\index{configuration space} and the ontological meaning of the many-body wave function defined in the space. This will be the main task of the next chapter.









\chapter{The ontology of quantum mechanics (II)}


I have analyzed the ontology of\index{quantum mechanics!ontology of} quantum mechanics for one-body quantum systems such as an electron.
The analysis suggests that the electron is a particle, and its motion is random and discontinuous. 
In this chapter, I will extend this picture of random discontinuous motion\index{random discontinuous motion of particles} of particles for one-body systems to many-body systems.
I will argue that the wave function of an $N$-body quantum system represents the state of random discontinuous motion\index{random discontinuous motion of particles} of $N$ particles in three-dimensional space, and in particular, the modulus squared of the wave function gives the probability density that the particles appear in every possible group of positions in space. 
Moreover, I will present a more detailed analysis of random discontinuous motion\index{random discontinuous motion of particles} of particles and the interpretation of the wave function in terms of it.








\section{Wave function realism?}







The wave function of a physical system is in general a mathematical object defined in a high-dimensional configuration space\index{configuration space}. For an $N$-body system, the configuration space\index{configuration space} in which its wave function is defined is $3N$-dimensional. 
Before presenting my analysis of the nature of configuration space\index{configuration space} and the meaning of the wave function, I will first examine a widely-discussed view, wave function realism\index{wave function!realism},\footnote{Since a realist interpretation of the wave function does not necessarily imply that the wave function describes a real, physical field in configuration space\index{configuration space}, the appellation ``wave function realism\index{wave function!realism}" seems misleading. But for the sake of convenience I will still use this commonly used appellation in the following disussion.} which regards the wave function as a description of a real, physical field in a fundamental high-dimensional space (Albert\index{Albert, David Z.}, 1996, 2013, 2015).

In recent years, wave function realism\index{wave function!realism} seems to become an increasingly popular position among philosophers of physics and metaphysicians (Ney\index{Ney, Alyssa} and Albert\index{Albert, David Z.}, 2013). This view is composed of two parts. The first part says that configuration space\index{configuration space} is a real, fundamental space. Albert\index{Albert, David Z.} (1996) writes clearly,

\begin{quote}
The space \emph{we} live in, the space in which any realistic interpretation of quantum mechanics is necessarily going to depict the history of the world as \emph{playing itself out} ... is \emph{configuration}-space. And whatever impression we have to the contrary (whatever impression we have, say, of living in a three-dimensional space, or in a four-dimensional space-time) is somehow flatly illusory. (Albert\index{Albert, David Z.}, 1996, p.277)
\end{quote}

\noindent The second part of this view states what kind of entity the wave function is in the configuration space\index{configuration space}. Again, according to Albert\index{Albert, David Z.} (1996),

\begin{quote}
The sorts of physical objects that wave functions \emph{are}, on this way of thinking, are (plainly) \emph{fields} - which is to say that they are the sorts of objects whose states one specifies by specifying the values of some set of numbers at every point in the space where they live, the sorts of objects whose states one specifies (in \emph{this} case) by specifying the values of two numbers (one of which is usually referred to as an \emph{amplitude}, and the other as a \emph{phase} at every point in the universe's so-called \emph{configuration} space. \\
\indent The values of the amplitude and the phase are thought of (as with all fields) as intrinsic properties of the points in the configuration space\index{configuration space} with which they are associated. (Albert\index{Albert, David Z.}, 1996, p.278)
\end{quote}

\noindent Note that configuration space\index{configuration space} conventionally refers to an abstract space that is used to represent possible configurations of particles in three-dimensional space, and thus when assuming wave function realism\index{wave function!realism!introduction of} 
it is not accurate to call the high-dimensional space in which the wave function exists ``configuration space\index{configuration space}".  For wave function realism\index{wave function!realism}, there are no particles and their configurations, and the high-dimensional space is also fundamental, whose dimensionality is defined in terms of the number of degrees of freedom needed to capture the wave function of the system. But I will still use the appellation ``configuration space\index{configuration space}" in my discussion of wave function realism\index{wave function!realism} for the sake of  convenience.
 
Here it is also worth noting that Bell (\index{Bell, John S.}1987) once gave a clear recognition of  the \emph{prima facie} argument for wave function realism\index{wave function!realism} (see also Lewis,\index{Lewis, Peter J.} 2004). Concerning Bohm's (1952) theory, Bell writes,

\begin{quote}
\emph{No one can understand this theory until he is willing to think of $\psi$ as a real objective field rather than just a `probability amplitude'. Even though it propagates not in 3-space but in $3N$-space.} (emphasis in original) (Bell,\index{Bell, John S.} 1987, p.128) 
\end{quote}

\noindent Concerning Ghirardi\index{Ghirardi, GianCarlo}, Rimini\index{Rimini, Alberto} and Weber\index{Weber, Tullio}'s (1986)\index{GRW theory} dynamical collapse theory, he also writes,

\begin{quote}
There is nothing in this theory but the wavefunction. It is in the wavefunction that we must find an image of the physical world, and in particular of the arrangement of things in ordinary three-dimensional space. But the wavefunction as a whole lives in a much bigger space, of $3N$-dimensions. (Bell,\index{Bell, John S.} 1987, p.204) 
\end{quote}

There are two main motivations for adopting wave function realism\index{wave function!realism!motivations of}.
The first, broader motivation is that it seems to be the simplest, most straightforward, and most flat-footed way of thinking about the wave function realistically (Albert\index{Albert, David Z.}, 1996). In Lewis's (2004) words,

\begin{quote}
The wavefunction figures in quantum mechanics in much the same way that particle configurations figure in classical mechanics; its evolution over time successfully explains our observations. So absent some compelling argument to the contrary, the prima facie conclusion is that the wavefunction should be accorded the same status that we used to accord to particle configurations. Realists, then, should regard the wavefunction as part of the basic furniture of the world... This conclusion is independent of the theoretical choices one might make in response to the measurement problem\index{measurement problem}; whether one supplements the wavefunction with hidden variables (Bohm 1952), supplements the dynamics with a collapse mechanism (Ghirardi\index{Ghirardi, GianCarlo}, Rimini\index{Rimini, Alberto} and Weber\index{Weber, Tullio} 1986\index{GRW theory}), or neither (Everett 1957), it is the wavefunction that plays the central explanatory and predictive role. (Lewis,\index{Lewis, Peter J.} 2004)
\end{quote}

\noindent The end result is then the assumption of a physical space with a geometrical structure isomorphic to the configuration space\index{configuration space} and a set of physical properties isomorphic to the amplitude and phase of the wave function.\index{wave function!realism!motivations of}

The second, more specific motivation for adopting wave function realism\index{wave function!realism} is entanglement\index{quantum entanglement} or nonseparability\index{nonseparability} of quantum mechanical states. 
In classical mechanics, the state of a system of $N$ particles can be represented as a point in a $3N$-dimensional configuration space\index{configuration space}, and the configuration space\index{configuration space} representation\index{wave function!configuration space\index{configuration space} representation of} is simply a convenient summary of the positions of all these particles. In quantum mechanics, however, 
the wave function of an entangled $N$-body system, which is defined in a $3N$-dimensional configuration space\index{configuration space}, cannot be broken down into individual  three-dimensional wave functions of its subsystems. 
Thus the configuration space\index{configuration space} representation\index{wave function!configuration space\index{configuration space} representation of} cannot be regarded as a convenient summary of the individual subsystem states in three-dimensional space; there are physical properties of an entangled $N$-body system that cannot be represented in terms of the sum of the properties of $N$ subsystems moving in three-dimensional space (Lewis,\index{Lewis, Peter J.} 2004, 2016). Therefore, wave function realism\index{wave function!realism} seems to be an inescapable consequence, according to which there exist a configuration space\index{configuration space} entity, the wave function, as a basic physical ingredient of the world.\index{wave function!realism!motivations of}

In the following, I will first analyze the above motivations for wave function realism\index{wave function!realism} and then analyze its potential problems.\index{wave function!realism!motivations of}
To begin with, the first broader motivation to adopt wave function realism\index{wave function!realism} is debatable, since the approach of reading off the nature of the physical entity represented directly from the structure of the mathematical representation is problematic (Maudlin\index{Maudlin, Tim}, 2013).
The problem has two aspects. \index{wave function!realism!critics of}
On the one hand, a physical theory may have different mathematical formulations. 
If one reads off the physical ontology of the theory directly from the mathematical structure used to formulate the theory, then one will in general obtain different, conflicting ontologies of the same theory.
This means that one at least needs to consider the reasons to choose one ontology rather than another.

Take standard\index{quantum mechanics!standard formulation of} quantum mechanics  as an example (Maudlin\index{Maudlin, Tim}, 2013). When the wave function is multiplied by a constant phase, the new wave function yields the same empirical predictions. Thus standard quantum mechanics has two empirically equivalent mathematical formulations: one in terms of vectors in Hilbert space\index{Hilbert space}, and the other in terms of elements of projective Hilbert space\index{Hilbert space}. Obviously, taking the mathematics of these two formulations at face value will lead to different physical ontologies.  Moreover, choosing the former, namely assuming wave function realism\index{wave function!realism}, is obviously inconsistent with the widely accepted view that gauge degrees of freedom are not physical, and thus one also needs to explain why the overall phase of the wave function, which is in principle unobservable, represents a real physical degree of freedom. In contrast, choosing the latter will avoid this thorny issue. As a result, even if it is reasonable to read off the physical ontology directly from the mathematical structure, it seems that one should not choose the Hilbert space\index{Hilbert space} formulation and assume wave function realism\index{wave function!realism}.\footnote{For a more detailed analysis of this problem see Maudlin\index{Maudlin, Tim} (2013).}\index{wave function!realism!critics of}


On the other hand, a mathematical representation may represent different physical ontologies, and one also  needs to explain why  choose one ontology rather than another. Although wave function realism\index{wave function!realism} is the straightforward way of thinking about the wave function realistically, there are also other possible realistic interpretations of the wave function. For example, as argued by Monton\index{Monton, Bradley} (2002, 2013) and Lewis\index{Lewis, Peter J.} (2013), the wave function of an $N$-body quantum system may represent the property of $N$ particles in three-dimensional space.\footnote{As I have argued in Section 2.2.4, however, Monton\index{Monton, Bradley}'s (2002) argument for this view based on the eigenstate-eigenvaue half link is problematic.} 
Moreover, according to their analysis, this interpretation is devoid of several potential problems of wave function realism\index{wave function!realism} (see later discussion). Thus, it seems that even if one chooses the Hilbert space\index{Hilbert space} formulation one should not assume wave function realism\index{wave function!realism} either.\index{wave function!realism!critics of}

 
To sum up, as Maudlin\index{Maudlin, Tim} (2013) concluded, studying only the mathematics in which a physical theory is formulated is not the royal road to grasp its ontology. As I have argued in the previous chapters, we also need to study the connection between theory and experience in order to grasp the ontology of a physical theory.


Next,  I will analyze the second concrete motivation to adopt wave function realism\index{wave function!realism}, the existence of quantum entanglement\index{quantum entanglement}. Indeed, quantum mechanical states are distinct from classical mechanical states in their nonseparability\index{nonseparability} or entanglement\index{quantum entanglement}. The wave function of an entangled $N$-body system, which is defined in a $3N$-dimensional space, cannot be decomposed of individual  three-dimensional wave functions of its subsystems. Schr\"{o}dinger\index{Schr\"{o}dinger, Erwin} took this as the defining feature of quantum mechanics (Schr\"{o}dinger\index{Schr\"{o}dinger, Erwin}, 1935b). \index{wave function!realism!critics of}
However, contrary to the claims of Ney\index{Ney, Alyssa} (2012), this feature does not necessarily imply that the $3N$-dimensional space is fundamental and real, and the wave function represents a real, physical field in this space. As pointed out by French\index{French, Steven} (2013), there is at least an alternative understanding of entanglement\index{quantum entanglement} in terms of the notion of ``nonsupervenient" relations holding between individual physical entities existing in three-dimensional space (Teller\index{Teller, Paul}, 1986; French\index{French, Steven} and Krause\index{Krause, D\'{e}cio}, 2006). In addition, as noted above, one can also interpret the wave function of an $N$-body quantum system as the property of $N$ particles in three-dimensional space (Monton\index{Monton, Bradley}, 2002, 2013; Lewis,\index{Lewis, Peter J.} 2013).\footnote{It is worth noting that there is also another possibility, namely the wave function is a multiple-field, a configuration of which assigns properties to sets of $N$ points in three-dimensional space (Forrest\index{Forrest, Peter}, 1988, Chapter 5; Belot\index{Belot, Gordon}, 2012).} The existence of these alternatives undoubtedly reduces the force of the second motivation to adopt wave function realism\index{wave function!realism}.

In the following, I will analyze the potential problems of wave function realism\index{wave function!realism}.
The most obvious problem is how to  explain our three-dimensional impressions, which has been called the ``problem of perception" (Sol\'{e}\index{Sol\'{e}, Albert}, 2013). Albert\index{Albert, David Z.} clearly realized this problem:\index{wave function!realism!critics of}

\begin{quote}
The particularly urgent question (again) is where, in this picture, all the tables, and chairs, and buildings, and people are. The particularly urgent question is how it can possibly have come to pass, on a picture like this one, that there appear to us to be \emph{multiple} particles moving around in a  \emph{three-dimensional} space (Albert\index{Albert, David Z.}, 2013, p.54). 
\end{quote}

The first possible solution to this problem is the so-called instantaneous solution, which attempts to extract an image of a three-dimensional world from the instantaneous $3N$-dimensional wave function (Lewis,\index{Lewis, Peter J.} 2004). However, as argued by Monton\index{Monton, Bradley} (2002),  this solution faces a serious objection. It is that a point in configuration space\index{configuration space} alone does not pick out  a unique arrangement of objects in three-dimensional space, as this solution obviously requires. A point in configuration space\index{configuration space} is given by specifying the values of $3N$ parameters, but nothing intrinsic to the space specifies which parameters correspond to which objects in three-dimensional space. Therefore, the $3N$-dimensional wave function at a particular instant cannot underpin our experiences of a three-dimensional world at that instant (except there is a preferred coordinatization of the $3N$-dimensional space that wave function realism\index{wave function!realism} seems to lack). 

The second possible solution is the so-called dynamical solution, which attempts to show that the dynamical evolution of the $3N$-dimensional wave function over time produce the illusion of $N$ particles moving around in a  three-dimensional space (Albert\index{Albert, David Z.}, 1996; Lewis,\index{Lewis, Peter J.} 2004).\index{wave function!realism!critics of}
The key is to notice that the Hamiltonian governing the evolution of the wave function takes a uniquely simple form for a particular grouping of the coordinates of the wave function  into ordered triples.
In a classical mechanical world, the same form of Hamiltonian provides a notion of three-dimensional inter-particle distance, which can play a natural physical role since it ``reliably measures the degree to which the particles in question can dynamically affect one another" (Albert\index{Albert, David Z.}, 1996). 
Thus inhabitants living in such a world will perceive the system as containing $N$ particles in  a  three-dimensional space. 
How about the quantum mechanical world then?
According to Albert\index{Albert, David Z.} (1996), if inhabitants in such a world don't look too closely, the appearances they encounter will also correspond with those of their classical counterparts. The reason is roughly that for a everyday object its true representation in terms of the evolution of a wave function can be approximated by the corresponding evolution of a point in a classical configuration space\index{configuration space} (Lewis,\index{Lewis, Peter J.} 2004).\footnote{However, as admited by Albert\index{Albert, David Z.} (2013, p.56), ``The business of actually filling in the details of these accounts is not an altogether trivial matter and needs to be approached separately, and anew, for each particular way of solving the measurement problem\index{measurement problem}, and requires that we attend carefully to exactly how it is that the things we call particles actually manifest themselves in our empirical experience of the world." For a further analysis of this issue see Albert\index{Albert, David Z.} (2015).} Therefore, even though wave function realism\index{wave function!realism} says that the world is $3N$-dimensional, we perceive it as having only three dimensions. 

However, as argued by Lewis (2004, 2013), this dynamical solution to the ``problem of perception" is also problematic.\index{wave function!realism!critics of}
The reason is that the dynamical laws are not invariant under the coordinate transformations of the $3N$-dimensional space, but  invariant under the coordinate transformations of a three-dimensional space. 
This indicates that the configuration space\index{configuration space} has a three-dimensional  structure. In other words, although the wave function is a function of $3N$ independent parameters, but the transformational properties of the Hamiltonian require that these parameters refer to only three different spatial directions. 
Then, according to Lewis (2004), Albert\index{Albert, David Z.}'s dynamical solution is not only impossible but also unnecessary. It is impossible because the Hamiltonian takes exactly the same form under every choice of coordinates in three-dimensional space, and thus no choice makes it particularly simple. It is unnecessary because the outcome that the coordinates are naturally grouped into threes is built into the structure of configuration space\index{configuration space}, and thus does not need to be generated as a mere appearance based on the simplicity of the dynamics.

It is worth emphasizing that Lewis's (2004) argument based on invariance of dynamical laws also poses a serious threat to wave function realism\index{wave function!realism}. It strongly suggests that the quantum mechanical configuration space\index{configuration space} is not a real, fundamental space, but a space of configurations of particles existing in a three-dimensional space, quite like the classical mechanical configuration space\index{configuration space}. Correspondingly, the wave function is not a physical field in the high-dimensional configuration space\index{configuration space} either. This conclusion is also supported by a recent analysis of Myrvold\index{Myrvold, Wayne C.} (2015). He argued that since quantum mechanics arises from a relativistic quantum field theory\index{quantum field theory}, and in particular, the wave function of quantum mechanics and the configuration space\index{configuration space} in which it is defined are constructed from field operators defined on ordinary spacetime,
the configuration space\index{configuration space} is not fundamental, but rather is derivative of structures defined in three-dimensional space, and the wave function is not like a physical field either. \index{wave function!realism!critics of}

However, as Lewis (2004) also admitted, the above analysis does not provide a direct interpretation of the wave function. In his view, the wave function is a distribution over three-dimensional particle configurations, while a distribution over particle configurations is not itself a particle configuration. 
Then, what is the state of distribution the wave function represents? Are there really particles existing in three-dimensional space? I will try to answer these intriguing questions in the next section.\index{wave function!realism!critics of}



\section{A new ontological analysis of the wave function}

A better way to analyze the relationship between the wave function and the physical entity it describes is not only analyzing the structure of the wave function itself, but also analyzing the whole Schr\"{o}dinger equation\index{Schr\"{o}dinger equation}, which governs the evolution of the wave function over time. The Schr\"{o}dinger equation\index{Schr\"{o}dinger equation} contains more information about a quantum system than the wave function of the system, an important piece of which is the mass and charge properties of the system that are responsible for the gravitational and electromagnetic interactions between it and other systems.\footnote{In this sense, the wave function is not a complete description of a physical system, since it contains no information about the mass and charge of the system.
Note that other authors have already analyzed the Hamiltonian that  governs the evolution of the wave function in analyzing the meaning of the wave function, but what they have analyzed is only the coordinates of the Hamiltonian, not the other important parameters of the Hamiltonian such as mass and charge and their relations with the coordinates.} 
Mainly based on an analysis of these properties, I will argue that what the wave function of an $N$-body system describes is not one physical entity, either a continuous field or a discrete particle, in a $3N$-dimensional space, but $N$ physical entities in our ordinary three-dimensional space.
Moreover, by a new ontological analysis of the entangled states\index{entangled states} of an $N$-body quantum system, I will further argue that these physical entities are not continuous fields but discrete particles, and the motion of these particles is discontinuous and random.

\subsection{Understanding configuration space\index{configuration space}}

In order to know the meaning of the wave function, we need to first know the meaning of the coordinates on the configuration space\index{configuration space} of a quantum system, in which the wave function of the system is defined. 
As we have seen above, there are already some analyses of this issue. Here I will give a new analysis.

One way to understand configuration space\index{configuration space} is to see how the wave function transforms under a Galilean transformation between two inertial frames. Consider the wave function of an $N$-body quantum system, $\psi(x_1,y_1,z_1, ..., x_N,y_N,z_N,t)$, in an inertial frame $S$ with coordinates $(x,y,z,t)$. The coordinates $(x_1,y_1,z_1, ..., x_N,y_N,z_N)$ are coordinates on the $3N$-dimensioanl configuration space\index{configuration space} of the system, and the wave function of the system is defined in this space. Now, in another inertial frame $S'$ with coordinates $(x',y',z',t')$, where $(x',y',z',t')=G(x,y,z,t)$ is the Galilean transformation, the wave function becomes $\psi^{'}(x_1^{'},y_1^{'},z_1^{'}, ..., x_N^{'},y_N^{'},z_N^{'},t')$, where $(x_i^{'},y_i^{'},z_i^{'},t')=G(x_i,y_i,z_i,t)$ for $i=1, ..., N$. Then the transformation of the arguments of the wave function already tells us the meaning of these arguments or the meaning of the coordinates on the configuration space\index{configuration space} of the system. It is that the $3N$ coordinates of a point in the configuration space\index{configuration space} of an $N$-body quantum system are $N$ groups of three position coordinates in three-dimensional space. Under the Galilean transformation between two inertial coordinate systems $S$ and $S'$, each group of three coordinates $(x_i, y_i, z_i)$ of the $3N$ coordinates on configuration space\index{configuration space} also transforms according to the Galilean transformation.

In addition, the interaction Hamiltonian\index{Hamiltonian!interaction} of an $N$-body quantum system says the same thing. For example,
the interaction Hamiltonian\index{Hamiltonian!interaction} of an $N$-body quantum system under an external potential $V(x,y,z)$ is $\sum_{i=1}^N{V(x_i, y_i, z_i)}$, and the corresponding term in the Schr\"{o}dinger equation\index{Schr\"{o}dinger equation} is $\sum_{i=1}^N{V(x_i, y_i, z_i)}\psi(x_1,y_1,z_1, ..., x_N,y_N,z_N,t)$. Obviously the arguments of the potential function are the three position coordinates in our ordinary three-dimensional space, so do the corresponding group of three arguments of the wave function. 

Another way to understand configuration space\index{configuration space} is to resort to experience. The Born rule\index{Born rule} (for projective measurements\index{projective measurements}) also tells us the meaning of the coordinates on the configuration space\index{configuration space} of a quantum system.
According to the Born rule\index{Born rule},  the modulus squared of the wave function of an $N$-body quantum system, $|\psi(x_1,y_1,z_1, ..., x_N,y_N,z_N,t)|^2$, represents the probability density that the first subsystem is detected in position $(x_1,y_1,z_1)$ in our three-dimensional space and the second subsystem is detected in position $(x_2,y_2,z_2)$ in our three-dimensional space and so on. Thus, each group of three coordinates $(x_i, y_i, z_i)$ of the $3N$ coordinates on configuration space\index{configuration space} are the three position coordinates in our ordinary three-dimensional space. Similarly, the result of a protective measurement on an $N$-body quantum system says the same thing. For example, the result of a protective measurement of the charge density of an $N$-body quantum system in position $(x,y,z)$ in our three-dimensional space is:

\begin{equation}
 \rho(x,y,z)=\sum_i{\int...\int{Q_i|\psi(x_1, ..., z_{i-1}, x, y, z, x_{i+1},... z_N, t)|^2dx_1 ...dz_{i-1}dx_{i+1}...dz_N}},
\end{equation}

\noindent where $Q_i$ is the charge of the $i$-th subsystem. This formula also tells us that each group of three arguments $(x_i, y_i, z_i)$ of the wave function are the three position coordinates in our three-dimensional space. 

In summary, I have argued that the $3N$ coordinates of a point in the configuration space\index{configuration space} of an $N$-body quantum system are $N$ groups of three position coordinates in our ordinary three-dimensional space. 

\subsection{Understanding subsystems}

In order to know the ontological meaning of the wave function of an $N$-body quantum system, we also need to understand the characteristics of the subsystems which constitute the  system.

First of all, the Schr\"{o}dinger equation\index{Schr\"{o}dinger equation} tells us something about subsystems. In the Schr\"{o}dinger equation\index{Schr\"{o}dinger equation} for an $N$-body quantum system, there are $N$ mass parameters $m_1$, $m_2$,  ..., $m_N$ (as well as $N$ charge parameters etc). These parameters are not natural constants, but properties of the system; they may be different for different systems. Moreover, each mass parameter describes the same mass property, and it may assume different values for different subsystems. Therefore, it is arguable that these $N$ mass parameters describe the same mass property of $N$ subsystems. In other words, an $N$-body quantum system contains $N$ subsystems or $N$ physical entities, each of which has its respective mass and charge properties, and the wave function of the system describes the state of these physical entities. This conclusion is obvious when the wave function of an $N$-body quantum system is a product state of $N$ wave functions. 

Moreover, these $N$ physical entities exist in our three-dimensional space, not in the $3N$-dimensional configuration space\index{configuration space}. The reason is that in the Schr\"{o}dinger equation\index{Schr\"{o}dinger equation} for an $N$-body quantum system, each mass parameter $m_i$ is \emph{only} correlated with each group of three coordinates $(x_i, y_i, z_i)$ of the $3N$ coordinates on configuration space\index{configuration space}, while these three coordinates $(x_i, y_i, z_i)$, according to the above analysis, are the three position coordinates in our three-dimensional space. 
Here it is also worth noting that the configuration space\index{configuration space} (as a fundamental space) cannot accommodate mass and charge distribution\index{charge distribution}s. For example, consider a product state of a two-body quantum system, in which there are point masses $m_1$ and $m_2$ in two positions in our three-dimensional space. This requires that in the corresponding \emph{single} position in the six-dimensional configuration space\index{configuration space} of the system there should exist \emph{two} distinguishable point masses $m_1$ and $m_2$. But  in every position in the configuration space\index{configuration space} there can only exist one total point mass (if the space is real and fundamental).

Secondly, the connection of quantum mechanics with experience\index{quantum mechanics!and experience} also tells us the same thing. Recall the Born rule\index{Born rule} says that the modulus squared of the wave function of an $N$-body quantum system, $|\psi(x_1,y_1,z_1..., x_N,y_N,z_N,t)|^2$, represents the probability density that the first subsystem with mass $m_1$ is detected in position $(x_1,y_1,z_1)$ and the second subsystem with mass $m_2$ is detected in position $(x_2,y_2,z_2)$ and so on. Thus the Born rule\index{Born rule} also says that each subsystem with its respective mass $m_i$ is  correlated with each group of three coordinates $(x_i, y_i, z_i)$ of the $3N$ coordinates on configuration space\index{configuration space}. Moreover, the results of the measurements of the masses and charges of an $N$-body quantum system also indicate that the system is composed of $N$ subsystems with their respective masses and charges.

To sum up, it is arguable that for an $N$-body quantum system, there are $N$ subsystems or $N$ physical entities with respective masses and charges in our three-dimensional space. 


\subsection{Understanding entangled states\index{entangled states}}

As I noted at the end of the last chapter, 
when the picture of random discontinuous motion\index{random discontinuous motion of particles} of a particle for one-body systems is extended to many-body systems, it needs to be determined whether the ontology for an $N$-body quantum system is $N$ particles in our three-dimension space or one particle in the configuration space\index{configuration space} of the system.
Now according to the above analysis, there are $N$ physical entities in our three-dimensional space for an $N$-body quantum system.
Therefore, the picture of random discontinuous motion\index{random discontinuous motion of particles!arguments for} for an $N$-body quantum system will be random discontinuous motion\index{random discontinuous motion of particles} of $N$ particles in our three-dimensional space.
In the following, I will present a new ontological analysis of the entangled states\index{entangled states} of an $N$-body quantum system. The analysis will provide further support for the existence of particles and their random discontinuous motion\index{random discontinuous motion of particles}.

Consider a two-body system whose wave function is defined in a six-dimensional configuration space\index{configuration space}. First of all, suppose the wave function of the system is localized in one position $(x_1, y_1, z_1, x_2, y_2, z_2)$ in the space at a given instant. This wave function can be decomposed into a product of two wave functions which are localized in positions $(x_1,y_1,z_1)$ and $(x_2,y_2,z_2)$ in the same three-dimensional space, our ordinary three-dimensional space, respectively. It is uncontroversial that this wave function describes two independent physical entities, which are localized in positions $(x_1,y_1,z_1)$ and $(x_2,y_2,z_2)$ in our three-dimensional space, respectively. Moreover, as I have argued previously, the Schr\"{o}dinger equation\index{Schr\"{o}dinger equation} that governs the evolution of the system further indicates that these two physical entities have respective masses such as $m_1$ and $m_2$ (as well as respective charges such as $Q_1$ and $Q_2$ etc). \index{random discontinuous motion of particles!arguments for}

Next, suppose the wave function of the two-body system is localized in two positions $(x_1, y_1, z_1, x_2, y_2, z_2)$ and $(x_3, y_3, z_3, x_4, y_4, z_4)$ in the six-dimensional configuration space\index{configuration space} at a given instant. This is an entangled state, which can be generated from a product state by the Schr\"{o}dinger\index{Schr\"{o}dinger, Erwin} evolution of the system. In this case, there are still two physical entities with the original masses and charges in three-dimensional space, since the Schr\"{o}dinger\index{Schr\"{o}dinger, Erwin} evolution does not create or annihilate physical entities,\footnote{In other words, when the state of the two physical entities evolves from a product state to an entangled state, the interaction between them does not annihilate any of them from the three-dimensional space.} and the mass and charge properties of the system do not change during its evolution either. 
According to the above analysis, the wave function of the two-body system being localized in position $(x_1, y_1, z_1, x_2, y_2, z_2)$ means that physical entity 1 with mass $m_1$ and charge $Q_1$ exists in position $(x_1,y_1,z_1)$ in three-dimensional space, and physical entity 2 with mass $m_2$ and charge $Q_2$ exists in position $(x_2,y_2,z_2)$ in three-dimensional space. Similarly, the wave function of the two-body system being localized in position $(x_3, y_3, z_3, x_4, y_4, z_4)$ means that physical entity 1 exists in position $(x_3, y_3, z_3)$ in three-dimensional space, and physical entity 2 exists in position $(x_4, y_4, z_4)$ in three-dimensional space. These are two ordinary physical situations. Then, when the wave function of these two physical entities is an entangled state, being localized in both positions $(x_1, y_1, z_1, x_2, y_2, z_2)$ and $(x_3, y_3, z_3, x_4, y_4, z_4)$, how do they exist in three-dimensional space? \index{random discontinuous motion of particles!arguments for}

Since the state of the physical entities described by the wave function is defined either at a precise instant or during an infinitesimal time interval around a given instant as the limit of a time-averaged state, there are two possible existent forms.\footnote{I have discussed these two possibilities when analyzing the origin of the charge distribution\index{charge distribution} of a quantum system in the last chapter.} One is that the above two physical situations exist at the same time at the precise given instant in three-dimensional space. This means that physical entity 1 exists in positions $(x_1,y_1,z_1)$ and $(x_3, y_3, z_3)$, and physical entity 2 exists in positions $(x_2,y_2,z_2)$ and $(x_4, y_4, z_4)$. Since there is no correlation between the positions of the two physical entities, the wave function that describes this existent form is not an entangled state but a product state, which is localized in four positions $(x_1, y_1, z_1, x_2, y_2, z_2)$, $(x_3, y_3, z_3, x_4, y_4, z_4)$, $(x_1, y_1, z_1, x_4, y_4, z_4)$, and $(x_3, y_3, z_3, x_2, y_2, z_2)$ in the six-dimensional configuration space\index{configuration space}. Thus this possiblity is excluded.


The other possible existent form, which is thus the actual existent form, is that the above two physical situations exist ``at the same time" during an arbitrarily short time interval or an infinitesimal time interval around the given instant in three-dimensional space.
Concretely speaking, the situation in which physical entity 1 is in position $(x_1,y_1,z_1)$ and physical entity 2 is in position $(x_2,y_2,z_2)$ exists in one part of the continuous time flow, and the situation in which physical entity 1 is in position $(x_3, y_3, z_3)$ and physical entity 2 is in position $(x_4, y_4, z_4)$ exists in the other part. The restriction is that the temporal part in which each situation exists cannot be a continuous time interval during an arbitrarily short time interval; otherwise the wave function describing the state in the time interval will be not the original superposition of two branches, but one of the branches. This means that the set of the instants at which each situation exists is not a continuous instant set but a discontinuous, dense instant set. 
At some discontinuous instants, physical entity 1 with mass $m_1$ and charge $Q_1$ exists in position $(x_1,y_1,z_1)$ and physical entity 2 with mass $m_2$ and charge $Q_2$ exists in position $(x_2,y_2,z_2)$, while at other discontinuous instants, physical entity 1 exists in position $(x_3, y_3, z_3)$ and physical entity 2 exists in position $(x_4, y_4, z_4)$. By this way of time division, the above two physical situations exist ``at the same time" during an arbitrarily short time interval or during an infinitesimal time interval around the given instant.\index{random discontinuous motion of particles!arguments for}

\begin{center} 
\begin{figure}[h]\label{schcat}

\includegraphics[scale=0.42]{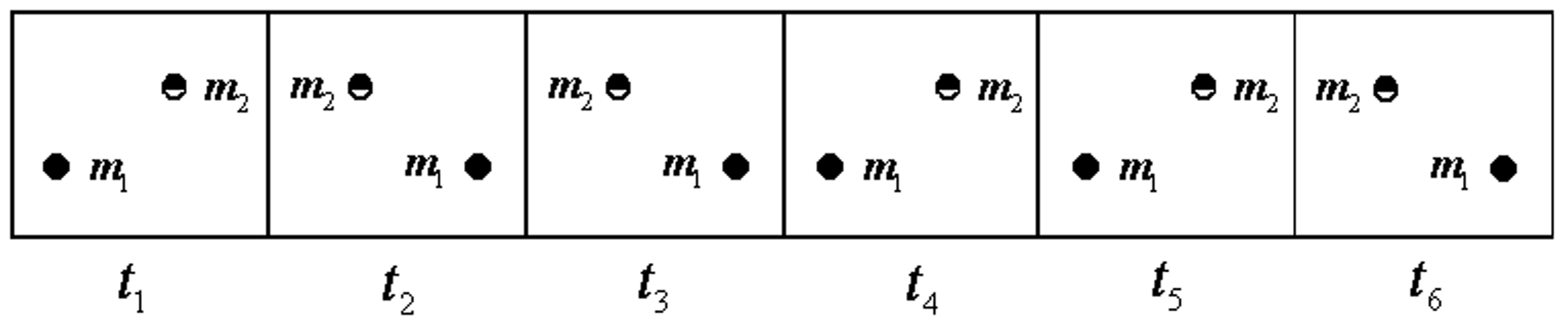}


\caption{Two entangled physical entities in space at six neighboring instants}

\end{figure}

 \end{center} 


This way of time division implies a picture of discontinuous motion for the involved physical entities, which is as follows. Physical entity 1 with mass $m_1$ and charge $Q_1$ jumps discontinuously between positions $(x_1,y_1,z_1)$ and $(x_3, y_3, z_3)$, and physical entity 2 with mass $m_2$ and charge $Q_2$ jumps discontinuously between positions $(x_2,y_2,z_2)$ and $(x_4, y_4, z_4)$. Moreover, they jump in a precisely simultaneous way. When physical entity 1 jumps from position $(x_1,y_1,z_1)$ to position $(x_3, y_3, z_3)$, physical entity 2 always jumps from position $(x_2,y_2,z_2)$ to position $(x_4, y_4, z_4)$, and vice versa.  In the limit case where position $(x_2,y_2,z_2)$ is the same as position $(x_4, y_4, z_4)$, physical entities 1 and 2 are no longer entangled, while physical entity 1 with mass $m_1$ and charge $Q_1$ still jumps discontinuously between positions $(x_1,y_1,z_1)$ and $(x_3, y_3, z_3)$. This means that the picture of discontinuous motion also exists for one-body systems. As argued before, since quantum mechanics does not provide further information about the positions of the physical entities at each instant, the discontinuous motion described by the theory is  essentially random too.

This result can be illustrated more vividly with the Schr\"{o}dinger's cat state\index{Schr\"{o}dinger's cat state}. For Schr\"{o}dinger's cat which is in a superposition of both alive and dead states, at each instant the cat is either alive or dead in a purely random way, while during a time interval the cat discontinuously jumps between the alive and dead states. The following figure depicts Schr\"{o}dinger's cat at six neighboring instants.\footnote{The cats in the figure are copied from Physics Today 23(9), 30-35 (1970), with the permission of the American Institute of Physics. DOI: http://dx.doi.org/10.1063/1.3022331.}

\begin{center} 
\begin{figure}[h]\label{schcat}

\includegraphics[scale=0.20]{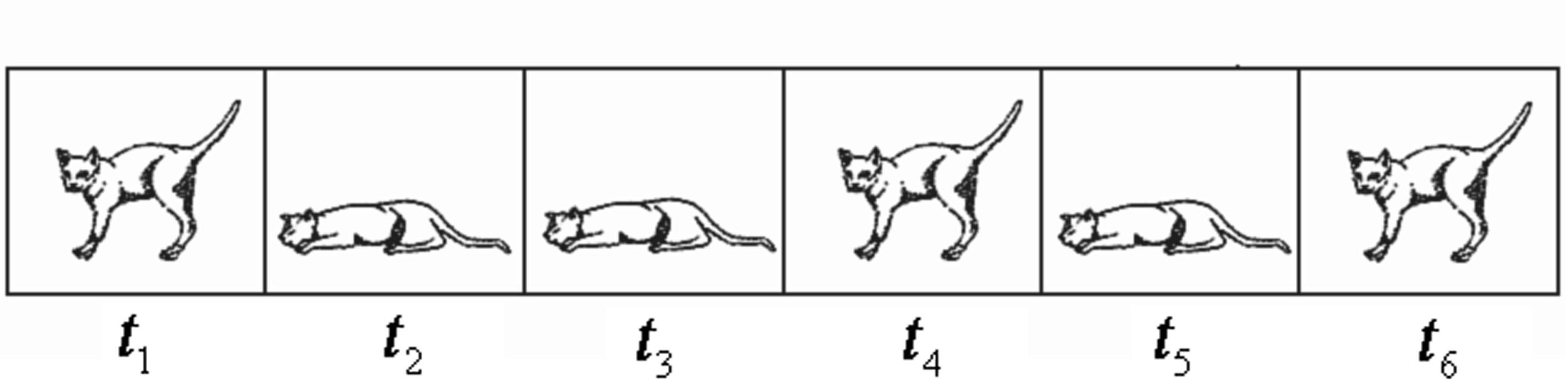}


\caption{Schr\"{o}dinger's cat at six neighboring instants}

\end{figure}

 \end{center} 


The above analysis may also tell us what these physical entities are. A physical entity in our three-dimensional space may be a continuous field or a discrete particle. For the above entangled state of a two-body system, since each physical entity is only in one position in space at each instant (when there are two positions it may occupy), it is not a continuous field but a localized particle.
In fact, there is a more general reason why these physical entities are not continuous fields in three-dimensional space.\index{random discontinuous motion of particles!arguments for}
It is that for an entangled state of an $N$-body system we cannot even define $N$ continuous fields in three-dimensional space which contain the whole information of the entangled state. 
 


Since a general position entangled state of a many-body system can be decomposed into a superposition of the product states of the position eigenstates of its subsystems, the above analysis applies to all entangled states\index{entangled states}.  
Therefore, it is arguable that an $N$-body quantum system is composed not of $N$ continuous fields but of $N$ discrete particles in our three-dimensional space. Moreover, the motion of these particles is not continuous but discontinuous and random in nature, and especially, the motion of entangled particles is precisely simultaneous.\footnote{Note that the analysis and its results given in this section also hold true for other instantaneous properties. I will discuss this point later.} \index{random discontinuous motion of particles!arguments for}



\section{The wave function as a description of random discontinuous motion\index{random discontinuous motion of particles!as a new interpretation of the wave function} of particles}

In classical mechanics, we have a clear physical picture of motion. It is well understood that the trajectory function $x(t)$ in the theory describes continuous motion of a particle. In quantum mechanics, the trajectory function $x(t)$ is replaced by a wave function $\psi(x,t)$. If the particle ontology is still viable in the quantum world, then it seems natural that the wave function should describe some sort of more fundamental motion of particles, of which continuous motion is only an approximation in the classical domain, as quantum mechanics is a more fundamental theory of the physical world, of which classical mechanics is an approximation. The previous analysis provides a strong support for this conjecture. It says that a quantum system is a system of particles that undergo random discontinuous motion\index{random discontinuous motion of particles}.\footnote{We may say that an electron is a quantum particle in the sense that its motion is not continuous motion described by classical mechanics, but random discontinuous motion\index{random discontinuous motion of particles!as a new interpretation of the wave function} described by quantum mechanics.} 
Here the concept of particle is used in its usual sense. A particle is a small localized object with mass and charge, and it is only in one position in space at each instant. 
As a result, the wave function in quantum mechanics can be regarded as a description of the more fundamental motion of particles, which is essentially discontinuous and random. In this section, I will give a more detailed analysis of random discontinuous motion\index{random discontinuous motion of particles} of particles and the ontological meaning of the wave function (Gao\index{Gao, Shan}, 1993, 1999, 2000, 2003, 2006b, 2008, 2011a, 2011b, 2014b, 2015a).\index{random discontinuous motion of particles!as a new interpretation of the wave function}


\subsection{A mathematical viewpoint}

Compared with continuous motion of particles, the picture of random discontinuous motion\index{random discontinuous motion of particles} of particles seems strange and unnatural for most people. 
This is not beyond expectations, since we are most familiar with the apparent continuous motion of objects in our everyday world.
However, it can be argued that random discontinuous motion\index{random discontinuous motion of particles} is more natural and logical than continuous motion from a mathematical point of view. 
Let us see why this is the case.

The motion of a particle can be described by a functional relation between each instant and its position at this instant in mathematics. In this way, continuous motion is described by continuous functions\index{continuous functions}, while discontinuous motion is described by discontinuous function\index{discontinuous function}s. 
The question is: Which sort of functions universally exist in the mathematical world? 
This question can also be put in another more appropriate way. 
Since motion does not exist at an instant, the state of motion of a particle at a given instant is defined not by its position at the precise instant, but by its positions during an infinitesimal time interval around the instant.
This means that the state of motion of a particle at a given instant will be described by a set of points\index{set of points} in space and time, in which each point represents the position of the particle at each instant during an infinitesimal time interval around the given instant.
Then the question is: What is the general form of such a set of points\index{set of points!general}? 
Is it a continuous line? Or is it a discontinuous set of points\index{set of points!discontinuous}? The former corresponds to continuous motion, while the latter corresponds to discontinuous motion.\index{random discontinuous motion of particles!as a new interpretation of the wave function}

\begin{center} 
\begin{figure}[h]\label{cmdm}

\includegraphics[scale=0.50]{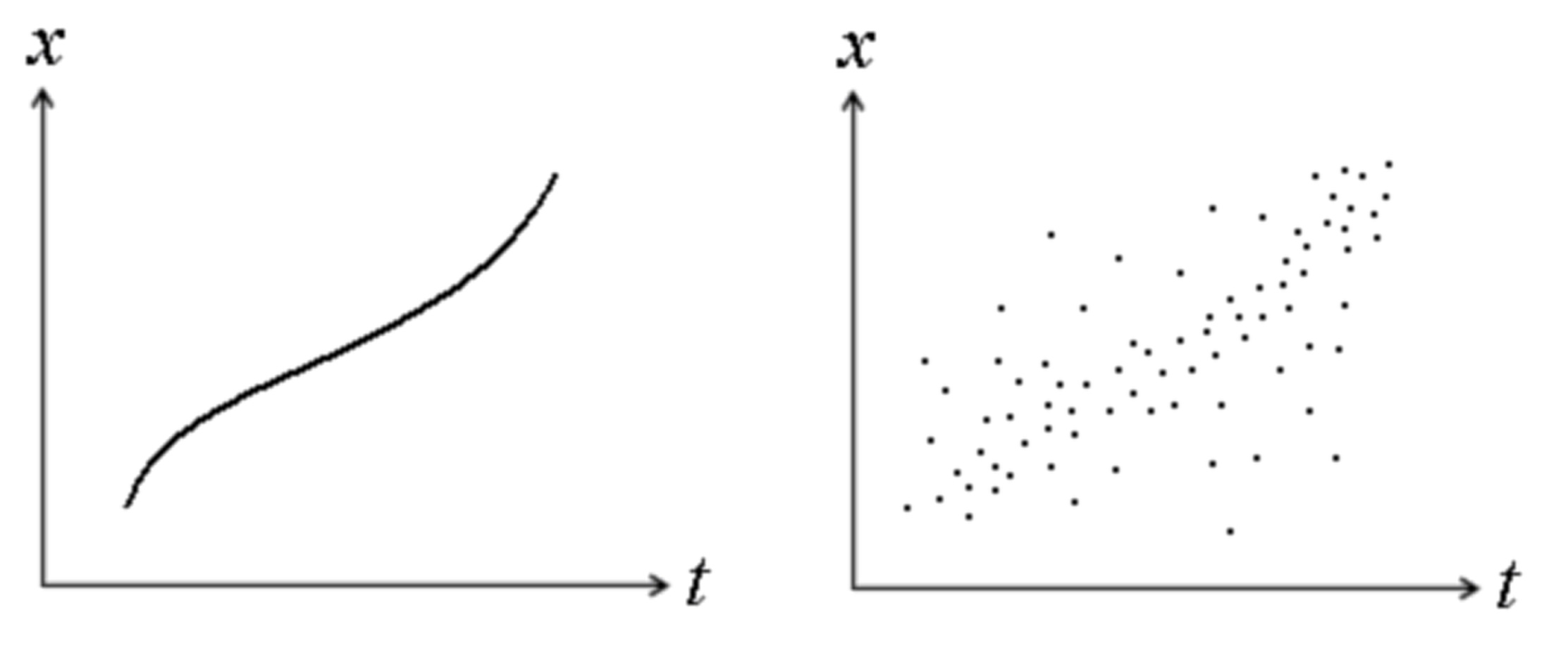}


\caption{Continuous motion vs. discontinuous motion}

\end{figure}

 \end{center} 


The right answers to these questions had not been found until the late 19th and early 20th centuries. Before then only continuous functions\index{continuous functions} were studied in mathematics. The existence of continuous functions\index{continuous functions} accords with everyday experience. In particular, the motion of macroscopic objects\index{macroscopic objects} is apparently continuous, and thus can be directly described by continuous functions\index{continuous functions}. However, mathematics became more and more dependent on logic rather than on experience in the second half of the 19th century. Many discontinuous function\index{discontinuous function}s were invented during that period. For example, a famous discontinuous function\index{discontinuous function} is that its value is zero at each rational point and is one at each irrational point.\index{random discontinuous motion of particles!as a new interpretation of the wave function}

At first, most elder mathematicians were very hostile to discontinuous function\index{discontinuous function}s. They called them pathological functions\index{pathological functions}. As Poincar\'{e}\index{Poincar\'{e}, Henri} remarked in 1899, 

\begin{quote}
Logic sometimes makes monsters. For half a century we have seen a mass of bizarre functions which appear to be forced to resemble as little as possible honest functions which serve some purpose. More of continuity, or less of continuity, more derivatives, and so forth... In former times when one invented a new function it was for a practical purpose; today one invents them purposely to show up defects in the reasoning of our fathers. (Quoted in Kline\index{Kline, Morris}, 1990, p. 973)
\end{quote}

\noindent However, several young mathematicians, notably Borel and Lebesgue, took discontinuous function\index{discontinuous function}s seriously. They discovered that these functions could also be strictly analyzed with the help of the set theory. As a result, they led a revolution in mathematical analysis, which transformed classical analysis into modern analysis.\footnote{It is well known that in the same period there also happened a revolution in physics, which transformed classical mechanics into quantum mechanics. I will argue below that there is a connection between modern analysis in mathematics and quantum mechanics in physics.}
A core notion introduced by them is measure.
A measure of a set is a generalization of the concepts of the length of a line, the area of a plane figure, and the volume of a solid, and it can be intuitively understood as the size of the set.
Length can only be used to describe continuous lines, which are very special sets of points, while measure\index{set of points!measure of} can be used to describe more general sets of points.
For example, the set of all rational points between 0 and 1 in a real line is a dense set of points\index{set of points!dense}, and its measure\index{set of points!measure of} is zero. Certainly, the measure of a line still equals to its length.\index{random discontinuous motion of particles!as a new interpretation of the wave function}

The above questions can now be answered by using the measure theory\index{measure theory}.
All functions form a set, whose measure\index{set of points!measure of} can be set to one.
Then according to the measure theory\index{measure theory}, the measure of the subset of all continuous functions\index{continuous functions} is zero, while the measure\index{set of points!measure of} of the subset of all discontinuous function\index{discontinuous function}s is one. 
This means that continuous functions\index{continuous functions} are extremely special functions, and nearly all functions are discontinuous function\index{discontinuous function}s. As Poincar\'{e}\index{Poincar\'{e}, Henri} also admitted,

\begin{quote}
Indeed, from the point of view of logic, these strange functions are the most general... If logic were the sole guide of the teacher, it would be necessary to begin with the most general functions. (Quoted in Kline\index{Kline, Morris}, 1990, p. 973)
\end{quote}

\noindent Similarly, a general set of points\index{set of points!general} in an infinitesimal region of space-time, which represents a general local state of motion of a particle, is a discontinuous, dense set of points\index{set of points!discontinuous}\index{set of points!dense}. And a continuous line in the region is an extremely special set of points\index{set of points!special}, and the measure of the subset of all continuous lines is zero.\index{random discontinuous motion of particles!as a new interpretation of the wave function}

Therefore, from a mathematical point of view, random discontinuous motion\index{random discontinuous motion of particles}, which is described by general discontinuous function\index{discontinuous function}s and discontinuous sets of points\index{set of points!discontinuous}, is more natural and logical than continuous motion, which is described by extremely special continuous functions\index{continuous functions} and continuous lines.
Moreover, if the great book of nature is indeed written in the language of mathematics, as Galileo\index{Galilei, Galileo} once put it, then it seems that it is discontinuous motion of particles, not continuous motion of particles, that universally exists in the physical world.



\subsection{Describing random discontinuous motion\index{random discontinuous motion of particles!description of} of particles}

In the following, I will give a strict description of random discontinuous motion\index{random discontinuous motion of particles} of particles based on the measure theory\index{measure theory}. For the sake of simplicity, I will mainly analyze one-dimensional motion. The results can be readily extended to the three-dimensional situation.\index{random discontinuous motion of particles!as a new interpretation of the wave function}

\begin{center} 
\begin{figure}[h]\label{rdm}

\includegraphics[scale=0.50]{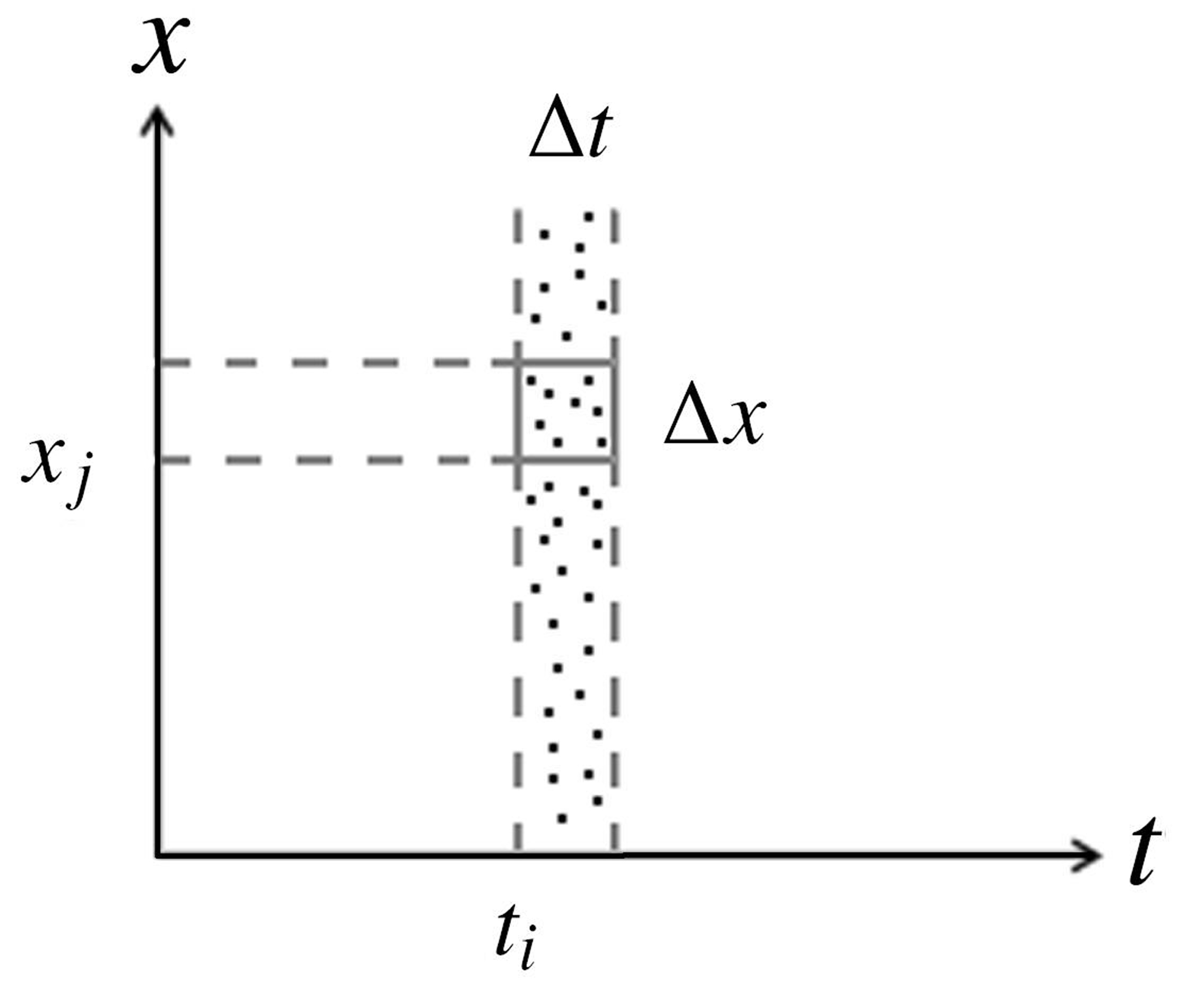}


\caption{Describing random discontinuous motion of a particle}

\end{figure}

\end{center} 


Consider the state of random discontinuous motion\index{random discontinuous motion of particles} of a particle in finite intervals $\Delta t$ and $\Delta x$ around a space-time point ($t_i$,$x_j$) as shown in Figure 7.2. The positions of the particle form a random, discontinuous trajectory in this square region.\footnote{Unlike deterministic continuous motion of particles, the discontinuous trajectory function, $x(t)$, no longer provides a useful description for random discontinuous motion\index{random discontinuous motion of particles} of particles. Recall that a trajectory function $x(t)$ is essentially discontinuous if it is not continuous at every instant $t$. A trajectory function $x(t)$ is continuous if and only if for every $t$ and every real number $\varepsilon >0$, there exists a real number $\delta >0$ such that whenever a point $t_0$ has distance less than $\delta$ to $t$, the point $x(t_0)$  has distance less than $\varepsilon$ to $x(t)$.}
We study the projection of this trajectory in the $t$-axis, which is a dense instant set in the time interval $\Delta t$. Let $W$ be the discontinuous trajectory of the particle and $Q$ be the square region $[x_j,x_j+\Delta x] \times [t_i,t_i+\Delta t] $. The dense instant set can be denoted by $\pi_t(W \cap Q) \in \Re$, where $\pi_t$ is the projection on the $t$-axis. According to the measure theory\index{measure theory}, we can define the Lebesgue measure\index{set of points!Lebesgue measure of}:\index{random discontinuous motion of particles!description of}\index{random discontinuous motion of particles!as a new interpretation of the wave function}

\begin{equation}
M_{\Delta x,\Delta t}(x_j,t_i)=\int_{\pi_t(W \cap Q) \in \Re}dt.  
\label{}
\end{equation}

\noindent Since the sum of the measures of the dense instant sets in the time interval $\Delta t$ for all $x_j$ is equal to the length of the continuous time interval $\Delta t$, we have: 

\begin{equation}
\sum_j {M_{\Delta x,\Delta t}(x_j,t_i)}=\Delta t.
\label{m}
\end{equation}

\noindent Then we can define the measure density\index{set of points!measure density of} as follows:

\begin{equation}
\rho(x,t)=\lim_{\Delta x, \Delta t \rightarrow 0} {M_{\Delta x,\Delta t}(x,t)/(\Delta x \cdot \Delta t)}.
\label{rho}
\end{equation}

\noindent We call $\rho(x,t)$ position measure density or position density in brief. \index{random discontinuous motion of particles!description of!position density}
This quantity provides a strict description of the position distribution of the particle in an infinitesimal space interval $dx$ around position $x$ during an infinitesimal interval $dt$ around instant $t$, and it satisfies the normalization relation $\int_{-\infty}^{+\infty}{\rho(x,t)dx}=1$ by (\ref{m}). 
Note that the existence of the above limit relies on the precondition that the probability density that the particle appears in each position $x$ at each instant $t$, which may be denoted by $\varrho(x,t)$, is differentiable with respect to both $x$ and $t$. It can be seen that $\rho(x,t)$ is determined by $\varrho(x,t)$, and there exists the relation $\rho(x,t)=\varrho(x,t)$.\index{random discontinuous motion of particles!description of}

Since the position density $\rho(x,t)$ changes with time in general, we may further define the position flux density $j(x,t)$ through the relation $j(x,t)=\rho(x,t)v(x,t)$, where $v(x,t)$ is the velocity of the local position density. It describes the change rate of the position density. Due to the conservation of measure, $\rho(x,t)$ and $j(x,t)$ satisfy the continuity equation: \index{random discontinuous motion of particles!description of!position flux density}

\begin{equation}
{{\partial \rho(x,t)}\over{\partial t}}+{{\partial j(x,t)}\over{\partial x}}=0.
\label{}
\end{equation}

\noindent The position density $\rho(x,t)$ and position flux density $j(x,t)$ provide a complete description of the state of random discontinuous motion\index{random discontinuous motion of particles} of a particle.\footnote{It is also possible that the position density $\rho(x,t)$ alone provides a complete description of the state of random discontinuous motion\index{random discontinuous motion of particles} of a particle. Which possibility is the actual one depends on the laws of motion\index{laws of motion}. As we will see later, quantum mechanics requires that a complete description of the state of random discontinuous motion\index{random discontinuous motion of particles} of particles includes both the position density and the position flux density.}\index{random discontinuous motion of particles!description of}\index{random discontinuous motion of particles!as a new interpretation of the wave function}

This description of the motion of a particle can be extended to the motion of many particles. At each instant a quantum system of $N$ particles can be represented by a point in an $3N$-dimensional configuration space\index{configuration space}.
During an arbitrarily short time interval or an infinitesimal time interval around each instant, these particles perform random discontinuous motion\index{random discontinuous motion of particles} in three-dimensional space, and correspondingly, this point performs random discontinuous motion\index{random discontinuous motion of particles} in the configuration space\index{configuration space}. 
Then, similar to the single particle case, the state of the system can be described by the position density $\rho(x_1,x_2,...x_N,t)$ and position flux density $j(x_1,x_2,...x_N,t)$ defined in the configuration space\index{configuration space}. There is also the relation $\rho(x_1,x_2,...x_N,t)=\varrho(x_1,x_2,...x_N,t)$, where $\varrho(x_1,x_2,...x_N,t)$ is the probability density that particle 1 appears in position $x_1$ and particle 2 appears in position $x_2$ ... and particle $N$ appears in position $x_N$. When these $N$ particles are independent with each other, the position density can be reduced to the direct product of the position density for each particle, namely $\rho(x_1,x_2,...x_N,t)=\prod_{i=1}^N\rho(x_i,t)$. \index{random discontinuous motion of particles!description of}\index{random discontinuous motion of particles!as a new interpretation of the wave function}

\subsection{Interpreting the wave function}

Although the motion of particles is essentially discontinuous and random, the discontinuity and randomness of motion are absorbed into the state of motion, which is defined during an infinitesimal time interval around a given instant and described by the position density and position flux density. Therefore, the evolution of the state of random discontinuous motion\index{random discontinuous motion of particles!as a new interpretation of the wave function} of particles may obey a deterministic continuous equation. 
By assuming the nonrelativistic equation of random discontinuous motion\index{random discontinuous motion of particles} is the Schr\"{o}dinger equation\index{Schr\"{o}dinger equation} and considering the form of the resulting continuity equation,\footnote{I have given a rigorous derivation of the free Schr\"{o}dinger equation\index{Schr\"{o}dinger equation!free} in Chapter 5. The derivation is based not on the position density and position flux density, but directly on the wave function. Thus it avoids Wallstrom\index{Wallstrom, Timothy C.}'s (1994) objections to Nelson\index{Nelson, Edward}'s stochastic mechanics (see also Bacciagaluppi\index{Bacciagaluppi, Guido}, 1999; Schmelzer\index{Schmelzer, Ilja}, 2011).} we can obtain the relationship between the position density $\rho(x,t)$, position flux density $j(x,t)$ and the wave function $\psi(x,t)$. $\rho(x,t)$ and $j(x,t)$ can be expressed by $\psi(x,t)$ as follows:\footnote{Note that the relation between $j(x,t)$ and $\psi(x,t)$ depends on the concrete form of the external potential under which the studied system evolves, and the relation given below holds true for an external scalar potential. In contrast, the relation $\rho(x,t)=|\psi(x,t)|^2$ holds true universally, independently of the concrete evolution of the studied system. }

\begin{equation}
\rho(x,t)=|\psi(x,t)|^2,
\label{}
\end{equation}

\begin{equation}
j(x,t)={\hbar \over{2mi}}[\psi^*(x,t){{\partial \psi(x,t)}\over{\partial x}}-\psi(x,t){{\partial \psi^*(x,t)}\over{\partial x}}].
\label{}
\end{equation}

\noindent Correspondingly, the wave function $\psi(x,t)$ can be uniquely expressed by $\rho(x,t)$ and $j(x,t)$ or $v(x,t)$ (except for an overall phase factor): 

\begin{equation}
\psi(x,t)=\sqrt{\rho(x,t)}e^{im\int_{-\infty}^{x}{v(x',t)}dx'/\hbar}.
\label{}
\end{equation}

\noindent In this way, the wave function $\psi(x,t)$ also provides a complete description of the state of random discontinuous motion\index{random discontinuous motion of particles!as a new interpretation of the wave function} of a particle. 
A similar one-to-one relationship between the wave function and position density, position flux density also exists for random discontinuous motion\index{random discontinuous motion of particles} of many particles. For the motion of many particles, the position density and position flux density are defined in a $3N$-dimensional configuration space\index{configuration space}, and thus the many-particle wave function, which is composed of these two quantities, also lives on the $3N$-dimensional configuration space\index{configuration space}.

It is well known that there are several ways to understand objective probability, such as frequentist, propensity, and best-system intepretations (H\'{a}jek\index{H\'{a}jek, Alan}, 2012). In the case of random discontinuous motion\index{random discontinuous motion of particles} of particles, the propensity interpretation seems more appropriate. 
This means that the wave function in quantum mechanics should be regarded not simply as a description of the state of random discontinuous motion\index{random discontinuous motion of particles} of particles, but more suitably as a description of the instantaneous property of the particles that determines their random discontinuous motion\index{random discontinuous motion of particles} at a deeper level.
In particular, the modulus squared of the wave function represents the propensity property of the particles that determines the probability density that they appear in every possible group of positions in space.\footnote{Note that the propensity here denotes single case propensity, as long run propensity theories fail to explain objective single-case probabilities. For a helpful analysis of the single-case propensity interpretation of probability in GRW\index{GRW theory} theory see Frigg \index{Frigg, Roman}and Hoefer\index{Hoefer, Carl} (2007). In addition, it is worth emphasizing that the propensities possessed by particles relate to their objective motion, not to the measurements on them. In contrast, according to the existing propensity interpretations of quantum mechanics, the propensities a quantum system has relate only to measurements; a quantum system possesses the propensity to exhibit a particular value of an observable if the observable is measured on the system (Su\'{a}rez\index{Su\'{a}rez, Mauricio}, 2004, 2007).} 
In contrast, the position density and position flux density, which are defined during an infinitesimal time interval around a given instant, are only a description of the state of the resulting random discontinuous motion\index{random discontinuous motion of particles!as a new interpretation of the wave function} of particles, and they are determined by the wave function. In this sense, we may say that the motion of particles is ``guided" by their wave function in a probabilistic way.

\subsection{On momentum, energy and spin}

I have been analyzing random discontinuous motion\index{random discontinuous motion of particles} of particles in position space. 
Does the picture of random discontinuous motion\index{random discontinuous motion of particles} exist for other observables such as momentum, energy and spin? Since there are also momentum wave functions etc in quantum mechanics, it seems tempting to assume that the above interpretation of the wave function in position space also applies to the wave functions in momentum space etc.
This means that when a particle is in a superposition of the eigenstates of an observable, it also undergoes random discontinuous motion\index{random discontinuous motion of particles} among the eigenvalues of this observable. 
For example, a particle in a superposition of momentum eigenstates also undergoes random discontinuous motion\index{random discontinuous motion of particles} among all momentum eigenvalues. At each instant the momentum of the particle is definite, randomly assuming one of the momentum eigenvalues with probability density given by the modulus squared of the wave function at this momentum eigenvalue, and during an infinitesimal time interval around each instant the momentum of the particle spreads throughout all momentum eigenvalues.

Indeed, it can be argued that if an observable has a definite value at each instant, then, like position, it will also undergo random discontinuous change over time. 
Recall that there are two arguments for the existence of random discontinuous motion\index{random discontinuous motion of particles} of particles. 
The first argument is based on an analysis of the origin of the charge distribution\index{charge distribution} of a one-body quantum system such as an electron (see Section 6.3), and the second argument is based on an analysis of the entangled states\index{entangled states} of a many-body system (see Section 7.2).
Although the first argument applies only to position in general,\footnote{The reason is that the interaction Hamiltonian\index{Hamiltonian!interaction} usually relates to position, not to momentum and energy, and thus the non-existence of electrostatic self-interaction of the charge distribution\index{charge distribution} of an electron only suggests the picture of ergodic motion of a particle in position space.} the second argument applies to all observables of a quantum system which have a definite value at each instant.


However, there is a well-known constraint on ascribing definite values to the observables of a quantum system.
It is the Kochen-Specker theorem\index{Kochen-Specker theorem}, which  proves the impossibility of ascribing sharp values to all observables of a quantum system simultaneously, while preserving the functional relations between commuting observables (Kochen and Specker, 1967\index{Kochen, Simon B.}\index{Specker, Ernst}).
There are two possible ways to deal with this constraint.
The first way is to still ascribe sharp values to all observables of a quantum system and take all of them as physical properties of the system. This requires that the functional relations between commuting observables should not be always preserved for the values assigned to them.\footnote{
Note that incompatible observables can also have sharp values simultaneously and thus have a joint probability distribution when the functional relations between these observables are not preserved for the values assigned to them.
The joint probability distribution of incompatible observables may be a product form, which fulfills the requirement that the marginal for every observable is the same given by quantum mechanics.
The proofs of nonexistence of such a joint probability distribution require that the functional relations between these noncommuting observables should be preserved for the values assigned to them, which is a problematic assumption (see, e.g. Nelson\index{Nelson, Edward}, 2001, p. 95).} 
There is an argument supporting this option.
As Vink\index{Vink, Jeroen C.} (1993) has noticed, it is not necessary to require that the functional relations between commuting observables should hold for the values assigned to them for an arbitrary wave function; rather,  the requirement  must only hold for post-measurement wave functions (which are effectively eigenstates of the measured observable) and for observables that commute with the measured observable, which is enough to avoid conflicts with the predictions of quantum mechanics.\footnote{Note that this strategy to circumvent the Kochen-Specker theorem\index{Kochen-Specker theorem} does not apply to the straightforward solution of the measurement problem\index{measurement problem} provided by the realist $\psi$-epistemic view\index{wave function!epistemic view of}, which explains the collapse of the wave function\index{wave function!collapse} merely as the effect of acquiring new information (see Section 2.2.2). For this solution, the functional relations between commuting observables must hold for the values assigned to them for an arbitrary wave function in order to avoid conflicts with the predictions of quantum mechanics, since these values do not change during the measurement process. In contrast, for the $\psi$-ontic solutions to the measurement problem\index{measurement problem}, the values of commuting observables may change during the measurement process.}
In this way, the picture of random discontinuous motion\index{random discontinuous motion of particles} of particles may exist for all observables including momentum, energy and spin observables along all directions. But the functional relations between commuting observables will fail to hold for the values assigned to them in general.

The second way to deal with the constraint of the Kochen-Specker theorem\index{Kochen-Specker theorem} is to ascribe sharp values to a finite number of observables of a quantum system and only take them as physical properties of the system.
In this way, the functional relations between commuting observables can be preserved for the values assigned to them, while the picture of random discontinuous motion\index{random discontinuous motion of particles} exists only for a finite number of observables.
The question is: Which observables?

First of all, since the proofs of the Kochen-Specker theorem\index{Kochen-Specker theorem} do not prohibit ascribing sharp values to the position, momentum and energy (and their functions) of a quantum system simultaneously, the picture of random discontinuous motion\index{random discontinuous motion of particles} may exist also for momentum and energy.
Next, it needs a more careful analysis whether the picture of random discontinuous motion\index{random discontinuous motion of particles} exists for spin.
Consider a free quantum system with spin one.
The Kochen-Specker theorem\index{Kochen-Specker theorem} requires that only the system's squared spin components along a finite number of directions  can be ascribed sharp values simultaneously (when the functional relations between commuting observables are preserved). 
However, it can be argued that there is only one special direction for the spin of the system.
It is the direction along which the spin of the system is definite.
Thus it seems more reasonable to assume that only the spin observable along this direction can be ascribed a  sharp value and taken as a physical property of the system.\footnote{This analysis of spin also applies to projection operators on a system's Hilbert space\index{Hilbert space} (of dimension greater than or equal to 3).
According to the Kochen-Specker theorem\index{Kochen-Specker theorem}, only a finite number of projection operators can be ascribed sharp values simultaneously when the functional relations between commuting projection operators are preserved for the values assigned to them.
Since there is no special direction for the set of orthogonal rays in the Hilbert space\index{Hilbert space}, all projection operators have the same status and no one is special.
Therefore, it is arguable that the number of the projection operators which can be ascribed sharp values simultaneously should be either zero or infinity, depending on whether the functional relations between commuting observables are preserved for the values assigned to them.
In particular, if this constraint for commuting observables is always satisfied, then projection operators on a system's Hilbert space\index{Hilbert space} do not correspond to properties of the system.}
Under this assumption, the picture of random discontinuous motion\index{random discontinuous motion of particles} does not exist for the spin of a free quantum system.
But if the spin state of a quantum system is entangled with its spatial state due to interaction and the branches of the entangled state are well separated in space, the system in different branches will have different spin, and it will also undergo random discontinuous motion\index{random discontinuous motion of particles} between these different spin states. This is the situation that usually happens during a spin measurement.

The key to decide which way is right is to determine whether the functional relations between commuting observables should be preserved for the values assigned to them for an arbitrary wave function.
In my view, although this requirement is not necessary as Vink\index{Vink, Jeroen C.} (1993) has argued, there is no compelling reason to drop it either.
Moreover, dropping this requirement and taking all observables including infinitely many related spin observables along various directions as physical properties of a quantum system seems to be superfluous in ontology and inconsistent with Occam's razor.
Therefore, I think the second way is the right way to deal with the constraint of the Kochen-Specker theorem\index{Kochen-Specker theorem}.

To sum up, I have argued that the picture of random discontinuous motion\index{random discontinuous motion of particles} also exists for some observables of a quantum system other than position, such as momentum and energy.
But spin is a distinct property; the spin of a free quantum system is always definite along a certain direction, and it does not undergo random discontinuous motion\index{random discontinuous motion of particles}.

\section{Similar pictures of motion through history}


Although the picture of random discontinuous motion\index{random discontinuous motion of particles} of particles seems very strange, similar ideas have been proposed for different purposes in history. In this section, I will briefly discuss these ideas.

\subsection{Epicurus's atomic swerve\index{Epicurus's idea of atomic swerve} and Al-Nazzam's leap motion\index{Al-Nazzam's leap theory}}

The ancient Greek philosopher Epicurus first considered the randomness of motion seriously. He presented the well-known idea of atomic ``swerve" on the basis of Democritus's atomic theory (Englert, 1987). Like Democritus, Epicurus also held that the elementary constituents of the world are atoms, which are indivisible microscopic bits of matter, moving in empty space. He, however, modified Democritus's strict determinism of elementary processes. Epicurus thought that occasionally the atoms swerve from their course at random times and places. Such swerves are uncaused motions. The main reason for introduing them is that they are needed to explain why there are atomic collisions. According to Democritus's atomic theory, the natural tendency of atoms is to fall straight downward at uniform velocity. If this were the only natural atomic motion, the atoms would never have collided with one another, forming macroscopic bodies. Therefore, Epicurus saw it necessary to introduce the random atomic swerves.

In order to solve Zeno's paradoxes, the 9th-century Islamic  theologian  Abu Ishaq Ibrahim Al-Nazzam proposed the idea of discontinuous motion, and called it the theory of ``leap" (Jammer\index{Jammer, Max}, 1974, p.258). He argued that if a finite distance cannot be subdivided into a finite number of fractions but is subject to infinite divisibility, then an object in motion has to perform a leap (tafra) since not every imaginable point in space can be touched. In this way, ``The mobile may occupy a certain place and then proceed to the third place without passing through the intermediate second place on the fashion of a leap." (Jammer\index{Jammer, Max}, 1974, p.259). Moreover, a leap from position $A$ to position $B$ consists of two interlocking sub-events; the original body in position $A$ ceases to exist, and an ``identical" body comes into being in position $B$. 

It can be seen that neither Epicurus's atomic swerve nor Al-Nazzam's leap motion is exactly the  random discontinuous motion\index{random discontinuous motion of particles} of particles; the  atomic swerve is random but lack of explicit discontinuity, while the leap motion is discontinuous but lack of explicit randomness. In a sense, the picture of random discontinuous motion\index{random discontinuous motion of particles} can be considered as an integration of Epicurus's random swerve and Al-Nazzam's discontinuous leap. However, the above reasons for introducing random and discontinuous motion are no longer valid as seen today.\index{Epicurus's idea of atomic swerve} \index{Al-Nazzam's leap theory}

\subsection{Bohr\index{Bohr, Niels}'s discontinuous quantum jumps}

The picture of motion, which was proposed in modern times and more like random discontinuous motion\index{random discontinuous motion of particles}, is Bohr\index{Bohr, Niels}'s discontinuous quantum jumps.

In 1913, Bohr\index{Bohr, Niels} proposed what is now called the Bohr\index{Bohr, Niels} model of the atom (Bohr\index{Bohr, Niels}, 1913). 
He postulated that electrons as particles undergo two kinds of motion in atoms; they either move continuously around the nucleus in certain stationary orbits or discontinuously jump between these orbits. These  discontinuous quantum jumps are supposed to be also truly random and uncaused. 
Indeed, Rutherford\index{Rutherford, Ernest} once raised the issue of causality. In a letter to Bohr\index{Bohr, Niels} dated 20 March 1913, he wrote:

 \begin{quote}  
There appears to me one grave difficulty in your hypothesis, which I have no doubt you fully realize, namely, how does an electron decide what frequency it is going to vibrate at when it passes from one stationary state to the other? It seems to me that you would have to assume that the electron knows beforehand where it is going to stop. (Quoted in Pais\index{Pais, Abraham}, 1991, p.152-153)
 \end{quote}

Bohr\index{Bohr, Niels} admitted that ``the dynamical equilibrium of the systems in the stationary states can be discussed by help of the ordinary mechanics, while the passing of the systems between different stationary states cannot be treated on that basis." (Bohr\index{Bohr, Niels}, 1913, p.7) However, he did not offer a further analysis of his discontinuous quantum jumps.

Although the electrons in Bohr\index{Bohr's atomic model}'s atomic model only undergo the random, discontinuous jumps occasionally (which is like Epicurus's atomic swerve), these jumps let us perceive the flavor of random discontinuous motion\index{random discontinuous motion of particles} of particles in the quantum world.

\subsection{Schr\"{o}dinger\index{Schr\"{o}dinger's snapshot description}'s snapshot description}

In his 1927 Solvay congress report, Schr\"{o}dinger\index{Schr\"{o}dinger, Erwin} gave a further visualizable explanation of his interpretation of the wave function in terms of charge density. He said, 


 \begin{quote}  
The classical system of material points does not really exist, instead there exists something that continuously 
fills the entire space and of which one would obtain a `snapshot' if one dragged the classical system, with the camera shutter open,  through {\em all} its configurations, the representative point in $q$-space spending in each volume element $d\tau$ a time that is proportional to the {\em instantaneous} value of $\psi\psi^*$.  (Bacciagaluppi\index{Bacciagaluppi, Guido} and Valentini\index{Valentini, Antony}, 2009, p.409)
 \end{quote} 

If this description in terms of a snapshot is understood literally, then it seems to suggest that a quantum system is composed of point particles, and the continuous charge distribution\index{charge distribution} of the system in space is formed by the ergodic motion of the system's point particles. 
Moreover, since the snapshot picture also requires that the spending time of the particles in each volume element is proportional to the  instantaneous value of $\psi\psi^*$ there, it seems that the ergodic motion of the particles should be not continuous but discontinuous (as I have argued in the last chapter).

Indeed, what Schr\"{o}dinger\index{Schr\"{o}dinger, Erwin} said later in his report seems to also support this analysis. He said,

 \begin{quote}  
The pure field theory is not enough, it has to be supplemented by performing a 
kind of {\em individualisation} of the charge densities coming from the single point charges of the
classical model, where however each single `individual' may be spread over the whole of space, so that they overlap. (Bacciagaluppi\index{Bacciagaluppi, Guido} and Valentini\index{Valentini, Antony}, 2009, p.415)
 \end{quote} 

However, it is arguable that this picture of ergodic motion of particles was not the actual picture in Schr\"{o}dinger\index{Schr\"{o}dinger, Erwin}'s mind then, and he did not even mean by the above statement that the snapshot description is real. The reason is that Schr\"{o}dinger\index{Schr\"{o}dinger, Erwin} later gave a clearer explanation of the above statement in the discussion of his report, which totally avoided the snapshot description. He said,\index{Schr\"{o}dinger's snapshot description}

 \begin{quote}  
It would seem that my description in terms of a snapshot was not very fortunate, since it has been misunderstood. Perhaps the following explanation is clearer. The interpretation of Born is well-known, who takes $\psi\psi^*d\tau$ to be the probability for the system being in the volume element $d\tau$ of the configuration space\index{configuration space}. Distribute a very large number $N$ of systems in the configuration space\index{configuration space}, taking the above probability as `frequency function'. Imagine these systems as superposed in real space, dividing however by $N$ the charge of  each point mass in each system. In this way, in the limiting case where $N=\infty$ one obtains the wave mechanical picture of the system. (Bacciagaluppi\index{Bacciagaluppi, Guido} and Valentini\index{Valentini, Antony}, 2009, p.423)
 \end{quote}  

Schr\"{o}dinger\index{Schr\"{o}dinger, Erwin} never referred to his snapshot description, which he called a ``somewhat naive but quite concrete idea", in his later publications.


It is interesting to note that like Schr\"{o}dinger\index{Schr\"{o}dinger, Erwin}'s snapshot description, Bell once gave a statement which may also suggest the picture of random discontinuous motion\index{random discontinuous motion of particles} of particles. In his well-known article ``Against Measurement", after discussing Schr\"{o}dinger\index{Schr\"{o}dinger, Erwin}'s interpretation of the wave function in terms of charge density, Bell wrote:\index{Schr\"{o}dinger's snapshot description}

\begin{quote}
Then came the Born interpretation. The wavefunction gives not the density of stuff\index{density of stuff}, but gives rather (on squaring its modulus) the density of probability. Probability of \emph{what} exactly? Not of the electron  \emph{being} there, but of the electron being  \emph{found} there, if its position is `measured'. 

Why this aversion to `being' and insistence on `finding'? The founding fathers were unable to form a clear picture of things on the remote atomic scale. (Bell,\index{Bell, John S.} 1990)
\end{quote}

It seems that by this statement Bell supported the following two claims: (1) the wave function should give (on squaring its modulus) the density of probability of  the electron being in certain position in space, and (2) the wave function should also give the density of stuff\index{density of stuff}.
The key is to understand the meaning of probability in the first claim. 
If the probability means subjective probability, then the electron will be in a definite position which we don't know with certainty. But this result contradicts the second claim. The wave function of the electron giving the density of stuff\index{density of stuff} means that the electron cannot be in only one position, but fill in the whole space in certain way with local density given by the modulus squared of its wave function.

On the other hand, if the above probability means objective probability, then, as I have argued in the last chapter, the two claims can be reconciled and the resulting picture will be random discontinuous motion\index{random discontinuous motion of particles} of particles. The electron appears in every position in space with objective probability density given by the modulus squared of its wave function there, and during an arbitrarily short time interval such random discontinuous motion\index{random discontinuous motion of particles} forms the density of stuff\index{density of stuff} everywhere, which is also proportional to the modulus squared of the wave function of the electron.


\subsection{Bell's Everett (?) theory\index{Bell's Everett (?) theory}}

Although the above analysis may be an overinterpretation of Bell's statement, it can be argued that Bell's Everett (?) theory\index{Bell's Everett (?) theory} indeed suggests the picture of random discontinuous motion\index{random discontinuous motion of particles} of particles.

According to Bell (\index{Bell, John S.}1981), Everett's theory can be regarded as Bohm's theory\index{Bohm's theory} without the continuous particle trajectories. Thus

\begin{quote}
instantaneous classical configuration $x$ are supposed to exist, and to be distributed in the comparison class of possible worlds with probability $|\psi|^2$. But no pairing of configuration at different times, as would be effected by the existence of trajectories, is supposed. (Bell,\index{Bell, John S.} 1987, p.133)
\end{quote}

Obviously, in Bell's Everett (?) theory\index{Bell's Everett (?) theory}, a quantum system is composed of particles, which have a definite position in space at each instant. Moreover, these particles jump among all possible configurations with probability $|\psi|^2$, and such jumps are random and discontinuous. This is clearly the picture of  random discontinuous motion\index{random discontinuous motion of particles} of particles. I will analyze Bell's Everett (?) theory\index{Bell's Everett (?) theory} in more detail in the next chapter.

\chapter{Implications for solving the measurement problem\index{measurement problem}}  \label{chap:SMP}

There are two fundamental problems in the conceptual foundations of quantum mechanics. The first one concerns the physical meaning of the wave function in the theory. The second one is the measurement problem\index{measurement problem}. 
I have analyzed the ontological status and meaning of the wave function, and suggested a new ontological  interpretation of the wave function in terms of random discontinuous motion\index{random discontinuous motion of particles} (RDM) of particles. 
In this chapter, I will further analyze possible implications of this new interpretation of the wave function for the solutions to the measurement problem\index{measurement problem}. That is, I will analyze how to solve the measurement problem\index{measurement problem} when assuming the suggested interpretation of the wave function is true.
The analysis will also answer whether the ontology described by the wave function is enough to account for our experience, e.g. whether additional ontology such as many worlds or hidden variables is needed to account for our definite experience.


\section{The measurement problem\index{measurement problem} revisited}


Before analyzing the implications of my new interpretation of the wave function for solving the measurement problem\index{measurement problem}, I will first give a new formulation of the problem and present an analysis of its solutions based on the formulation.

\subsection{A new formulation}
Quantum mechanics assigns a wave function to an appropriately prepared physical system and specifies that the evolution of the wave function is governed by the Schr\"{o}dinger equation\index{Schr\"{o}dinger equation}. However, when assuming the wave function is a complete description of the system, the linear dynamics is apparently incompatible with the existence of definite results of measurements on the system. This leads to the measurement problem\index{measurement problem}.\footnote{The measurement problem\index{measurement problem} was originally fomulated in the context of \index{quantum mechanics!standard formulation of}standard quantum mechanics. The theory does not define measurement precisely, and as a result, it is unclear when the collapse postulate\index{quantum mechanics!standard formulation of!collapse postulate in} can be applied during the evolution of the wave function (Albert\index{Albert, David Z.}, 1992).} 

According to Maudlin\index{Maudlin, Tim}'s (1995a) formulation, the measurement problem\index{measurement problem!Maudlin's formulation of} originates from the incompatibility of the following three claims:

(C1). the wave function of a physical system is a complete description of the system; \\
\indent (C2). the wave function always evolves in accord with a linear dynamical equation, e.g. the Schr\"{o}dinger equation\index{Schr\"{o}dinger equation}; \\
\indent  (C3). each measurement has a definite result (which is one of the possible measurement results whose probability distribution satisfies the Born rule\index{Born rule}). 

The proof of the inconsistency of these three claims is familiar. 
Suppose a measuring device $M$ measures the $x$-spin of a spin one-half system $S$ that is in a superposition of two different $x$-spins $1/\sqrt{2}(\ket{up}_S+\ket{down}_S)$. If (C2) is correct, then the state of the composite system after the measurement must evolve into the superposition of $M$ recording $x$-spin up and $S$ being $x$-spin up and $M$ recording $x$-spin down and $S$ being $x$-spin down:

\begin{equation}
1/\sqrt{2}(\ket{up}_S \ket{up}_M+\ket{down}_S \ket{down}_M).
\label{ds}
\end{equation}
\noindent The question is what kind of state of the measuring device this represents.\index{measurement problem!Maudlin's formulation of}
If (C1) is also correct, then this superposition must specify every physical fact about the measuring device. But by symmetry of the two terms in the superposition, this superposed state cannot describe a measuring device recording either $x$-spin up or $x$-spin down. Thus if (C1) and (C2) are correct, (C3) must be wrong.

It can be seen that there are three basic approaches to solving the measurement problem\index{measurement problem} thus formulated.\footnote{It has been debated whether the consistent histories approach to quantum mechanics can solve the measurement problem (Griffiths\index{Griffiths, Robert B.}, 1984, 2002, 2013, 2015; \index{Omn\`{e}s, Roland}Omn\`{e}s, 1988; 1999;  Hartle\index{Hartle, James} and Gell-Mann\index{Gell-Mann, Murray}, 1993; Dowker\index{Dowker, Fay} and Kent\index{Kent, Adrian}, 1995, 1996; Okon\index{Okon, Elias} and Sudarsky\index{Sudarsky, Daniel}, 2014a, 2014b, 2015). I will not discuss this approach here.}\index{measurement problem!solutions to, \emph{see} Bohm's theory, Everett's theory, and collapse theories} 
The first approach is to deny the claim (C1), and add some additional variables and corresponding dynamics to explain definite measurement results. A well-known example is Bohm's theory\index{Bohm's theory} (Bohm,\index{Bohm, David} 1952; Goldstein\index{Goldstein, Sheldon}, 2013).\footnote{Two other interesting examples are the modal interpretations (Lombardi and Dieks\index{Lombardi, Olimpia}\index{Dieks, Dennis}, 2012) and the two-state-vector formalism (Aharonov\index{Aharonov, Yakir} and Vaidman\index{Vaidman, Lev}, 2008). In the two-state-vector formalism, the ontology is composed of the normal wave function and another backwards evolving wave function, which can be taken as a hidden variable. It has been demonstrated that a special final boundary condition on the universe may propagate backwards in time to account for all measurement results and draw the line between classical and quantum mechanics (Aharonov\index{Aharonov, Yakir}\index{Cohen, Eliahu} et al, 2014).}
The second approach is to deny the claim (C2), and revise the Schr\"{o}dinger equation\index{Schr\"{o}dinger equation} by adding certain nonlinear and stochastic evolution terms to explain definite measurement results. Such theories are called collapse theories\index{collapse theories} (J\'{a}nossy\index{J\'{a}nossy, Lajos}, 1952; Ghirardi\index{Ghirardi, GianCarlo}, 2016).
The third approach is to deny the claim (C3), and assume the existence of many equally real worlds to accommodate all possible results of measurements (Everett,\index{Everett, Hugh, III} 1957; Barrett and\index{Barrett, Jonathan} Byrne, 2012; Barrett\index{Barrett, Jeffrey A.}, 2014; Vaidman\index{Vaidman, Lev}, 2016). In this way, it may also explain definite measurement results in each world including our own world. This approach is called Everett's interpretation of quantum mechanics or Everett's theory\index{Everett's theory}.\index{measurement problem!Maudlin's formulation of}


It has been realized that the measurement problem\index{measurement problem} in fact has two levels: the physical level and the mental level, and it is essentially the determinate-experience problem (Barrett, 1999, 2005\index{Barrett, Jeffrey A.}). \index{determinate-experience problem, \emph{see} measurement problem}
The problem is not only to explain how the linear dynamics can be compatible with the existence of  definite measurement results obtained by physical devices, but also, and more importantly,  to explain how the linear dynamics can be compatible with the existence of definite experience of conscious observers. 
However, the mental aspect of the measurement problem\index{measurement problem} is ignored in Maudlin\index{Maudlin, Tim}'s (1995a) formulation. \index{measurement problem!Maudlin's formulation of}
Here I will suggest a new formulation of the measurement problem\index{measurement problem} which lays more stress on the psychophysical connection. In this formulation, the measurement problem\index{measurement problem} originates from the incompatibility of the following two assumptions:

(A1). the mental state of an observer supervenes on her wave function; \\ 
\indent (A2). the wave function always evolves in accord with a linear dynamical equation, e.g. the Schr\"{o}dinger equation\index{Schr\"{o}dinger equation}. \\



The proof of the inconsistency of these two claims is similar to the above proof. \index{measurement problem!a new formulation of}
Suppose an observer $M$ measures the $x$-spin of a spin one-half system $S$ that is in a superposition of two different $x$-spins, $1/\sqrt{2}(\ket{up}_S+\ket{down}_S)$. If (A2) is correct, then the physical state of the composite system after the measurement will evolve into the superposition of $M$ recording $x$-spin up and $S$ being $x$-spin up and $M$ recording $x$-spin down and $S$ being $x$-spin down:

\begin{equation}
1/\sqrt{2}(\ket{up}_S \ket{up}_M+\ket{down}_S \ket{down}_M).
\label{os}
\end{equation}

\noindent If (A1) is also correct, then the mental state of the observer $M$ will supervene on this superposed wave function. Since the mental states corresponding to the physical states  $\ket{up}_M$ and $\ket{down}_M$ differ in their mental content, the observer M being in the superposition (\ref{os}) will have a conscious experience different from the experience of M being in each branch of the superposition by the symmetry of the two branches. In other words, the record that M is consciously aware of is neither x-spin up nor x-spin down when she is physically in the superposition (\ref{os}). This is inconsistent with experimental observations. Therefore, (A1) and (A2) are incompatible.\index{measurement problem!a new formulation of}


By this new formulation of the measurement problem\index{measurement problem}, we can look at the three major solutions of the problem from a new angle.\index{measurement problem!solutions to, \emph{see} Bohm's theory, Everett's theory, and collapse theories}
First of all, the solution to the measurement problem\index{measurement problem} must deny either the claim (A1) or the claim (A2).\index{measurement problem!a new formulation of}
If (A1) is correct (as usually thought), then (A2) must be wrong. In other words, if the mental state of an observer supervenes on her wave function, then the Schr\"{o}dinger equation\index{Schr\"{o}dinger equation} must be revised and the solution to the  measurement problem\index{measurement problem} will be along the direction of collapse theories\index{collapse theories}.
On the other hand,  if (A2) is correct, then (A1) must be wrong. This means that if the wave function always evolves in accord with the Schr\"{o}dinger equation\index{Schr\"{o}dinger equation}, then the mental state of an observer cannot supervene on her wave function.
There are two other forms of psychophysical supervenience.\index{measurement problem!and psychophysical supervenience}
One is that when a wave function corresponds to many observers, the mental state of each observer supervenes on one branch of the wave function,\footnote{If an observer always has a unique mental state and the mental state supervenes only on a certain branch of her wave function, then the psychophysical supervenience will be obviously violated. This is the case of the single-mind theory (Albert\index{Albert, David Z.} and Loewer\index{Loewer, Barry}, 1988; Barrett, 1999\index{Barrett, Jeffrey A.}). Although I will not discuss this theory below, some of my analyses also apply to it.} and the other is that  the mental state of an observer supervenes on other additional variables. 
The first form corresponds to Everett's theory\index{Everett's theory}, and the second form corresponds to Bohm's theory\index{Bohm's theory}.

To sum up,  the three major solutions to the measurement problem\index{measurement problem} correspond to three different forms of psychophysical supervenience. \index{measurement problem!and psychophysical supervenience}
In fact, there are only three types of physical states on which the mental state of an observer may supervene, which are (1) the wave function, (2) certain branches of the wave function, and (3) other additional variables.
The question is: Exactly what physical state does the mental state of an observer supervene on?
It can be expected that an analysis of this question will help solve the measurement problem\index{measurement problem}.\index{measurement problem!a new formulation of}

\subsection{Everett's theory\index{Everett's theory}}

I will first analyze Everett's theory\index{Everett's theory}.\index{Everett's theory!psychophysical supervenience in}
The theory claims that for the above post-measurement state (\ref{os}) there are two observers, and each of them is consciously aware of a definite record, either x-spin up or x-spin down.\footnote{Note that in Wallace\index{Wallace, David}'s (2012) latest formulation of Everett's theory\index{Everett's theory} the number of the emergent observers after the measurement is not definite due to the imperfectness of decoherence. I will discuss this point later. } 

There are in general two ways of understanding the notion of multiplicity in Everett's  theory.\index{Everett's theory!multiplicity in}
One is the strong form which claims that there are two \emph{physical} observers (in material content) after the quantum measurement (see, e.g. DeWitt\index{DeWitt, Bryce S.} and Graham\index{Graham, R. Neill}, 1973). The resulting theory is called many-worlds theory\index{many-worlds theory}.\index{Everett's theory!strong form of, \emph{see also} many-worlds theory}
In this theory, a physical observer always has a unique mental state, and the mental state also supervenes on the whole physical state of the observer, although which may be only a branch of the post-measurement superposition.  \index{Everett's theory!strong form of, \emph{see also} many-worlds theory!inconsistency problem of}\index{Everett's theory!strong form of, \emph{see also} many-worlds theory!mass-energy non-conservation problem of}
Thus this theory is consistent with the common assumption of psychophysical supervenience, according to which the mental state of a physical observer supervenes on her whole physical state.\index{Everett's theory!psychophysical supervenience in}
As is well known, however, this theory has serious problems such as violation of mass-energy conservation and inconsistency with the dynamical equations (Albert\index{Albert, David Z.} and Loewer\index{Loewer, Barry}, 1988). The problem of inconsistency can also be seen as follows. The existence of many worlds is only relative to decoherent observers, not relative to non-decoherent observers, who can measure the whole superposition corresponding to the many worlds (e.g. by protective measurements\index{protective measurements}) and confirm that there is no increase in the total mass-energy and number of particles.

The other way of understanding the notion of multiplicity is the weak form\index{Everett's theory!weak form of, \emph{see also} many-minds theory} which claims that there is one \emph{physical} observer (in material content), but there are two \emph{mental} observers or two mental states of the same physical observer, after the quantum measurement (see, e.g. Zeh\index{Zeh, H. Dieter}, 1981, 2016a, 2016b).\index{Everett's theory!psychophysical supervenience in}
Wallace\index{Wallace, David}'s (2012) formulation of Everett's theory\index{Everett's theory} is arguably this view in nature (see also Kent\index{Kent, Adrian}, 2010).\footnote{In Wallace\index{Wallace, David}'s formulation, it is claimed that there are more than one emergent physical observers after the quantum measurement, but their existence is only in the sense of branch structure (i.e. the structure of certain parts of the whole physical state), not in the sense of material content. Therefore, strictly speaking, there is only one physical observer with her whole physical state after the quantum measurement. This means that Wallace\index{Wallace, David}'s formulation of Everett's theory\index{Everett's theory!Wallace's formulation of} is essentially a weak form of the theory. If such a theory is also regarded as a many-worlds theory\index{many-worlds theory}, then these worlds should be not physical worlds but mental worlds. In my opinion, it is the failure to clearly distinguish between the weak form and the strong form of Everett's theory\index{Everett's theory} or between physical worlds and mental worlds in the literature that causes much confusion in understanding the theory. In addition, it is worth noting that even if assuming the functionalist approach to consciousness and considering decoherence, the existence of many structures in a wave function does not necessarily lead to the appearance of many emergent worlds including emergent observers (Gao, 2016).} 
In order to derive the multiplicity prediction of the weak form\index{Everett's theory!weak form of, \emph{see also} many-minds theory} of Everett's theory\index{Everett's theory}, 
a physical observer cannot always have a unique mental state, and when she has more than one mental states, each mental state does not supervene on her whole physical state either.
For example, when the observer is in one of the two physical states  $\ket{up}_M$ and $\ket{down}_M$, she has a unique mental state and the mental state also supervenes on her whole physical state. While when she is in a superposition of these two physical states such as (\ref{os}), she has two mental states but each mental state does not supervene on her whole physical state; rather, each mental state supervenes only on a part of the whole physical state, such as one of the two terms in the superposition (\ref{os}).
Therefore, different from the strong form\index{Everett's theory!strong form of, \emph{see also} many-worlds theory} of Everett's theory\index{Everett's theory},  the weak form\index{Everett's theory!weak form of, \emph{see also} many-minds theory} of Everett's theory\index{Everett's theory!psychophysical supervenience in} obviously violates the common assumption of psychophysical supervenience.\footnote{The common assumption of psychophysical supervenience is arguably reasonable. A whole physical state is independent, while any two parts of the state are not independent; once one part is selected, the other part will be also fixed. Since a mental state is usually assumed to be autonomous, it is arguable that a mental state should supervene on a whole physical state, not on any part of the state.}

One may object that the psychophysical supervenience is not really violated here, since the sum of all mental states of a physical observer do supervene on her whole physical state.
However, this objection seems invalid.\index{Everett's theory!psychophysical supervenience in}
First of all, since these mental states may be incompatiable with each other, the sum of these mental states is arguably not a valid mental state. For instance, the combination of the mental state of seeing a cat and the mental state of not seeing a cat seems meaningless. 
Thus, strictly speaking, this form of supervenience is not really a form of psychophysical supervenience.
Next, it can be further argued that if one wants psychophysical supervenience, then one presumably wants the mental state that determines one's experience and the only mental state to which one has epistemic access to supervene on one's physical state (Barrett, 1999, p.196). But in the weak form\index{Everett's theory!weak form of, \emph{see also} many-minds theory} of Everett's theory\index{Everett's theory}, when a physical observer has many mental states, each mental observer can only have epistemic access to her own mental state, and thus the sum of all these mental states is certainly not a mental state that satisfies this requirement of psychophysical supervenience. 

In fact, it can be shown by a concrete example that in the weak form\index{Everett's theory!weak form of, \emph{see also} many-minds theory} of Everett's theory\index{Everett's theory}, although the total wave function of a physical observer does not change after a unitary evolution, the mental state of each mental observer, which supervenes on each branch of the wave function, may change. 
Consider again the post-measurement superposition (\ref{os}). 
There exists a unitary time evolution operator, which changes the first branch of the superposition to its second branch and changes the second branch to the first branch. It is similar to the NOT gate for a single q-bit. 
Then after the evolution the total superposition does not change. 
However, after the evolution the mental state which supervenes on either branch of the superposition will change; the mental state supervening on the first branch will change from being aware of x-spin up to being aware of x-spin down, and  the mental state supervening on the second branch will change from being aware of x-spin down to being aware of x-spin up.s
Therefore,  even though the sum of all mental states of a physical observer supervenes on her whole physical state, it seems that the psychophysical supervenience is also violated by the weak form\index{Everett's theory!weak form of, \emph{see also} many-minds theory} of Everett's theory\index{Everett's theory}. 



Certainly, one may also insist that the common assumption of psychophysical supervenience is not universally valid in the quantum domain. 
But even though this is true, one still needs to explain why, in particular, why this assumption applies to the physical states  $\ket{up}_M$ and $\ket{down}_M$, but not to any superposition of them. This is similar to the preferred basis problem.\index{Everett's theory!weak form of, \emph{see also} many-minds theory!preferred basis problem of}
It seems that the only difference one can think is that being in the superposition the physical observer has no definite mental state which contains a definite conscious experience about the measurement result, while being in each branch of the superposition, $\ket{up}_M$ or $\ket{down}_M$, she has a definite mental state which contains a definite conscious experience about the measurement result.
However, it has been argued that the common assumption of psychophysical supervenience can be applied to any physical state, and moreover, under this assumption a physical observer being in a post-measurement superposition such as (\ref{os}) also has a definite mental state which contains a definite conscious experience about the measurement result (Gao\index{Gao, Shan}, 2016).
Note that these analyses also apply to the many-minds theory\index{many-minds theory}, which is similar to the weak form\index{Everett's theory!weak form of, \emph{see also} many-minds theory} of Everett's theory\index{Everett's theory} in many aspects (Albert\index{Albert, David Z.} and Loewer\index{Loewer, Barry}, 1988; Barrett, 1999\index{Barrett, Jeffrey A.}).\index{Everett's theory!psychophysical supervenience in}

Finally, I will give a brief comment on the relationship between Everett's theory\index{Everett's theory!and decoherence} and decoherence. It is usually thought that the appearance of many observers after a quantum measurement is caused by decoherence. 
However, even if this claim is true for the strong form\index{Everett's theory!strong form of, \emph{see also} many-worlds theory} of Everett's theory\index{Everett's theory}, it seems that it cannot be true for the weak form\index{Everett's theory!weak form of, \emph{see also} many-minds theory} of the theory. 
The reason is that the generation of a superposed state of a physical observer (e.g. a superposition of two physical states  $\ket{up}_M$ and $\ket{down}_M$), as well as the psychophysical supervenience, have nothing to do with decoherence. In addition, resorting to decoherence seems to cause a further difficulty for the application of the doctrine of psychophysical supervenience. Since decoherence is never perfect, there will be no definite physical states on which the mental states can supervene.\footnote{This problem is more serious when assuming that the mental states cannot be reduced to the physical states.} 


\subsection{Bohm's theory\index{Bohm's theory}}
Let us now turn to Bohm's theory\index{Bohm's theory!psychophysical supervenience in}. In  this theory, there are two suggested forms of psychophysical supervenience. 
The first one is that the mental state supervenes on the branch of the wave function occupied by Bohmian particle\index{Bohm's theory!Bohmian particles in}s, and the second one is that the mental state supervenes only on the (relative) positions of Bohmian particle\index{Bohm's theory!Bohmian particles in}s.


The first form of psychophysical supervenience has been the standard view until recently, according to which the mental state of an observer being in a post-measurement superposition like (\ref{os}) supervenes only on the branch of the superposition occupied by her Bohmian particle\index{Bohm's theory!Bohmian particles in}s. 
Indeed, Bohm initially assumed this form of psychophysical supervenience. He said: ``the packet entered by the apparatus [hidden] variable...  determines the actual result of the measurement, which the observer will obtain when she looks at the apparatus." (Bohm,\index{Bohm, David} 1952, p.182). 
In this case, the role of the Bohmian particle\index{Bohm's theory!Bohmian particles in}s is merely to select the branch from amongst the other non-overlapping branches of the superposition.\index{Bohm's theory!psychophysical supervenience in}

The first form of psychophysical supervenience is also called Bohm's result assumption (Brown\index{Brown, Harvey. R.} and Wallace\index{Wallace, David}, 2005), and it has been widely argued to be problematic (Stone\index{Stone, Abraham D.}, 1994; Brown\index{Brown, Harvey. R.}, 1996; Zeh\index{Zeh, H. Dieter}, 1999; Brown\index{Brown, Harvey. R.} and Wallace\index{Wallace, David}, 2005; Lewis,\index{Lewis, Peter J.} 2007). 
For example, according to Brown\index{Brown, Harvey. R.} and Wallace\index{Wallace, David} (2005), in the general case each of the non-overlapping branches in the post-measurement superposition has the same credentials for representing a definite measurement result as the single branch does in the predictable case (i.e. the case in which the measured system is in an eigenstate of the measured observable). The fact that only one of them carries the Bohmian particle\index{Bohm's theory!Bohmian particles in}s does nothing to remove these credentials from the others, and adding the particles to the picture does not interfere destructively with the empty branches either.\index{Bohm's theory!psychophysical supervenience in}

In my view, the main problem with the first form of psychophysical supervenience is that the empty branches and the occupied branch have the same qualification to be supervened by the mental state.
Moreover, although it is imaginable that the Bohmian particle\index{Bohm's theory!Bohmian particles in}s may have influences on the occupied branch, e.g. disabling it from being supervened by the mental state, it is hardly conceivable that the Bohmian particle\index{Bohm's theory!Bohmian particles in}s have influences on all other empty branches, e.g. disabling them from being supervened by the mental state. 

In view of the first form of psychophysical supervenience being problematic, most Bohmians today seem to support the second form of psychophysical supervenience (e.g. Holland\index{Holland, Peter R.}, 1993, p.334), although they sometimes do not state it explicitly (e.g. Maudlin\index{Maudlin, Tim}, 1995b). 
If assuming this form of psychophysical supervenience, namely assuming the mental state supervenes only on the (relative) positions of Bohmian particle\index{Bohm's theory!Bohmian particles in}s, then the above problems can be avoided.\index{Bohm's theory!psychophysical supervenience in}
However, it has been argued that this form of psychophysical supervenience is inconsistent with the popular functionalist approach to consciousness (Brown\index{Brown, Harvey. R.} and Wallace\index{Wallace, David}, 2005; see also Bedard, 1999). The argument can be summarized as follows. If the functionalist assumption is correct, for consciousness to supervene on the Bohmian particle\index{Bohm's theory!Bohmian particles in}s but not the wave function, the Bohmian particle\index{Bohm's theory!Bohmian particles in}s must have some functional property that the wave function do not share. But the functional behaviour of the Bohmian particle\index{Bohm's theory!Bohmian particles in}s is arguably identical to that of the branch of the wave function in which they reside. 
Moreover, it has been argued that this form of psychophysical supervenience also leads to another problem of allowing superluminal signaling\index{superluminal signaling}\index{Bohm's theory!superluminal signaling in} (Brown\index{Brown, Harvey. R.} and Wallace\index{Wallace, David}, 2005; Lewis,\index{Lewis, Peter J.} 2007). If the mental state supervenes on the positions of Bohmian particle\index{Bohm's theory!Bohmian particles in}s, then an observer can in principle know the configuration of the Bohmian particle\index{Bohm's theory!Bohmian particles in}s in her brain with a greater level of accuracy than that defined by the wave function. This will allow superluminal signaling\index{superluminal signaling} and lead to a violation of the no-signaling theorem\index{quantum mechanics!no-signaling theorem of!violation of} (Valentini\index{Valentini, Antony}, 1992).\footnote{In my view, this problem is not as serious as usually thought, since the existence of such superluminal signaling\index{superluminal signaling} is not inconsistent with experience, and superluminal signaling\index{superluminal signaling} may also exist in other theories such as collapse theories\index{collapse theories}\index{collapse theories!superluminal signaling in} (Gao\index{Gao, Shan}, 2004, 2014d).} \index{Bohm's theory!psychophysical supervenience in}


A more serious problem with the second form of psychophysical supervenience, in my view, is that it seems inconsistent with the Born rule\index{Born rule}. 
Consider again an observer being in the post-measurement superposition (\ref{os}).
According to the Born rule\index{Born rule}, the modulus squared of the amplitude of each branch of this superposition represents the probability of obtaining the measurement result corresponding to the branch.
For example, the modulus squared of the amplitude of the branch $\ket{up}_M$ represents the probability of obtaining the x-spin up result.
This means that the Born rule\index{Born rule} requires that the quantities that represent the measurement results should be correlated with these branches of the superposition.\footnote{Note that this requirement is independent of whether the wave function is ontic or not.}
Then, in order that the measurement result is represented by the relative positions of Bohmian particle\index{Bohm's theory!Bohmian particles in}s as required by the second form of psychophysical supervenience, there must exist a correspondence between different branches of the superposition and different relative positions of the Bohmian particle\index{Bohm's theory!Bohmian particles in}s, and in particular, the relative positions of the Bohmian particle\index{Bohm's theory!Bohmian particles in}s corresponding to different branches of the superposition must be different.\index{Bohm's theory!psychophysical supervenience in}

However, Bohm's theory\index{Bohm's theory} does not give such a corresponding relationship, and thus it is at least incomplete when assuming the second form of psychophysical supervenience.
Moreover, it can be argued that the corresponding relationship may not exist. 
The probability of the Bohmian particle\index{Bohm's theory!Bohmian particles in}s appearing at a location in configuration space\index{configuration space} is equal to the modulus squared of the amplitude of the wave function at the location.
It is permitted by the linear dynamics that one branch of the post-measurement superposition (\ref{os}) is a spatial translation of the other branch, e.g. the spatial part of $\ket{down}_M$ is $\psi(x_1,y_1,z_1,..., x_N,y_N,z_N,t)$, and the spatial part of $\ket{up}_M$ is $\psi(x_1+a,y_1,z_1,..., x_N+a,y_N,z_N,t)$.
Then if a relative configuration of the Bohmian particle\index{Bohm's theory!Bohmian particles in}s appears in the region of one branch in configuration space\index{configuration space}, it may also appear in the region of the other branch in configuration space\index{configuration space}. Moreover, the probability densities that the configuration appears in both regions are the same. 
This means that a relative configuration of the Bohmian particle\index{Bohm's theory!Bohmian particles in}s can correspond to either branch of the superposition, and there does not exist a corresponding relationship between different branches of the superposition and different relative configurations of the Bohmian particle\index{Bohm's theory!Bohmian particles in}s.\index{Bohm's theory!psychophysical supervenience in}

Since the Born rule\index{Born rule} requires that there should exist such a corresponding relationship when assuming the second form of psychophysical supervenience, the non-existence of the corresponding relationship then means that the second form of psychophysical supervenience is inconsistent with the Born rule\index{Born rule}. This can be seen more clearly as follows. If assuming the second form of psychophysical supervenience, namely if the measurement result is represented by the relative positions of Bohmian particle\index{Bohm's theory!Bohmian particles in}s, then no matter which branch of the post-measurement superposition the Bohmian particle\index{Bohm's theory!Bohmian particles in}s reside in, the measurement result will be the same. This is obviously inconsistent with the Born rule\index{Born rule}.\index{Bohm's theory!psychophysical supervenience in}

Finally, I note that the above analysis of psychophysical supervenience also raises a doubt about the whole strategy of Bohm's theory\index{Bohm's theory} to solve the measurement problem\index{measurement problem}. Why add hidden variables such as positions of Bohmian particle\index{Bohm's theory!Bohmian particles in}s to quantum mechanics? It has been thought that adding these variables which have definite values at every instant is enough to ensure the definiteness of measurement results and further solve the measurement problem\index{measurement problem}.\index{Bohm's theory!psychophysical supervenience in}
However, if the mental state cannot supervene on these additional variables, then even though these variables have definite values at every instant, they are unable to account for our definite experience and thus do not help solve the measurement problem (see also Barrett, 2005)\index{measurement problem}.



\subsection{Collapse theories}

I have argued that one will meet some serious difficulties if assuming the mental state of an observer supervenes either on certain branches of her wave function or on other additional variables, and thus it seems that Everett's and Bohm's theories are not promising solutions to the measurement problem\index{measurement problem}.\index{collapse theories!psychophysical supervenience in}
This also suggests that the mental state of an observer supervenes directly on her wave function, and 
collapse theories\index{collapse theories} may be in the right direction to solve the measurement problem\index{measurement problem}.

However, it has been known that collapse theories\index{collapse theories} are still plagued by a few serious problems such as the tails problem\index{collapse theories!tails problem} (Albert\index{Albert, David Z.} and Loewer\index{Loewer, Barry}, 1996). In particular, the structured tails problem\index{collapse theories!tails problem!structured} has not been solved in a satisfactory way (see McQueen\index{McQueen, Kelvin J.}, 2015 and references therein).
The problem is essentially that collapse theories\index{collapse theories} such as the GRW\index{GRW theory} theory predicts that the post-measurement state is still a superposition of different outcome branches with similar structure (although the modulus squared of the coefficient of one branch is close to one), and they need to explain why high modulus-squared values are macro-existence determiners (McQueen\index{McQueen, Kelvin J.}, 2015).
In my view, the key to solving the structured tails problem\index{collapse theories!tails problem!structured} is not to analyze the connection between high modulus-squared values and macro-existence, but to analyze the connection between these values and our experience of macro-existence, which requires us to further analyze how the mental state of an observer supervenes on her wave function.

Admittedly this is an unsolved, difficult issue. I will give a few speculations here.\index{collapse theories!tails problem!a solution to structured}
I conjecture that the mental content of an observer being in a post-measurement superposition like (\ref{os}) is composed of the mental content corresponding to every branch of the superposition, and in particular, the modulus squared of the amplitude of each branch determines the vividness of the mental content corresponding to the branch (Gao\index{Gao, Shan}, 2016).
Under this assumption, when the modulus squared of the amplitude of a branch is close to zero, the mental content corresponding to the branch will be the least vivid.
It is conceivable that below a certain threshold of vividness an ordinary observer or even an ideal observer will not be consciously aware of the corresponding mental content.
Then even though in collapse theories\index{collapse theories} the post-measurement state of an observer is still a superposition of different outcome branches with similar structure, the observer can only be consciously aware of the  mental content corresponding to the branch with very high amplitude, and the mental content corresponding to the branches with very low amplitude will not appear in the whole mental content of the observer.
This may solve the structured tails problem\index{collapse theories!tails problem!structured} of collapse theories\index{collapse theories}.

\section{Can RDM of particles\index{random discontinuous motion of particles} directly solve the measurement problem\index{measurement problem}?}

According to the suggested interpretation of the wave function in terms of RDM of particles\index{random discontinuous motion of particles}, 
a quantum system is composed of particles, and the wave function of the system represents an instantaneous property of the particles of the system that determines their random discontinuous motion\index{random discontinuous motion of particles}. 
It is obvious that the underlying RDM of particles\index{random discontinuous motion of particles} adds an additional variable, the particle configuration, and its random dynamics to quantum mechanics. 
In this sense, the resulting theory may be regarded as a stochastic ``hidden" variables theory.
The question is: Can RDM of particles\index{random discontinuous motion of particles} provide a direct solution to the measurement problem\index{measurement problem}? 


Interestingly, Bell (\index{Bell, John S.}1981) already gave a positive answer to this question when analyzing his Everett (?) theory, which interprets Everett's theory\index{Everett's theory} as Bohm's theory\index{Bohm's theory} without trajectories.
The reason is that the picture of RDM of particles\index{random discontinuous motion of particles} is arguably the same as the picture given by Bell's Everett (?)\index{Bell's Everett (?) theory} theory (see Section 7.4.4), and it has been argued that Bell's Everett (?)\index{Bell's Everett (?) theory} theory can account for our definite experience and be consistent with the predictions of quantum mechanics (Bell,\index{Bell, John S.} 1981; Barrett, 1999\index{Barrett, Jeffrey A.}).
In the following, I will explain how the RDM of particles\index{random discontinuous motion of particles} can directly solve the measurement problem\index{measurement problem}, and  why the solution is unsatisfactory due to its potential problems.\footnote{My explanation basically follows Barrett's (1999) analysis of Bell's Everett (?)\index{Bell's Everett (?) theory} theory.}

Consider a simple $x$-spin measurement, in which an observer $M$ measures the $x$-spin of a spin one-half system $S$ that is in a superposition of two different $x$-spins. According to the linear Schr\"{o}dinger equation\index{Schr\"{o}dinger equation}, the state of the composite system after the measurement will be the superposition of $M$ recording $x$-spin up and $S$ being $x$-spin up and $M$ recording $x$-spin down and $S$ being $x$-spin down:

\begin{equation}
\alpha \ket{up}_S \ket{up}_M+\beta \ket{down}_S \ket{down}_M,
\label{}
\end{equation}

\noindent  where $\alpha$ and $\beta$ are not zero and satisfy the normalization condition $|\alpha|^2+|\beta|^2=1$.

According to the picture of RDM of particles\index{random discontinuous motion of particles}, the positions of the particles representing the measurement record of $M$ are definite at each instant. Moreover, these particles discontinuously and randomly jump between the two states $\ket{up}_M$ and $\ket{down}_M$ over time, and the probability of they being in these two states at each instant are $|\alpha|^2$ and $|\beta|^2$, respectively.
Then when assuming that an observer's mental state is well-defined at an instant, and it also supervenes on the particle configuration of the observer at the instant,
the observer $M$ will at each instant have a determinate record corresponding to one of the two terms in the above superposition, 
that is, at each instant $M$'s particle configuration will effectively select one of the two terms in the superposition as actual and thus $M$'s mental state will be the state with the determinate record $x$-spin up or the determinate record $x$-spin down.
Moreover, which particle configuration $M$ ends up with, and thus which determinate record he gets, is randomly determined at the instant, and the probability of $M$ getting a particular record is equal to the modulus squared of the wave function associated with the record, namely the probability of $M$ ending up with a configuration recording $x$-spin up is $|\alpha|^2$ and the probability of $M$ ending up with a configuration recording  $x$-spin down is $|\beta|^2$.
This means that the Born probabilities come from the objective probabilities inherent in the RDM of particles\index{random discontinuous motion of particles} directly.

Obviously, due to the essential discontinuity and randomness of RDM, the observer $M$'s measurement record will change in a random and discontinuous way over time and thus be unreliable as a record of what actually happened.
As Bell (\index{Bell, John S.}1981) argued, however, that there is no association of the particular present with any particular past does not matter. ``For we have no access to the past. We have only our `memories' and 
`records'. But these memories and records are in fact present phenomena. The theory should account for the present correlations between these present phenomena. And in this respect we have seen it to agree with ordinary quantum mechanics, in so far as the latter is unambiguous." (Bell,\index{Bell, John S.} 1981)

Here is a more detailed explanation of Bell's argument. 
Suppose the observer $M$ gets the result $x$-spin up for the result of her first measurement. When she repeats her measurement, the state of the composite system after this second measurement will be

\begin{equation}
\alpha \ket{up}_S \ket{up,up}_M+\beta \ket{down}_S \ket{down,down}_M
\label{}
\end{equation}

\noindent by the linear Schr\"{o}dinger\index{Schr\"{o}dinger, Erwin} evolution. Now, according to the picture of RDM of particles\index{random discontinuous motion of particles},  there is a probability of $|\beta|^2$ that $M$ will end up with a configuration recording $x$-spin down for the second result even though he recorded $x$-spin up for the first result. Thus it appears that there is a probability of $|\beta|^2$ that $M$'s two measurement results will disagree.
However, if $M$ does get $x$-spin down for her second measurement, her configuration will be the one associated with the second term of the above state. This means that $M$'s actual memory configuration will record $x$-spin down for her first result, and thus for $M$ the two measurements in fact yield the same result. 

Although the predictions of the above theory are consistent with experience, the unreliability of an observer's memories will lead to an empirical incoherence problem (Barrett, 1999\index{Barrett, Jeffrey A.}, p.126).
The problem is that although one can test the instantaneous empirical predictions of the theory (i.e. the way that measurement records are correlated at a particular instant), one cannot test its dynamical law that governs the time evolution of the particle configuration because one's memories of measurement records are unreliable.
In other words, even if the dynamical law of the theory were correct, one could not have an empirical justification for accepting that it is correct.

The more serious problem of the above theory, in my opinion, is that the assumption of the existence of instantaneous mental states is implausible (see also Butterfield\index{Butterfield, Jeremy}, 1996; Donald, 1996).
A large number of neuroscience experiments show that the appearance and persistence of a conscious experience in our brain involves a dynamical process of many neuron groups in the brain, which certainly requires a duration.
Note that the limit of our brain is relevant here, because it is we human beings who do the quantum experiments. 
If a mental state has a duration, no matter how short the duation is, then the above theory will be wrong.
In this case, the doctrine of psychophysical supervenience requires that the mental state of an observer  supervenes on her physical state in the duration. 


To sum up, it is arguable that the RDM of particles\index{random discontinuous motion of particles} cannot solve the measurement problem\index{measurement problem} in a satisfactory way. 
This may be not against expectation, since the RDM of particles\index{random discontinuous motion of particles} provides only an ontological interpretation of the wave function.
As I will argue below, however, this ontological interpretation of the wave function, if it is true, may also have implications for solving the measurement problem\index{measurement problem}.


\section{The origin of the Born probabilities}

An important aspect of the measurement problem\index{measurement problem} is to explain the origin of the Born probabilities or the probabilities of measurement results.\footnote{The Born probabilities are usually given the frequency interpretation, which is neutral with respect to the issue whether the probabilities are fundamental or due to ignorance. Although there have been attempts to derive the Born rule\index{Born rule} from more fundamental postulates of quantum mechanics, it is still controversial whether these arguments really derive the Born rule\index{Born rule} or they are in fact circular (see Landsman\index{Landsman, Nicolaas P.}, 2009 and references therein). In my opinion, only after finding the origin of the Born probabilities, can one find the correct derivation of the Born rule\index{Born rule!derivations of} (see Section 8.4).} 
According to the suggested interpretation of the wave function in terms of RDM of particles\index{random discontinuous motion of particles}, the ontological meaning of the modulus squared of the wave function of an electron in a given position is that it represents the probability density that the electron as a particle appears in this position, while according to the Born rule\index{Born rule}, the modulus squared of the wave function of the electron in the position also gives the probability density that the electron is found there. It is hardly conceivable that these two probabilities, which are both equal to the modulus squared of the wave function, have no connection. On the contrary, it seems natural to assume that the origin of the Born probabilities is the RDM of particles\index{random discontinuous motion of particles}. \index{random discontinuous motion of particles!and the origin of the Born probabilities}

There is a further argument supporting this assumption.
According to the picture of RDM of particles\index{random discontinuous motion of particles}, the wave function of a quantum system represents an instantaneous property of the particles of the system that determines their motion. However, the wave function is not a complete description of the instantaneous state of the particles. The instantaneous state of the particles at a given instant also includes their random positions, momenta and energies at the instant. Although the probability of the particles being in each random instantaneous state is completely determined by the wave function, their stay in the state at each instant is independent of the wave function.\index{random discontinuous motion of particles!and the origin of the Born probabilities}
Therefore, it seems natural to assume that the random stays of the particles may have certain stochastic influences on the evolution of the system, which are manifested in the measurement process, e.g. including generating the random result during a measurement. I will discuss this point in more detail in Section 8.4.1.

If the assumption that the origin of the Born probabilities is the RDM of particles\index{random discontinuous motion of particles} turns out to be true, then it will have significant implications for solving the measurement problem\index{measurement problem}, because the existing solutions have not taken this assumption into account.\footnote{Note that Bell's Everett (?)\index{Bell's Everett (?) theory} theory or the theory discussed in the last section is an exception, where the Born probabilities directly come from the objective probabilities inherent in the RDM of particles\index{random discontinuous motion of particles}.}
In Bohm's theory\index{Bohm's theory}, the dynamics is deterministic and the Born probabilities are epistemic in nature. \index{Bohm's theory!origin of the Born probabilities in}
In the latest formulation of Everett's theory\index{Everett's theory}\index{Everett's theory!origin of the Born probabilities in} (Wallace\index{Wallace, David}, 2012), the Born probabilities are subjective in the sense that it is defined via decision theory.
In collapse theories\index{collapse theories}, although the Born probabilities are objective, it is usually assumed that the randomness originates from a classical noise source independent of the wave function of the studied system. \index{random discontinuous motion of particles!and the origin of the Born probabilities}
In short, none of these major solutions to the measurement problem\index{measurement problem} assumes that the Born probabilities originate from the wave function itself. 
Therefore, if the Born probabilities originate from the objective probabilities inherent in the RDM of particles\index{random discontinuous motion of particles} described by the wave function, then all these realistic alternatives to\index{quantum mechanics!alternatives to, \emph{see} Bohm's theory, Everett's theory, and collapse theories}  quantum mechanics need to be reformulated so that they can be consistent with this result. 
The reformulation may be easier for some alternatives, but  more difficult or even impossible for others. In the following, I will give an analysis of this issue.\index{random discontinuous motion of particles!and the origin of the Born probabilities}

In order that the Born probabilities originate from the RDM of particles\index{random discontinuous motion of particles}, the instantaneous particle configuration of a measuring device must have efficacy in determining the generation of the measurement result during the measurement process.\index{random discontinuous motion of particles!and the origin of the Born probabilities} 
In other words, besides the deterministic quantum dynamics,
there must exist an additional random dynamics which results from the RDM of particles\index{random discontinuous motion of particles} and which results in the appearance of a random measurement result.
As noted before, a necessary condition for a certain variable to be able to represent a measurement result is that the mental state of an observer supervenes on this variable.
This variable may be the instantaneous particle configuration or the wave function or certain branches of the wave function or additional variables.
Then if the RDM of particles\index{random discontinuous motion of particles} is the origin of the Born probabilities, then the instantaneous particle configuration must influence the evolution of this variable by a random dynamics during the measurement process.\index{random discontinuous motion of particles!and the origin of the Born probabilities}

There are two possible ways to realize the random dynamics. One way is that the instantaneous particle configuration of a measuring device or an observer represents the measurement result, and correspondingly the RDM of particles\index{random discontinuous motion of particles}  provides the random dynamics.
This requires that the mental state of an observer is well-defined at an instant and it also supervenes on the instantaneous particle configuration of the observer. 
This is the case in Bell's Everett (?)\index{Bell's Everett (?) theory} theory, in which the Born probabilities directly come from the objective probabilities inherent in the RDM of particles\index{random discontinuous motion of particles}.

The other way is that the instantaneous particle configuration, which does not represent the measurement result, influences the variable representing the measurement result by a new random dynamics.
The variable representing the measurement result may be the wave function or certain branches of the wave function or additional variables, corresponding to collapse theories\index{collapse theories} or Everett's theory\index{Everett's theory} or Bohm's theory\index{Bohm's theory}.
In collapse theories\index{collapse theories}, it is required that the instantaneous particle configuration influences the evolution of the wave function by a random dynamics to generate the right Born probabilities. It has been shown by a concrete collapse model that this is possible (Gao\index{Gao, Shan}, 2013a). \index{random discontinuous motion of particles!and the origin of the Born probabilities}
In Everett's theory\index{Everett's theory}, it is required that the instantaneous particle configuration influences the evolution of certain branches of the wave function by a random dynamics to generate the right Born probabilities.
This seems impossible since  the evolution of  the wave function is always governed by the deterministic Schr\"{o}dinger equation\index{Schr\"{o}dinger equation} according to the theory.
In Bohm's theory\index{Bohm's theory}, it is required that the instantaneous particle configuration (which undergoes RDM) influences the evolution of the positions of Bohmian particle\index{Bohm's theory!Bohmian particles in}s (which undergoes continuous motion) by a random dynamics to generate the right Born probabilities.
This seems possible, but the picture of two kinds of particles can hardly be satisfactory.

In the following sections, I will analyze these implications for the major solutions to the measurement problem\index{measurement problem} in detail. In addition, I will also present some other objections to these solutions based on the picture of RDM of particles\index{random discontinuous motion of particles}.

\subsection{Everett's theory\index{Everett's theory}} 

I will first analyze Everett's theory\index{Everett's theory}. According to this theory, there is only the wave function physically, and its evolution is always governed by the Schr\"{o}dinger equation\index{Schr\"{o}dinger equation}.\index{Everett's theory!origin of the Born probabilities in}
Moreover, the theory claims that after a quantum measurement with many possible results there will be many observers, and each of them is consciously aware of one of these definite results.
For example, when an observer $M$ measures the $x$-spin of a spin one-half system $S$ and the post-measurement state of the composite system is the superposition of $M$ recording $x$-spin up and $S$ being $x$-spin up and $M$ recording $x$-spin down and $S$ being $x$-spin down:

\begin{equation}
\alpha \ket{up}_S \ket{up}_M+\beta \ket{down}_S \ket{down}_M,
\label{ss}
\end{equation}

\noindent  where $\alpha$ and $\beta$ are not zero and satisfy the normalization condition $|\alpha|^2+|\beta|^2=1$, there will be two observers after the measurement, and each of them is consciously aware of a definite record, either x-spin up or x-spin down. 

As I noted before, there are in general two ways of understanding the notion of multiplicity in Everett's theory\index{Everett's theory}.\index{Everett's theory!origin of the Born probabilities in}\index{random discontinuous motion of particles!and Everett's theory}
One is the strong form\index{Everett's theory!strong form of, \emph{see also} many-worlds theory} which claims that there are many \emph{physical} observers (in material content) after a quantum measurement with many possible results, and the other is the weak form\index{Everett's theory!weak form of, \emph{see also} many-minds theory} which claims that there is one physical observer (in material content), but there are many \emph{mental} observers or many mental states of the same physical observer after the quantum measurement.
It is obvious that the strong form\index{Everett's theory!strong form of, \emph{see also} many-worlds theory} is not consistent with the picture of RDM of particles\index{random discontinuous motion of particles}, according to which the number of particles is not changed when the wave function of a quantum system evolves over time, e.g. from $ (\alpha \ket{up}_S+\beta \ket{down}_S)\ket{ready}_M$ to $\alpha \ket{up}_S \ket{up}_M+\beta \ket{down}_S \ket{down}_M$. In other words, there is only one physical world according to the picture of RDM of particles\index{random discontinuous motion of particles}.

However, a more detailed analysis is needed to determine whether the weak form\index{Everett's theory!weak form of, \emph{see also} many-minds theory} of Everett's theory\index{Everett's theory} is consistent with the picture of RDM of particles\index{random discontinuous motion of particles!and Everett's theory}.
To begin with, if assuming that an observer's mental state is well-defined at an instant, and it supervenes on the  physical state of the observer at the instant, then the resulting theory will be Bell's Everett (?)\index{Bell's Everett (?) theory} theory. The theory is a one-world and one-mind theory. In this case, the weak form\index{Everett's theory!weak form of, \emph{see also} many-minds theory} of Everett's theory\index{Everett's theory} is not consistent with the picture of RDM of particles\index{random discontinuous motion of particles}.\index{Everett's theory!origin of the Born probabilities in}
Next, assume an observer's mental state is well-defined in a finite (continuous) time interval, and it supervenes on the physical state of the observer in the time interval. This is a well-accepted assumption. 
According to the picture of RDM of particles\index{random discontinuous motion of particles}, when an observer is in a post-measurement superposition such as (\ref{ss}),  her physical state during any finite time interval is the whole superposition, not any of its branches.
Each branch of the superposition exists not in a continuous time interval, but only in a discontinuous, dense instant set, and the combination of all these instant sets forms a continuous time interval. 
Then according to the above assumption, when an observer is in a post-measurement superposition such as (\ref{ss}),\index{Everett's theory!origin of the Born probabilities in} 
her mental state supervenes on the whole superposition, not on any of its branches, and thus there is still one world and one mind. This means that the weak form\index{Everett's theory!weak form of, \emph{see also} many-minds theory} of Everett's theory\index{Everett's theory} is not consistent with the picture of RDM of particles\index{random discontinuous motion of particles} either in this case.

In order that the weak form\index{Everett's theory!weak form of, \emph{see also} many-minds theory} of Everett's theory\index{Everett's theory} is consistent with the picture of RDM of particles\index{random discontinuous motion of particles}, one must assume that when an observer is in a post-measurement superposition such as  (\ref{ss}), she has many mental states, each of which  supervenes on one branch of the superposition, which exists in a discontinuous, dense instant set.\index{random discontinuous motion of particles!and Everett's theory}
This requires that an observer's mental states are well-defined even in a discontinuous, dense instant set, and each of them  supervenes on her partial physical state in the discontinuous, dense instant set.
However, this requirement seems \emph{ad hoc}, only for the purpose of being consistent with the predictions of the weak form\index{Everett's theory!weak form of, \emph{see also} many-minds theory} of Everett's theory\index{Everett's theory}. Moreover, it will lead to a very strange picture, namely that when a physical observer has many minds, these minds do not exist at the same time at any precise instant, but exist at different instants and interlace with each other in time.

In the following, I will analyze the origin of the Born probabilities in the weak form\index{Everett's theory!weak form of, \emph{see also} many-minds theory} of Everett's theory\index{Everett's theory}.
I will argue that even if the theory is consistent with the picture of RDM of particles\index{random discontinuous motion of particles}, it cannot accommodate the assumption that the origin of the Born probabilities is the RDM of particles\index{random discontinuous motion of particles}.
In the weak form\index{Everett's theory!weak form of, \emph{see also} many-minds theory} of Everett's theory\index{Everett's theory}, measurement records are represented by certain branches of the wave function of the observer on which her mental states supervene.\index{Everett's theory!origin of the Born probabilities in} 
In order that the Born probabilities originate from the RDM of particles\index{random discontinuous motion of particles}, the instantaneous particle configuration of the observer must influence the evolution of these branches of her wave function. \index{random discontinuous motion of particles!and Everett's theory}
This means that besides the deterministic quantum dynamics, there must exist an additional random dynamics which results from the RDM of particles\index{random discontinuous motion of particles} and results in the appearance of random measurement records.
However, according to the weak form\index{Everett's theory!weak form of, \emph{see also} many-minds theory} of Everett's theory\index{Everett's theory}, there is only the wave function, and its evolution is always governed by the Schr\"{o}dinger equation\index{Schr\"{o}dinger equation}. Therefore, in the theory, the instantaneous particle configuration of an observer can have no influence on the evolution of any branches of her wave function, and as a result, the Born probabilities cannot originate from the objective probability inherent in the RDM of particles\index{random discontinuous motion of particles}.
Note that Everett's theory\index{Everett's theory} cannot be reformulated by adding the required random dynamics  for the wave function either; otherwise it will become a collapse theory.

There is another possibility which deserves to be considered. \index{Everett's theory!origin of the Born probabilities in}\index{random discontinuous motion of particles!and Everett's theory}
If the doctrine of psychophysical supervenience is violated, then even though the evolution of the wave function is always governed by the Schr\"{o}dinger equation\index{Schr\"{o}dinger equation},  the instantaneous particle configuration of an observer may also influence the evolution of her mental states (which supervene on certain branches of her wave function), and thus it may be still possible that the Born probabilities originate from the RDM of particles\index{random discontinuous motion of particles}.
In the single-mind theory, this possibility indeed exists.
In the many-minds theory\index{many-minds theory}, however, this possibility does not exist.
Since  the instantaneous particle configuration is the same for the mental states of all minds, its random influence on the evolution of the mental state of each mind will be also the same. 
This means that the measurement record for every mind will be the same, which is certainly inconsistent with the predictions of the many-minds theory\index{many-minds theory}.
Therefore, at least for some minds the origin of the Born probabilities is not the RDM of particles\index{random discontinuous motion of particles} in the many-minds theory\index{many-minds theory}.

\subsection{Bohm's theory\index{Bohm's theory}}

In this section, I will analyze Bohm's theory\index{Bohm's theory}. In particular, I will examine whether it is possible that the Born probabilities originate from the RDM of particles\index{random discontinuous motion of particles} described by the wave function in Bohm's theory\index{Bohm's theory}\index{Bohm's theory!origin of the Born probabilities in}.\index{random discontinuous motion of particles!and Bohm's theory}

According to Bohm's theory\index{Bohm's theory}, a complete realistic description of a quantum system is provided by the configuration defined by the positions of its Bohmian particle\index{Bohm's theory!Bohmian particles in}s together with its wave function. The wave function follows the linear Schr\"{o}dinger equation\index{Schr\"{o}dinger equation} and never collapses. The Bohmian particle\index{Bohm's theory!Bohmian particles in}s are guided by the wave function via the guiding equation\index{Bohm's theory!guiding equation of} to undergo deterministic continuous motion.
As noted before, in Bohm's theory\index{Bohm's theory}, there are two suggested forms of psychophysical supervenience, and for either form the result of a measurement is determined by the positions of the Bohmian particle\index{Bohm's theory!Bohmian particles in}s of the measuring device. 
Thus, in order that the Born probabilities originate from the RDM of particles\index{random discontinuous motion of particles!and Bohm's theory}, the instantaneous RDM particle configuration must influence the evolution of the positions of Bohmian particle\index{Bohm's theory!Bohmian particles in}s by a random dynamics.\index{random discontinuous motion of particles!and Bohm's theory}
This means that the guiding equation\index{Bohm's theory!guiding equation of} needs to be added a stochastic evolution term, which originates from the RDM of particles\index{random discontinuous motion of particles} and is required to generate the right measurement results in accordance with the Born rule\index{Born rule}.
In the following, I will show that this is in principle possible. I use Vink\index{Vink, Jeroen C.}'s (1993) discrete dynamics for a one-body system to illustrate this possibility.\footnote{Vink\index{Vink, Jeroen C.}'s dynamics is also called the Bohm-Bell-Vink\index{Vink, Jeroen C.} dynamics (Barrett, 1999\index{Barrett, Jeffrey A.}).}

The continuity equation in the discrete position representation $\ket{x_n}$ is:

\begin{equation}
\hbar \partial P_n(t)/\partial t = \sum_m J_{nm}(t),
\end{equation}

\noindent where 

\begin{equation}
P_n(t) = |\langle{x_n}\ket{\psi(t)}|^2,
\end{equation}

\begin{equation}
J_{nm}(t) = 2 Im(\langle{\psi(t)}\ket{x_n}\bra{x_n}H\ket{x_m}\langle{x_m}\ket{\psi(t)}),
\end{equation}

\noindent where $\ket{\psi(t)}$ is the wave function of the system, and $H$ is the Hamiltonian of the system.

In Vink\index{Vink, Jeroen C.}'s dynamics, the position jumps of the Bohmian particle\index{Bohm's theory!Bohmian particles in} of the system are governed by a transition probability $T_{mn}dt$ which gives the probability to go from position $x_n$ to $x_m$ in the time interval $dt$. The transition matrix $T$ gives rise to a time-dependent probability distribution $x_n$ (for an ensemble of identically prepared systems), $P_n(t) $, which has to satisfy the master equation:

\begin{equation}
\partial P_n(t)/\partial t = \sum_m (T_{nm}P_m-T_{mn}P_n).
\end{equation}

\noindent Then when the transition matrix $T$ satisfies the following equation:\index{random discontinuous motion of particles!and Bohm's theory}

\begin{equation}
J_{nm}/\hbar = T_{nm}P_m-T_{mn}P_n.
\end{equation}

\noindent  the above continuity equation can be satisfied.

Vink\index{Vink, Jeroen C.} (1993) showed that when choosing Bell's simple solution where for $n \neq m$\footnote{The probability $T_{nn}dt$ follows from the normalization relation $ \sum_m{T_{nm}dt}=1$ for a time discretization step $dt$.}

\begin{equation}
T_{nm}= 
\begin{cases} 
{J_{nm}/\hbar P_m},& J_{nm} \geq 0
\\
0,&J_{nm}<0,
\end{cases}
\end{equation}

\noindent the dynamics reduces to the (first-order) guiding equation\index{Bohm's theory!guiding equation of} of Bohm's theory\index{Bohm's theory} in the continuum limit.\index{random discontinuous motion of particles!and Bohm's theory}

Besides the Bohmian particle\index{Bohm's theory!Bohmian particles in} undergoing Vink\index{Vink, Jeroen C.}'s discrete position jumps, the system is also composed of a particle which undergoes RDM according to the picture of RDM of particles\index{random discontinuous motion of particles}.
If the RDM of the particle influences the position jumps of the Bohmian particle\index{Bohm's theory!Bohmian particles in}, then there must exist a stochastic evolution term in the solution $T_{nm}$, which depends on the instantaneous Bohmian particle\index{Bohm's theory!Bohmian particles in} configuration $x_k$ and may be denoted by $S_{m}(k)$. It can be seen that one can add to $T_{nm}$ defined above any solution $T^0$ of the homogeneous  equation:

\begin{equation}
T^0_{nm}P_m-T^0_{mn}P_n=0.
\end{equation}

\noindent This requires that  $S_{m}(k)$ should satisfy the following relation:

\begin{equation}
S_{m}(k)P_m-S_{n}(k)P_n=0.
\end{equation}

\noindent This result shows that when this relation is satisfied the random influences of the RDM of particles\index{random discontinuous motion of particles} can keep the probability distribution of the positions of the Bohmian particle\index{Bohm's theory!Bohmian particles in}s of an ensemble of identically prepared systems unchanged.\index{random discontinuous motion of particles!and Bohm's theory}
Then under the quantum equilibrium hypothesis\index{Bohm's theory!quantum equilibrium hypothesis of},\footnote{The quantum equilibrium hypothesis\index{Bohm's theory!quantum equilibrium hypothesis of} provides the initial conditions for the guidance equation which make Bohm's theory\index{Bohm's theory} obey Born's rule in terms of position distributions.}  the probabilities of measurement results predicted by the revised theory are still the same as the Born probabilities.

In this revised theory, however, it can hardly say that the Born probabilities originate from the RDM of particles\index{random discontinuous motion of particles}. In the theory, the quantum equilibrium hypothesis\index{Bohm's theory!quantum equilibrium hypothesis of} still plays a major role, while the random influences of the RDM of particles\index{random discontinuous motion of particles} are only a side effect.
In order that the Born probabilities really originate from the RDM of particles\index{random discontinuous motion of particles}, 
the random influences of the RDM of particles\index{random discontinuous motion of particles} on the evolution of the positions of Bohmian particle\index{Bohm's theory!Bohmian particles in}s are required to generate the right measurement results in accordance with the Born rule\index{Born rule} no matter what the initial probability distribution of the positions of these Bohmian particle\index{Bohm's theory!Bohmian particles in}s is (i.e. without resorting to the quantum equilibrium hypothesis\index{Bohm's theory!quantum equilibrium hypothesis of}).\index{random discontinuous motion of particles!and Bohm's theory}
However, it can be argued by a concrete example that this is impossible.

Consider an electron in a superposition of two energy eigenstates in two boxes.
Due to the restriction of box walls, the Bohmian particle\index{Bohm's theory!Bohmian particles in} of the electron cannot \emph{continuously} move from one box to another.
Then no matter what the random influence of the RDM of the electron is,
the probability distribution of the positions of the Bohmian particle\index{Bohm's theory!Bohmian particles in}s of an ensemble of identical prepared electrons in the two boxes keeps unchanged.\footnote{Here it is implicitly assumed that the motion of Bohmian particle\index{Bohm's theory!Bohmian particles in}s keeps continuous under the random influence of the RDM of particles\index{random discontinuous motion of particles}. If the motion of the Bohmian particle\index{Bohm's theory!Bohmian particles in} of the electron becomes discontinuous, then it seems that its state of motion will be also described by a wave function, and as a result, there will be two electrons. This leads to a contradiction.}
Therefore, if the initial probability distribution of the positions of these Bohmian particle\index{Bohm's theory!Bohmian particles in}s is different from the Born probabilities, e.g. the initial positions of these Bohmian particle\index{Bohm's theory!Bohmian particles in} are all in one box,\index{random discontinuous motion of particles!and Bohm's theory}
then the probabilities of position measurement results will be different from the Born probabilities (cf. Valentini\index{Valentini, Antony} and Westman\index{Westman, Hans}, 2005)). 

One may object that this example is only an unrealistic situation. But even for such situations the revised theory is also required to be consistent with the Born rule\index{Born rule}.
It can be imagined that for realistic situations the motion of each Bohmian particle\index{Bohm's theory!Bohmian particles in} under the random influences of the RDM of particles\index{random discontinuous motion of particles} may be ergodic, during which the probability of each Bohmian particle\index{Bohm's theory!Bohmian particles in} being in any position is in accordance with the Born probabilities.
Then the probabilities of position measurement results will be the same as the Born probabilities. Moreover, these probabilities also originate from the RDM of particles\index{random discontinuous motion of particles}.

However, there are at least two problems with this picture of ergodic motion of Bohmian particle\index{Bohm's theory!Bohmian particles in}s.\index{random discontinuous motion of particles!and Bohm's theory}
The first problem concerns the time scale of forming the ergodic motion.
The random influences of  the RDM of particles\index{random discontinuous motion of particles} need a certain time to make the initial non-ergodic motion of Bohmian particle\index{Bohm's theory!Bohmian particles in}s become ergodic motion. But when the ergodic motion is formed, the original wave function already evolves to a new one. If only the time scale of forming the ergodic motion is finite, then there is always a finite discrepency between the probabilities of measurement results and the Born probabilities, although it is possible that current experiments cannot yet detect the difference.
The second and more serious problem is that this picture of ergodic motion is inconsistent with the predictions  of quantum mechanics.
Since the wave function never collapses in the theory, even though the Bohmian particle\index{Bohm's theory!Bohmian particles in}s are initially trapped in one branch of the post-measurement state after a measurement, the random influences of the RDM of particles\index{random discontinuous motion of particles} (which is described by the post-measurement state) will soon make the motion of the Bohmian particle\index{Bohm's theory!Bohmian particles in}s ergodic, which means that the Bohmian particle\index{Bohm's theory!Bohmian particles in}s will move throughout the whole region the post-measurement state spreads. Then a subsequent measurement may obtain a result different from the result of the initial measurement. This contradicts the predictions of quantum mechanics.\index{random discontinuous motion of particles!and Bohm's theory}

Finally, I will give a few comments on the reality of the wave function\index{wave function!reality of} and Bohmian particle\index{Bohm's theory!Bohmian particles in}s in Bohm's theory\index{Bohm's theory}.  
It is usually thought that in Bohm's theory\index{Bohm's theory} the wave function cannot be directly measured and thus is more hidden than Bohmian particle\index{Bohm's theory!Bohmian particles in}s. For example, Bell once wrote, 

\begin{quote}
the most hidden of all variables, in the pilot wave picture, is the wavefunction, which manifests itself to us only by its influence on the complementary variables [i.e. positions of Bohmian particle\index{Bohm's theory!Bohmian particles in}s]. (Bell,\index{Bell, John S.} 1987, p.201)
\end{quote}

\noindent Maudlin\index{Maudlin, Tim} made this point more clearly, he said: 

\begin{quote}
no experiment can directly \emph{reveal} the quantum state of any system: our only clues to the quantum state are inferences from the behavior of the Primary Ontology [i.e. Bohmian particle\index{Bohm's theory!Bohmian particles in}s]. (Maudlin\index{Maudlin, Tim}, 2013, p.147)
\end{quote}

\noindent  However, this is a misunderstanding. Although an unknown quantum state or wave function cannot be measured, a known wave function can be directly measured by a series of protective measurements\index{protective measurements} (Aharonov\index{Aharonov, Yakir} and Vaidman\index{Vaidman, Lev}, 1993; Aharonov\index{Aharonov, Yakir}, Anandan\index{Anandan, Jeeva S.} and Vaidman\index{Vaidman, Lev}, 1993; Gao\index{Gao, Shan}, 2014a) (see also Section 1.3 for a brief introduction). 
In contrast,  it is the exact positions of Bohmian particle\index{Bohm's theory!Bohmian particles in}s that cannot be measured even in principle. 
As D\"{u}rr\index{D\"{u}rr, Detlef}, Goldstein\index{Goldstein, Sheldon} and Zangh\`{i}\index{Zangh\`{i}, Nino} stated explicitly, 

\begin{quote}
in a universe governed by Bohmian mechanics it is in principle impossible to know more about the configuration of any subsystem than what is expressed by (4.1) [i.e. $|\psi|^2$]. (D\"{u}rr\index{D\"{u}rr, Detlef}, Goldstein\index{Goldstein, Sheldon} and Zangh\`{i}\index{Zangh\`{i}, Nino}, 1992)
\end{quote}

\noindent Therefore, Bohmian particle\index{Bohm's theory!Bohmian particles in}s are more hidden than the wave function in Bohm's theory\index{Bohm's theory}.

This result also raises a doubt on the reality of Bohmian particle\index{Bohm's theory!Bohmian particles in}s.\footnote{The reality of the trajectories of the Bohmian particle\index{Bohm's theory!Bohmian particles in}s has been questioned based on analysis of weak measurement and protective measurement (Englert\index{Englert, Berthold-Georg}, Scully\index{Scully, Marlan O.}, S\"{u}ssmann and Walther, 1992; Aharonov\index{Aharonov, Yakir} and Vaidman\index{Vaidman, Lev}, 1996; Aharonov\index{Aharonov, Yakir}, Englert\index{Englert, Berthold-Georg} and Scully, 1999; Aharonov\index{Aharonov, Yakir}, Erez and Scully\index{Scully, Marlan O.}, 2004). However, these objections may be answered by noticing that what the protective measurement measures is the wave function, not the Bohmian particle\index{Bohm's theory!Bohmian particles in}s (see also Drezet\index{Drezet, Aurelien}, 2006). For another answer to these objections see Hiley\index{Hiley, Basil J.}, Callaghan and Maroney\index{Maroney, Owen J. E.} (2000).}
It seems not wholly satisfactory to directly claim reality for certain mathematical objects in a physical theory, e.g. directly claiming reality for Bohmian particle\index{Bohm's theory!Bohmian particles in}s and their trajectories in Bohm's theory\index{Bohm's theory}.\footnote{For example, an opponent of Bohm's theory\index{Bohm's theory} would say: ``This kind of reality is entirely in a lofty Platonic world, and we are merely invited to imagine that we live in such a world." (Werner\index{Werner, Reinhard F.}, 2015)}
This point of view is also supported by a recent research program on the reality of the wave function\index{wave function!reality of} (Harrigan\index{Harrigan, Nicholas} and Spekkens\index{Spekkens, Robert W.}, 2010; Pusey,\index{Pusey, Matthew F.} Barrett and\index{Barrett, Jonathan} Rudolph,\index{Rudolph, Terry} 2012; Leifer\index{Leifer, Matthew S.}, 2014a; Gao\index{Gao, Shan}, 2015b).
To determine whether or not a mathematical object in a physical theory directly describes the physical reality, a more satisfactory strategy is to analyze the connection between them via experience.
A physical theory is a theory connecting with experience after all.

There are in general two ways to realize this strategy.
One way is to assume the existence of the underlying state of  reality and its connection with experience or results of measurements. A typical example is the ontological models framework\index{ontological models framework} (Harrigan\index{Harrigan, Nicholas} and Spekkens\index{Spekkens, Robert W.}, 2010), which assumes that when a measurement is performed, the behaviour of the measuring device is determined by the ontic state of the measured system, along with the physical properties of the measuring device.
The other way is via a criterion of reality\index{criterion of reality} related to experience. A well-known example is the EPR criterion of reality\index{criterion of reality!EPR} (Einstein\index{Einstein, Albert}, Podolsky and Rosen\index{Rosen, Nathan}\index{Podolsky, Boris}, 1935). I have also suggested an improved criterion of reality\index{criterion of reality} before (in Section 4.4). 

As I have argued in Chapter 4, with the help of protective measurements\index{protective measurements} one can prove the reality of the wave function\index{wave function!reality of} by either realization of the above strategy. 
This provides a convincing reason for including the wave function in the ontology of Bohm's theory\index{Bohm's theory}\index{Bohm's theory!ontology of}. In contrast,  the reality of Bohmian particle\index{Bohm's theory!Bohmian particles in}s cannot be proved by either realization of the above strategy.
The main reason is still that the exact positions of Bohmian particle\index{Bohm's theory!Bohmian particles in}s cannot be measured in principle,\footnote{Even the approximate positions of Bohmian particle\index{Bohm's theory!Bohmian particles in}s can only be measured by a strong measurement which disturbs the measured system greatly. Moreover, the result of a strong position measurement does not reflect the actual position of the measured Bohmian particle\index{Bohm's theory!Bohmian particles in}, but only reflect the disturbed position of the Bohmian particle\index{Bohm's theory!Bohmian particles in} immediately before the result appears.}
 or in other words, there is no connection between Bohmian particle\index{Bohm's theory!Bohmian particles in}s and experience.\footnote{It seems that the connection between Bohmian particle\index{Bohm's theory!Bohmian particles in}s and experience may be directly established by assuming that the mental state of an observer supervenes on the relative positions of Bohmian particle\index{Bohm's theory!Bohmian particles in}s. As I have argued in the first section of this chapter, however, this assumption is problematic.}
Although the lack of proof cannot exclude the possibility that Bohmian particle\index{Bohm's theory!Bohmian particles in}s are real, it does raise a doubt about the reality of Bohmian particle\index{Bohm's theory!Bohmian particles in}s.

In addition, my previous analysis of the meaning of the wave function further increases the doubt.
I have argued that the wave function is neither nomological nor a physical field in a high-dimensional space; rather, it represents the state of RDM of particles\index{random discontinuous motion of particles} in our three-dimensional space. 
If this interpretation of the wave function turns out to be true, then there are already particles in Bohm's theory\index{Bohm's theory}.
These particles are more like real particles than Bohmian particle\index{Bohm's theory!Bohmian particles in}s; they have mass and charge, and their state of motion, which is described by the wave function, can also be measured.
Morover, there are already additional variables besides the wave function. They are the definite positions, momenta and energies of these particles at each instant. Although the evolution of these variables over time is not continuous and deterministic, their random dynamics might just lead to the stochastic collapse of the wave function\index{wave function!collapse} and further account for the appearance of random measurement results. I will analyze this possibility in detail later on.
Then the motivation to introduce additional, unobservable Bohmian particle\index{Bohm's theory!Bohmian particles in}s will be largly reduced.
Last but not least, introducing Bohmian particle\index{Bohm's theory!Bohmian particles in}s will also lead to a very clumsy and unnatural picture. For example, an electron will contain two particles, one undergoing RDM, and the other undergoing deterministic continuous motion.

To sum up, I have argued that Bohm's theory\index{Bohm's theory} can hardly accommodate the assumption that the Born probabilities originate from the RDM of particles\index{random discontinuous motion of particles} described by the wave function. Moreover, the existence of the RDM of particles\index{random discontinuous motion of particles} itself also reduces the necessity of introducing additional Bohmian particle\index{Bohm's theory!Bohmian particles in}s in the first place.


\subsection{Collapse theories}

I have analyzed Everett's theory\index{Everett's theory} and Bohm's theory\index{Bohm's theory}, which are two major solutions to the measurement problem\index{measurement problem}.
It is argued that these two theories can hardly be consistent with the suggested interpretation of the wave function in terms of RDM of particles\index{random discontinuous motion of particles}.\index{random discontinuous motion of particles!and collapse theories} 
Moreover, even if they can be reformulated to be consistent with the new interpretation,  they can hardly accommodate the assumption that the Born probabilities originate from the RDM of particles\index{random discontinuous motion of particles}. 
If there are no additional variables (that represent definite measurement results) besides the wave function, then the state of a quantum system including a measuring device will be represented by its wave function. If there are no many worlds either, then a definite measurement result, which is usually denoted by a definite position of the pointer of a measuring device, will be represented by a local wavepacket of the pointer, rather than by a superposition of many local wavepackets. As a consequence, the generation of definite measurement results can only be achieved by the collapse of the wave function\index{wave function!collapse}. In other words, wavefunction collapse will be a real physical process.
Therefore, the previous analyses strongly suggest that the picture of RDM of particles\index{random discontinuous motion of particles} favors collapse theories\index{collapse theories}, the other major solution to the measurement problem\index{measurement problem} besides Everett's theory\index{Everett's theory} and Bohm's theory\index{Bohm's theory}.

Admittedly, the existing collapse theories\index{collapse theories} are still phenomenological models, and they are also plagued by a few serious problems, such as energy non-conservation\index{collapse theories!energy non-conservation problem of} problem (Pearle\index{Pearle, Philip}, 2007, 2009). In particular, the physical origin of wavefunction collapse, including the origin of the randomness of the collapse process, is still unknown, although there are already several interesting conjectures (see, e.g. Di\'{o}si\index{Di\'{o}si, Lajos}, 1989; Penrose\index{Penrose, Roger}, 1996).
In the next section, I will try to solve these problems and propose a new collapse model in terms of RDM of particles\index{random discontinuous motion of particles!and collapse theories}. 
It will be shown that the picture of RDM of particles\index{random discontinuous motion of particles} can be readily combined with the picture of wavefunction collapse. \index{random discontinuous motion of particles!and collapse theories}
On the one hand, the dynamical collapse of the wave function\index{wave function!collapse} can release the randomness of the RDM of particles\index{random discontinuous motion of particles}, and on the other hand, the RDM of particles\index{random discontinuous motion of particles} can provide an appropriate noise source that collapses the wave function. Especially, it will be shown by the new collapse model that the Born probabilities can indeed originate from the objective probabilities inherent in the RDM of particles\index{random discontinuous motion of particles}.


\section{A model of wavefunction collapse in terms of RDM of particles\index{random discontinuous motion of particles}}

It is well known that a `chooser' and a `choice' are needed to bring the required dynamical collapse of the wave function\index{wave function!collapse} (Pearle\index{Pearle, Philip} 1999). The chooser is the noise source that collapses the wave function, and the choices are the states toward which the collapse tends. In this section, I will first analyze these two problems and 
propose a new model of energy-conserved\index{collapse models!energy-conserved} wavefunction collapse in terms of RDM of particle. Then I will investigate the consistency of the model and experiments. Finally, I will also give a few speculations on the physical origin of wavefunction collapse. 

\subsection{The chooser in discrete time\index{discrete time}}

To begin with, I will analyze the chooser problem. In the existing collapse models, the chooser is usually assumed to be an unknown classical noise field independent of the collapsed wave function (Pearle\index{Pearle, Philip}, 2007, 2009). If what the wave function describes is the RDM of particles\index{random discontinuous motion of particles}, then it seems natural to assume that the random motion of particles is the appropriate noise source to collapse the wave function. This assumption has four merits at least. First, the noise source and its properties are already known. For example, the probability density that the particles of a quantum system appear in certain positions at each instant is given by the modulus squared of the wave function of the system in these positions at the instant. Next, this noise source is not a classical field, and thus the model can avoid the problems introduced by the field such as the problem of infinite energy etc (Pearle\index{Pearle, Philip}, 2009). 
Thirdly, as I have argued before, the RDM of particles\index{random discontinuous motion of particles} can explain the origin of the Born probabilities.
Last but not least, the RDM of particles\index{random discontinuous motion of particles} can also manifest itself in the laws of motion\index{laws of motion} by introducing the collapse evolution of the wave function. In the following, I will give a more detailed analysis of this assumption.




According to the picture of RDM of particles\index{random discontinuous motion of particles}, the wave function of a quantum system represents an instantaneous property of the particles of the system that determines their random motion. Besides the wave function, the instantaneous state of the particles at a given instant also includes their random positions, momenta and energies at the instant. 
Although the probability of the particles being in each random instantaneous state is completely determined by (the modulus squared of) their wave function, their stay in the state at each instant is independent of the wave function.
Therefore, it seems natural to assume that the random stays of the particles may have certain physical efficacy that manifests in the evolution of the system.\footnote{This is distinct from the case of continuous motion. For the latter, the position of a particle at each instant is \emph{completely} determined by its intrinisic velocity at the instant (and its initial position), and thus the position of the particle has no influence on its velocity.}
Since the motion of the particles is essentially random, their stay at an instant does not influence their stays at other instants in any direct way. Then the random stays of the particles can only influence the evolution of their wave function.\footnote{In fact, since the  random stays of the particles as one part of the instantaneous state of the system are completely random, the complete equation of motion for the instantaneous state is only about the evolution of the wave function. Therefore,  the random stays of the particles  can only manifest themselves in the complete equation of motion by their stochastic influences on the evolution of the wave function.}
Moreover, since the quantity that directly determines the probability of the random stays is the modulus squared of the wave function, it seems more appropriate to assume that it is not the wave function, but its modulus squared, that is influenced by the random stays of the particles.
This forms a feedback; the modulus squared of the wave function of a quantum system determines the probabilities of the random stays of its particles in certain positions, momenta and energies, while these random stays also influence the evolution of  the modulus squared of the wave function in a stochastic way.

However, the existence of the stochastic influences on the evolution of the wave function seems to rely on an important precondition: the discreteness of time. If time is continuous and instants are durationless, then the accumulated influences of the random stays during an arbitrarily short time interval, even if they exist, will contain no randomness. The reason is that the discontinuity and randomness of motion exist only at each durationless instant, and they don't exist during an arbitrarily short time interval or an infinitesimal time interval.\footnote{For example, the state of RDM of particles\index{random discontinuous motion of particles} in real space, which is defined during an infinitesimal time interval at a given instant, is described by the position density and position flux density, and they are continuous quantities that contain no discontinuity and randomness.} Concretely speaking, the integral of the influences of the random stays during an infinitesimal time interval contains no randomness inherent in the random stays, no matter how the influence at each instant is. The integral can be formulated as $\int_t^{t+dt}{\rho(X,t)N(X,t)dt}$, where $X=X(t)$ is a random variable that describes the random stays, $\rho(X,t)$ is the probability density function, and $N(X,t)$ is a general influence function which is essentially discontinuous. Note that this integral is Lebesgue integrable when $\rho(X,t)$ is integrable and $N(X,t)$ is finite for any $X$ and $t$. It can be shown by a simple example that this integral as a function of time contains no randomness. Suppose the random variable $X$ only assumes two values $0$ and $1$, and $N(X,t)=X(t)$. Then we have $\int_t^{t+dt}\rho(X,t)N(X,t)dt=\rho(1,t)dt$. This integral is a continuous function of time, and its evolution with time contains no randomness. In contrast, if time is discrete and instants are not zero-sized but finite-sized, the integral during a finite time interval will  be obviously a random function of time.\footnote{In some sense, the discreteness of time prevents particles from jumping from their present instantaneous state to another instantaneous state and makes the particles stay in the present instantaneous state all through during each finite-sized instant.}

There is also another argument for the discrete stochastic evolution of the wave function. It has been widely argued that the existence of a minimum \emph{observable} interval of space and time, the Planck scale, is a model-independent result of the proper combination of quantum field theory\index{quantum field theory} and general relativity (see, e.g. Garay, 1995 for a review).\footnote{Note that the existing arguments do not imply but only suggest that spacetime is discrete in the ontological sense. Moreover, the meanings and realization of discrete spacetime are also different in the existing models of quantum gravity\index{quantum gravity}.} The existence of a minimum observable interval of time or the Planck time means that any physical change during a time interval shorter than the Planck time is unobservable, or in other words, a physically observable change can only happen during a time interval not shorter than the Planck time. Since the above stochastic influences on the evolution of the wave function depend not only on time duration but also on the wave function itself in general, the influences can always be observable for some wave functions during an arbitrarily short time interval (at the ensemble level). However, the existence of a minimum observable Planck time demands that all observable processes should happen during a time interval not shorter than the Planck time, and thus each tiny stochastic influence must happen during one unit of Planck time or more.\footnote{This means that the minimum duration of the random stay of a particle in a definite position, momentum and energy is always a discrete instant.}
Moreover, if there are many possible instantaneous states where the stochastic influences can happen at each time (e.g. for a general wave function), the duration of each tiny stochastic influence will be exactly one unit of Planck time for most time; when the time interval is longer than one unit of Planck time the stochastic influence will happen in other instantaneous states with probability almost equal to one.

To sum up, I have argued that the manifestation of the randomness and discontinuity of RDM of particles\index{random discontinuous motion of particles} in the laws of motion\index{laws of motion} requires that time is discrete. In the following analysis, I will assume that time is indeed discrete, and the size of each discrete instant is the Planck time.\footnote{It has been conjectured that a fundamental theory of physics may be formulated by three natural constants: the Planck time ($t_P$), the Planck length ($l_P$), and the Planck constant ($\hbar$), and all other physical constants are expressed by the combinations of them (Gao\index{Gao, Shan}, 2006b). For example, the speed of light is $c=l_P/t_P$, and the Einstein\index{Einstein, Albert} gravitational constant is $\kappa = 8\pi l_P t_P /\hbar $. In this sense, the RDM of particles\index{random discontinuous motion of particles} in discrete space and time, represented by the above three constants, is more fundamental than the phenomena described by the special and general theory of relativity. However, even if this conjecture is true, it is still a big challenge how to work out the details (see Gao\index{Gao, Shan}, 2014c for an initial attempt).} In discrete time\index{discrete time}, the particles of a quantum system randomly stay in an instantaneous state with definite positions, momenta and energies at each discrete instant with probability determined by the modulus squared of the wave function of the system at the instant. Each random, finite stay of the particles may have a finite influence on the evolution of the wave function. As I will demonstrate later, the accumulation of such discrete and random influences may lead to the right collapse of the wave function\index{wave function!collapse}, which can then explain the generation of the definite measurement result. Accordingly, the evolution of the wave function will be governed by a revised Schr\"{o}dinger equation\index{Schr\"{o}dinger equation}, which includes the normal linear terms and a stochastic nonlinear term that describes the discrete collapse dynamics.\footnote{Note that the wave function (as an instantaneous property of particles) also exists in discrete time\index{discrete time}, which means that the evolution of the wave function including the linear Schr\"{o}dinger\index{Schr\"{o}dinger, Erwin} evolution is also discrete in nature.}

\subsection{Energy conservation and the choices}

Now I will analyze the choice problem. The random stay of the particles of a quantum system at each discrete instant may have a stochastic influence on the evolution of the wave function of the system at the instant. If the stochastic influences accumulate and result in the collapse of the wave function\index{wave function!collapse}, then what are the states toward which the collapse tends? This is the choice problem or preferred basis problem. It may be expected that the stochastic influences of the RDM of particles\index{random discontinuous motion of particles} on the wave function should not be arbitrary but be restricted by certain fundamental principles. In particular, it seems reasonable to assume that the resulting dynamical collapse of the wave function\index{wave function!collapse} should also satisfy the principle of conservation of energy\index{conservation of energy!principle of}.\footnote{It is worth noting that there might also exist a possibility that the principle of conservation of energy\index{conservation of energy!principle of} is not universal and indeed violated by wavefunction collapse. A hint is that the usual proof that spacetime translation invariance\index{laws of motion!spacetime translation invariance of} leads to the conservation of energy and momentum relies on the linearity of quantum dynamics, and it does not apply to nonlinear quantum dynamics such as wavefunction collapse. I will not consider this possibility here.}  If this assumption is true, then the choices or preferred bases will be the energy eigenstates of the total Hamiltonian of the studied system.\footnote{For superpositions of degenerate energy eigenstates of a many-body system, a further collapse rule is needed. I will discuss this issue later on.} In the following, I will give a detailed analysis of the consequences of this assumption. Its possible physical basis will be investigated in the last subsection.

As is well known, for a deterministic evolution of the wave function such as the linear Schr\"{o}dinger\index{Schr\"{o}dinger, Erwin} evolution, the requirement of conservation of energy\index{conservation of energy!principle of} may apply to a single isolated system. However, for a stochastic evolution of the wave function such as the dynamical collapse process, the requirement of conservation of energy cannot apply to a single system in general but only apply  to an ensemble of identically prepared systems.  
 It can be proved that only when the preferred bases are energy eigenstates of the total Hamiltonian for each identical system in an ensemble, can energy be conserved at the ensemble level \index{conservation of energy!for wavefunction collapse}for wavefunction collapse (see Pearle\index{Pearle, Philip}, 2000 for a more detailed analysis). Note that for the linear Schr\"{o}dinger\index{Schr\"{o}dinger, Erwin} evolution under a time-independent external potential, energy is conserved but momentum is not conserved even at the ensemble level, and thus it is not conservation of momentum but conservation of energy\index{conservation of energy!principle of} that is a more universal restriction for wavefunction collapse\index{conservation of energy!for wavefunction collapse}.

The conservation of energy\index{conservation of energy!principle of} can not only solve the preferred basis problem, but also further determine the law of dynamical collapse to a large extent.
For each system in the same quantum state in an ensemble, in order that the probability distribution of energy eigenvalues keeps constant at all times for the whole ensemble (i.e. energy is conserved at the ensemble level), the random change of the  energy probability distribution of the system, which results from the random stay of the system at each discrete instant, must satisfy a certain restriction. Concretely speaking, the random stay in an energy eigenvalue $E_i$ will increase the probability of the energy eigenstate $|E_i>$ and decrease the probabilities of all other energy eigenstates pro rata. Moreover, the increasing amplitude must be proportional to the total probability of all other energy eigenstates, and the coefficient is related to the energy uncertainty of the state. 
I will prove these results in the next subsection.

A more important question is whether this energy-conserved\index{collapse models!energy-conserved} collapse model can explain definite measurement results. At first sight the answer appears negative. For example, the energy eigenstates being preferred bases seems apparently inconsistent with the localization of macroscopic objects\index{macroscopic objects} including the pointers of measuring devices. However, a detailed analysis given later will demonstrate that the model can be consistent with experiments. The key is to realize that the energy uncertainty driving the collapse of the entangled state of a many-body system is not the uncertainty of the total energy of all sub-systems, but the sum of the absolute energy uncertainty of every sub-system. As a result, the preferred bases are the product states of the energy eigenstates of the Hamiltonian of each sub-system for a non-interacting or weakly-interacting many-body system. This gives a further collapse rule for superpositions of degenerate energy eigenstates of a many-body system.

\subsection{A discrete model of energy-conserved\index{collapse models!energy-conserved} wavefunction collapse}

In this subsection, I will present a concrete model of energy-conserved\index{collapse models!energy-conserved} wavefunction collapse based on the above analysis (see also Gao, 2013a).

Consider a multi-level system with a constant Hamiltonian. Its initial state is:  

\begin{equation}
\ket{\psi(0)}=\sum_{i=1}^m{c_i(0) \ket{E_i}},
\label{}
\end{equation}

\noindent  where $\ket{E_i}$ are the energy eigenstates of the Hamiltonian of the system, $E_i$ is the corresponding energy eigenvalue, and the coefficients $c_i(0)$ satisfy the normalization relation $\sum_{i=1}^m{|c_i(0)|^2}=1$. 

Based on the previous analysis, I assume that this superposition of energy eigenstates will collapse to one of the eigenstates after a discrete dynamical process, and the collapse evolution satisfies the conservation of energy\index{conservation of energy!at the ensemble level} at the ensemble level. The physical picture of the dynamical collapse process is as follows. At the initial discrete instant $t_0=t_P$ (where $t_P$ is the Planck time), the system randomly stays in a branch $\ket{E_{i}}$ with probability $P_i(0) \equiv |c_i(0)|^2$.
This finite stay slightly changes the probability of the staying branch, and correspondingly the probabilities of all other branches are also changed pro rata. Similarly, at any discrete instant $t=n t_P$ the system randomly stays in a branch $\ket{E_{i}}$ with probability $P_i(t) \equiv |c_i(t)|^2$, and the random stay also changes the probabilities of the branches slightly. Then during a finite time interval much larger than $t_P$, the probability of each branch will undergo a discrete and stochastic evolution. In the end, the probability of one branch will be close to one, and the probabilities of other branches will be close to zero. In other words, the initial superposition will randomly collapse to one of the energy branches in the superposition effectively.\footnote{I have suggested a possible solution to the tails problem\index{collapse theories!tails problem} in Section 8.1.3.}

Now I will give a concrete analysis of this dynamical collapse process. Since the linear Schr\"{o}dinger\index{Schr\"{o}dinger, Erwin} evolution does not change the energy probability distribution, we may only consider the influence of dynamical collapse on the energy probability distribution. Suppose the system stays in branch $\ket{E_{i}}$ at instant $t=n t_P$, and the stay changes the probability of this branch, $P_{i}(t)$, to

\begin{equation}
P_{i}^i (t+t_P)=P_{i}(t)+\Delta P_{i},
\label{}
\end{equation} 

\noindent where the superscript $i$ denotes the staying branch, and $\Delta P_i$ is a functional of $P_{i}(t)$. Due to the conservation of probability, the increase of the probability of one branch can only come from the scale-down of the probabilities of all other branches. This means that the probability of another branch $P_{j}(t)$ ($j \neq i$) correspondingly changes to

\begin{equation}
P_{j}^i (t+t_P)=P_{j}(t)-{{P_{j}(t) \Delta P_{i}} \over {1-P_{i}(t)}},
\label{}
\end{equation} 

\noindent where the superscript $i$ still denotes the staying branch.\footnote{This result can also be obtained by first increasing the probability of the staying branch and then normalizing the probabilities of all branches. This means that $P_{i}^i(t+t_P)={{P_{i}(t)+\Delta }\over {1+\Delta }}$ and $P_{j}^i(t+t_P)={{P_{j}(t)}\over {1+\Delta}}$ for any $j \neq i$. In this way, we have $\Delta P_{i}={{\Delta}\over{1+\Delta}}(1-P_{i}(t))$ and $\Delta P_{j}=-{{\Delta}\over{1+\Delta}}P_{j}(t)$ for any $j \neq i$. } The probability of this random stay at the instant is $p(E_i,t)=P_{i}(t)$. Then we can work out the diagonal density matrix elements of the evolution: \footnote{The density matrix here describes the ensemble of states which arise from all possible random evolution (Pearle\index{Pearle, Philip}, 1999).}

\begin{eqnarray}
\rho_{ii}(t+t_P)&=&\sum_{j=1}^m p(E_j,t) P_{i}^j (t+t_P)\nonumber
\\&=&P_{i}(t)[P_{i}(t)+\Delta P_{i}]+\sum_{j \neq i} P_{j}(t)[P_{i}(t)-{{P_{i}(t) \Delta P_{j}(t)} \over {1-P_{j}(t)}}]\nonumber
\\ &=&\rho_{ii}(t)+P_{i}(t)[\Delta P_{i}-\sum_{j \neq i} P_{j}(t){{\Delta P_{j}(t)} \over {1-P_{j}(t)}}].
\label{}
\end{eqnarray}

Here I will introduce the first rule of dynamical collapse, which says that the probability distribution of energy eigenvalues for an ensemble of identically prepared systems is constant during the dynamical collapse process. As I have argued in the last subsection, this rule is required by the principle of conservation of energy\index{conservation of energy!principle of} at the ensemble level. By this rule, we have $\rho_{ii}(t+t_P)=\rho_{ii}(t)$ for any $i$. This leads to the following set of equations:

\begin{eqnarray}
\Delta P_{1}(t)-\sum_{j \neq 1}{{P_{j}(t)\Delta P_{j}(t)} \over {1-P_{j}(t)}}=0, \nonumber
\\
\Delta P_{2}(t)-\sum_{j \neq 2}{{P_{j}(t)\Delta P_{j}(t)} \over {1-P_{j}(t)}}=0, \nonumber
\\
...\nonumber
\\
\Delta P_{m}(t)-\sum_{j \neq m}{{P_{j}(t)\Delta P_{j}(t)} \over {1-P_{j}(t)}}=0.
\end{eqnarray}

\noindent By solving this equations set (e.g. by subtracting each other), we can find the following relation for any $i$:

\begin{equation}
{{\Delta P_{i}} \over {1-P_{i}(t)}}=k,
\label{MK}
\end{equation}

\noindent where $k$ is an undetermined dimensionless quantity that relates to the state $\ket{\psi(t)}$.

By using (\ref{MK}), we can further work out the non-diagonal density matrix elements of the evolution. But it is more convenient to calculate the following variant of non-diagonal density matrix elements:

\begin{eqnarray}
  \rho_{ij}(t+t_P)&=&\sum_{l=1}^m p(E_l,t) P_{i}^l (t+t_P)P_{j}^l (t+t_P)\nonumber
\\&=& \sum_{l \neq i,j} P_{l}(t)[P_{i}(t)-kP_{i}(t)][P_{j}(t)-kP_{j}(t)]\nonumber
\\ 
& & + P_{i}(t)[P_{i}(t)+k(1-P_{i}(t))][P_{j}(t)-kP_{j}(t)]\nonumber
\\ 
& & + P_{j}(t)[P_{j}(t)+k(1-P_{j}(t))][P_{i}(t)-kP_{i}(t)]\nonumber
\\ 
&=& (1-k^2)\rho_{ij}(t).
\end{eqnarray}

\noindent  Since the collapse time, $\tau_c$, is usually defined by the relation $\rho_{ij}(\tau_c)={1 \over 2}\rho_{ij}(0)$, we may use an appropriate approximation, where $k$ is assumed to be the same as its initial value during the time interval $[0, \tau_c]$, to simplify the calculation of the collapse time. Then we have:

\begin{equation}
\rho_{ij}(t)\approx (1-k^2)^{n}\rho_{ij}(0).
\label{}
\end{equation}

\noindent The corresponding collapse time is in the order of:

\begin{equation}
\tau_c \approx {1 \over {k^2}}t_P,
\label{TC}
\end{equation}

In the following, I will analyze the formula of $k$ defined by (\ref{MK}). To begin with, the probability restricting condition $0\leqslant P_{i}(t)\leqslant 1$ for any $i$ requires that $0\leqslant k \leqslant 1$. When $k=0$, no collapse happens, and when $k=1$, collapse happens instantaneously. Note that $k$ cannot be smaller than zero, as this will lead to the negative value of $P_{i}(t)$ in some cases. For instance, when $k$ is negative and $P_{i}(t)< {{|k|}\over{1+|k|}}$, $P_{i}^i(t+t_P)=P_{i}(t)+k[1-P_{i}(t)]$ will be negative and violate the probability restricting condition. That $k$ is positive indicates that each random stay increases the probability of the staying branch and decreases the probabilities of other branches.

Next, it can be argued that $k$ is proportional to the energy uncertainty of the state. When the energy uncertainty is zero, i.e., when the state is an energy eigenstate, no collapse happens. When the energy uncertainty is not zero, collapse happens. The energy uncertainty can be defined with the common RMS (mean square root) uncertainty:

\begin{equation}
\Delta E =\sqrt{  \sum_{i=1}^{m} {P_i(E_{i}-\overline{E})^2}  },
\label{EMUL}
\end{equation}

\noindent where $\overline{E}=\sum_{i=1}^{m} {P_iE_i}$ is the average energy. For the simplest two-level system, we have

\begin{equation}
\Delta E =\sqrt{P_1P_2}|E_1-E_2|.
\label{ETWO}
\end{equation}

Thirdly, in order to cancel out the dimension of energy, the dimensionless quantity $k$ should also include a constant with the dimension of time.
This constant is arguably the Planck time $t_P$. In continuous time where $t_P=0$, no stochastic influence exists and no collapse happens. In discrete time\index{discrete time} where $t_P \neq 0$, collapse happens. 

Then after omitting a coefficient in the order of unity, we can get the formula of $k$ in the first order:

\begin{equation}
k \approx \Delta E t_P/\hbar.
\label{K}
\end{equation}

\noindent This is the second rule of dynamical collapse. By inputting (\ref{K}) into (\ref{TC}), we can further get the collapse time formula:

\begin{equation}
\tau_c \approx {\hbar E_P \over {(\Delta E)^2}},
\label{CT}
\end{equation}

\noindent where $E_P=h/t_P$ is the Planck energy, and $\Delta E$ is the energy uncertainty of the initial state.

Here it is worth pointing out that $k$ must contain the first order term of $ \Delta E$. The reason is that the only existence of the second order or higher order term of $ \Delta E$ will lead to much longer collapse time for some common measurement situations, which contradicts experiments (Gao\index{Gao, Shan}, 2006a, 2006b). In addition, a similar analysis of the consistency with experiments may also provide a further support for the energy-conserved\index{collapse models!energy-conserved} collapse model in which the preferred bases are energy eigenstates. First of all, if the preferred bases are not energy eigenstates but momentum eigenstates, then the energy uncertainty will be replaced by momentum uncertainty in the collapse time formula (\ref{CT}), namely $\tau_c \approx {\hbar E_P \over {(\Delta p c)^2}}$. As a result, the collapse time will be too short to be consistent with experiments for some situations. For example, for the ground state of hydrogen atom the collapse time will be a few days. Note that the second order or higher order term of $ \Delta p$ will also lead to much longer collapse time for some common measurement situations, which contradicts experiments.

Next, if the preferred bases are position eigenstates,\footnote{In continuous space and time, a position eigenstate has infinite average energy and cannot be physically real. But in discrete space and time, position eigenstates will be the states whose spatial dimension is about the Planck length, and they may exist. } then the collapse time formula (\ref{CT}) will be replaced by something like $\tau_c \approx {l^2 t_P \over {(\Delta x)^2}}$, where $l$ is a certain length scale which may relate to the studied system, such as the Compton wavelength of the system. No matter what length scale $l$ is, the collapse time of a momentum eigenstate will be zero since its position uncertainty is infinite. This means that the momentum eigenstates of a quantum system will instantaneously collapse  to one of its position eigenstates and thus cannot exist. Moreover, the wave functions with very small momentum uncertainty will also collapse very quickly even for microscopic particles. These predictions can hardly be consistent with experiments.

The results of the above analysis can be summarized as follows. Suppose the state of the studied multi-level system at instant $t=nt_P$ is:

\begin{equation}
\ket{\psi(t)}=\sum_{i=1}^m{c_i(t) e^{-iE_it/\hbar}\ket{E_i}}.
\label{}
\end{equation}

\noindent  The equation of discrete collapse dynamics for $P_i(t) \equiv |c_i(t)|^2$ is:

\begin{equation}
P_i(t+t_P) = P_i(t) +{\Delta E \over E_P} [\delta_{E_sE_i}-P_i(t)],
\label{EDC}
\end{equation}

\noindent  where $\Delta E$ is the energy uncertainty of the state at instant $t$ defined by (\ref {EMUL}), $E_s$ is a random variable representing the random stay of the system, and its probability of assuming $E_i$ at instant $t$ is $P_i(t)$. When $E_s=E_i$, $\delta_{E_sE_i}=1$, and when $E_s \neq E_i$, $\delta_{E_sE_i}=0$.

This equation of dynamical collapse can be extended to the entangled states\index{entangled states} of a many-body system. The difference only lies in the definition of the energy uncertainty $\Delta E$. For a non-interacting or weakly-interacting many-body system in an entangled state, for which the energy uncertainty of each sub-system can be properly defined, $\Delta E$ is the sum of the absolute energy uncertainty of all sub-systems, namely

\begin{equation}
\Delta E =\sqrt{\sum_{j=1}^n  \sum_{i=1}^{m} {P_i(E_{ji}-\overline{E_j})^2}  },
\label{E}
\end{equation}

\noindent where $n$ is the total number of the entangled sub-systems, $m$ is the total number of energy branches in the entangled state, $E_{ji}$ is the energy of the \emph{j}-th sub-system in the \emph{i}-th energy branch of the state, and $\overline{E_j}$ is the average energy of the \emph{j}-th sub-system. Correspondingly, the preferred bases are the product states of the energy eigenstates of the Hamiltonian of each sub-system. 

It is worth pointing out that the above $\Delta E$ is not defined as the uncertainty of the total energy of all sub-systems as in the energy-driven collapse models\index{collapse models!energy-driven} (see, e.g. Percival\index{Percival, Ian C.}, 1995, 1998a; Hughston\index{Hughston, Lane P.}, 1996). The reason is that  each sub-system has its own energy uncertainty that drives its collapse, and the total driving ``force" for the whole entangled state should be the sum of the driving ``forces" of all sub-systems, at least in the first order approximation. Although these two kinds of energy uncertainty are equal in numerical values in some situations (e.g. for a strongly-interacting many-body system),  there are also some situations where they are not equal. For example, for a superposition of degenerate energy eigenstates of a non-interacting many-body system, which may arise during a common measurement process, the uncertainty of the total energy of all sub-systems is exactly zero, but the absolute energy uncertainty of each sub-system and their sum may be not zero. As a result, the superpositions of degenerate energy eigenstates of a many-body system may also collapse. As we will see later, it is this distinct feature of my model that makes it be able to avoid Pearle\index{Pearle, Philip}'s (2004) serious objections to the energy-driven collapse models\index{collapse models!energy-driven}.

It can be seen that the equation of dynamical collapse, Eq.(\ref{EDC}), has an interesting property, scale invariance. After a  discrete instant $t_P$, the probability increase of the staying branch $\ket{E_i}$ is $\Delta P_i = k (1- P_i)$, and the probability decrease of the neighboring branch $\ket{E_{i+1}}$ is $\Delta P_{i+1} = k P_{i+1}$. Then the probability increase of these two branches is

\begin{equation}
\Delta (P_i+P_{i+1}) = k [1-(P_i+P_{i+1})].
\label{}
\end{equation}

\noindent Similarly, the equation $\Delta P = k (1- P)$ holds true for the total probability of arbitrarily many branches (one of which is the staying branch). This property of scale invariance may simplify the calculation in many cases. For instance, for a superposition of two wavepackets with energy difference, $\Delta E_{12}$, much larger than the energy uncertainty of each wavepacket, $\Delta E_1=\Delta E_2$ , we can calculate the collapse dynamics in two steps. First, we use Eq.(\ref{EDC}) and Eq.(\ref{ETWO}) with $|E_1-E_2|=\Delta E_{12}$ to calculate the time of the superposition collapsing into one of the two wavepackets. Here we need not to consider the infinitely many energy eigenstates constituting each wavepacket and their probability distribution. Next, we use Eq.(\ref{EDC}) with $\Delta E=\Delta E_1$ to calculate the time of the wavepacket collapsing into one of its energy eigenstates. In general, this collapse process is so slow that its effect can be neglected. 


There is another point that needs to be clarified. As I have argued before, the discontinuity of motion requires that the collapse dynamics should be discrete in nature, and moreover, the preferred bases should be energy sssssssseigenstates in order that the collapse dynamics satisfies the conservation of energy at the ensemble level\index{conservation of energy!at the ensemble level}. As a consequence, the energy eigenvalues must be also discrete for any quantum system. This result seems to contradict the predictions of quantum mechanics. But when considering that our accelerating universe has a finite event horizon, the momentum and energy eigenvalues of any quantum system in the universe may be indeed discrete.\footnote{It has been suggested that certain quantum fluctuations of discrete spacetime within a finite event horizon may be a possible form of dark energy, and their existence may explain the observed cosmic acceleration (Gao\index{Gao, Shan}, 2005, 2013c). In addition, it is worth noting that the existence of discrete energy levels for a quantum system limited in our universe is also supported by the holographic principle, which implies that the total information within a universe with a finite event horizon is finite. If the energy of a quantum system is continuous, then the information contained in the system will be infinite.} The reason is that all quantum systems in our universe are limited by the finite horizon, and thus no free quantum systems exist in the strict sense. For example, the energy of a massless particle (e.g. photon) can only assume discrete values $E_{n} = n^2 {hc \over 4 R_U}$, and the minimum energy is $E_{1} ={hc \over 4 R_U} \approx 10^{-33}eV$, where $R_U \approx 10^{25}m$ is the radius of the event horizon of our universe.\footnote{Note that the current upper bound on the photon mass is about $m_{\gamma}< 10^{-18}eV/c^2$ (Nakamura et al, 2010).} In addition, for a  particle with mass $m_0$, its energy also assumes discrete values $E_n=n^2 {h^2 \over 32m_0R_U^2}$. For instance, the minimum energy is $E_{1} \approx 10^{-72}eV$ for electrons, which is much smaller than the minimum energy of photons.


Finally, it will be interesting to see whether the discreteness of energy still keeps the collapse dynamics smooth. Suppose the energy uncertainty of a quantum system is $\Delta E \approx 1eV$, and its energy ranges between the minimum energy $E_1$ and $1eV$. Then we can get the maximum energy level $l_{max} \approx \sqrt{1eV\over {10^{-33}eV}} \approx 10^{16}$. The probability of most energy eigenstates in the superposition will be about $P \approx 10^{-16}$. During each discrete instant $t_P$, the probability increase of the staying energy branch is $\Delta P \approx {\Delta E \over E_P}(1- P) \approx 10^{-28}$. This indicates that the probability change during each random stay is still very tiny. However, it can be seen that  when the energy uncertainty is larger than $10^{8}eV$,  the probability change during each random stay will be abrupt. 

\subsection{On the consistency of the model and experiments}

 
In this subsection, I will analyze whether the proposed model of energy-conserved\index{collapse models!energy-conserved} wavefunction collapse is consistent with experiments. Note that Adler\index{Adler, Stephen L.} (2002) has already given a detailed consistency analysis in the context of energy-driven collapse models\index{collapse models!energy-driven}, and as we will see below, some of his analyses also apply to my model.

\subsubsection{Maintenance of coherence}

First of all, it can be shown that the model satisfies the constraint of predicting the maintenance of coherence when this is observed. Since the energy uncertainty of the state of a microscopic particle is very small in general, its collapse will be too slow to have any detectable effect in present experiments on these particles. For example, the energy uncertainty of a photon emitted from an atom is in the order of $10^{-6}eV$, and the corresponding collapse time is $10^{25}s$ according to Eq. (\ref{CT}) of our collapse model, which is much longer than the age of the universe, $10^{17}s$. This means that the final states of collapse (i.e. energy eigenstates) are never reached for a quantum system with small energy uncertainty even during a time interval as long as the age of the universe. 

As another example, consider the SQUID experiment of Friedman et al (2000), where the coherent superpositions of macroscopic states consisting of oppositely circulating supercurrents are observed.\footnote{Note that the possibility of using the SQUID experiments to test the collapse theories has been discussed in great detail by Rae\index{Rae, Alastair I. M.} (1990) and Buffa, Nicrosini\index{Nicrosini, Oreste}\index{Rimini, Alberto} and Rimini (1995). See also Leggett\index{Leggett, Anthony J.} (2002) for a helpful review of SQUID experiments as tests of the limits of quantum mechanics.} In the experiment, each circulating current corresponds to the collective motion of about $10^9$ Cooper pairs, and the energy uncertainty is about $8.6 \times 10^{-6}eV$. Eq. (\ref{CT})  predicts a collapse time of $10^{23}s$, and thus maintenance of coherence is expected despite the macroscopic structure of the state.

A more interesting example is provided by certain long-lived nuclear isomers, which have large energy gaps from their ground states (see Adler\index{Adler, Stephen L.}, 2002 and references therein). For example, the metastable isomer of $^{180}$Ta, the only nuclear isomer to exist naturally on earth, has a half-life of more than $10^{15}$ years and an energy gap of $75keV$ from the ground state. According to Eq. (\ref{CT}), a coherent superposition of the ground state and metastable isomer of $^{180}$Ta will spontaneously collapse to either the isomeric state or the ground state, with a collapse time of order 20 minutes. It will be a promising way to test my collapse model by examining the maintenance of coherence of such a superposition.

\subsubsection{Rapid localization in measurement situations}

In the following, I will show that my model of energy-conserved\index{collapse models!energy-conserved} wavefunction collapse can explain definite measurement results. 

Consider a typical measurement process in quantum mechanics. According to the standard von Neumann\index{von Neumann, John} procedure, measuring an observable $A$ in a quantum state $\ket{\psi}$ involves an interaction Hamiltonian\index{Hamiltonian!interaction}

\begin{equation}
H_I = g(t)PA
\label{H_i}
\end{equation} 

\noindent coupling the measured system to an appropriate measuring device, where $P$ is the conjugate momentum of the pointer variable. The time-dependent coupling strength $g(t)$ is a smooth function normalized to $\int dt g(t)=1$ during the interaction interval $\tau$, and $g(0)=g(\tau)=0$. The initial state of the pointer is supposed to be a Gaussian wavepacket\index{Gaussian wavepacket} centered at initial position $0$, denoted by $|\phi(0)\rangle$.

For a conventional projective measurement, the interaction $H_I$  is of very short duration and so strong that it dominates the rest of the Hamiltonian (i.e. the effect of the free Hamiltonians of the measuring device and the measured system can be neglected). Then the state of the combined system at the end of the interaction can be written as

\begin{equation}
\ket{t=\tau} = e^{-{i\over\hbar} P A } \ket{\psi}  \ket{\phi(0)}.
\end{equation}

\noindent By expanding $\ket{\psi}$  in the eigenstates of $A$, $\ket{a_i}$, we obtain
 
\begin{equation}
\ket{t=\tau} = \sum_{i} e^{-{i\over\hbar} P a_i } c_i \ket{a_i} \ket{\phi(0)},
\end{equation}

\noindent where $c_i$ are the expansion coefficients. The exponential term shifts the center of the pointer by $a_i$:

\begin{equation}
\ket{t=\tau} = \sum_{i} c_i \ket{a_i} \ket{\phi(a_i)}.
\end{equation}

\noindent  This is an entangled state, where the eigenstates of $A$ with eigenvalues $a_i$ get correlated to macroscopically distinguishable states of the measuring device in which the pointer is shifted by these values $a_i$ (but the width of the pointer wavepacket is not changed). According to the collapse postulate\index{quantum mechanics!standard formulation of!collapse postulate in}, this state will instantaneously and randomly collapse into one of its branches $\ket{a_i} \ket{\phi(a_i)}$. Correspondingly, the measurement will obtain a definite result, $a_i$, which is one of the eigenvalues of the measured observable.

Let us see whether my energy-conserved\index{collapse models!energy-conserved} collapse model can explain the definite measurement results. At first sight, the answer seems negative. As emphasized by Pearle\index{Pearle, Philip} (2004), each device state in the above entangled superposition has precisely the same energy spectrum for an ideal measurement.\footnote{According to Pearle\index{Pearle, Philip} (2004), when considering environmental influences, each device/environment state in the superposition also has precisely the same energy spectrum.} Then it seems that the superposition will not collapse according to the energy-conserved\index{collapse models!energy-conserved} collapse model. However, this is not the case. The key is to realize that different eigenstates of the measured observable are generally measured in different parts of the measuring device, and they interact with different groups of atoms or molecules in these parts. Therefore, we should rewrite the device states explicitly as $\ket{\phi(0)} = \prod_j \ket{\varphi_j(0)}$ and $\ket{\phi(a_i)}=\ket{\varphi_i(1)}\prod_{j\neq i}\ket{\varphi_j(0)}$, where $ \ket{\varphi_j(0)}$ denotes the initial state of the device in part $j$, and $\ket{\varphi_i(1)}$ denotes the result state of the device in part $i$. Then we have

\begin{equation}
\sum_{i} c_i \ket{a_i} \ket{\phi(a_i)} =  \sum_{i} c_i \ket{a_i} \ket{\varphi_i(1)}\prod_{j\neq i}\ket{\varphi_j(0)}.
\end{equation}

\noindent Since there is always some kind of measurement amplification from the microscopic state to the macroscopic post-measurement state in the measurement process, there is a large energy difference between the states $\ket{\varphi_i(1)}$ and $\ket{\varphi_i(0)}$ for any $i$.\footnote{Since each result state of the measuring device has the same energy spectrum, the energy difference between the states $\ket{\varphi_i(1)}$ and $\ket{\varphi_i(0)}$ is the same for any $i$.} As a result, the total energy uncertainty, which is approximately equal to the energy difference according to Eq. (\ref{E}), is also very large, and it will result in a rapid collapse of the above superposition into one of its branches according to my energy-conserved\index{collapse models!energy-conserved} collapse model.\footnote{Since the uncertainty of the total energy of the whole entangled system is still zero, the energy-driven collapse models\index{collapse models!energy-driven} (e.g. Percival\index{Percival, Ian C.}, 1995; Hughston\index{Hughston, Lane P.}, 1996) will predict that no wavefunction collapse happens and no definite measurement result appears for the above measurement process (Pearle\index{Pearle, Philip}, 2004).}

Let me give a more realistic example, a photon being detected via photoelectric effect. In the beginning of the detection, the spreading spatial wave function of the photon is entangled with the states of a large number of surface atoms of the detector. In each local branch of the entangled state, the total energy of the photon is wholly absorbed by the electron in the local atom interacting with the photon.\footnote{In more general measurement situations, the measured particle (e.g. an electron) is not annihilated by the detector. However, in each local branch of the entangled state of the whole system, the particle also interacts with a single atom of the detector by an ionizing process, and energy is also conserved during the interaction. Due to this important property, although the measured particle is detected locally in a detector (the size of the local region is in the order of the size of an atom), its wave function does not necessarily undergo position collapse as assumed by the GRW\index{GRW theory} and CSL model\index{CSL model}s (Ghirardi\index{Ghirardi, GianCarlo}, 2016), and especially, energy can still be conserved (even at the individual level) during the localization process according to my model.} This is clearly indicated by the term $\delta (E_f-E_i-\hbar \omega)$ in the transition rate of photoelectric effect. The state of the ejecting electron is a (spherical) wavepacket moving outward from the local atom, whose average direction and momentum distribution are determined by the momentum and polarization of the photon.

This microscopic effect of ejecting electron is then amplified (e.g. by an avalanche process of atoms) to form a macroscopic signal such as the shift of the pointer of a measuring device. During the amplification process, the energy difference is constantly increasing between the branch in which the photon is absorbed and the branch in which the photon is not absorbed near each atom interacting with the photon. This large energy difference will soon lead to the collapse of the whole superposition into one of the local branches, and thus the photon is only detected locally.\footnote{In a similar way, a spherically symmetric wave function will be detected as a linear track in a cloud chamber (cf. Mott\index{Mott, Nevill F. }, 1929).} Take the single photon detector - avalanche photodiode as a concrete example.\footnote{Take the widely-used Geiger counter as another illustration of the amplification process during a measurement. A Geiger counter is an instrument used to detect particles such as $\alpha$ particles, $\beta$ particles and $\gamma$ rays etc. It consists of a glass envelope containing a low-pressure gas (usually a mixture of methane with argon and neon) and two electrodes, with a cylindrical mesh being the cathode and a fine-wire anode running through the centre of the tube. A potential difference of about $10^3V$ relative to the tube is maintained between the electrodes, therefore creating a strong electric field near the wire. The counter works on the mechanism of gas multiplication. Ionization in the gas is caused by the entry of a particle. The ions are attracted to their appropriate electrode, and they gain sufficient energy to eject electrons from the gas atoms as they pass through the gas. This further causes the atoms to ionize. Therefore, electrons are produced continuously by this process and rapid gas multiplication takes place (especially in the central electrode because of its strong electric field strength). Its effect is that more than $10^6$ electrons are collected by the central electrode for every ion produced in the primary absorption process. These ``electron avalanches" create electric pulses which then can be amplified electronically and counted by a meter to calculate the number of initial ionization events. In this way, a Geiger counter can detect low-energy radiation because even one ionized particle produces a full pulse on the central wire. It can be estimated that the introduced energy difference during a detection is $\Delta E \approx 10^{9}eV$, and the corresponding collapse time is $\tau_c \approx 10^{-5}s$ according to my collapse model.} Its energy consumption is sharply peaked in a very short measuring interval. One type of avalanche photodiode operates at $10^5$ cps and has a mean power dissipation of 4mW (Gao\index{Gao, Shan}, 2006b). This corresponds to an energy consumption of about $2.5\times 10^{11}eV$ per measuring interval $10^{-5}s$. By using the collapse time formula (\ref{CT}), where the energy uncertainty is $\Delta E \approx 2.5\times 10^{11}eV$, we find the collapse time is  $\tau_c \approx 1.25\times 10^{-10}s$, which is much shorter than the measuring interval.

\subsubsection{Emergence of the classical world}

In this subsection, I will show that my model of energy-conserved\index{collapse models!energy-conserved} wavefunction collapse can also account for the emergence of the classical world.

At first glance, it seems that there is an apparent inconsistency between the predictions of my model and our macroscopic experience. According to the model, when there is a superposition of a macroscopic object in an identical physical state (an approximate energy eigenstate) at two different, widely separated locations, the superposition does not collapse, since there is no energy difference between the two branches of the superposition. But our experience tells us that macroscopic objects\index{macroscopic objects} are localized. This inconsistency problem has been basically solved by Adler\index{Adler, Stephen L.} (2002). The crux of the matter lies in the influences of environment. The collisions and especially the accretions of environmental particles will quickly increase the energy uncertainty of the entangled state of the whole system including the object and environmental particles, and thus the initial superposition will soon collapse to one of the localized branches according to my model. Accordingly, macroscopic objects\index{macroscopic objects} are always localized due to environmental influences. It is worth emphasizing again that the energy uncertainty here denotes the sum of the absolute energy uncertainty of each sub-system in the entangled state as defined by my model.\footnote{The uncertainty of the total energy of the whole system is still very small even if the environmental influences are counted. Thus no observable collapse happens for the above situation according to the energy-driven collapse models\index{collapse models!energy-driven} (Pearle\index{Pearle, Philip}, 2004).}

As a typical example, consider a dust particle of radius $a \approx 10^{-5}cm$ and mass $m \approx 10^{-7}g$. It is well known that localized states of macroscopic objects\index{macroscopic objects} spread very slowly under the free Schr\"{o}dinger\index{Schr\"{o}dinger, Erwin} evolution. For instance, for a Gaussian wavepacket\index{Gaussian wavepacket} with initial width $\Delta$, the wavepacket will spread so that the width doubles in a time $t=2m \Delta^2 / \hbar $. This means that the double time is almost infinite for a macroscopic object. If the dust particle had no interactions with environment and its initial state is a Gaussian wavepacket\index{Gaussian wavepacket} with width $\Delta \approx 10^{-5}cm$, the doubling time would be about the age of our universe. However, if the dust particle interacts with environment, the situation turns out to be very different. Although the different components that couple to the environment will be individually incredibly localised, collectively they can have a spread that is many orders of magnitude larger. In other words, the state of the dust particle and the environment will be a superposition of zillions of very well localised terms, each with slightly different positions, and which are collectively spread over a macroscopic distance (Bacciagaluppi\index{Bacciagaluppi, Guido}, 2008). According to Joos\index{Joos, Eric} and Zeh\index{Zeh, H. Dieter} (1985), the spread in an environment full of thermal radiation only is proportional to mass times the cube of time for large times, namely $(\Delta x) ^2 \approx \Lambda m\tau^3$, where $\Lambda$ is the localization rate depending on the environment, defined by the evolution equation of density matrix $\rho_t(x,x')=\rho_0(x,x')e^{-\Lambda t (x-x')^2}$. For example, if the above dust particle interacts with thermal radiation at $T=300K$, the localization rate is $\Lambda = 10^{12}$, and the overall spread of its state is of the order of $10m$ after a second (Joos\index{Joos, Eric} and Zeh\index{Zeh, H. Dieter}, 1985). If the dust particle interacts with air molecules, e.g. floating in the air, the spread of its state will be much faster. 

Let us see whether the energy-conserved\index{collapse models!energy-conserved} wavefunction collapse in my model can prevent the above spreading. Suppose the dust particle is in a superposition of two identical localized states that are separated by $10^{-5}cm$ in space. The particle floats in the air, and its average velocity is about zero. At standard temperature and pressure, one nitrogen molecule accretes in the dust particle, whose area is $10^{-10}cm^2$, during a time interval of $10^{-14}s$ in average (Adler\index{Adler, Stephen L.}, 2002). Since the mass of the dust particle is much larger than the mass of a nitrogen molecule, the change of the velocity of the particle is negligible when compared with the change of the velocity of the nitrogen molecules during the process of accretion. Then the kinetic energy difference between an accreted molecule and a freely moving molecule is about $\Delta E = {3 \over 2}kT \approx 10^{-2}eV$. When one nitrogen molecule accretes in one localized branch of the dust particle (the molecule is freely moving in the other localized branch), it will increase the energy uncertainty of the total entangled state by $\Delta E \approx 10^{-2}eV$. Then after a time interval of $10^{-4}s$, the number of accreted nitrogen molecules is about $10^{10}$, and the total energy uncertainty is about $10^{8}eV$. According to Eq. (\ref{CT}) of my collapse model, the corresponding collapse time is about $10^{-4}s$. 

In my energy-conserved\index{collapse models!energy-conserved} collapse model, the final states of collapse are energy eigenstates, and in particular, they are nonlocal momentum eigenstates for free quantum systems. Thus it is somewhat counterintuitive that the energy-conserved\index{collapse models!energy-conserved} wavefunction collapse can make the states of macroscopic objects\index{macroscopic objects} local. As I have argued above, this is due to the constant influences of environmental particles. When the spreading of the state of a macroscopic object becomes larger, its interaction with environmental particles will introduce a larger energy difference between its different local branches, and this will collapse the spreading state again into a more localized state.
 As a result, the states of macroscopic objects\index{macroscopic objects} in an environment will not reach the final states of collapse, namely momentum eigenstates, although they do continuously undergo the energy-conserved\index{collapse models!energy-conserved} dynamical collapse. 

In a word, according to my energy-conserved\index{collapse models!energy-conserved} collapse model, there are two opposite processes for a macroscopic object constantly interacting with environmental particles. One is the spreading process due to the linear Schr\"{o}dinger\index{Schr\"{o}dinger, Erwin} evolution, and the other is the localization process due to the energy-conserved\index{collapse models!energy-conserved} collapse evolution. The interactions with environmental particles not only make the spreading more rapidly but also make the localization more frequently. In the end these two processes will reach an approximate equilibrium. The state of a macroscopic object will be a wavepacket narrow in both position and momentum, and this narrow wavepacket will approximately follow Newtonian trajectories by Ehrenfest's theorem\index{Ehrenfest's theorem} (if the external potential is uniform enough along the width of the packet). This may explain the emergence of the classical world around us.


\subsubsection{Definite conscious experiences of observers}

Ultimately, my energy-conserved\index{collapse models!energy-conserved} collapse model should be able to account for the definite conscious experiences of us as observers. 
If the observed system is the pointer of a measuring device or another macroscopic object, which is already in a definite state, then it will be easy to explain the definite conscious experiences of observers. 
But if the observed system is a microscopic system which can trigger a conscious perception of the observer, then we will need a careful analysis of the process of observation. 
For example, a small number of photons entering into the eyes of an observer from direction $A$ may trigger a visual perception $v_A$ of the observer, and the same photons from direction $B$ may trigger another different visual perception $v_B$ of the observer. Then, what visual perception of the observer will a superposition of these two input states trigger? It is required that a qualified observer should be like a measuring device and her visual perception should be either  $v_A$ or $v_B$ in this case. In the following, I will show that we are indeed qualified observers according to my energy-conserved\index{collapse models!energy-conserved} collapse model.\footnote{Note that a serious analysis of human perceptions such as visual perception in terms of collapse theories such as the GRW\index{GRW theory} model was first given by Aicardi\index{Aicardi, Francesca} et al (1991) and Ghirardi\index{Ghirardi, GianCarlo} (1999).}.

According to the neuroscience literature, the generation of a conscious perception such as a visual perception in human brains involves a large number of neurons changing their states from a resting state (resting potential) to a firing state (action potential). In each neuron, the main difference of these two states lies in the motion of $10^6$ $Na^+$s passing through the neuron membrane. Since the membrane potential is in the order of $10^{-2}V$, the energy difference between the firing state and the resting state is $\Delta E \approx 10^{4}eV$.\footnote{Since there are also other contributions to the energy difference from environmental particles, the energy difference will be larger and the collapse time will be shorter.} According to Eq. (\ref{CT}) of my energy-conserved\index{collapse models!energy-conserved} collapse model, the collapse time of a  superposition of these two states of a neuron is $\tau_c \approx 10^{5}$s. When considering the number of neurons that can form a  conscious perception is usually in the order of $10^7$, the collapse time of the quantum superposition of two different conscious perceptions is $\tau_c  \approx 10^{-9}$s. Since the normal conscious time of a human being is in the order of several hundred milliseconds, the collapse time is much shorter than the normal conscious time. This demonstrates that we human beings are qualified observers and our conscious experiences are always definite according to my energy-conserved\index{collapse models!energy-conserved} collapse model.

\subsection{In search of a deeper basis}

In this last subsection, I will give a few speculations about the physical basis of my energy-conserved\index{collapse models!energy-conserved} collapse model.

As I have pointed out before, the requirement of \index{conservation of energy!for wavefunction collapse}conservation of energy for wavefunction collapse is for an ensemble of identically prepared systems, not for a single system. 
Since each system in an ensemble does not ``know" the other systems and the whole ensemble (see, however, Smolin\index{Smolin, Lee}, 2012), it seems that there must exist certain underlying mechanism for each system which can ensure the conservation of energy for an ensemble\index{conservation of energy!at the ensemble level}. This means that the conservation of energy for an ensemble\index{conservation of energy!at the ensemble level} of identically prepared systems can be more appropriately understood as a result of the laws of motion\index{laws of motion} for individual systems in the ensemble. 
Here is a possible underlying mechanism. First of all, energy is conserved for the evolution of individual energy eigenstates. 
Next, a superposition of energy eigenstates will dynamically collapse to one of these energy eigenstates, and the probability of the collapse result satisfies the Born rule\index{Born rule}. Then the wavefunction collapse will satisfy the conservation of energy for an ensemble\index{conservation of energy!at the ensemble level} of identically prepared systems. In the following, I will give a more detailed analysis of this underlying mechanism.



According to the picture of RDM of particles\index{random discontinuous motion of particles}, for a particle in a superposition of energy eigenstates, the particle stays in an instantaneous state with a definite energy eigenvalue at a discrete instant, and at another instant it may jump to another instantaneous state with a different energy eigenvalue. It seems to be a reasonable assumption that the particle has both the tendency to jump among the instantaneous states with different energies and the tendency to stay in the instantaneous states with the same energy, and their relative strength is determined by the energy uncertainty of the superposition. This assumption seems natural and comprehensible, since there should exist two opposite tendencies in general, and their relative strength is determined by a certain condition. In some sense, the two tendencies of a particle are related to the two parts of its instantaneous state, respectively; the jumping tendency is related to the wave function, and it is needed to manifest the superposition of different energy eigenstates, while the staying tendency is related to the random stays. These two opposite tendencies together constitute the complete ``temperament" of a particle.

It can be argued that the tendency to stay in the same energy for individual particles may be the physical origin of the energy-conserved\index{collapse models!energy-conserved} wavefunction collapse. For a particle in a superposition of energy eigenstates, the particle stays in an instantaneous state with an energy eigenvalue at a discrete instant, and the staying tendency of the particle will increase its probability of being in the instantaneous states with the present energy, so that it can stay in the same energy with higher probability at the next instant (which manifests its staying tendency).
In other words, the random stay of a particle in an instantaneous state with an energy eigenvalue will increase the probability of the energy eigenvalue (and correspondingly decrease the probabilities of other energy eigenvalues pro rata). Moreover, the increase of probability is arguably proportional to the energy uncertainty of the particle; when  the energy uncertainty is zero, the probability does not change, while when the energy uncertainty is not zero, the probability increases.

It can be further argued that the probability distribution of energy eigenvalues should remain constant during the random evolution of an ensemble of identically prepared systems, and thus the resulting wavefunction collapse will satisfy the Born rule\index{Born rule}. The reason is as follows.
At a deeper level, it is very likely that the laws of nature permit nature to manifest itself, or else we will be unable to find the laws of nature and verify them by experiments, and our scientific investigations will be also pointless. This may be regarded as a meta-law. 
By this meta-law, when an initial superposition of energy eigenstates undergoes the energy collapse process, which is essentially random and irreversible, the probability distribution of energy eigenvalues should manifest itself through the collapse results for an ensemble of identically prepared systems.
This means that the diagonal density matrix elements for the ensemble should be precisely the same as the initial probability distribution at every step of the evolution. 
Otherwise the probability distribution of the collapse results in the ensemble cannot reflect the initial probability distribution, or in other words, the probability information contained in the initial state will be completely lost due to the random and irreversible wavefunction collapse.\footnote{Note that the reversible Schr\"{o}dinger\index{Schr\"{o}dinger, Erwin} evolution conserves the information even for individual isolated systems.} As a consequence, the collapse evolution will conserve energy at the ensemble level, and the probabilities of collapse results will also satisfy the Born rule\index{Born rule}.

Certainly, even if the above argument is valid, there is still a question that needs to be answered. Why energy? Why not position or momentum?
If there is only one property that undergoes RDM, then the above tendency argument for the unique property may be satisfying. But if there are many properties that undergoes RDM, then we need to answer why the tendency argument applies only to energy. A possible answer is that energy is the property that determines the linear Schr\"{o}dinger\index{Schr\"{o}dinger, Erwin} evolution of the state of motion, and thus it seems natural and uniform that energy also determines the nonlinear collapse evolution. Moreover, energy eigenstates are the states of motion that no longer evolve (except an absolute phase) for the linear evolution. Then by analogy, it is likely that energy eigenstates are also the states that no longer evolve for the nonlinear collapse evolution, i.e., that energy eigenstates are the preferred bases. I must admit that these arguments are very speculative, and the physical origin of wavefunction collapse is still an unsolved issue. 

\section{An analysis of other collapse models}

In this section, I will give a critical analysis of other collapse models. These models can be sorted into two categories.\footnote{For a helpful analysis of the properties of collapse models in a more general formalism see Weinberg\index{Weinberg, Steven} (2012). In addition, it is worth noting that  the requirement of no-faster-than-light signaling implies that the dynamics of the density matrix must be linear for collapse models (Gisin\index{Gisin, Nicolas}, 1989, 1990; Adler\index{Adler, Stephen L.} and Bassi\index{Bassi, Angelo}, 2009; Bassi and Hejazi, 2015).} The first category may be called spontaneous collapse models\index{collapse models!spontaneous}, in which the dynamical collapse of the wave function\index{wave function!collapse} is assumed to happen even for an isolated system. They include the gravity-induced\index{collapse models!gravity-induced} collapse model (Di\'{o}si\index{Di\'{o}si, Lajos}, 1989; Penrose\index{Penrose, Roger}, 1996), the GRW\index{GRW theory} model (Ghirardi\index{Ghirardi, GianCarlo}, Rimini\index{Rimini, Alberto} and Weber\index{Weber, Tullio}, 1986) etc.\footnote{The GRW\index{GRW theory} model was originally referred to as QMSL (Quantum Mechanics with Spontaneous Localizations). In this model, it is assumed that each elementary constituent of any physical system is subjected, at random times, to random and spontaneous localization processes (or hittings) around appropriate positions. The random hittings happen much less frequently for a microscopic system, e.g. an electron undergoes a hitting, on average, every hundred million years. If these hittings are assumed to be brought about by an external system, then the GRW\index{GRW theory} model should be regarded not as a spontaneous collapse model but as an interaction-induced\index{collapse models!interaction-induced} collapse model.} The second category may be called interaction-induced\index{collapse models!interaction-induced} collapse models, which assume that the dynamical collapse of the wave function\index{wave function!collapse} of a given system results from its particular interaction with a noise field. A typical example is the CSL (Continuous Spontaneous Localization) model (Pearle\index{Pearle, Philip}, 1989; Ghirardi\index{Ghirardi, GianCarlo}, Pearle\index{Pearle, Philip} and Rimini\index{Rimini, Alberto}, 1990).\footnote{If the involved noise field in the CSL model\index{CSL model} is not taken as real, then the model should be regarded as a spontaneous collapse model\index{collapse models!spontaneous}.} In the following, I will mainly analyze Penrose\index{Penrose, Roger}'s gravity-induced\index{collapse models!gravity-induced} collapse model and the CSL model\index{CSL model}.

\subsection{Penrose\index{Penrose, Roger}'s gravity-induced\index{collapse models!gravity-induced}  collapse model}

It seems very natural to guess that the collapse of the wave function\index{wave function!collapse} is induced by gravity. The reasons include: (1) gravity is the only universal force being present in all physical interactions; (2) gravitational effects grow with the size of the objects concerned, and it is in the context of macroscopic objects\index{macroscopic objects} that linear superpositions may be violated. The gravity-induced\index{collapse models!gravity-induced} wavefunction collapse conjecture can be traced back to Feynman\index{Feynman, Richard} (1995). In his \emph{Lectures on Gravitation}, Feynman\index{Feynman, Richard} considered the philosophical problems in quantizing macroscopic objects\index{macroscopic objects} and contemplates on a possible breakdown of quantum theory.\footnote{It is worth noting that Feynman\index{Feynman, Richard} considered this conjecture even earlier at the 1957 Chapel Hill conference (see DeWitt\index{DeWitt, Cecile M.} and Rickles\index{Rickles, Dean}, 2011, ch.22).} He said, ``I would like to suggest that it is possible that quantum mechanics fails at large distances and for large objects, …it is not inconsistent with what we do know. If this failure of quantum mechanics is connected with gravity, we might speculatively expect this to happen for masses such that $GM^2/\hbar c=1$, of $M$ near $10^{-5}$ grams." (Feynman\index{Feynman, Richard}, 1995)

Feynman\index{Feynman, Richard}'s suggestion was later investigated by several authors (e.g. K\'{a}ro lyh\'{a}zy, 1966; K\'{a}rolyh\'{a}zy\index{K\'{a}rolyh\'{a}zy, F.}, Frenkel\index{Frenkel, Andor} and Luk\'{a}cs\index{Luk\'{a}cs, B\'{e}la}, 1986; Di\'{o}si\index{Di\'{o}si, Lajos}, 1984, 1987, 1989; Penrose\index{Penrose, Roger}, 1981, 1986, 1989, 1994, 1996, 1998, 2000, 2002, 2004). In particular, Penrose\index{Penrose, Roger} (1996) proposed a detailed gravity-induced\index{collapse models!gravity-induced} wavefunction collapse argument, and the proposal is a `minimalist' one in the sense that it does not aspire to a more complete dynamics. The argument is based on a fundamental conflict between the superposition principle of quantum mechanics and the principle of general covariance of general relativity. The conflict can be seen by considering the superposition of a static mass distribution in two different locations, say position A and position B. On the one hand, according to quantum mechanics, the valid definition of such a superposed state requires the existence of a definite space-time background, in which position A and position B can be distinguished. On the other hand, according to general relativity, the space-time geometry, including the distinguishability of position A and position B, cannot be predetermined, and must be dynamically determined by the superposed state. Since the different position states in the superposition determine different space-time geometries, the space-time geometry determined by the whole superposition is indefinite, and as a result, the superposed state and its evolution cannot be consistently defined. In particular, the definition of the time-translation operator for the superposed space-time geometries involves an inherent ill-definedness, leading to an essential uncertainty in the energy of the superposed state. Then by analogy Penrose\index{Penrose, Roger} argued that this superposition, like an unstable particle in quantum mechanics, is also unstable, and it will decay or collapse into one of the two states in the superposition after a finite lifetime. 

Moreover, Penrose\index{Penrose, Roger} (1996) suggested that the essential energy uncertainty in the Newtonian limit is proportional to the gravitational self-energy $E_{\Delta}$ of the difference between the two mass distributions,\footnote{Penrose\index{Penrose, Roger}'s Newtonian expression for the energy uncertainty has been generalized to an arbitrary quantum superposition of relativistic, but weak, gravitational fields (Anandan\index{Anandan, Jeeva S.}, 1998).} and the collapse time, analogous to the half-life of an unstable particle, is 

\begin{equation}
T \approx \hbar/E_{\Delta}.
\label{PCT}
\end{equation}

\noindent This criterion is very close to that put forward by Di\'{o}si\index{Di\'{o}si, Lajos} (1989) earlier,\footnote{In Di\'{o}si\index{Di\'{o}si, Lajos}'s (1989) collapse model, the increase of energy induced by wavefunction collapse is too large to be consistent with experiments. This problem was pointed out and solved by Ghirardi\index{Ghirardi, GianCarlo}, Grassi\index{Grassi, Renata} and Rimini\index{Rimini, Alberto} (1990).} and it is usually called the Di\'{o}si\index{Di\'{o}si, Lajos}-Penrose\index{Penrose, Roger} criterion\index{Di\'{o}si-Penrose\index{Penrose, Roger} criterion}. Later, Penrose\index{Penrose, Roger} (1998) further suggested that the preferred bases (i.e. the states toward which the collapse tends) are the stationary solutions of the so-called Schr\"{o}dinger-Newton equation within Newtonian approximation.

Now I will examine Penrose\index{Penrose, Roger}'s gravity-induced\index{collapse models!gravity-induced} wavefunction collapse argument in detail (see also Gao\index{Gao, Shan}, 2013b). The crux of the argument is whether the conflict between quantum mechanics and general relativity requires that a quantum superposition of two space-time geometries must collapse after a finite time. I will argue in the following that the answer seems negative. First of all, although it is widely acknowledged that there exists a fundamental conflict between the superposition principle of quantum mechanics and the principle of general covariance of general relativity, it is still a controversial issue what the exact nature of the conflict is and how to resolve it. The problem is often referred to as the `problem of time' in various approaches to quantum gravity\index{quantum gravity} (Kucha\v r, 1992\index{Kucha\v r, Karel V.}; Isham, 1993; Isham\index{Isham, Christopher J.} and Butterfield\index{Butterfield, Jeremy}, 1999; Kiefer\index{Kiefer, Claus}, 2007; Anderson\index{Anderson, Edward}, 2012). It seems not impossible that the conflict may be solved by reformulating quantum mechanics in a way that does not rely on a definite space-time background (see, e.g. Rovelli\index{Rovelli, Carlo}, 2004, 2011).

Secondly, Penrose\index{Penrose, Roger}'s argument by analogy seems too weak to establish a necessary connection between wavefunction collapse and the conflict between general relativity and quantum mechanics. Even though there is an essential uncertainty in the energy of the superposition of different space-time geometries, this kind of energy uncertainty is different in nature from the energy uncertainty of unstable particles or unstable states in quantum mechanics (Gao\index{Gao, Shan}, 2010). The former results from the ill-definedness of the time-translation operator for the superposed space-time geometries, while the latter exists in a definite space-time background, and there is a well-defined time-translation operator for the unstable states. Moreover, the decay of an unstable state (e.g. an excited state of an atom) is a natural result of the linear quantum dynamics, and the process is not random but deterministic. In particular, the decay process is not spontaneous but caused by the background field constantly interacting with the unstable state, e.g. the state may not decay at all when being in a very special background field with bandgap (Yablonovitch\index{Yablonovitch, Eli}, 1987). In contrast, the hypothetical decay or collapse of the superposed space-time geometries is spontaneous, nonlinear and random. In short, there exists no convincing analogy between a superposition of different space-time geometries and an unstable state in quantum mechanics. Accordingly, one cannot argue for the collapse of the superposition of different space-time geometries by this analogy. Although an unstable state in quantum mechanics may decay after a very short time, this does not \emph{imply} that a superposition of different space-time geometries should also decay - and, again, sometimes an unstable state does not decay at all under special circumstances. To sum up, Penrose\index{Penrose, Roger}'s argument by analogy has a very limited force, and it is not strong enough to establish a necessary connection between wavefunction collapse and the conflict between quantum mechanics and general relativity.\footnote{In my opinion, Penrose\index{Penrose, Roger} also realized the limitation of the analogy and only considered it as a plausibility argument.}

Thirdly, it can be further argued that the conflict between quantum mechanics and general relativity does not necessarily lead to wavefunction collapse. The key is to realize that the conflict also needs to be resolved before the wavefunction collapse finishes, and when the conflict has been resolved, the wavefunction collapse will lose its physical basis relating to the conflict. As argued by Penrose\index{Penrose, Roger} (1996), a quantum superposition of different space-time geometries and its evolution are both ill-defined due to the fundamental conflict between the principle of general covariance of general relativity and the superposition principle of quantum mechanics, and the ill-definedness requires that the superposition must collapse into one of the definite space-time geometries, which has no problem of ill-definedness. However, the wavefunction collapse seems too late to save the superposition from the ``suffering" of the ill-definedness during the collapse. In the final analysis, the conflict or the problem of ill-definedness needs to be solved \emph{before} defining a quantum superposition of different space-time geometries and its evolution. In particular, the hypothetical collapse evolution of the superposition also needs to be consistently defined, which again indicates that wavefunction collapse does not solve the problem of ill-definedness. On the other hand, once the problem of ill-definedness is solved and a consistent description obtained, wavefunction collapse will lose its connection with the problem. Therefore, contrary to Penrose\index{Penrose, Roger}'s expectation, it seems that the conflict between quantum mechanics and general relativity does not entail the existence of wavefunction collapse.

Even though Penrose\index{Penrose, Roger}'s gravity-induced\index{collapse models!gravity-induced} wavefunction collapse argument may be problematic, it is still possible that wavefunction collapse is a real physical process as I have argued in previous sections. Moreover, Penrose\index{Penrose, Roger}'s collapse time formula can also be assumed as it is, and numerical estimates based on the formula for life-times of superpositions indeed turn out to be realistic (Penrose\index{Penrose, Roger}, 1994, 1996). Therefore, Penrose\index{Penrose, Roger}'s suggestions for the collapse time formula and the preferred basis also deserve to be examined as some aspects of a phenomenological model.

To begin with, I will analyze Penrose\index{Penrose, Roger}'s collapse time formula, Eq. (\ref{PCT}), according to which the collapse time of a superposition of two mass distributions is inversely proportional to the gravitational self-energy of the difference between the two mass distributions. As I have argued above, there does not exist a precise analogy between such a superposition and an unstable state in quantum mechanics, and gravity does not necessarily induce wavefunction collapse either. Thus this collapse time formula, which is originally based on a similar application of Heisenberg's uncertainty principle to unstable states, will lose its original physical basis. In particular, the appearance of the gravitational self-energy term in the formula is in want of a reasonable explanation (see below). In fact, it has been shown that this gravitational self-energy term does not represent the ill-definedness of time-translation operator in the strictly Newtonian regime (Christian\index{Christian, Joy}, 2001). In this regime, the time-translation operator can be well defined, but the gravitational self-energy term is obviously not zero. Moreover, as Di\'{o}si\index{Di\'{o}si, Lajos} (2007) pointed out, the microscopic formulation of Penrose\index{Penrose, Roger}'s collapse time formula also meets the cut-off difficulty.

Next, I will analyze Penrose\index{Penrose, Roger}'s choice of the preferred basis. According to Penrose\index{Penrose, Roger} (1998), the preferred bases are the stationary solutions of the Schr\"{o}dinger-Newton equation\index{Schr\"{o}dinger-Newton equation}:

\begin{equation}
i\hbar {\partial \psi(x,t) \over \partial t}=-{\hbar^2 \over 2m}\nabla^2 \psi(x,t)-Gm^2\int{{|\psi(x',t)|^2 \over |x-x'|}d^3 x'}\psi(x,t)+V\psi(x,t),
\label{SN}
\end{equation}

\noindent  where $m$ is the mass of a quantum system, $V$ is an external potential, $G$  is Newton's gravitational constant. This equation describes the gravitational self-interaction of a single quantum system, in which the mass density $m|\psi(x,t)|^2$ is the source of the classical gravitational potential. However, there is an obvious objection to the Schr\"{o}dinger\index{Schr\"{o}dinger, Erwin}-Newton equation\index{Schr\"{o}dinger-Newton equation} (see also \index{Giulini, Domenico}Giulini and Gro$\beta$ardt, 2012). Since charge accompanies mass for a charged particle such as an electron, the existence of the gravitational self-interaction, although which is too weak to be excluded by present experiments (Salzman\index{Salzman, P. J.} and Carlip\index{Carlip, Steven}, 2006; Giulini and Gro$\beta$ardt\index{Gro$\beta$ardt, Andr\'{e}}, 2011),\footnote{Note that Salzman and Carlip\index{Carlip, Steven} (2006) overestimated the influence of gravitational self-interaction on the dispersion of wavepackets by about 6 orders of magnitude. This was pointed out and corrected by Giulini and Gro$\beta$ardt (2011).} entails the existence of a remarkable electrostatic self-interaction of the charged particle, which, as I have argued before, is incompatible with experimental observations.\footnote{Since the Schr\"{o}dinger\index{Schr\"{o}dinger, Erwin}-Newton equation\index{Schr\"{o}dinger-Newton equation} is the nonrelativistic realization of the typical model of semiclassical gravity\index{semiclassical gravity}, in which the source term in the classical Einstein\index{Einstein, Albert} equation is taken as the expectation of the energy momentum operator in the quantum state (Rosenfeld\index{Rosenfeld, L\'{e}on}, 1963), the above analysis also presents a serious objection to the approach of semiclassical gravity\index{semiclassical gravity}. Note that although the existing arguments against the semiclassical gravity\index{semiclassical gravity} models seem very strong, they are not conclusive (Carlip\index{Carlip, Steven}, 2008; Boughn, 2009).} For example, for the electron in the hydrogen atom, the potential of the electrostatic self-interaction is of the same order as the Coulomb potential produced by the nucleus, and thus it is impossible that the revised Schr\"{o}dinger equation\index{Schr\"{o}dinger equation} with such an electrostatic self-interaction term, like the Schr\"{o}dinger equation\index{Schr\"{o}dinger equation}, gives predictions of the hydrogen spectra that agree with experiment.

On the other hand, it is worth noting that protective measurements\index{protective measurements} show that a charged quantum system such as an electron does have mass and charge distribution\index{charge distribution}s in space, and the mass and charge density in each position is also proportional to the modulus squared of the wave function of the system there (Aharonov\index{Aharonov, Yakir} and Vaidman\index{Vaidman, Lev}, 1993; Aharonov\index{Aharonov, Yakir}, Anandan\index{Anandan, Jeeva S.} and Vaidman\index{Vaidman, Lev}, 1993). However, as I have argued in Chapter 6, the distributions do not exist throughout space at the same time, for which there are gravitational and electrostatic self-interactions of the distributions. Rather, they are effective, formed by the ergodic motion of a point-like particle with the total mass and charge of the system. In this case, there will exist no gravitational and electrostatic self-interactions of the distributions. This is consistent with the predictions of quantum mechanics and experimental observations.

Finally, I will briefly discuss another two potential problems of Penrose\index{Penrose, Roger}'s collapse model. 
The first problem is the origin of the randomness of collapse results. 
Since wavefunction collapse is spontaneous in Penrose\index{Penrose, Roger}'s collapse model, the randomness of collapse results can  only come from the studied quantum system itself. Yet the gravitational field of the studied quantum system, even if it does induce wavefunction collapse, seems to fail to account for the origin of the randomness. 
The second problem is energy non-conservation\index{collapse theories!energy non-conservation problem of}. Although Penrose\index{Penrose, Roger} did not give a concrete model of wavefunction collapse, he thought that the energy uncertainty $E_{\Delta}$ may cover such a potential non-conservation, leading to no actual violation of energy conservation\index{conservation of energy!violation of} (Penrose\index{Penrose, Roger}, 2004). However, this is still a controversial issue. For instance, Di\'{o}si\index{Di\'{o}si, Lajos} (2007) pointed out that the von Neumann\index{von Neumann, John}-Newton equation\index{von Neumann-Newton equation}, which may be regarded as a realization of Penrose\index{Penrose, Roger}'s scheme, does not conserve energy. 

In conclusion, Penrose\index{Penrose, Roger}'s proposal that gravity induces wavefunction collapse seems debatable. However, as I have argued in previous sections, it is very likely that wavefunction collapse is a real physical process as Penrose\index{Penrose, Roger} thinks, although its origin remains a deep mystery. Moreover, relating the process with gravity is still an extremely crucial problem which deserves a lot of attention, and approaches that are not fully satisfactory may also give hints concerning where to go or how to proceed.



\subsection{The CSL model\index{CSL model}}

Different from Penrose\index{Penrose, Roger}'s gravity-induced\index{collapse models!gravity-induced}  collapse model, the CSL model\index{CSL model} is a typical interaction-induced\index{collapse models!interaction-induced} collapse model. In the model, it is assumed that the collapse of the wave function\index{wave function!collapse} of a quantum system is caused by its interaction with a classical scalar field, $w(x, t)$. Moreover, the preferred bases are the eigenstates of the smeared mass density operator, and the mechanism leading to the suppression of the superpositions of macroscopically different states is fundamentally governed by the integral of the squared differences of the mass densities associated to the superposed states. It may be expected that the introduction of the noise field can help solve the problems plagued by the spontaneous collapse models\index{collapse models!spontaneous}, e.g. the problems of energy non-conservation\index{collapse theories!energy non-conservation problem of} and the origin of randomness. However, one must first answer what field the noise field is and especially why it can collapse the wave functions of all quantum systems. The validity of the CSL model\index{CSL model} strongly depends on the existence of this hypothetical noise field. In the following, I will mainly analyze this important legitimization problem of the CSL model\index{CSL model}.\footnote{As admitted by Pearle\index{Pearle, Philip} (2009), ``When, over 35 years ago, ... I had the idea of introducing a randomly fluctuating quantity to cause wave function collapse, I thought, because there are so many things in nature which fluctuate randomly, that when the theory is better developed, it would become clear what thing in nature to identify with that randomly fluctuating quantity. Perhaps ironically, this problem of legitimizing the phenomenological CSL collapse description by tying it in a natural way to established physics remains almost untouched." Related to this legitimization problem\index{CSL model!legitimization problem of} is that the two parameters which specify the model are ad hoc (Pearle\index{Pearle, Philip}, 2007). These two parameters, which were originally introduced by Ghirardi\index{Ghirardi, GianCarlo}, Rimini\index{Rimini, Alberto} and Weber\index{Weber, Tullio} (1986\index{GRW theory}), are a distance scale, $a \approx 10^{-5}cm$, characterising the distance beyond which the collapse becomes effective, and a time scale,  $\lambda^{-1} \approx 10^{16}$sec, giving the rate of collapse for a microscopic system. If wavefunction collapse is a fundamental physical process related to other fundamental processes, the parameters should be able to be written in terms of other physical constants.}

Whatever the nature of the noise field $w(x, t)$ is, it cannot be quantum in the usual sense since its coupling to a quantum system is not a standard coupling between two quantum systems\index{CSL model!origin of the noise field in}. The coupling is anti-Hermitian, and the equation of the resulting dynamical collapse is not the standard Schr\"{o}dinger equation\index{Schr\"{o}dinger equation} with a stochastic potential either. According to our current understandings, the gravitational field is the only universal field that might be not quantized, although this possibility seems extremely small in the view of most researchers. Therefore, it seems natural to assume that this noise field is the gravitational field, and the randomness of collapse results originates from the fluctuations of the gravitational field (see, e.g. K\'{a}rolyh\'{a}zy\index{K\'{a}rolyh\'{a}zy, F.}, Frenkel\index{Frenkel, Andor} and Luk\'{a}cs\index{Luk\'{a}cs, B\'{e}la}, 1986; Di\'{o}si\index{Di\'{o}si, Lajos}, 1989, 2007; Adler\index{Adler, Stephen L.}, 2016).
In fact, it has been argued that in the CSL model\index{CSL model} the $w$-field energy density must have a gravitational interaction with ordinary matter (Pearle\index{Pearle, Philip} and Squires\index{Squires,  Euan J.}, 1996; Pearle\index{Pearle, Philip}, 2009). The argument of Pearle\index{Pearle, Philip} and Squires\index{Squires,  Euan J.} (1996) can be summarized as follows.\footnote{Pearle\index{Pearle, Philip} (2009) further argued that compatibility with general relativity requires a gravitational force exerted upon matter by the $w$-field.  However, as Pearle\index{Pearle, Philip} (2009) also admitted, no convincing connection (for example, identification of metric fluctuations, dark matter or dark energy with $w(x, t)$) has yet emerged, and the legitimization problem (i.e. the problem of endowing physical reality to the noise field) is still in its infancy.}

There are two equations which characterize the CSL model\index{CSL model}\index{CSL model!origin of the noise field in}. The first equation is a modified Schr\"{o}dinger equation\index{Schr\"{o}dinger equation}, which expresses the influence of an arbitrary field $w(x, t)$ on the studied  quantum system. The second equation is a probability rule whichs gives the probability that nature actually chooses a particular $w(x, t)$. This probability rule can also be interpreted as expressing the influence of the quantum system on the field. As a result, $w(x, t)$ can be written as follows:

\begin{equation}
w(x, t) = w_0(x, t) + \exptt{A(x, t)},
\label{}
\end{equation}

\noindent where $A(x, t)$ is the mass density operator smeared over the GRW scale $a$, $\exptt{A(x, t)}$ is its expectation value, and $w_0(x, t)$ is a Gaussian randomly fluctuating field with zero drift, temporally white noise in character and with a particular spatial correlation function. Then the scalar field $w(x, t)$ that causes wavefunction collapse can be interpreted as the gravitational curvature scalar with two sources, the expectation value of the smeared mass density operator and an independent white noise fluctuating source.\footnote{Recently Adler\index{Adler, Stephen L.} (2016) gave a new conjecture on the physical origin of the noise field, according to which the noise field comes from a rapidly fluctuating complex part of the classical gravitational metric.} This indicates that the CSL model\index{CSL model} is based on the semiclassical gravity\index{semiclassical gravity}, and the smeared mass density is the source of the gravitational potential.\footnote{Note that Ghirardi\index{Ghirardi, GianCarlo}, Grassi\index{Grassi, Renata} and Benatti (1995) and Ghirardi\index{Ghirardi, GianCarlo} (1997) explicitly introduced the mass density ontology\index{GRW theory!mass density ontology of} in the context of collapse theories\index{collapse theories} (see also Allori\index{Allori, Valia} et al, 2008; Bedingham\index{Bedingham, Daniel J.} et al, 2014). According to Ghirardi\index{Ghirardi, GianCarlo} (2016), ``what the theory is about, what is real `out there' at a given space point $x$, is just a field, i.e. a variable $m(x,t)$ given by the expectation value of the mass density operator $M(x)$ at $x$ obtained by multiplying the mass of any kind of particle times the number density operator for the considered type of particle and summing over all possible types of particles."}

According to my previous analysis in Chapter 6, however,assuming the existence of smeared mass density is arguably debatable.\index{CSL model!origin of the noise field in}
First, protective measurements\index{protective measurements} show that the mass density of a quantum system is proportional to the modulus squared of its wave function and thus it is not smeared. In other words,  the assumed existence of the smeared mass density in the CSL model\index{CSL model} contradicts the results of protective measurements\index{protective measurements}.
Note that it is crucial that the mass density of a quantum system should be smeared over the  GRW scale $a$ in the CSL model\index{CSL model}; without such a smearing the energy excitation of particles undergoing collapse would be beyond experimental constraints (Pearle\index{Pearle, Philip} and Squires\index{Squires,  Euan J.}, 1996).
Next, my previous analysis of Schr\"{o}dinger\index{Schr\"{o}dinger, Erwin!charge density hypothesis of}'s charge density hypothesis also applies to the mass density ontology\index{GRW theory!mass density ontology of}.
According to the analysis, the mass density of a quantum system, even if it is smeared, is not real but effective; it is formed by the ergodic motion of point-like particles with masses.
Therefore, it is arguable that the quantum ontology of the CSL model\index{CSL model} is not fields, but particles.\index{CSL model!and semiclassical gravity}
Finally, although the approach of semiclassical gravity\index{semiclassical gravity} may be consistent in the context of collapse models (Pearle\index{Pearle, Philip} and Squires\index{Squires,  Euan J.}, 1996; Ghirardi\index{Ghirardi, GianCarlo}, 2016), it may have been excluded as I have argued in the last subsection. 
In conclusion, it seems that the noise field introduced in the CSL model\index{CSL model} cannot have a gravitational origin, and this may raise a strong doubt about the reality of the field.\index{CSL model!origin of the noise field in}

On the other hand, even though the approach of semiclassical gravity\index{semiclassical gravity} is viable and the noise field in the CSL model\index{CSL model} can be the gravitational field, one still needs to answer why the gravitational field has the very ability to collapse the wave functions of all quantum systems as required by the model. It is worth noting that the randomly fluctuating field in the model, $w_0(x, t)$, is not the gravitational field of the studied quantum system but the background gravitational field. Therefore, Penrose\index{Penrose, Roger}'s gravity-induced\index{collapse models!gravity-induced} wavefunction collapse argument, even if it is valid, does not apply to the CSL model\index{CSL model}. The fluctuations of the background gravitational field can readily lead to the decoherence of the wave function of a quantum system, but it seems that they have no magical ability to cause the collapse of the wave function\index{wave function!collapse}.
Moreover, since the Schr\"{o}dinger equation\index{Schr\"{o}dinger equation} is purely deterministic, it seems that the random quantum fluctuations must also result from the collapse of the wave function\index{wave function!collapse} in such collapse models. If this is true, then these models will be based on circular reasonings.

Finally, I will briefly discuss another two problems of the CSL model\index{CSL model}.\footnote{Pearle\index{Pearle, Philip} (2007, 2009), Bassi\index{Bassi, Angelo} (2007) and Ghirardi\index{Ghirardi, GianCarlo} (2016) gave a very detailed analysis of the problems of the CSL model\index{CSL model} and the present status of the investigations of them.} The first problem is energy non-conservation\index{collapse theories!energy non-conservation problem of}\index{CSL model!energy non-conservation problem of}. The collapse in the model narrows the wave function in position space, thereby producing an increase of energy.\footnote{Note that with appropriate choice for the parameters in the CSL model\index{CSL model}, such a violation of energy conservation\index{conservation of energy!violation of} is very tiny and hardly detectable by present day technology. For a recent proposal of experimental test see Di\'{o}si\index{Di\'{o}si, Lajos} (2015).} A possible solution to this problem is that the conservation laws may be satisfied when the contributions of the noise field $w(x, t)$ to the conserved quantities are taken into account. It has been shown that the total mean energy can be conserved (Pearle\index{Pearle, Philip}, 2004), and the energy increase can also be made finite when revising the coupling between the noise field and the studied quantum system (Bassi\index{Bassi, Angelo}, Ippoliti and Vacchini, 2005). But a complete solution has not been found yet, and it is still unknown whether such a solution indeed exists.

The second problem is to make a relativistic quantum field theory\index{quantum field theory} which describes wavefunction collapse (Pearle\index{Pearle, Philip}, 2009; Ghirardi\index{Ghirardi, GianCarlo}, 2016). Notwithstanding a good deal of effort, a satisfactory theory has not been obtained.\footnote{For example, Tumulka\index{Tumulka, Roderich} (2006, 2009) proposed a relativistic version of the GRW\index{GRW theory} flash theory. But it is still debatable whether the ontology of the theory, which is known as the GRW flash ontology\index{GRW theory!flash ontology of}, is satisfactory (Esfeld\index{Esfeld, Michael} and Gisin\index{Gisin, Nicolas}, 2014).} The main difficulty is that the hypothetical interaction responsible for collapse will produce too many particles out of the vacuum, amounting to infinite energy per second per volume, in the relativistic extension of these interaction-induced\index{collapse models!interaction-induced} collapse models.
It has been suggested that the problem of infinities\index{CSL model!problem of infinities of} may be solved by smearing out the point interactions. For example, Nicrosini\index{Nicrosini, Oreste}\index{Rimini, Alberto} and Rimini (2003) showed that this is possible when including a locally preferred frame. More recently, Bedingham\index{Bedingham, Daniel J.} (2011) introduced a new relativistic field responsible for mediating the collapse process, and showed that his model can fulfill the aim of smearing the interactions whilst preserving Lorentz covariance and frame independence. Whether this promising model is wholly satisfactory needs to be further studied.
Note that spontaneous collapse models\index{collapse models!spontaneous} without collapse interaction (e.g. my energy-conserved\index{collapse models!energy-conserved} collapse model) do not have the problem of infinities. I will analyze the problem of compatibility between wavefunction collapse and the principle of relativity in the next chapter.


\chapter{Quantum ontology and relativity}


In this chapter, I will give a primary analysis of how special relativity\index{special relativity} influences the suggested ontology of quantum mechanics\index{quantum mechanics!ontology of}, namely the RDM of particles\index{random discontinuous motion of particles}, as well as how the quantum ontology influences special relativity\index{special relativity} reciprocally.

It is well known that there are two important conceptual issues concerning the unification of quantum mechanics and special relativity\index{special relativity}. First, the apparent incompatibility between wavefunction collapse (and the resulting quantum nonlocality) and the principle of relativity has been an unsolved problem since the founding of quantum mechanics (see, e.g. Bell\index{Bell, Mary} and Gao\index{Gao, Shan}, 2016 and references therein). For example, it is still debatable whether a preferred Lorentz frame\index{preferred Lorentz frame} is needed to solve the incompatibility problem. Second, although the combination of the linear quantum dynamics and special relativity\index{special relativity} has been obtained in quantum field theory\index{quantum field theory!ontology of}, it is still a controversial issue what the ontology of the theory really is (see, e.g. Cao, 1999\index{Cao, Tian Yu}; Kuhlmann\index{Kuhlmann, Meinard}, 2015, section 5). Is it fields or particles? Or it is other physical entities?
In the following sections, I will analyze how to solve these problems in terms of RDM of particles\index{random discontinuous motion of particles}. 
The analysis may not only give the picture of quantum ontology in the relativistic domain, but also suggest how special relativity\index{special relativity} should be revised to accommodate the quantum ontology.

\section{The picture of motion distorted by the Lorentz transformation\index{Lorentz transformation}}

Let us first see how the picture of RDM of particles\index{random discontinuous motion of particles} is changed by the Lorentz transformation\index{Lorentz transformation}.

\subsection{Picture of motion of a single particle}

For the RDM of a particle, the particle has a tendency to be in any possible position at a given instant, and the probability density that the particle appears in each position $x$ at a given instant $t$ is given by the modulus squared of its wave function, namely $\rho(x,t)=|\psi(x,t)|^2$. The physical picture of the motion of the particle is as follows. At a discrete instant the particle randomly stays in a position, and at the next instant it will still stay there or randomly appear in another position, which is probably not in the neighborhood of the previous position. In this way, during a time interval much longer than the duration of a discrete instant, the particle will move discontinuously and randomly throughout the whole space with position probability density $\rho(x,t)$. Since the distance between the locations occupied by the particle at two neighboring instants may be very large, this jumping process is obviously nonlocal. In the nonrelativistic domain where time is absolute, the nonlocal jumping process is the same in every inertial frame. But in the relativistic domain, the jumping process will look different in different inertial frames due to the Lorentz transformation\index{Lorentz transformation}.

Suppose a particle is in position $x_1$ at instant $t_1$ and in position $x_2$ at instant $t_2$ in an inertial frame $S$. In another inertial frame $S'$ with velocity $v$ relative to $S$, the Lorentz transformation\index{Lorentz transformation} leads to:

\begin{equation}
t_1^{'}={{t_1-x_1v/c^2} \over \sqrt{1-v^2/c^2} },
\label{}
\end{equation}

\begin{equation}
t_2^{'}={{t_2-x_2v/c^2} \over \sqrt{1-v^2/c^2} },
\label{}
\end{equation}

\begin{equation}
x_1^{'}={{x_1-vt_1} \over \sqrt{1-v^2/c^2} },
\label{}
\end{equation}

\begin{equation}
x_2^{'}={{x_2-vt_2} \over \sqrt{1-v^2/c^2} }.
\label{}
\end{equation}

\noindent Since the jumping process of the particle is nonlocal, the two events ($t_1$, $x_1$) and ($t_2$, $x_2$) may readily satisfy the spacelike separation condition $|x_2-x_1|>c|t_2-t_1|$. Then we can always select a possible velocity $v<c$ that leads to $t_1^{'}=t_2^{'}$:

\begin{equation}
v={ {t_2-t_1} \over {x_2-x_1}}c^2.
\label{}
\end{equation}

\noindent But obviously the two positions of the particle in frame $S'$, namely $x_1^{'}$ and $x_2^{'}$, are not equal. This means that in frame $S'$ the particle will be in two different positions $x_1^{'}$ and $x_2^{'}$ at the same time at instant $t_1^{'}$. In other words, it seems that there are two identical particles at instant $t_1^{'}$ in frame $S'$. Note that the velocity of $S'$ relative to $S$ may be much smaller than the speed of light, and thus the appearance of the two-particle picture is irrelevant to the high-energy processes described by relativistic quantum field theory\index{quantum field theory}, e.g. the creation and annihilation of particles. 

The above result shows that for any pair of events in frame $S$ that satisfies the spacelike separation condition, there always exists an inertial frame in which the two-particle picture will appear. Since the jumping process of the particle in frame $S$ is essentially random, it can be expected that the two-particle picture will appear in infinitely many inertial frames with the same probability. Then during an arbitrary finite time interval, in each inertial frame the measure of the set of the instants at which there are two particles in appearance, which is equal to the length of the time interval divided by the total number of the frames that is infinite, will be zero. Moreover, there may also exist situations where the particle is at arbitrarily many positions at the same time at an instant in an inertial frame, although the appearing probability of these situations is also zero. Certainly, at nearly all instants in a time interval, whose measure is equal to the length of the time interval, the particle is still in one position at an instant in all inertial frames. Therefore, the many-particle appearance of the RDM of a particle cannot be measured in principle.

However, for the RDM of a particle, in any inertial frame different from $S$, the Lorentz transformation\index{Lorentz transformation} will generally make the time order of the random stays of the particle in $S$ reversal, since the discontinuous motion of the particle is nonlocal and most neighboring random stays are spacelike separated events. In other words, the time order is not Lorentz invariant. Moreover, the set of the instantsin a time interval  at which the time order of the random stays of the particle is reversed has non-zero measure, which may be close to the length of the time interval. As we will see below, this reversal of time order will lead to more distorted pictures for quantum entanglement\index{quantum entanglement} and wavefunction collapse.

\subsection{Picture of quantum entanglement\index{quantum entanglement}}

Now I will analyze the motion of two particles in an entangled state. For the RDM of two particles in an entangled state, the two particles have a joint tendency to be in any two possible positions, and the probability density that the two particles appear in each position pair $x_1$ and $x_2$ at a given instant $t$ is determined by the modulus squared of their wave function at the instant, namely $\rho(x_1,x_2,t)=|\psi(x_1,x_2,t)|^2$.

Suppose two particles are in an entangled state ${1 \over \sqrt{2}}(\psi_u\varphi_u+\psi_d\varphi_d$), where $\psi_u$ and $\psi_d$ are two spatially separated states of particle 1, $\varphi_u$ and $\varphi_d$ are two spatially separated states of particle 2. The physical picture of this entangled state is as follows. At an instant, particles 1 and 2 randomly stay in two positions in the region where the state $\psi_u\varphi_u$ or $\psi_d\varphi_d$ spreads. At the next instant, they will still stay there or jump to another two positions in the region where the state $\psi_u\varphi_u$ or $\psi_d\varphi_d$ spreads.  During a very short time interval, the two particles will discontinuously and randomly move throughout the two regions where the states $\psi_u\varphi_u$ and $\psi_d\varphi_d$ spread with the same probability $1 \over 2$. In this way, the two particles jump in a precisely simultaneous way. At an arbitrary instant, if particle 1 is in the region of $\psi_u$ or $\psi_d$, then particle 2 must be in the region of $\varphi_u$ or $\varphi_d$, and vice versa. Moreover, when particle 1 jumps from the region of $\psi_u$ to the region of $\psi_d$ or from the region of $\psi_d$ to the region of $\psi_u$, particle 2 must simultaneously jump from the region of $\varphi_u$ to the region of $\varphi_d$ or from the region of $\varphi_d$ to the region of $\varphi_u$, and vice versa. Note that this kind of  random synchronicity between the motion of particle 1 and the motion of particle 2 is irrelevant to the distance between them, and it can only be accounted for by the existence of joint tendency of the two particles as a whole.

The above picture of quantum entanglement\index{quantum entanglement} is supposed to exist in one inertial frame. It can be expected that when observed in another inertial frame, this perfect picture will be distorted in a similar way as the single particle picture. Let me give a concrete analysis. Suppose in an inertial frame $S$, at instant $t_a$ particle 1 is in position $x_{1a}$ in the region of $\psi_u$ and particle 2 is in position $x_{2a}$ in the region of  $\varphi_u$, and at instant $t_b$ particle 1 is in position $x_{1b}$ in the region of $\psi_d$ and particle 2 is in position $x_{2b}$  in the region of $\varphi_d$. Then according to the Lorentz transformation\index{Lorentz transformation}, in another inertial frame $S'$ with velocity $v'$ relative to $S$, where $v'$ satisfies:

\begin{equation}
v'={ {t_a-t_b} \over {x_{1a}-x_{2b}}}c^2,
\label{V'}
\end{equation}

\noindent the instant at which particle 1 is in position $x'_{1a}$ in the region of $\psi'_u$ is the same as the instant at which particle 2 is in position $x'_{2b}$ in the region of $\varphi'_d$, namely

\begin{equation}
t'_{1a}=t'_{2b}={1 \over \sqrt{1-v'^2/c^2}}  { {x_{1a}t_b-x_{2b}t_a} \over {x_{1a}-x_{2b}}}.
\label{}
\end{equation}

\noindent This means that in $S'$ there exists an instant at which particle 1 is in the region of $\psi'_u$ but particle 2 is in the region of $\varphi'_d$. Similarly, in another inertial frame $S''$ with velocity $v''$ relative to $S$, there also exists an instant $t''$ at which particle 1 is in the region of $\psi''_d$ but particle 2 is in the region of $\varphi''_u$, where $v''$ and $t''$ satisfy the following relations:

\begin{equation}
v''={ {t_a-t_b} \over {x_{2a}-x_{1b}}}c^2,
\label{V''}
\end{equation}

\begin{equation}
t''={1 \over \sqrt{1-v''^2/c^2}} { {x_{2a}t_b-x_{1b}t_a} \over {x_{2a}-x_{1b}}}.
\label{}
\end{equation}

\noindent Note that since the two particles are well separated in space, the above two velocities can readily satisfy the restricting conditions $v'<c$ and $v''<c$ when the time interval $|t_a-t_b|$ is very short. 

Since the motion of particles is essentially random, in any inertial frame $S'$ which is different from $S$, the original correlation between the motion of the two particles in $S$ can only keep half the time for the above entangled state, and the correlation will be reversed for the other half of time, during which the two particles will be in state $\psi'_u\varphi'_d$ or $\psi'_d\varphi'_u$ at each instant. For a general entangled state $a\psi_u\varphi_u+b\psi_d\varphi_d$, where $|a|^2+|b|^2=1$, the proportion of correlation-reversed time will be $2|ab|^2$, and the proportion of correlation-kept time will be $|a|^4 + |b|^4$. Moreover, the instants at which the original correlation is kept or reversed are discontinuous and random. This means that the synchronicity between the jumps of the two particles is destroyed too. 

To sum up, the above analysis indicates that the instantaneous correlation and synchronicity between the motion of two entangled particles in one inertial frame is destroyed in other frames due to the Lorentz transformation\index{Lorentz transformation}.\footnote{Certainly, in these frames there are still correlations and synchronicity between the jumps of the two particles at different instants. As noted above, however, since these instants are discontinuous and random, such correlations and synchronicity can hardly be identified.} As we will see below, however, this distorted picture of quantum entanglement\index{quantum entanglement} cannot be measured either.

\subsection{Picture of wavefunction collapse}

I have shown that the picture of the instantaneous motion of particles is distorted by the Lorentz transformation\index{Lorentz transformation} due to the nonlocality and randomness of motion. In the following, I will further show that 
the nonlocal and random collapse evolution of the state of motion (defined during an infinitesimal time interval around a given instant) will be influenced more seriously by the Lorentz transformation\index{Lorentz transformation}.

Consider a particle being in a superposition of two Gaussian wavepacket\index{Gaussian wavepacket}s ${1 \over \sqrt{2}}(\psi_1+\psi_2)$ in an inertial frame $S$. The centers of the two wavepackets are located in $x_1$ and $x_2$ ($x_1<x_2$), respectively, and the width of each wavepacket is much smaller than the distance between them. 
Suppose after an appropriate measurement, this superposed state randomly collapses to $\psi_1$ or $\psi_2$ with the same probability $1 \over 2$, and the collapse happens in different positions at the same time in $S$. This means that when the superposition collapses to the branch $\psi_1$ near position $x_1$, the other branch $\psi_2$ near position  $x_2$ will disappear simultaneously. The simultaneity\index{simultaneity} of wavefunction collapse ensures that the sum of the probabilities of the particle being in all branches is 1 at every instant.

According to the picture of RDM of particles\index{random discontinuous motion of particles}, the above collapse process can be described as follows. Before the collapse of the superposition ${1 \over \sqrt{2}}(\psi_1+\psi_2)$, the particle jumps between the two branches $\psi_1$ and  $\psi_2$ or the two regions near $x_1$ and $x_2$ in a discontinuous and random way. At each instant, the particle is either in a position near $x_1$ or in a position near $x_2$, and its probability of being in each region is the same $1 \over 2$. This means that at every instant there is always one particle, which spends half the time near $x_1$ and half the time near $x_2$.  After the superposition collapses to one of its branches, e.g. $\psi_1$, the particle only jumps in the region near $x_1$ in a discontinuous and random way, and its probability of being in this region is 1. This means that at every instant there is always one particle being in a position inside the region. 

Now let us see the picture of the above collapse process in another inertial frame $S'$ with velocity $v$ relative to $S$.  Suppose the superposition ${1 \over \sqrt{2}}(\psi_1+\psi_2)$ collapses to the branch $\psi_1$ near position $x_1$ at instant $t$ in $S$. This process contains two events happening simultaneously in two spatially separated regions. One event is the disappearance of the branch ${1 \over \sqrt{2}}\psi_2$ near position $x_2$ at instant $t$, and the other event is the change of the branch ${1 \over \sqrt{2}}\psi_1$ to $\psi_1$ happening near position $x_1$ at instant $t$.\footnote{Strictly speaking, since the collapse time is  finite, these events happen not at a precise instant but during a very short time, which may be much shorter than the time of light propagating between $x_1$ and $x_2$.} According to the Lorentz transformation\index{Lorentz transformation}, the times of occurrence of these two events in $S'$ are

\begin{equation}
t'_1={{t-x_1v/c^2} \over \sqrt{1-v^2/c^2} },
\label{}
\end{equation}

\begin{equation}
t'_2={{t-x_2v/c^2} \over \sqrt{1-v^2/c^2} }.
\label{}
\end{equation}

\noindent It can be seen that $x_1<x_2$ leads to $t'_1>t'_2$. Then during the period between $t'_1$ and $t'_2$, the branch ${1 \over \sqrt{2}}\psi'_2$ near position $x'_2$ has already disappeared, but the branch ${1 \over \sqrt{2}}\psi'_1$  near position $x'_1$ has not changed to $\psi'_1$. This means that at any instant between $t'_1$ and $t'_2$, there is only a non-normalized state ${1 \over \sqrt{2}}\psi'_1$. According to the picture of  RDM of particles\index{random discontinuous motion of particles}, for a particle being in the state ${1 \over \sqrt{2}}\psi'_1$, the probability of the particle being in the branch $\psi'_1$ or in the region near $x_1$ is $1 \over 2$. 
In other words, at each instant the particle either exists in a position near $x_1$ or disappears in the whole space with the same probability $1 \over 2$. This result indicates that in $S'$ the particle only exists half the time during the period between $t'_1$ and $t'_2$. In contrast, the particle always exists in a certain position in space at any time in $S$.

Similarly, if the superposition ${1 \over \sqrt{2}}(\psi_1+\psi_2)$  collapses to the branch $\psi_2$ near position $x_2$ at instant $t$ in $S$, then in $S'$, during the period between $t'_1$ and $t'_2$, the branch ${1 \over \sqrt{2}}\psi'_2$ near position $x'_2$ has already changed to $\psi'_2$, while the branch ${1 \over \sqrt{2}}\psi'_1$ near position $x'_1$ has not disappeared and is still there. Therefore, there is only a non-normalized state ${1 \over \sqrt{2}}\psi'_1+\psi'_2$ at any instant between $t'_1$ and $t'_2$. According to the picture of RDM of particles\index{random discontinuous motion of particles}, this means that during the period between $t'_1$ and $t'_2$, there is more than one particle in $S'$: the first particle is in the branch $\psi'_2$ all the time, and the second identical particle exists half the time in the branch $\psi'_1$ (and it exists nowhere in space half the time).

However, although the state of the particle in $S'$ is not normalized, the total probability of \emph{finding} the particle in the whole space is still 1, not $1 \over 2$ or $3 \over 2$, in the frame. In other words, although the collapse process is seriously distorted in $S'$, the distortion  cannot be measured. The reason is that in $S'$ the collapse resulting from measurement happens at different instants in different positions,\footnote{Concretely speaking, the time order of the collapses happening in different positions in $S'$ is connected with the time order of the corresponding collapses in $S$ by the Lorentz transformation\index{Lorentz transformation}.} and the superposition of the branches in these positions and at these instants is always normalized. In the following, I will give a more detailed explanation.

As noted above, in $S'$ the collapse first happens at $t'_2$ for the branch ${1 \over \sqrt{2}}\psi'_2$ near position $x'_2$, and then happens at $t'_1$ for the branch ${1 \over \sqrt{2}}\psi'_1$ near position $x'_1$ after a delay. If we measure the branch ${1 \over \sqrt{2}}\psi'_2$, then the resulting collapse will influence the other branch ${1 \over \sqrt{2}}\psi'_1$ only after a delay of $\Delta t'={{|x_1-x_2|v/c^2} \over \sqrt{1-v^2/c^2} }$, while if we measure the branch ${1 \over \sqrt{2}}\psi'_1$, then the resulting collapse will influence the other branch ${1 \over \sqrt{2}}\psi'_2$ in advance by the same time interval $\Delta t'$, and the influence is backward in time. Now suppose we make a measurement on the branch ${1 \over \sqrt{2}}\psi'_2$ near position $x'_2$ and detect the particle there (i.e., the state after collapse is $\psi'_2$). Then before the other branch ${1 \over \sqrt{2}}\psi'_1$ disappears, which happens after a delay of $\Delta t'$, we can make a second measurement on the branch ${1 \over \sqrt{2}}\psi'_1$ near position $x'_1$. It seems that the probability of finding the particle there is not 0 but $1 \over 2$, and thus the total probability of finding the particle in the whole space is great than one and it is possible that we can detect two particles. However, this is not the case. Although the second measurement on the branch ${1 \over \sqrt{2}}\psi'_1$ near position $x'_1$ is made later than the first measurement, it is the second measurement that collapses the superposition  ${1 \over \sqrt{2}}(\psi'_1+\psi'_2)$ to $\psi'_2$ near position $x'_2$; the local branch ${1 \over \sqrt{2}}\psi'_1$ near position $x'_1$ disappears immediately after the measurement, while the influence of the resulting collapse on the other branch ${1 \over \sqrt{2}}\psi'_2$ near position $x'_2$ is backward in time and happens before the first measurement on this branch. Therefore, the second measurement near position $x'_1$ must obtain a null result, and the reason the first measurement detects the particle near position $x'_2$ is that the superposition already collapses to $\psi'_2$ near position $x'_2$ before the measurement due to the second measurement.

By a similar analysis, it can also be shown that the measurements on an entangled state of two particles, e.g. $\psi_u\varphi_u+\psi_d\varphi_d$, can only obtain correlated results in every inertial frame. If a measurement on particle 1 obtains the result $u$ or $d$, indicating the state of the particle collapses to the state $\psi_u$ or $\psi_d$ after the measurement, then a second measurement on particle 2 can only obtain the result $u$ or $d$, indicating the state of particle 2 collapses to the state $\varphi_u$ or $\varphi_d$ after the measurement. Accordingly, although the instantaneous correlation and synchronicity between the motion of two entangled particles is destroyed in all but one inertial frame, the distorted picture of quantum entanglement\index{quantum entanglement} cannot be measured.

\section{Simultaneity: relative or absolute?}

The above analysis clearly demonstrates the apparent conflict between the RDM of particles\index{random discontinuous motion of particles} and the Lorentz transformation\index{Lorentz transformation} in special relativity\index{special relativity}. The crux of the matter lies in the relativity of simultaneity\index{simultaneity!relativity of}. If simultaneity\index{simultaneity} is relative as manifested by the Lorentz transformation\index{Lorentz transformation}, then the picture of RDM of particles\index{random discontinuous motion of particles} will be seriously distorted except in a preferred frame, although the distortion is unobservable in principle. Only when simultaneity\index{simultaneity} is absolute, can the picture of RDM of particles\index{random discontinuous motion of particles} be kept perfect in every inertial frame. 


Although the relativity of simultaneity\index{simultaneity!relativity of} has been often regarded as one of the essential concepts of special relativity\index{special relativity}, it is not necessitated by experimental facts but a result of the choice of standard synchrony\index{special relativity!standard synchrony in} (see, e.g. Reichenbach\index{Reichenbach, Hans}, 1958; Gr\"{u}nbaum\index{Gr\"{u}nbaum, Adolf}, 1973).\footnote{For a comprehensive discussion of this issue see Janis\index{Janis, Allen} (2014) and references therein.} As Einstein\index{Einstein, Albert} (1905) already pointed out in his first paper on special relativity\index{special relativity}, whether or not two spatially separated events are simultaneous depends on the adoption of a convention in the framework of special relativity\index{special relativity}. In particular, the choice of standard synchrony\index{special relativity!standard synchrony in}, which is based on the constancy of the one-way speed of light and results in the relativity of simultaneity\index{simultaneity!relativity of}, is only a convenient convention.\footnote{The standard synchrony\index{special relativity!standard synchrony in} can be described in terms of the following thought experiment. There are two spatial locations $A$ and $B$ in an inertial frame. Let a light ray, traveling in vacuum, leave $A$ at time $t_1$ (as measured by a clock at rest there), and arrive at $B$ coincident with the event $E$ at $B$. Let the ray be instantaneously reflected back to $A$, arriving at time $t_2$ (as measured by the same clock at rest there). Then the standard synchrony\index{special relativity!standard synchrony in} is defined by saying that $E$ is simultaneous with the event at A that occurred at time $(t_1+t_2)/2$. This definition is equivalent to the requirement that the one-way speeds of light are the same on the two segments of its round-trip journey between $A$ and $B$. Here one may argue that the definition of standard synchrony\index{special relativity!standard synchrony in} makes use only of the relation of equality (of the one-way speeds of light in different directions), so that simplicity dictates its choice. However, even in the framework of special relativity\index{special relativity}, since the equality of the one-way speeds of light is a convention this choice does not simplify the postulational basis of the theory but only gives a symbolically simpler representation (Gr\"{u}nbaum\index{Gr\"{u}nbaum, Adolf}, 1973). On the other hand, as I have demonstrated in the last section, when going beyond the framework of special relativity\index{special relativity} and considering the RDM of particles\index{random discontinuous motion of particles} and its collapse evolution, standard synchrony\index{special relativity!standard synchrony in} is not simple but complex and it will lead to serious distortions in describing the dynamical collapse of the wave function\index{wave function!collapse}.} Strictly speaking, the speed constant $c$ in special relativity\index{special relativity} is two-way speed, not one-way speed, and as a result, the general space-time transformation required by the constancy of the two-way speed of light is not the Lorentz transformation\index{Lorentz transformation} but the Edwards-Winnie transformation\index{Edwards-Winnie transformation} (Edwards\index{Edwards, W. F.}, 1963; Winnie\index{Winnie, John A.}, 1970): 

\begin{equation}
x'=\eta(x-vt),
\label{}
\end{equation}

\begin{equation}
t'=\eta[1+\beta (k+k')]t+\eta [\beta (k^2-1)+k-k']x/c,
\label{}
\end{equation}

\noindent where $x,t$ and $x',t'$ are the coordinates of inertial frames $S$ and $S'$, respectively, $v$ is the velocity of $S'$ relative to $S$, $c$ is the invariant two-way speed of light, $\beta=v/c$, and $\eta=1/\sqrt{(1+\beta k)^2-\beta^2}$. $k$ and $k'$ represent the directionality of the one-way  speed of light in $S$ and $S'$, respectively, and they satisfy $-1 \leqslant  k, k' \leqslant 1$. Concretely speaking, the one-way  speeds of light along $x$ and $-x$  directions in $S$ are $c_x={c \over {1-k}}$ and $c_{-x}={c \over {1+k}}$, respectively, and the one-way  speeds of light along $x'$ and $-x'$ directions in $S'$ are $c_{x'}={c \over {1-k'}}$ and $c_{-x'}={c \over {1+k'}}$, respectively. 

If we adopt the convention of standard synchrony\index{special relativity!standard synchrony in}, namely assuming that the one-way speed of light is isotropic and constant in every inertial frame, then $k,k' =0$ and the Edwards-Winnie transformation\index{Edwards-Winnie transformation}  reduces to the Lorentz transformation\index{Lorentz transformation}, which leads to the relativity of simultaneity\index{simultaneity!relativity of}. Alternatively, one can  adopt the convention of nonstandard synchrony\index{special relativity!standard synchrony in} that makes simultaneity\index{simultaneity} absolute. In order to do this, one may first synchronize the clocks at different locations in an arbitrary inertial frame by Einstein\index{Einstein, Albert}'s standard synchrony\index{special relativity!standard synchrony in}, that is, one may assume the one-way speed of light is isotropic in this frame, and then let the clocks in other frames be directly regulated by the clocks in this frame when they coincide in space. The corresponding space-time transformation can be derived as follows. Let $S$ be the preferred Lorentz frame\index{preferred Lorentz frame} in which the one-way speed of light is isotropic, namely, let $k = 0$. Then we get

\begin{equation}
k'= \beta(k^2 -1) + k = -\beta. 
\label{}
\end{equation}

\noindent Since the synchrony convention leads to the absoluteness of simultaneity\index{simultaneity}, we also have in the Edwards-Winnie transformation\index{Edwards-Winnie transformation}:

\begin{equation}
\beta (k^2-1)+k-k'=0.
\label{}
\end{equation}

\noindent Thus the space-time transformation that restores absolute simultaneity\index{simultaneity!absolute} is:

\begin{equation}
x'={1 \over \sqrt{1-v^2/c^2} } (x-vt),
\label{}
\end{equation}

\begin{equation}
t'=\sqrt{1-v^2/c^2} t,
\label{}
\end{equation}

\noindent where  $x,t$ are the coordinates of the preferred Lorentz frame\index{preferred Lorentz frame} $S$, $x',t'$ are the coordinates of another inertial frame $S'$, and $v$ is the velocity of this frame relative to the preferred frame. In $S'$, the one-way  speed of light along the $x'$ and $-x'$ direction is $c_{x'}={c^2 \over {c-v}}$ and $c_{-x'}={c^2 \over {c+v}}$, respectively.

The above analysis demonstrates the possibility of keeping simultaneity\index{simultaneity} absolute within the framework of special relativity\index{special relativity}. One can adopt the convention of standard synchrony\index{special relativity!standard synchrony in} that leads to the relativity of simultaneity\index{simultaneity!relativity of}, and one can also adopt the convention of nonstandard synchrony\index{special relativity!standard synchrony in} that restores the absoluteness of simultaneity\index{simultaneity}.
It is often thought that if there is a causal influence connecting two distinct events, then the claim that they are not simultaneous will have a nonconventional basis (Reichenbach\index{Reichenbach, Hans}, 1958; Gr\"{u}nbaum\index{Gr\"{u}nbaum, Adolf}, 1973; Janis\index{Janis, Allen}, 2014). In particular, if there is an arbitrarily fast causal influence connecting two spacelike separated events, then these two events will be simultaneous and simultaneity\index{simultaneity} will be nonconventional. 
In my opinion, this view is problematic, since it depends on a debatable concept of causality.
It can be argued that a causal influence may be instantaneous and its transmission requires no time, and moreover, a cause may not precede its effect either, namely there may exist retrocausality\index{retrocausality} (Price\index{Price, Huw}, 2008; Faye\index{Faye, Jan}, 2015; Price\index{Price, Huw} and Wharton\index{Wharton, Ken}, 2016).\footnote{In my view, the time order of two events, causally related or otherwise, is not necessarily invariant, and it may be physically meaningless. } Then even if there is a causal influence connecting two distinct events, one cannot make sure that  these two  events are not simultaneous, and one event must precede the other event.
In the final analysis, it seems that one still needs a certain way of synchrony to determine the transmission speed and direction of a causal influence. 

Although the concept of causality may not provide a nonconventional basis of simultaneity\index{simultaneity}, some other requirements may do.
For example, if it is required that the number of particles is constant in every inertial frame, then the RDM of particles\index{random discontinuous motion of particles} and its collapse evolution will provide a nonconventional basis for the absoluteness of simultaneity\index{simultaneity}.
As I have shown in the last section, for the RDM of particles\index{random discontinuous motion of particles}, 
if adopting standard synchrony\index{special relativity!standard synchrony in} that leads to the relativity of simultaneity\index{simultaneity!relativity of}, 
then even if there is always one particle in a preferred frame, there may exist two particles at some instants in  other inertial frames.
Similarly, during the collapse evolution of the RDM of particles\index{random discontinuous motion of particles}, if adopting standard synchrony\index{special relativity!standard synchrony in} that leads to the relativity of simultaneity\index{simultaneity!relativity of}, then even if there is always one particle in the preferred frame, there may exist two particles or no particles at some instants in other inertial frames.
Therefore, if the number of particles is required to be constant in every inertial frame, then the existence of RDM of particles\index{random discontinuous motion of particles} and its collapse evolution will require that simultaneity\index{simultaneity} is absolute, thus providing a nonconventional basis for the absoluteness of simultaneity\index{simultaneity}.
However, one may still object that this requirement is not necessary. For instance, one may think the picture of RDM of particles\index{random discontinuous motion of particles} is only real in the preferred frame, and the appearance of the two-particle picture etc in other inertial frames is a mere illusion. Then the nonconventional basis of simultaneity\index{simultaneity} provided by the requirement is also debatable.

In any case, whether simultaneity\index{simultaneity} is relative or absolute, there is always a preferred Lorentz frame\index{preferred Lorentz frame!existence of} for the RDM of particles\index{random discontinuous motion of particles} and its collapse evolution in the relativistic domain. 
If the invariance of the one-way speed of light or standard synchrony\index{special relativity!standard synchrony in} is assumed, then the collapse evolution of the RDM of particles\index{random discontinuous motion of particles} will not happen simultaneously at different locations in space in all but one Lorentz frame, and thus it will  single out  a preferred Lorentz frame\index{preferred Lorentz frame}, in which the collapse of the wave function\index{wave function!collapse} happens simultaneously at different locations in space.
Alternatively, if restoring absolute simultaneity\index{simultaneity!absolute} and assuming the collapse of the wave function\index{wave function!collapse} happens simultaneously at different locations in space in every inertial frame, then the one-way speed of light will be not isotropic in all but one Lorentz frame, and thus the non-invariance of the one-way speed of light will also single out a preferred Lorentz frame\index{preferred Lorentz frame}, in which the one-way speed of light is isotropic.\index{preferred Lorentz frame!existence of}
In the final analysis, the existence of a preferred Lorentz frame\index{preferred Lorentz frame} is an inevitable result of the combination of the constancy of the two-way speed of light and the existence of RDM of particles\index{random discontinuous motion of particles} and its collapse evolution. Therefore, no matter which assumption is adopted, the preferred Lorentz frame\index{preferred Lorentz frame} can always be defined as the Lorentz frame in which the one-way speed of light is isotropic and the collapse of the wave function\index{wave function!collapse} happens simultaneously in the whole space.


\section{Collapse dynamics and preferred Lorentz frame}

If a preferred Lorentz frame\index{preferred Lorentz frame} as defined above indeed exists, then it will be natural to ask whether the frame can be detected. It is usually thought that the answer to this question is negative (see, e.g. Maudlin\index{Maudlin, Tim}, 2002). For example, although Bohm's theory\index{Bohm's theory}\index{Bohm's theory} and certain collapse theories\index{collapse theories}\index{collapse theories} assume the existence of a preferred Lorentz frame\index{preferred Lorentz frame}, the frame is undetectable in these theories. In this section, I will first give a few arguments supporting the detectability of the preferred Lorentz frame\index{preferred Lorentz frame!detectability of}, and then show that the frame can be detected in my model of energy-conserved\index{collapse models!energy-conserved} wavefunction collapse.

First of all, it is worth noting that the detectability of the preferred Lorentz frame\index{preferred Lorentz frame} is not prohibited. \index{preferred Lorentz frame!detectability of}
It is usually thought that in order to detect the preferred Lorentz frame\index{preferred Lorentz frame} one must be able to measure the collapse of an individual wave function.
If the collapse of an individual wave function can be measured, then when adopting standard synchrony\index{special relativity!standard synchrony in} 
one will be able to detect the preferred Lorentz frame\index{preferred Lorentz frame} by measuring the time order of the collapses of the wave function happening at different locations in space. Only in the preferred Lorentz frame\index{preferred Lorentz frame}, the collapse of the wave function\index{wave function!collapse} happens simultaneously at different locations in space.
But the measurability of the collapse of an individual wave function will lead to superluminal signaling\index{superluminal signaling} and thus is prohibited by the no-signaling theorem of quantum mechanics\index{quantum mechanics!no-signaling theorem of}. 
As a result, the preferred Lorentz frame\index{preferred Lorentz frame} cannot be detected.

There are two loopholes in this argument. \index{preferred Lorentz frame!detectability of!arguments for}
One loophole is that the no-signaling theorem may be not universally true. It has been argued that when the measuring device is replaced with a conscious observer superluminal\index{collapse theories!superluminal signaling in}\index{superluminal signaling} signaling may be achieved in principle in collapse theories\index{collapse theories} (Squires\index{Squires,  Euan J.}, 1992; Gao\index{Gao, Shan}, 2004, 2014d).
The other loophole is that there are  other possible methods of detecting the preferred Lorentz frame\index{preferred Lorentz frame}, which do not depend on the measurability of individual wavefunction collapse and the existence of superluminal signaling\index{superluminal signaling} (see below). Therefore, even if the no-signaling theorem is universally true, it does not prohibit the detectability of the preferred Lorentz frame\index{preferred Lorentz frame}.\index{preferred Lorentz frame!detectability of!arguments for}\index{quantum mechanics!no-signaling theorem of}

Next, the detectability of the preferred Lorentz frame\index{preferred Lorentz frame} is also supported by one of our basic scientific beliefs, the so-called minimum ontology\index{minimum ontology}. It says that if a certain thing cannot be detected in principle, then it does not exist, whereas if a certain thing does exist, then it can be detected. According to this view, the preferred Lorentz frame\index{preferred Lorentz frame} should be detectable in principle if it exists. Imagine that there exists some kind of fundamental particles around us, but they cannot be detected in principle. How unbelievable this is! \index{preferred Lorentz frame!detectability of!arguments for}

However, it seems that there are two common objections to this view. First, one might refute this view by resorting to the fact that the objects beyond the event horizon of an observer cannot be detected by the observer. My answer is that although these objects cannot be detected by the observer, they can be detected locally by other observers. Second, one may refute this view based on the fact that an unknown quantum state cannot be measured. This objection seems to have a certain force, and presumably it makes some people believe in the undetectability of the preferred Lorentz frame\index{preferred Lorentz frame}. But this objection is arguably invalid too. To begin with, a preferred Lorentz frame\index{preferred Lorentz frame} is a classical system, not a quantum system, and its state of motion is described by a definite velocity, not by a superposed quantum state, while the unknown velocity of a classical system can be measured. Next, although the unknown state of a quantum system cannot be measured, the other definite properties of the system, such as its mass and charge, can still be detected by gravitational and electromagnetic interactions. Moreover, an unknown quantum state being not measurable does not mean that a known quantum state is not measurable either. A known quantum state of a single system can be measured by a series of protective measurements\index{protective measurements}, and even an unknown nondegenerate energy eigenstate can also be measured by protective measurements\index{protective measurements} (Aharonov\index{Aharonov, Yakir} and Vaidman\index{Vaidman, Lev}, 1993; Aharonov\index{Aharonov, Yakir}, Anandan\index{Anandan, Jeeva S.} and Vaidman\index{Vaidman, Lev}, 1993; Gao\index{Gao, Shan}, 2014a). Thus the fact that an unknown quantum cannot be measured does not refute the the minimum ontology\index{minimum ontology}. \index{preferred Lorentz frame!detectability of!arguments for}

In the\index{preferred Lorentz frame!collapse dynamics and} following, I will demonstrate that the preferred Lorentz frame\index{preferred Lorentz frame} can be detected by measuring the (average) collapse time of the wave function in my model of energy-conserved\index{collapse models!energy-conserved} wavefunction collapse. According to the model, the collapse time formula for a superposition of energy eigenstates is\index{preferred Lorentz frame!method of detecting}

\begin{equation}
\tau_c \approx {\hbar^2 \over {t_P (\Delta E)^2}},
\label{CT1}
\end{equation}

\noindent where $t_P$ is the Planck time, $\Delta E$ is the energy uncertainty of the state. It is assumed here that  this collapse time formula is still valid in a Lorentz frame in the relativistic domain. This assumption seems reasonable, as the collapse time formula already contains the speed of light $c$ via the Planck time $t_P$.\footnote{In contrast, the collapse theories\index{collapse theories}\index{collapse theories} in which the collapse time formula does not contain $c$ are not directly applicable in the relativistic domain.} Since the formula is not relativistically invariant, its relativistically invariant form must contain a term relating to the velocity of the experimental frame relative to a preferred Lorentz frame\index{preferred Lorentz frame}. In other words, there must exist a preferred Lorentz frame\index{preferred Lorentz frame} according to the collapse model. 

The preferred Lorentz frame\index{preferred Lorentz frame}, denoted by $S_0$, can be defined as the Lorentz frame in which the above formula is valid. Then in another Lorentz frame the collapse time will depend on the velocity of the frame relative to $S_0$. According to the Lorentz transformation\index{Lorentz transformation}, in a Lorentz frame $S'$ with velocity $v$ relative to the frame $S_0$ we have\index{preferred Lorentz frame!method of detecting}

\begin{equation}
\tau'_c={1 \over \sqrt{1-v^2/c^2} } \tau_c,
\label{}
\end{equation}

\begin{equation}
t'_P={1 \over \sqrt{1-v^2/c^2} } t_P,
\label{}
\end{equation}

\begin{equation}
\Delta E' \approx {{1-v/c} \over \sqrt{1-v^2/c^2} } \Delta E.
\label{E'E}
\end{equation}

\noindent Here I only consider the situation where the particle has very high energy, namely $E \approx pc$, and thus Eq. (\ref{E'E}) holds. Further, I assume that the Planck time $t_P$ is the minimum time (i.e., the duration of a discrete instant) in the preferred Lorentz frame\index{preferred Lorentz frame}, and in another Lorentz frame the minimum time is connected with the Planck time $t_P$ by the time dilation formula required by special relativity\index{special relativity}. Then by inputting these equations into Eq. (\ref{CT1}), we can obtain the relativistic collapse time formula for an arbitrary experimental frame with velocity $v$ relative to the frame $S_0$:\index{preferred Lorentz frame!method of detecting}

\begin{equation}
\tau_c \approx (1+v/c)^{-2}{\hbar^2 \over {t_P (\Delta E)^2}}.
\label{CT2}
\end{equation}

\noindent This formula contains a factor relating to the velocity of the experimental frame relative to the preferred Lorentz frame\index{preferred Lorentz frame}.\footnote{Note that in order to be relativistic invariant the nonrelativistic equation of collapse dynamics, Eq. (\ref{EDC}), must contain a velocity term as follows: $P_i(t+t_P) = P_i(t) + f(v){\Delta E \over E_P} [\delta_{E_sE_i}-P_i(t)]$, where $f(v) \approx 1+v/c$ when $E \approx pc$, and $v$ is the velocity of the experimental frame relative to the preferred Lorentz frame\index{preferred Lorentz frame}. The above relativistic collapse time formula can be derived from this relativistic equation of collapse dynamics.}\index{preferred Lorentz frame!collapse dynamics and}

Therefore, according to my energy-conserved\index{collapse models!energy-conserved} collapse model, the collapse time of a given wave function will differ in different Lorentz frames. For example, considering the maximum difference of the speed of revolution  of the Earth with respect to the Sun is $\Delta v \approx 60 $km/s, the maximum difference of the collapse times measured at different times (e.g., spring and fall) on the Earth will be $\Delta \tau_c \approx 4 \times 10^{-4} \tau_c $. As a result, the collapse dynamics will single out a preferred Lorentz frame\index{preferred Lorentz frame} in which the collapse time of a given wave function is longest, and the frame can also be determined by comparing the collapse times of a given wave function in different Lorentz frames.\footnote{In general, we can measure the collapse time of a wave function by measuring the change of the interference between the corresponding collapse branches for an ensemble of identically prepared systems. The main technical difficulty of realizing such a measurement is to exclude the influence of environmental decoherence (cf. Marshall\index{Marshall, William} et al, 2003).} It may be expected that this preferred Lorentz frame\index{preferred Lorentz frame!CMB-frame as} is the CMB-frame in which the cosmic background radiation is isotropic, and the one-way speed of light is also isotropic.\index{preferred Lorentz frame!method of detecting}\index{preferred Lorentz frame!collapse dynamics and}


\section{Particle ontology for quantum field theory\index{quantum field theory!particle ontology for}}

It is well known that although the combination of linear quantum dynamics and special relativity\index{special relativity} has been obtained in quantum field theory\index{quantum field theory!ontology of}, it is still a controversial issue how to understand the ontology of the theory. Does quantum field theory\index{quantum field theory} really describe a world composed of physical fields? Or it describes a world composed of particles or other new entities? 
I have argued in the previous chapters that the onology of (nonrelativistic) quantum mechanics is particles which undergo random discontinuous motion\index{random discontinuous motion of particles} in our three-dimensional space.
If this is true, then it seems that the ontology of (relativistic) quantum field theory\index{quantum field theory!particle ontology for} will be still these particles, since the Lorentz transformation\index{Lorentz transformation} in special relativity\index{special relativity} does not change the existent form of particles as shown in the previous sections.
In the following, I will give a more detailed argument supporting this answer.

The picture of particles appears from my analysis of the charge density of a quantum system such as an electron, which is given in Chapter 6.
First, I argue with the help of protective measurements\index{protective measurements} that a quantum system has a well-defined charge distribution\index{charge distribution} in space, in exactly the same sense that a classical system has a well-defined charge distribution\index{charge distribution} in space. Moreover, protective measurements\index{protective measurements} also show that the charge of a quantum system such as an electron is distributed throughout space, and the charge density in each position is equal to the modulus squared of the wave function of the system there multiplied by the charge of the system. 
Second, I argue that the charge distribution\index{charge distribution} of an electron is effective, that is, it is formed by the ergodic motion of a localized particle with the total charge of the electron.
If the charge distribution\index{charge distribution} of an electron is continuous and exists throughout space at the same time, then any two parts of the distribution, like two electrons, will arguably have electrostatic interaction too. The existence of such electrostatic self-interaction for an electron not only contradicts the superposition principle of quantum mechanics but also is incompatible with experimental observations. In contrast, if the charge distribution\index{charge distribution} of an electron is effective, namely if there is only a localized particle at every instant, then it is understandable that there exists no such electrostatic self-interaction for the effective charge distribution\index{charge distribution!effective} formed by the motion of the particle. This is consistent with the superposition principle of quantum mechanics and experimental observations. 

Now I will argue that the above analysis in quantum mechanics also holds true in quantum field theory\index{quantum field theory}.
Consider a quantum system in the low-energy regime such as the electron being in the ground state in a Hydrogen atom which is at rest in a Lorentz frame.
First, it can be argued that the electron still has a charge distribution\index{charge distribution} throughout the whole space according to quantum field theory\index{quantum field theory!particle ontology for}.\footnote{It is interesting to note that in relativistic quantum mechanics and quantum field theory\index{quantum field theory!charge density interpretation in} there is already the charge density interpretation for charged particles such as electrons, although it is not clear whether it means what it should mean.}
The reason is that the principle of protective measurements\index{protective measurements} is still valid in quantum field theory\index{quantum field theory}; when the electron is measured by a protective measurement during which its state is not changed, the measurement result is still the expectation value of the measured observable in the measured state.
Then quantum field theory\index{quantum field theory} also predicts that the charge of the electron is distributed throughout space, although the charge density in each position is somewhat different from that predicted by quantum mechanics due to the correction from quantum field theory\index{quantum field theory}.
This result is also guaranteed by an analysis of experimental observations.
Since the predictions of quantum mechanics are consistent with experimental observations to a very high precision in the low-energy regime, a series of actual protective measurements\index{protective measurements} will also confirm that the charge of the electron is distributed throughout space.

Next, it can be argued that the above analysis of the physical origin of the charge distribution\index{charge distribution} of an electron is still valid in quantum field theory\index{quantum field theory}.
The reason is that the superposition principle also holds true in quantum field theory\index{quantum field theory}.
Then if the charge distribution\index{charge distribution} of an electron is continuous and exists throughout space at the same time, the  resulting electrostatic self-interaction of the distribution will contradict the superposition principle of quantum field theory\index{quantum field theory}.
Moreover, the resulting electrostatic self-interaction of the distribution will be incompatible with experimental observations too.
Therefore, the charge distribution\index{charge distribution} of an electron is still formed by the ergodic motion of a localized particle with the total charge of the electron.
This means that an electron is still a particle in the low-energy regime in the relativistic domain.

Now consider a quantum system in the high-energy regime such as an electron moving close to the speed of light in a Lorentz frame.
This electron can be generated by a Lorentz boost from an electron being at rest in another Lorentz frame in the low-energy regime.
As I have shown in the previous sections, the Lorentz transformation\index{Lorentz transformation} does not change the existent form of particles, although it distorts the picture of motion when assuming the relativity of simultaneity\index{simultaneity!relativity of}. 
Thus an electron is still a particle in the high-energy regime in the relativistic domain.

Although the ontology of (relativistic) quantum field theory\index{quantum field theory!ontology of} is still particles, the theory does introduce a pair of new processes for the motion of particles, which are the creation and annihilation of particles.
In quantum mechanics, the number of particles is conserved and the existence of a particle is eternal. 
In quantum field theory\index{quantum field theory}, however, a particle can be created and annihilated.
This will lead to a new form of RDM of particles\index{random discontinuous motion of particles}.
For example, there will be superpositions of states with different particle numbers, such as $\alpha \ket{0} + \beta \ket{1}$. In this superposition, a particle only exists in the branch $\ket{1}$, which, according to the picture of RDM of particles\index{random discontinuous motion of particles}, means that the particle does not exist in the whole continuous time flow, but only exists part the time at certain discontinuous and random instants.

Finally, it is worth emphasizing that the above picture of particles is independent of whether the state of motion of particles can be localized or not.
For example, the fact that there are no Lorentz invariant localized states poses no threat to the existence of such particles.
According to the picture of RDM of particles\index{random discontinuous motion of particles}, the nature of a particle lies in that it exists only in one position at an instant, and the existence of such a particle at an instant is at a deeper level than its state of motion which is defined during an infinitesimal time interval around the instant.
Moreover, whether the state of motion of a particle can be localized or not is determined by the laws of motion\index{laws of motion},  and it does not depend on the existence of the particles defined in the above picture.
Similarly, it seems that whether there exist local and total number operators in quantum field theory\index{quantum field theory} does not poses a threat to the picture of RDM of particles\index{random discontinuous motion of particles} either. In my view, the current particle versus field debate in the philosophy of quantum field theory\index{quantum field theory!philosophy of} (see, e.g. Fraser\index{Fraser, Doreen}, 2008; Baker\index{Baker, David J.}, 2009; Bain\index{Bain, Jonathan}, 2011) may be mostly irrelevant, as the definition of particle there is essentially different from the definition of particle given here. But a more careful analysis is still needed and will be given in future work.


\chapter*{Epilogue} 


In this book, I have argued that the wave function in quantum mechanics is real, and it represents the state of random discontinuous motion\index{random discontinuous motion of particles} (RDM) of particles in three-dimensional space. Moreover, this picture of quantum ontology is complete in accounting for our definite experience, but it requires that the quantum dynamics be revised to include a stochastic, nonlinear evolution term resulting from the RDM of particles\index{random discontinuous motion of particles}. 
Obviously I took a road less traveled by researchers in the foundations of quantum mechanics.
In order to convince more readers that this road is deserved to be taken, I will review the main results of this book and think about them in a broader context in the epilogue.




The starting point of my road is protective measurements\index{protective measurements} (Chapter 1).
In 1993, this new method of measurement in quantum mechanics was discovered (Aharonov\index{Aharonov, Yakir} and Vaidman\index{Vaidman, Lev}, 1993; Aharonov\index{Aharonov, Yakir}, Anandan\index{Anandan, Jeeva S.} and Vaidman\index{Vaidman, Lev}, 1993).
Distinct from conventional projective measurements\index{projective measurements}, protective measurement is a method for measuring the expectation value of an observable on a single quantum system. By a series of protective measurements\index{protective measurements}, one can even measure the wave function of a single quantum system.
As thus, besides the Born rule\index{Born rule} for projective measurements\index{projective measurements}, protective measurements\index{protective measurements} provide another more direct connection between the wave function and results of measurements, which is not probabilistic but definite.
It can be expected that a definite connection between the wave function and results of measurements is more important for understanding the meaning of the wave function and searching for the ontology of quantum mechanics.
However, it seems that this connection provided by protective measurements\index{protective measurements} is still less well-known today, and its significance has not been fully realized by most researchers either.

The first stop on my road is the reality of the wave function\index{wave function!reality of} (Chapter 4).
It is correct to say that protective measurements\index{protective measurements} alone do not imply the reality of the wave function\index{wave function!reality of}.
An additional connection between the state of reality and results of measurements is needed.
This connection is provided by the second assumption of the ontological models framework\index{ontological models framework} (Spekkens\index{Spekkens, Robert W.}, 2005; Harrigan\index{Harrigan, Nicholas} and Spekkens\index{Spekkens, Robert W.}, 2010; Pusey,\index{Pusey, Matthew F.} Barrett and\index{Barrett, Jonathan} Rudolph,\index{Rudolph, Terry} 2012), which says that when a measurement is performed, the behaviour of the measuring device is only determined by the ontic state of the system, along with the physical properties of the measuring device. For protective measurements\index{protective measurements}, this  means that the ontic state  of a physical system determines the definite result of a protective measurement on the system.
When combined with this connection between the state of reality and results of measurements, protective measurements\index{protective measurements} will imply the reality of the wave function\index{wave function!reality of}.
Protective measurements can measure the wave function of a single quantum system, while the measurement results, which are represented by the wave function, are determined by the ontic state of the system. Therefore, the wave function is also real, representing a physical property of a single quantum system.

In fact, protective measurements\index{protective measurements} can provide a stronger argument for the reality of the wave function\index{wave function!reality of}, which may even persuade people who assume an anti-realist view of quantum mechanics.
Quantum mechanics, like classical mechanics, is also composed of a mathematical formalism and a rule of connection between the formalism and our experience at its core.
Then why assume a realist view of classical mechanics but assume an anti-realist view of quantum mechanics?
Presumaly the main reason is that the connection between quantum mechanics and experience\index{quantum mechanics!and experience} by the Born rule\index{Born rule} is probabilistic, while the connection between classical mechanics and experience is definite.
But the existence of protective measurements\index{protective measurements} in quantum mechanics makes this reason invalid; it also provides a definite connection between quantum mechanics and experience\index{quantum mechanics!and experience}.

On the other hand, it is arguable that a universal approach is needed to investigate the ontological content of all physical theories.
The familiar ontology of classical mechanics should not be given \emph{a priori} either, but be derived, like any other physical theories, through the universal approach.
Since a physical theory already specifies a connection between its mathematical formalism and our experience, we only need another connection between the state of reality and our experience to derive its ontological content.\footnote{This approach provides a strong support for the realist conception of scientific progress, which asserts that later science improves on earlier science by approaching closer to the truth. 
By this approach, if a physical theory is more successful in its empirical predictions, then it is also more accurate in depicting reality, and thus it is closer to truth. I will analyze the implications of this approach for the philosophy of science in detail in future work.} 
As noted above, the second assumption of the ontological models framework\index{ontological models framework} just provides such a connection.
Alternatively, the ontological content of a physical theory may also be directly derived from a reasonable criterion of reality\index{criterion of reality!and ontology of a physical theory}. 
The essential point here is not that these approaches must be valid and the ontological content thus derived must be true and complete, but that if they can be applied to classical mechanics and macroscopic objects\index{macroscopic objects} to derive the anticipant classical ontology, then they should also be applied  to quantum mechanics and microscopic objects\index{microscopic objects} to derive the quantum ontology, no matter how unexpected and strange it is.


In a word, due to the existence of protective measurements\index{protective measurements}, quantum mechanics is not different from classical mechanics when considering its reality. If we assume a realist view of classical mechanics, we should also assume a realist view of quantum mechanics for consistency. In particular, the wave function in quantum mechanics, like the trajectory function in classical mechanics, is also real, representing the physical state of a single quantum system (although the representation may be incomplete).


The second stop on my road is a new ontological interpretation of the wave function in terms of RDM of particles\index{random discontinuous motion of particles} (Chapters 6,7).
If the wave function in quantum mechanics represents the physical state of a single system, then what physical state does the wave function represent? The answer to this question will lead us to the ontology of quantum mechanics.
The most popular position among philosophers of physics and metaphysicians seems to be wave function realism\index{wave function!realism}, according to which the wave function represents a real, physical field on configuration space\index{configuration space} (Albert\index{Albert, David Z.}, 1996, 2013).
This is the simplest and most straightforward way of thinking about the wave function realistically.
However, it is well known that this interpretation is plagued by the problem of how to explain our three-dimensional impressions, while a satisfying solution to this problem seems still missing (see Albert\index{Albert, David Z.}, 2015 for a recent attempt).
This motivates a few authors to suggest that the wave function represents a property of particles in three-dimensional space (e.g. Monton\index{Monton, Bradley}, 2013; Lewis,\index{Lewis, Peter J.} 2013, 2016), although they do not give a concrete ontological picture of these particles in space and time and specify what property the property is.

In some sense, my idea of RDM of particles\index{random discontinuous motion of particles} can be regarded as a further development of this suggestion, although it already came to my mind more than 20 years ago (Gao\index{Gao, Shan}, 1993).
The picture of RDM of particles\index{random discontinuous motion of particles} clearly shows that interpreting the multi-dimensional wave function as representing the state of motion of particles in three-dimensional space \emph{is} possible. 
In particular, the RDM of particles\index{random discontinuous motion of particles} can explain quantum entanglement\index{quantum entanglement} in a more vivid way, which thus reduces the force of the main motivation to adopt wave function realism\index{wave function!realism}.
In any case, this new interpretion of the wave function in terms of RDM of particles\index{random discontinuous motion of particles} provides an alternative to wave function realism\index{wave function!realism}. \index{random discontinuous motion of particles!arguments for}

Furthermore, the picture of RDM of particles\index{random discontinuous motion of particles} has more explanatory power than wave function realism\index{wave function!realism}.
It can readily explain our three-dimensional impressions, which is still a difficult task for wave function realism\index{wave function!realism}.
Moreover, it can also explain many fundamental features of the Schr\"{o}dinger equation\index{Schr\"{o}dinger equation} that governs the evolution of the wave function, which seem puzzling for wave function realism\index{wave function!realism}.
For example, the existence of $N$ particles in three-dimensional space for an $N$-body quantum system can readily explain why there are $N$ mass parameters that are needed to describe the system, and why each mass parameter is \emph{only} correlated with each group of three coordinates of the $3N$ coordinates on the configuration space\index{configuration space} of the system, and why each group of three coordinates of the $3N$ coordinates transforms under the Galilean transformation between two inertial coordinate systems in our three-dimensional space, etc.

Finally, it can be expected that the difference between particle ontology and field ontology may also result in different predictions that may be tested with experiments, and thus the two interpretations of the wave function can be distinguished in physics.\index{random discontinuous motion of particles!arguments for}
The main difference  between a particle  and a field  is that a particle exists only in one position in space at each instant, while a field exists throughout the whole space at each instant.
It is a fundamental assumption in physics that a physical entity being at an instant has no interactions with itself being at another instant, while two physical entities may have interactions with each other. 
Therefore, a particle at an instant has no interactions with the particle at another instant, while any two parts of a field in space (as two local physical entities) may have interactions with each other. 
In particular, if a field is massive and charged, then any two parts of the field in space will have gravitational and electromagnetic interactions with each other.

Now consider a charged one-body quantum system such as an electron being in a superposition of two separated wavepackets. 
Since each wavepacket of the electron has gravitational and electromagnetic interactions with another electron, it is arguable that it is also massive and charged. 
But some people may be not convinced by this heuristic argument. Why?
A common reason, I guess, may be that if each wavepacket of the electron is massive and charged, then the two wavepackets will have gravitational and electromagnetic interactions with each other, but this is inconsistent with the superposition principle of quantum mechanics and experimental observations. But this reason is not valid. For if each wavepacket of the electron is not massive and charged, then how can it have gravitational and electromagnetic interactions with another electron? Rather, there should be a deeper reason why there are no interactions between the two wavepackets of the electron.

Besides the above heuristic argument, protective measurements\index{protective measurements} provide a more convincing argument for the existence of the mass and charge distribution\index{charge distribution}s of an electron in space.
This can be considered as a further implication of protective measurements\index{protective measurements} for the ontological meaning of the wave function.
As I have argued above, when combined with a reasonable connection between the state of reality and results of measurements, protective measurements\index{protective measurements} imply the reality of the wave function\index{wave function!reality of}.
For example, for an electron whose wave function is $\psi(x)$ at a given instant, we can measure the density $|\psi(x)|^2$ in each position $x$ in space by a protective measurement, and by the connection the density $|\psi(x)|^2$ is a physical property of the electron. Then, what density is the density $|\psi(x)|^2$?
Since a measurement must always be realized by a certain physical interaction between the measured system and the measuring device, the density must be, in the first place, the density of a certain interacting charge. For instance, if the measurement is realized by an electrostatic interaction between the electron and the measuring device, then the density multiplied by the charge of the electron, namely $-|\psi(x)|^2e$, will be the charge density of the electron in position $x$. This means that an electron has mass and charge distribution\index{charge distribution}s throughout space, and two separated wavepackets of an electron are both massive and charged.

Let us now see how the difference between particle ontology and field ontology may result in different empirical predictions.\index{random discontinuous motion of particles!arguments for}
According to the above analysis, if the wave function of an electron represents the state of RDM of a particle, then although two separated wavepackets of an electron are massive and charged, they will have no interactions with each other. This is consistent with the superposition principle of quantum mechanics and experimental observations.
On the other hand, if the wave function of an electron represents a physical field, then since this field is massive and charged, any two parts of the field will have gravitational and electromagnetic interactions with each other, which means that two separated wavepackets of an electron will have gravitational and electromagnetic interactions with each other. This is inconsistent with the superposition principle of quantum mechanics and experimental observations.
Therefore, it is arguable that the interpretation of the wave function in terms of motion of particles, rather than wave function realism\index{wave function!realism}, is supported by quantum mechanics and experience\index{quantum mechanics!and experience}.\index{random discontinuous motion of particles!arguments for}

The third stop on my road is the reality of wavefunction collapse (Chapter 8).
Admitting the validity of the above interpretation of the wave function in terms of RDM of particles\index{random discontinuous motion of particles}, the next question is whether this picture of quantum ontology is complete in accounting for our definite conscious experiences. 
This requires us to further analyze the measurement problem\index{measurement problem}.
The conventional research program is to first find a solution to the measurement problem\index{measurement problem}, such as Bohm's theory\index{Bohm's theory} or Everett's theory\index{Everett's theory} or collapse theories\index{collapse theories}, and then try to make sense of the wave function in the solution.
By such an approach, the meaning of the wave function will have no implications for solving the measurement problem\index{measurement problem}.
However, this approach is arguably problematic.
The reason is that the meaning of the wave function (in the Schr\"{o}dinger equation\index{Schr\"{o}dinger equation}) is independent of how to solve the measurement problem\index{measurement problem}, while the solution to the measurement problem\index{measurement problem} relies on the meaning of the wave function. For example, if assuming the operationalist $\psi$-epistemic view\index{wave function!epistemic view of}, then the measurement problem\index{measurement problem} will be dissolved.

My point is that even when assuming the $\psi$-ontic view\index{wave function!ontic view of}, the ontological meaning of the wave function also has implications for solving the measurement problem\index{measurement problem}. 
In particular, it can be argued that the RDM of particles\index{random discontinuous motion of particles} not only provides an ontological interpretation of the wave function, but also provides more resources for solving the measurement problem\index{measurement problem}.
The key is to notice that the modulus squared of the wave function of an electron not only gives the probability density that the electron is  \emph{found} in a certain position (according to the Born rule\index{Born rule}), but also gives the probability density that the electron as a particle \emph{is} in the position according to the picture of RDM of particles\index{random discontinuous motion of particles}. 
This should not be simply regarded as a coincidence, but be considered as a strong evidence for the existence of a deep connection.
In other words, it is natural to assume that the origin of the Born probabilities is the RDM of particles\index{random discontinuous motion of particles}. 
If this assumption turns out to be true, then it will have significant implications for solving the measurement problem\index{measurement problem}.
Since the existing solutions to the measurement problem\index{measurement problem}, including Bohm's theory\index{Bohm's theory}, Everett's theory\index{Everett's theory} and collapse theories\index{collapse theories}, have not taken this assumption into account, they need to be reformulated in order to be consistent with the assumption.
The reformulation may be easier for some, but more difficult or even impossible for others.

In order that the Born probabilities originate from the RDM of particles\index{random discontinuous motion of particles}\index{RDM of particles\index{random discontinuous motion of particles}|see{random discontinuous motion of particles}}, 
there must exist an additional random dynamics besides the linear, deterministic dynamics, which results from the RDM of particles\index{random discontinuous motion of particles} and results in the appearance of a random measurement result.
Since Everett's theory\index{Everett's theory} only admits the linear, deterministic dynamics, it cannot accommodate the above assumption. In other words, the Born probabilities cannot originate from the RDM of particles\index{random discontinuous motion of particles} in Everett's theory\index{Everett's theory}.
In Bohm's theory\index{Bohm's theory}, since the motion of the Bohmian particle\index{Bohm's theory!Bohmian particles in}s is not ergodic, the Born probabilities cannot wholly originate from the RDM of particles\index{random discontinuous motion of particles} either, and they must also depend on the the initial probability distribution of the positions of the Bohmian particle\index{Bohm's theory!Bohmian particles in}s.
Moreover, if there is an additional random dynamics responsible for generating the measurement result, then the guiding equation\index{Bohm's theory!guiding equation of}, which is added for the same purpose, will be redundant.
Last but not least, the existence of RDM of particles\index{random discontinuous motion of particles} itself already reduces the necessity of introducing additional Bohmian particle\index{Bohm's theory!Bohmian particles in}s in the first place; otherwise the theory will be clumsy and unnatural, since an electron will contain two particles, one undergoing random discontinuous motion\index{random discontinuous motion of particles}, and the other undergoing deterministic continuous motion.

Compared with Bohm's theory\index{Bohm's theory} and Everett's theory\index{Everett's theory}, collapse theories\index{collapse theories} seem to be the most appropriate framework to accommodate the above assumption about the origin of the Born probabilities, since a random dynamics  responsible for generating the measurement result is just what these theories need. Moreover, the RDM of particles\index{random discontinuous motion of particles} may also provide more resources for formulating a dynamical collapse theory, e.g. it already provides an appropriate noise source to collapse the wave function. Certainly, the resulting theory will be different from the existing collapse theories\index{collapse theories}, in which the Born probabilities do not originate from the RDM of particles\index{random discontinuous motion of particles}. A concrete model of wavefunction collapse in terms of RDM of particles\index{random discontinuous motion of particles} has been given in Section 8.4.

Besides this analysis of the measurement problem\index{measurement problem} in terms of RDM of particles\index{random discontinuous motion of particles}, I have also given a new formulation of the measurement problem\index{measurement problem} which lays more stress on psychophysical connection, and analyzed whether the major solutions to the problem can satisfy the restriction of psychophysical supervenience and thus can indeed solve the measurement problem\index{measurement problem}. The analysis also favors collapse theories\index{collapse theories} and disfavors Bohm's theory\index{Bohm's theory} and Everett's theory\index{Everett's theory}. In my view, the measurement problem\index{measurement problem} is essentially the determinate-experience problem\index{determinate-experience problem, \emph{see} measurement problem} (Barrett, 1999\index{Barrett, Jeffrey A.}), and the underlying ontology (e.g. RDM of particles\index{random discontinuous motion of particles}) and the psychophysical connection are the two extremes that should be understood fully in the first place when trying to solve the problem; the underlying ontology is at the lowest level, and the psychophysical connection is at the highest level.

These are the three main results obtained in this book. In addition to them, I have also shown that  the free Schr\"{o}dinger equation\index{Schr\"{o}dinger equation!free} can be derived in a rigorous way based on spacetime translation invariance\index{laws of motion!spacetime translation invariance of} and relativistic invariance\index{laws of motion!relativistic invariance of}. This may help explain the origin of the wave function in quantum mechanics.
Moreover, I have also given a primary analysis of how special relativity\index{special relativity} may influence the suggested ontology of quantum mechanics, as well as how the quantum ontology may influence special relativity\index{special relativity} reciprocally. Although these analyses are more speculative, they may inspire other researchers to find the right road.

Ninety years ago, Schr\"{o}dinger\index{Schr\"{o}dinger, Erwin} wrote in his second paper on wave mechanics: 

\begin{quote}
... it has even been doubted whether what goes on in an atom can be described within a scheme of space and time. From a philosophical standpoint, I should consider a conclusive decision in this sense as equivalent to a complete surrender. For we cannot really avoid our thinking in terms of space and time, and what we cannot comprehend within it, we cannot comprehend at all. (Moore\index{Moore, Walter J.}, 1989, p.208)
\end{quote}

\noindent However, the meaning of the wave function has not been fully understood until today, and we are still searching for the ontology of quantum mechanics.
Hopefully the suggested picture of quantum ontology, namely the RDM of particles\index{random discontinuous motion of particles} in space and time, may provide a satisfying description of what goes on in an atom and further help us understand the mysterious quantum world.




  \backmatter
  \appendix
  \endappendix

  \addtocontents{toc}{\vspace{\baselineskip}}


  \cleardoublepage


 \printindex



\end{document}